\documentclass[11pt]{article}
\usepackage{amsfonts,amssymb,graphicx}

\usepackage{fullpage}
\usepackage[tight]{subfigure}
\usepackage{rotating}
\usepackage{sidecap}
\pdfoutput=1
\usepackage{ifpdf}
\ifpdf
  \usepackage{color}
  \definecolor{darkblue}{rgb}{0.3,0.3,0.6}
    \definecolor{darkgreen}{rgb}{0,0.6,0}
  \usepackage[pdftitle={RigidCycles},pdfauthor={Jill Eckers Gabriele Honecker, Wieland Staessens},pdfsubject={String theory},pdfkeywords={string theory},colorlinks=true,linkcolor=black,urlcolor=darkblue,filecolor=darkblue,citecolor=darkblue,menucolor=darkblue,breaklinks=true,bookmarks=true,anchorcolor=black,unicode=true]{hyperref}
\else
  
\fi
\usepackage[numbers,compress]{natbib}
\usepackage{a4wide}
\usepackage{longtable}
\usepackage{graphicx,cancel}
\usepackage{subfigure}
\usepackage[centertags]{amsmath}
\usepackage{multirow}
\usepackage{slashed}
\usepackage{pdflscape}
\usepackage{rotating}
\usepackage{tikz}
\usetikzlibrary{arrows,positioning}
\usepackage{stmaryrd}
\usepackage{color}
 
\newcommand{\bCentering}{\centering}
\newcommand{\bCaption}{\caption}
\newcommand{\sgn}{{\rm sgn}}
\newcommand{\unity}{{\footnotesize\mbox{1\!\!I}}}

\def\muc{\multicolumn}

\def\Z{\mathbb{Z}}
\def\Q{\mathbb{Q}}
\def\R{\mathbb{R}}

\def\unity{1\!\!{\rm I}}
\def\ov{\overline}
\def\N{\mathbf{N}}
\def\Sym{\mathbf{Sym}}
\def\Anti{\mathbf{Anti}}
\def\Adj{\mathbf{Adj}}
\def\Tr{\text{Tr}}

\def\ov{\overline}
\def\1{{\bf 1}}
\def\2{{\bf 2}}
\def\3{{\bf 3}}
\def\4{{\bf 4}}
\def\6{{\bf 6}}
\def\8{{\bf 8}}
\def\OR{\Omega\mathcal{R}}

\def\pp{\uparrow\uparrow}

\def\targ#1#2{\genfrac{[}{]}{0pt}{}{#1}{#2}}
\def\tarh#1#2{\genfrac{(}{)}{0pt}{}{#1}{#2}}
\def\targ2#1#2{\genfrac{}{}{0pt}{}{#1}{#2}}

\definecolor{blus}{rgb}{0.1,0.1,0.8}
\definecolor{GreenYellow}{cmyk}{0.15,0,0.69,0}
\definecolor{Yellow}{cmyk}{0,0,1,0}
\definecolor{Goldenrod}{cmyk}{0,0.10,0.84,0}
\definecolor{Dandelion}{cmyk}{0,0.29,0.84,0}
\definecolor{Apricot}{cmyk}{0,0.32,0.52,0}
\definecolor{Peach}{cmyk}{0,0.50,0.70,0}
\definecolor{Melon}{cmyk}{0,0.46,0.50,0}
\definecolor{YellowOrange}{cmyk}{0,0.42,1,0}
\definecolor{Orange}{cmyk}{0,0.61,0.87,0}
\definecolor{BurntOrange}{cmyk}{0,0.51,1,0}
\definecolor{Bittersweet}{cmyk}{0,0.75,1,0.24}
\definecolor{RedOrange}{cmyk}{0,0.77,0.87,0}
\definecolor{Mahogany}{cmyk}{0,0.85,0.87,0.35}
\definecolor{Maroon}{cmyk}{0,0.87,0.68,0.32}
\definecolor{BrickRed}{cmyk}{0,0.89,0.94,0.28}
\definecolor{Red}{cmyk}{0,1,1,0}
\definecolor{OrangeRed}{cmyk}{0,1,0.50,0}
\definecolor{RubineRed}{cmyk}{0,1,0.13,0}
\definecolor{WildStrawberry}{cmyk}{0,0.96,0.39,0}
\definecolor{Salmon}{cmyk}{0,0.53,0.38,0}
\definecolor{CarnationPink}{cmyk}{0,0.63,0,0}
\definecolor{Magenta}{cmyk}{0,1,0,0}
\definecolor{VioletRed}{cmyk}{0,0.81,0,0}
\definecolor{Rhodamine}{cmyk}{0,0.82,0,0}
\definecolor{Mulberry}{cmyk}{0.34,0.90,0,0.02}
\definecolor{RedViolet}{cmyk}{0.07,0.90,0,0.34}
\definecolor{Fuchsia}{cmyk}{0.47,0.91,0,0.08}
\definecolor{Lavender}{cmyk}{0,0.48,0,0}
\definecolor{Thistle}{cmyk}{0.12,0.59,0,0}
\definecolor{Orchid}{cmyk}{0.32,0.64,0,0}
\definecolor{DarkOrchid}{cmyk}{0.40,0.80,0.20,0}
\definecolor{Purple}{cmyk}{0.45,0.86,0,0}
\definecolor{Plum}{cmyk}{0.50,1,0,0}
\definecolor{Violet}{cmyk}{0.79,0.88,0,0}
\definecolor{RoyalPurple}{cmyk}{0.75,0.90,0,0}
\definecolor{BlueViolet}{cmyk}{0.86,0.91,0,0.04}
\definecolor{Periwinkle}{cmyk}{0.57,0.55,0,0}
\definecolor{CadetBlue}{cmyk}{0.62,0.57,0.23,0}
\definecolor{CornflowerBlue}{cmyk}{0.65,0.13,0,0}
\definecolor{MidnightBlue}{cmyk}{0.98,0.13,0,0.43}
\definecolor{NavyBlue}{cmyk}{0.94,0.54,0,0}
\definecolor{RoyalBlue}{cmyk}{1,0.50,0,0}
\definecolor{Blue}{cmyk}{1,1,0,0}
\definecolor{Cerulean}{cmyk}{0.94,0.11,0,0}
\definecolor{Cyan}{cmyk}{1,0,0,0}
\definecolor{ProcessBlue}{cmyk}{0.96,0,0,0}
\definecolor{SkyBlue}{cmyk}{0.62,0,0.12,0}
\definecolor{Turquoise}{cmyk}{0.85,0,0.20,0}
\definecolor{TealBlue}{cmyk}{0.86,0,0.34,0.02}
\definecolor{Aquamarine}{cmyk}{0.82,0,0.30,0}
\definecolor{BlueGreen}{cmyk}{0.85,0,0.33,0}
\definecolor{Emerald}{cmyk}{1,0,0.50,0}
\definecolor{JungleGreen}{cmyk}{0.99,0,0.52,0}
\definecolor{SeaGreen}{cmyk}{0.69,0,0.50,0}
\definecolor{Green}{cmyk}{1,0,1,0}
\definecolor{ForestGreen}{cmyk}{0.91,0,0.88,0.12}
\definecolor{PineGreen}{cmyk}{0.92,0,0.59,0.25}
\definecolor{LimeGreen}{cmyk}{0.50,0,1,0}
\definecolor{YellowGreen}{cmyk}{0.44,0,0.74,0}
\definecolor{SpringGreen}{cmyk}{0.26,0,0.76,0}
\definecolor{OliveGreen}{cmyk}{0.64,0,0.95,0.40}
\definecolor{RawSienna}{cmyk}{0,0.72,1,0.45}
\definecolor{Sepia}{cmyk}{0,0.83,1,0.70}
\definecolor{Brown}{cmyk}{0,0.81,1,0.60}
\definecolor{Tan}{cmyk}{0.14,0.42,0.56,0}
\definecolor{Gray}{cmyk}{0,0,0,0.50}
\definecolor{Black}{cmyk}{0,0,0,1}
\definecolor{White}{cmyk}{0,0,0,0}

\definecolor{mygr}{rgb}{0,0.6,0}
\definecolor{mygrey}{rgb}{0,0.1,0.2}
\definecolor{myblue}{rgb}{0,0.5,0.9}
\definecolor{myblue2}{rgb}{0,0.5,0.5}
\definecolor{myorange}{rgb}{1,0.5,0}
\definecolor{mypurple}{rgb}{0.6,0,1}
\definecolor{mygolden}{rgb}{1,0.8,0.2}
\definecolor{mycyan}{rgb}{0,1,1}
\definecolor{mymagenta}{rgb}{1,0,1}

\newcommand{\bCaptionfonts}{\small}
\makeatletter
\long\def\@makecaption#1#2{%
  \vskip\abovecaptionskip
  \sbox\@tempboxa{{\bCaptionfonts #1: #2}}%
  \ifdim \wd\@tempboxa >\hsize
    {\bCaptionfonts #1: #2\par}
  \else
    \hbox to\hsize{\hfil\box\@tempboxa\hfil}%
  \fi
  \vskip\belowcaptionskip}
\makeatother

\makeatletter
\let\ORIGINALlatex@openbib@code=\@openbib@code
\renewcommand{\@openbib@code}{\ORIGINALlatex@openbib@code\setlength{\itemsep}{1ex plus.5ex minus.5ex}\setlength{\parsep}{0pt}}
\makeatother

\def\mathtab#1#2#3{\begin{table}[th]\bCentering$#1$\bCaption{#3}\label{tab:#2}\end{table}}
\def\mathtabfix#1#2#3{\begin{table}[th]\bCentering\resizebox{\linewidth}{!}{$#1$}\bCaption{#3}\label{tab:#2}\end{table}}

\allowdisplaybreaks[4]

\begin{document}
\begin{center}
\begin{flushright}
{\small MITP/15-066\\ 
IFT-UAM/CSIC-15-094\\
\today}

\end{flushright}

%\title{Model Building on $\Z_2 \times \Z_6$}
\vspace{25mm}
{\Large\bf D6-Brane Model Building on $\Z_2 \times \Z_6$: \\MSSM-like and Left-Right Symmetric Models}

\vspace{12mm}
{\large Jill Ecker${}^{a,\clubsuit}$, Gabriele Honecker${}^{a,\heartsuit}$ and Wieland Staessens${}^{b,\spadesuit}$
}

\vspace{8mm}
{
\it $^a$PRISMA Cluster of Excellence,Mainz Institute for Theoretical Physics (MITP),\\  \& Institut f\"ur Physik  (WA THEP), \\Johannes-Gutenberg-Universit\"at, D-55099 Mainz, Germany\\
$^b$Instituto de F\'isica Te\'orica UAM-CSIC, Cantoblanco, 28049 Madrid, Spain\\
\;$^{\clubsuit}${\tt eckerji@uni-mainz.de},~$^{\heartsuit}${\tt Gabriele.Honecker@uni-mainz.de} \\ $^{\spadesuit}${\tt wieland.staessens@csic.es}}

\vspace{15mm}{\bf Abstract}\\[2ex]\parbox{140mm}{
We perform a systematic search for globally defined MSSM-like and left-right symmetric models on D6-branes on the $T^6/(\Z_2 \times \Z_6 \times \OR)$ orientifold with discrete torsion.
Our search is exhaustive for models that are independent of the value of the one free complex structure modulus.
Preliminary investigations suggest that there exists one prototype of visible sector for MSSM-like and another for left-right symmetric models with differences arising from various hidden sector completions to global models. 
For each prototype, we provide the full matter spectrum, as well as the Yukawa and other three-point couplings needed to render vector-like matter states massive. This provides us with tentative explanations for the mass hierarchies within the quark and lepton sectors. We also observe that the MSSM-like models correspond to explicit realisations of the supersymmetric DFSZ axion model, and that the left-right symmetric models allow for global completions with either completely decoupled hidden sectors or with some messenger states charged under both visible and hidden gauge groups.
}
\end{center}

\thispagestyle{empty}
\clearpage 

\tableofcontents

%\newpage
\setlength{\parskip}{1em plus1ex minus.5ex}
%%%%%%%%%%%%%%%%%%%%%%%%%%%%%%%%%%%%%%%%%%%%%%%%%%%%%%%%%%%%%%%%%%%%%%%%%%%%%%%%%%%
%%%%%%%%%%%%%%%%%%%%%%%%%%%%%%%%%%%%%%%%%%%%%%%%%%%%%%%%%%%%%%%%%%%%%%%%%%%%%%%%%%%
%%%%%%%%%%%%%%%%%%%%%%%%%%%%%%%%%%%%%%%%%%%%%%%%%%%%%%%%%%%%%%%%%%%%%%%%%%%%%%%%%%%
%%%%%%%%%%%%%%%%%%%%%%%%%%%%%%%%%%%%%%%%%%%%%%%%%%%%%%%%%%%%%%%%%%%%%%%%%%%%%%%%%%%
\section{Introduction}\label{S:intro}

String theory is arguably the most promising framework for a unified description of all fundamental interactions. 
However, the question how the experimentally observed particle spectrum and interactions of the Standard Model of Particle Physics or some extension thereof arise from string theory
remains open to date, see e.g.~\cite{Candelas:1985hv,Ibanez:1987sn} for early works within the heterotic $E_8 \times E_8$ theory.
While a large fraction of today's efforts focusses on the construction of the {\it chiral} Standard Model or some GUT spectrum within the non-perturbative F-theory regime (see e.g. the lecture notes~\cite{Weigand:2010wm} and 
some very recent works~\cite{Lin:2014qga,Cvetic:2015txa,Krippendorf:2015kta,Mayrhofer:2014laa,Marchesano:2015dfa,Martucci:2015dxa,Martucci:2015oaa,Lin:2015qsa} and references therein), condoning the lack of control over the low-energy effective action, compactifications of Type II string theory at special points in moduli space provide for a very well controlled testing ground in the perturbative regime, where not only the full spectrum but (at least in principle) all interactions are computable.
The specific corner in the landscape of string compactifications, namely Type IIA orientifolds on toroidal orbifolds, which this article makes use of, relies on the combined power of using topology and algebraic geometry to describe the positions of D-branes and the chiral matter localised at their intersections, and Conformal Field Theory (CFT) techniques to compute the vector-like spectrum as well as gauge, Yukawa and higher $m$-point couplings~\cite{Blumenhagen:2006ci,Ibanez:2012zz}.\footnote{For further constructions at special points in moduli space see e.g. the review article~\cite{Schellekens:2013bpa}.}

The $T^6/(\Z_2 \times \Z_6 \times \OR)$ orbifold is chosen in this article since, based on earlier works with similar toroidal orbifold backgrounds, we expect it to be the most fertile one 
with a plethora of globally defined Type IIA/$\OR$ string vacua providing the phenomenologically most appealing spectra without flat directions allowing for any continuous gauge symmetry breaking.
Most notably, by trial and error we found in earlier works that the existence of some $\Z_3$ subsymmetry is favourable for providing three particle generations. For example, on the $T^6/(\Z_4 \times \OR)$~\cite{Blumenhagen:2002gw} and $T^6/(\Z_2 \times \Z_4 \times \OR)$~\cite{Honecker:2003vq,Honecker:2003vw,Honecker:2004np,Forste:2010gw} backgrounds, it is impossible to construct globally defined supersymmetric D6-brane models with three quark generations, while on $T^6/(\Z_6 \times \OR)$~\cite{Honecker:2004kb,Honecker:2004np,Gmeiner:2007we,Gmeiner:2009fb} and $T^6/(\Z_6' \times \OR)$~\cite{Bailin:2006zf,Gmeiner:2007zz,Bailin:2007va,Gmeiner:2008xq,Bailin:2008xx,Gmeiner:2009fb} such models have been obtained. For the latter, diverse investigations of the related low-energy effective 
field theory were performed in~\cite{Gmeiner:2009fb,Honecker:2011sm,Honecker:2011hm,Honecker:2012jd,Honecker:2012fn}, discrete remnants of gauge symmetries were first investigated in~\cite{Honecker:2013hda}, and the relation to some Peccei-Quinn symmetry and axions was studied in~\cite{Honecker:2013mya,Honecker:2015ela}. 
However, both types of $\Z_6^{(\prime)}$ orbifolds face the draw-back of containing matter in the adjoint representation as remnants of the underlying ${\cal N}=2$ supersymmetry in the gauge sector,
whose flat directions lead to continuous breakings of the non-Abelian gauge groups to subgroups of equal rank.
The sector of $\Z_2$-twisted three-cycles on the factorisable $T^6/(\Z_6 \times \OR)$ orientifold on the $SU(3)^3$ lattice can be viewed as one of three identical $\Z_2^{(i)}$-twisted sectors on $T^6/(\Z_2 \times \Z_6^{\prime} \times \OR)$ with discrete torsion. An exhaustive scan for globally defined, phenomenologically appealing, supersymmetric D-brane models on the latter only yielded Pati-Salam models~\cite{Honecker:2012qr}. 
 In a similar way, the $\Z_2$-twisted sector of the other factorisable $T^6/(\Z_6^{\prime} \times \OR)$ background can be viewed as occurring twice as subsector of the $T^6/(\Z_2 \times \Z_6 \times \OR)$ orientifold with discrete torsion, with the third $\Z_2$-twisted sector different and arising from $(T^2 \times (T^4/\Z_6))/\OR$. In our recent article~\cite{Ecker:2014hma}, we were able to exclude $SU(5)$ GUT models with three particle generations  (and no chiral exotic ${\bf 15}$-representations),  and we classified Pati-Salam models. This article is devoted to extending the search to left-right symmetric and MSSM-like spectra. A full systematic search has to be performed for any allowed value of the complex structure parameter $\varrho$ on the first two-torus of the $SU(2)^2 \times SU(3)^2$ background, as we will briefly comment on in section~\ref{Ss:IntersectSummary}. However, the present search focusses on D-brane configurations that are supersymmetric for arbitrary values of $\varrho$.

This article is organized as follows: in section~\ref{S:Recollections}, we review basic model building ingredients such as RR tadpole cancellation and supersymmetry on $T^6/(\Z_2 \times \Z_6 \times \OR)$ with discrete torsion,  where we also discuss the most basic phenomenological constraints, such as no exotic matter in the adjoint or symmetric representation of the $QCD$ D6-brane stack, and the resulting severe limitations on the corresponding three-cycle.  
In section~\ref{S:D6branePheno}, we discuss the missing global consistency conditions, namely the K-theory constraints, in the context of discrete $\Z_n$ gauge symmetry remnants from massive $U(1)$s in the 
low-energy effective field theory,  before we  present the results of a systematic computer search for MSSM-like and left-right symmetric models in sections~\ref{S:MSSM-pheno} and~\ref{S:LRsymModels-pheno}, respectively.
For each kind of attainable visible spectrum, we provide some prototype example of a globally defined D-brane configuration, for which we first compute the remnant discrete $\Z_n$ symmetries and/or surviving  massless $U(1)$ symmetries and then provide all Yukawa and other three-point couplings needed to render vector-like matter states massive.
Our conclusions are given in section~\ref{S:conclu}, and in appendix~\ref{A:ChanPatonMethod} we briefly review the method of Chan-Paton labels needed to determine the localisations of matter states for computing Yukawa couplings. Our focus lies here on the r\^{o}le of discrete Wilson lines not discussed before in the literature. In appendix~\ref{A:5StackLRS}, we present an example of a {\it semi-local} model, where all RR tadpoles are cancelled, but  some of the K-theory constraints are violated. Last but not least, appendix~\ref{A:ProtoIIModels} contains further prototype matter spectra for globally consistent left-right symmetric models with different hidden sectors.

%%%%%%%%%%%%%%%%%%%%%%%%%%%%%%%%%%%%%%%%%%%%%%%%%%%%%%%%%%%%%%%%%%%%%%%%%%%%%%%%%%%
%%%%%%%%%%%%%%%%%%%%%%%%%%%%%%%%%%%%%%%%%%%%%%%%%%%%%%%%%%%%%%%%%%%%%%%%%%%%%%%%%%%
%%%%%%%%%%%%%%%%%%%%%%%%%%%%%%%%%%%%%%%%%%%%%%%%%%%%%%%%%%%%%%%%%%%%%%%%%%%%%%%%%%%
%%%%%%%%%%%%%%%%%%%%%%%%%%%%%%%%%%%%%%%%%%%%%%%%%%%%%%%%%%%%%%%%%%%%%%%%%%%%%%%%%%%
\section{Recollections of the Orientifold $T^6/(\Z_2\times \Z_6\times \OR)$}\label{S:Recollections}
To fully appreciate the phenomenological aspects of intersecting D6-brane models on the toroidal orientifold $T^6/(\Z_2\times \Z_6\times \OR)$ with discrete torsion, a proper understanding of the background geometry is essential. Our starting point is thus a brief summary of indispensable geometric aspects related to the toroidal orientifold and its fractional three-cycles. For a more detailed account of these aspects we refer to~\cite{Forste:2010gw,Ecker:2014hma}. 
To prepare the systematic search and classification of {\it global} MSSM-like and left-right symmetric models in sections~\ref{S:MSSM-pheno} and~\ref{S:LRsymModels-pheno},
the second part of this section then reviews some results about the search for appropriate {\it local} rigid D6-brane configurations allowing for three chiral quark generations with a minimal amount of undesired exotic matter, as first presented in~\cite{Ecker:2014hma}. 

%%%%%%%%%%%%%%%%%%%%%%%%%%%%%%%%%%%%%%%%%%%%%%%%%%%%%%%%%%%%%%%%%%%%%%%%%%%%%%%%%%%
%%%%%%%%%%%%%%%%%%%%%%%%%%%%%%%%%%%%%%%%%%%%%%%%%%%%%%%%%%%%%%%%%%%%%%%%%%%%%%%%%%%
\subsection{Geometry and Fractional Three-Cycles}\label{Ss:Geometry}
\begin{figure}[ht]
\begin{center}
\begin{tabular}{c@{\qquad}c@{\hspace{-0.2in}}c}
\vspace{-0.7in}\includegraphics[width=4cm]{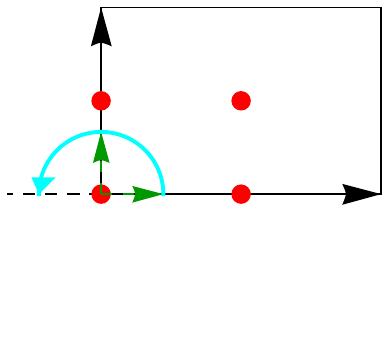} \begin{picture}(0,0) \put(-100,55){\bf \color{mygr} a} \put(-50,110){$R_1$} \put(0,75){$R_2$}  \put(-95,35){\bf \color{red} \bf 1} \put(-100,75){\bf \color{red} \bf 4} \put(-45,35){\bf \color{red} \bf 2} \put(-42,75){\bf \color{red} \bf 3}
\put(-15,35){$\pi_{1}$} \put(-105,105){$\pi_2$} \put(-70,65){\bf \color{mycyan}$\theta$}   \end{picture} &&\\
\vspace{-1.2in}  & \includegraphics[width=6cm]{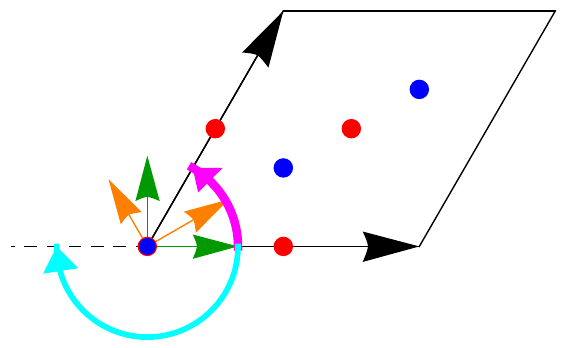} \begin{picture}(0,0) \put(-60,110){$r_2$} \put(-135,65){\color{mygr} {\bf A}} \put(-155,55){\color{myorange} {\bf B}} \put(-135,15){\color{red} \bf 1}  \put(-90,15){\color{red} \bf 4}  \put(-120,65){\color{red} \bf 5}  \put(-62,62){\color{red} \bf 6}  \put(-85,50){\color{blue} \bf 2}  \put(-42,75){\color{blue} \bf 3} \put(-60,15){$\pi_{3}$} \put(-105,105){$\pi_4$}  \put(-105,0){\bf \color{mycyan}$\theta$}   \put(-100,40){\bf \color{mymagenta}$\omega$}  \end{picture}  
&\includegraphics[width=6cm]{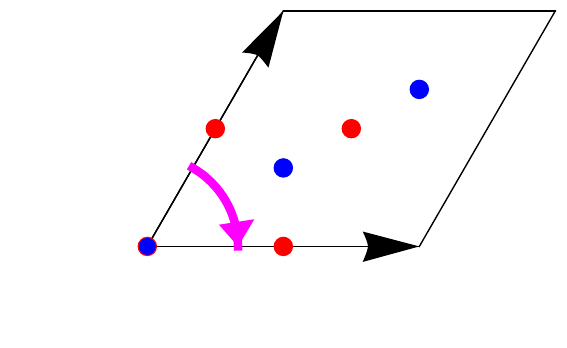} \begin{picture}(0,0) \put(-60,110){$r_3$}  \put(-135,15){\color{red} \bf 1}  \put(-90,15){\color{red} \bf 4}  \put(-120,65){\color{red} \bf 5}  \put(-62,62){\color{red} \bf 6}  \put(-85,50){\color{blue} \bf 2}  \put(-42,75){\color{blue} \bf 3} \put(-60,15){$\pi_{5}$} \put(-105,105){$\pi_6$}  \put(-107,52){\bf \color{mymagenta}$\omega$}  \end{picture} \\
\includegraphics[width=3.cm]{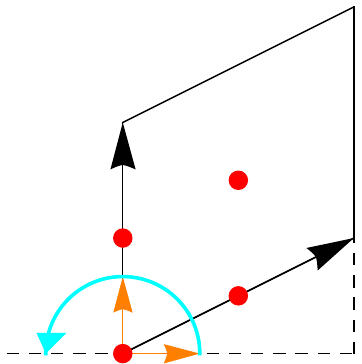} \begin{picture}(0,0) \put(-72,8){\bf \color{myorange} b}  \put(-40,-10){$R_1$} \put(0,55){$R_2$} \put(-68,-10){\bf \color{red} \bf 1} \put(-72,28){\bf \color{red} \bf 4} \put(-28,6){\bf \color{red} \bf 2} \put(-27,40){\bf \color{red} \bf 3}  \put(0,25){$\pi_{1}$} \put(-75,58){$\pi_2$}  \put(-50,22){\bf \color{mycyan}$\theta$}  \end{picture}  & &\\
\end{tabular}
\caption{The $SU(2)^2 \times SU(3) \times SU(3)$ compactification lattice for the $T^6/(\Z_2 \times \Z_6 \times \OR)$ orientifold defined by the action in equation~(\ref{Eq:Z2Z6action}) with complex structure modulus $\varrho \equiv \sqrt{3}{R_2/R_1}$ on $T^2_{(1)}$. The $\Z_2$ fixed points are labeled by the red points $(1, 2, 3, 4)$ on the first two-torus and $(1, 4, 5, 6)$ on the second and third two-torus. The $\Z_6$ action is trivial on the first torus and cyclically permutes three $\Z_2$ fixed points on the second and third torus: $1\stackrel{e^{\pi i/3}}{\circlearrowleft},4\stackrel{e^{\pi i/3}}{\to}5\stackrel{e^{\pi i/3}}{\to}6\stackrel{e^{\pi i/3}}{\to}4$.
The blue points denote the $\Z_3$ fixed points with $2\stackrel{e^{\pi i/3}}{\leftrightarrow} 3$. Invariance under the anti-holomorphic orientifold involution~(\ref{Eq:AntiHoloInvolution}) permits an untilted {\color{mygr}{\bf a}}-type (with $2  \stackrel{\cal R}{\circlearrowleft}, 3  \stackrel{\cal R}{\circlearrowleft}, 4 \stackrel{\cal R}{\circlearrowleft}$) or a tilted {\color{myorange}{\bf b}}-type lattice for $T_{(1)}^2$ (with $4 \stackrel{\cal R}{\circlearrowleft}, 2 \stackrel{\cal R}{\leftrightarrow} 3$), and two orientation choices for the other two-tori as well: {\color{mygr}{\bf A}}-type (with $4 \stackrel{\cal R}{\circlearrowleft}, 5 \stackrel{\cal R}{\leftrightarrow} 6$) and {\color{myorange}{\bf B}}-type (with $4 \stackrel{\cal R}{\leftrightarrow} 5, 6 \stackrel{\cal R}{\circlearrowleft}$). 
\label{Fig:LatticesZ2Z6}}
\end{center}
\end{figure}

Considering the factorisable six-torus $T_{(1)}^2 \times T_{(2)}^2 \times T_{(3)}^2$, the action of the point group $\Z_2\times \Z_6$ is described by a rotation of the complex coordinate $z^{k}$ parametrising the two-torus $T_{(k)}^2$ with $k \in \{1,2,3\}$:
\begin{equation}\label{Eq:Z2Z6action}
\theta^m \omega^n : z^k \rightarrow e^{2 \pi i (m v_k + n w_k)} z^k, \qquad \text{ with } \quad \vec{v} = \frac{1}{2} (1,-1,0), \quad \vec{w} = \frac{1}{6} (0,1,-1).
\end{equation}
In this expression, $\theta^m \omega^n$ corresponds to a generic element of the point group, where $\theta$ generates the $\Z_2$ part of the orbifold group acting on $T^2_{(1)} \times T^2_{(2)}$, and the $\Z_6$ part generated by $\omega$ acts only on the four-torus $T_{(2)}^2 \times T_{(3)}^2$. As an immediate consequence  of the $\Z_6$-action, the lattices of the factorisable four-torus $T_{(2)}^2 \times T_{(3)}^2$ take the shape of $SU(3)$ root lattices, while the complex structure modulus of the first two-torus $T_{(1)}^2$ remains unconstrained, see figure \ref{Fig:LatticesZ2Z6}. The $T_{(1)}^2$ lattice is thus given by a $SU(2)^2$ root lattice. Various combinations of the generators $\theta$ and $\omega$ generate additional $\Z_N$ subgroups with accompanying fixed points and/or fixed lines:  $\Z_6'$ symmetries are generated by $(\theta \omega, \theta \omega^2)$, a $\Z_3$ symmetry by $\omega^2$, and $\Z_2$ symmetries by $(\theta, \omega^3, \theta \omega^3)$.

Given that the orbifold group consists of a direct product of two Abelian factors, the $\Z_2$ generator $\theta$ can act on the ($\Z_6$)  $\omega$-twisted sectors with a phase $\eta = \pm 1$ and vice versa as discussed in detail in~\cite{Forste:2010gw}. For $\eta=-1$ we say that the orientifold has `discrete torsion', and the presence or absence of discrete torsion has a non-trivial impact on the Hodge numbers counting the two- and three-cycles in the twisted sectors, as discussed for $T^6/(\Z_2 \times \Z_6)$ e.g. in~\cite{Forste:2010gw,Ecker:2014hma}. For instance, the Hodge numbers associated to the three $\Z_2$ twisted sectors are given by:
\begin{equation}
\mbox{\resizebox{0.34\textwidth}{!}{%
$
\left( \begin{array}{c}
h_{11}^{\Z_2} \\ h_{12}^{\Z_2}
\end{array}
\right) = \left\{ \begin{array}{cc} 
\left(\begin{array}{c}
6 + 2 \times 8 \\
0
\end{array}\right)  & \eta =1,
\\
\left(\begin{array}{c}
 0 \\
 6 + 2 \times 4
\end{array}\right) & \eta =-1.
\end{array}\right.$}}
\end{equation} 
As we will review later on, the intersecting D6-brane models considered in this article are supposed to wrap exceptional three-cycles stuck at $\Z_2$ fixed points on $T^6/(\Z_2\times \Z_6)$. The absence of such three-cycles for $\eta=1$ implies that we should focus our attention on the toroidal orbifold with discrete torsion, $\eta=-1$, from now onwards.
%
%%%%%%%%%%%%%%%%%%%%%%%%
%%%%%%%%%%%%%%%%%%%%%%%%
%%%%%%%%%%%%%%%%%%%%%%%%%%%%%%%%%%%%%%%%

The orbifold group is extended by an orientifold projection $\OR(-)^{F_L}$, consisting of the worldsheet parity $\Omega$, the projection involving the left-moving fermion number $F_L$, and the anti-holomorphic involution ${\cal R}$ acting on the coordinates as:
\begin{equation}\label{Eq:AntiHoloInvolution}
{\cal R}: z^k \rightarrow \ov{z}^k.
\end{equation}
Besides reducing the amount of four-dimensional spacetime supersymmetry to ${\cal N} = 1$ for Type IIA string theory, the orientifold projection also constrains the shape of the two-torus $T_{(1)}^2$ to be rectangular ({\bf a}-lattice) or tilted ({\bf b}-lattice) and reduces the complex structure parameter on $T_{(1)}^2$ to one real parameter captured by the ratio $\varrho \equiv \sqrt{3} R_2/R_1$. The tiltedness of $T_{(1)}^2$ will be denoted by a discrete parameter $b\in \{0, \frac{1}{2}\}$, where the {\bf b}-type lattice configuration corresponds to $b=\frac{1}{2}$. The lattices for the two-tori $T_{(2)}^2$ and $T^2_{(3)}$, which are always tilted, only admit two orientations w.r.t.~the orientifold invariant direction: an {\bf A}-type lattice or a {\bf B}-type lattice orientation as depicted in figure~\ref{Fig:LatticesZ2Z6}. A priori one expects six different $\OR$-invariant lattice configurations, {\bf a/bAA}, {\bf a/bAB} and {\bf a/bBB}, but only the first two, {\bf aAA} and {\bf bAA}, 
are truly physically independent as shown in \cite{Ecker:2014hma}. More explicitly, non-supersymmetric rotations among the lattices relate the lattices {\bf aAB} and {\bf aBB} to the lattice {\bf aAA} on the one hand, and the lattices {\bf bAB} and {\bf bBB} to the lattice {\bf bAA} on the other hand.\footnote{The full equivalence between the lattices was shown at the level of all explicitly computable quantities, such as the splitting of the Hodge number $h_{11}$ into orientifold-even and -odd parts $(h_{11}^+, h_{11}^-)$ counting the closed string vectors and K\"ahler moduli, respectively, the RR tadpole cancellation and supersymmetry conditions, the massless matter spectrum, vacuum amplitudes, and the one-loop gauge threshold corrections.} Hence, it suffices to limit investigations and discussions to the two lattices {\bf a/bAA} in the remainder of the paper. 

The O6-planes are grouped into four inequivalent orbits under the $\Z_6$-action, denoted as the $\OR$- and $\OR\Z_2^{(k=1,2,3)}$-invariant orbits. Each of the four O6-plane orbits carries RR charges, and the sign of their RR charges is denoted by $\eta_{\OR}$ and $\eta_{\OR\Z_2^{(k)}}$, respectively. Worldsheet consistency of the Klein bottle amplitude relates~\cite{Blumenhagen:2005tn,Forste:2010gw} these charges to the discrete torsion parameter $\eta$:
\begin{equation}\label{Eq:eta-for-ORZ2s}
\eta = \eta_{\OR} \prod_{k=1}^3 \eta_{\OR\Z_2^{(k)}}, \qquad \text{ where } \;  \eta_{\OR},  \eta_{\OR\Z_2^{(k)}} \in \{\pm 1\}.
\end{equation} 
This relation indicates that one of the O6-plane orbits has to be `exotic' with positive RR charges ($\eta_{\OR(\Z_2^{(k)})} = -1$) in the presence of discrete torsion $\eta=-1$. As pointed out in~\cite{Ecker:2014hma}, configurations with three exotic O6-plane orbits on $T^6/(\Z_2 \times \Z_6 \times \OR)$ are excluded based on supersymmetry requirements and bulk RR tadpole cancellation conditions.

In order to cancel the RR charges of the O6-planes, we introduce supersymmetric D6-branes whose RR charges compensate those of the O6-planes. For model building purposes we consider D6-branes wrapping fractional three-cycles on the toroidal orientifold. Such fractional three-cycles $\Pi^{\rm frac}_a$ are constructed as linear combinations of bulk three-cycles $\Pi^{\rm bulk}_a$ and exceptional three-cycles $\Pi^{\Z_2^{(i)}}_a$:
\begin{equation}\label{Eq:Z2Z6FractCycles}
\mbox{\resizebox{0.92\textwidth}{!}{
$
\begin{aligned}\hspace{-4mm}
\Pi^{\rm frac}_a =&  \frac{1}{4} \Pi^{\rm bulk}_a  + \frac{1}{4} \sum_{i=1}^3 \Pi^{\Z_2^{(i)}}_a\\
=& \frac{1}{4} \left( P_a \rho_1 + Q_a  \rho_2 + U_a \rho_3 + V_a \rho_4 \right) + \frac{1}{4} \sum_{\alpha=0}^5 \left( x^{(1)}_{\alpha, a} \,
    \varepsilon_\alpha^{(1)} + y ^{(1)}_{\alpha, a} \, \tilde{\varepsilon}_\alpha^{(1)}
  \right) 
+ \frac{1}{4} \sum_{l=2,3} \sum_{\alpha=1}^4 \left( x^{(l)}_{\alpha, a} \, \varepsilon_\alpha^{(l)} + y ^{(l)}_{\alpha, a} \, \tilde{\varepsilon}_\alpha^{(l)} \right),
\end{aligned}
$}}
\end{equation}
where we decomposed the bulk and exceptional three-cycles with respect to an orbifold-invariant basis 
in the second line. More concretely, the integers $(P_a,Q_a,U_a,V_a)$ correspond to the bulk wrapping numbers expressed in terms of the $b_3^{\rm bulk} = 2 + 2 h_{21}^{\rm bulk} = 4$ dimensional basis of bulk three-cycles $\rho_{i\in\{1, 2, 3, 4\}}$ with non-vanishing intersection numbers~\cite{Forste:2010gw}:
\begin{equation}\label{Eq:BulkIntersectionNumbers}
\begin{array}{l}
\rho_1 \circ \rho_3 = \rho_2 \circ \rho_4 = 8,\\
\rho_1 \circ \rho_4 = \rho_2 \circ \rho_3 = 4.
\end{array}
\end{equation}
For factorisable three-cycles, the bulk wrapping numbers can be written out explicitly in terms of the torus wrapping numbers $(n_a^i,m_a^i)_{i=1,2,3}$:
\begin{equation}\label{Eq:Def-PQUV-via-nm}
\begin{array}{l@{\hspace{0.4in}}l}
P_a \equiv n_a^1 \left( n_a^2 n_a^3 - m_a^2 m_a^3 \right),& Q_a \equiv n_a^1 \left( n_a^2 m_a^3 + m_a^2 n_a^3 + m_a^2 m_a^3 \right),\\
U_a \equiv m_a^1 \left( n_a^2 n_a^3 - m_a^2 m_a^3 \right),& V_a \equiv m_a^1 \left( n_a^2 m_a^3 + m_a^2 n_a^3 + m_a^2 m_a^3 \right).
\end{array}
\end{equation}
Note that the torus wrapping numbers do transform non-trivially under the $\Z_6$-action:
\begin{equation}\label{Eq:1-cycle-orbits}
\left(\begin{array}{cc}
n^1_a & m^1_a \\ n^2_a & m^2_a \\ n^3_a & m^3_a 
\end{array}\right) \stackrel{\omega}{\rightarrow}
\left(\begin{array}{cc}
n^1_a & m^1_a \\ m^2_a & -(n^2_a +m^2_a) \\ -(n^3_a+m^3_a) & n^3_a 
\end{array}\right) \stackrel{\omega}{\rightarrow}
\left(\begin{array}{cc}
n_a^1& m^1_a\\ -(n^2_a+m^2_a) & n^2_a \\ m^3_a & -(n^3_a+m^3_a) 
\end{array}\right),
\end{equation}
whereas the bulk wrapping numbers are $\Z_6$-invariant quantities inherent to a $\omega$-orbit and thus independent of the choice of toroidal representant.

The integers $(x^{(k)}_{\alpha, a}, y^{(k)}_{\alpha, a})$ are the so-called exceptional wrapping numbers expressed in terms of the $b_3^{\Z_2} = 2 h_{21}^{\Z_2} = 28$ dimensional basis of exceptional three-cycles $(\varepsilon^{(1)}_{\alpha}, \tilde \varepsilon^{(1)}_{\alpha})_{\alpha\in\{0, \ldots, 5\}}$ and $(\varepsilon_{\alpha}^{(l)}, \tilde \varepsilon_{\alpha}^{(l)})^{l=2,3}_{\alpha\in\{1,\ldots, 4\}}$ with intersection form given by~\cite{Forste:2010gw}:
\begin{equation}
\begin{array}{lll}
\varepsilon^{(1)}_0 \circ \tilde \varepsilon^{(1)}_0 = -12, & \varepsilon^{(1)}_\alpha \circ \tilde \varepsilon^{(1)}_\beta= - 4\, \delta_{\alpha \beta}, & \alpha,\beta \in \{1,2,3,4,5\},\\ 
\varepsilon^{(l)}_\alpha \circ \tilde \varepsilon^{(l)}_\beta =  - 4\, \delta_{\alpha \beta} & \text{with }\,  l  = 2,3 & \alpha, \beta \in \{1,2,3,4\} .
\end{array}
\end{equation}
The exceptional wrapping numbers of factorisable three-cycles can be written for each $\Z_{2}^{(k)}$-twisted sector in terms of linear combinations of the torus wrapping numbers $(n_a^k, m_a^k)$ of the one-cycle along the $\Z_2^{(k)}$-invariant two-torus $T_{(k)}^2$, where the linear combination is determined  by a set of eight independent discrete parameters, e.g. 
$(x^{(k)}_{\alpha, a}, y^{(k)}_{\alpha, a}) = (\pm m^k_a\, , \, \mp (n^k_a+m^k_a))$ for one of the cases where the given index $\alpha$ receives the sole contribution from a single ($\Z_6$ orbit of a) $\Z_2^{(k)}$ fixed point. 
The eight discrete parameters can be divided into three types, with each type representing a different geometric characteristic of the exceptional divisor located at the $\Z_2^{(k)}$ fixed points on the four-torus $T^4_{(k)}\equiv T^2_{(i)} \times T^2_{(j)}$:  
\begin{itemize}
\item[(i)] three discrete displacement parameters $(\vec{\sigma}_a)$: the `bulk' cycle can pass through the origin ($\sigma_a^i=0$) of the two-torus $T_{(i)}^2$, or it can be shifted by one-half of a lattice vector ($\sigma_a^i=1$);
\item[(ii)] two independent $\Z_2^{(k)}$ eigenvalues $(-)^{\tau^{\Z_2^{(k)}}_a}$: such a parameter indicates the orientation with which the D-brane wraps the exceptional divisor at a reference fixed point on the four-torus $T^4_{(k)}$; we will loosely speaking say that a three-cycle encircles a fixed point  `clockwise' ($\tau_a^{\Z_2^{(k)}}=0$) or `counter-clockwise' ($\tau_a^{\Z_2^{(k)}}=1$). Note that only two $\Z_2^{(k)}$ eigenvalues are truly independent, due to the relation $(-)^{\tau_a^{\Z_2^{(3)}}} = (-)^{\tau_a^{\Z_2^{(1)}}+ \tau_a^{\Z_2^{(2)}}}$;
\item[(iii)] three discrete Wilson lines $(\vec{\tau}_a)$: a parameter $\tau_a^i$ loosely speaking describes how the exceptional divisor encircles a second $\Z_2^{(k)}$ fixed point on the two-torus $T_{(i)}^2$, namely with the same orientation ($\tau_a^i=0$) or opposite orientation ($\tau_a^i=1$) as the divisor at the reference point.
\end{itemize} 
The interplay of displacements $(\vec{\sigma}_a)$ and Wilson lines $(\vec{\tau}_a)$ is summarised in table~\ref{tab:Z2Z6SignAssignment} of appendix~\ref{A:ChanPatonMethod} along the relevant four-torus  $T^4_{(1)} \equiv T^2_{(2)} \times T^2_{(3)}$ for the $\varrho$-independent global models discussed in this article. 
More details regarding the construction of the orbifold-invariant basis of three-cycles and the explicit expressions for the exceptional wrapping numbers $(x^{(k)}_{\alpha, a}, y^{(k)}_{\alpha, a})$ in terms of the torus wrapping numbers $(n_a^k, m_a^k)$ can be found in~\cite{Forste:2010gw}, and we display the result here in table~\ref{tab:Z2Z6ExceptionalWrappingNumbers}.
%%%%%%%%%%%%%%%%%%%%%%%%%%%%
\mathtabfix{
\begin{array}{|c|c||c|c|}
\hline \multicolumn{4}{|c|}{\text{\bf Exceptional wrapping numbers } (x^{(k)}_{\alpha, a}, y^{(k)}_{\alpha, a})\, \text{\bf on $T^6/(\Z_2 \times \Z_6 \times \OR)$ in terms of torus wrapping numbers } (n_a^i, m_a^i) }\\
\hline
\hline
\multicolumn{2}{|c||}{\Z_2^{(1)} \text{\bf twisted sector}} & \multicolumn{2}{|c|}{\Z_2^{(l)} \text{\bf twisted sector with } l=2,3 }\\
\hline
\hline
{\rm I.} & {\rm II.} & {\rm I.} & {\rm II.}\\
\hline
\hline
 ( z_{\alpha, a}^{(1)}\, n^1_a, z_{\alpha, a}^{(1)}\, m^1_a) &   ( \hat z_{\alpha, a}^{(1)}\, n^1_a, \hat z_{\alpha, a}^{(1)}\, m^1_a) & 
 \begin{array}{c}
 (\zeta^{(l)}_{\alpha,a} \; n^l_a \;, \; \zeta^{(l)}_{\alpha,a} \, m^l_a) \\
 (\zeta^{(l)}_{\alpha,a} \; m^l_a \;, \; -\zeta^{(l)}_{\alpha,a} \; (n^l_a+m^l_a) )\\
  (-\zeta^{(l)}_{\alpha,a} \; (n^l_a+m^l_a)  \; , \; \zeta^{(l)}_{\alpha,a} \; n^l_a)
 \end{array}
 &
  \begin{array}{c}
  \left(-\zeta^{(l)}_{\alpha,a} \;  n^l_a+ (\hat \zeta^{(l)}_{\alpha,a}-\zeta^{(l)}_{\alpha,a}) \, m^l_a \; , \, (\zeta^{(l)}_{\alpha,a}-\hat \zeta^{(l)}_{\alpha,a}) \; n^l_a -\hat \zeta^{(l)}_{\alpha,a} \, m^l_a \right)\\
 \left( (\zeta^{(l)}_{\alpha,a}-\hat \zeta^{(l)}_{\alpha,a}) \; n^l_a -\hat \zeta^{(l)}_{\alpha,a} \; m^l \; , \;  \zeta^{(l)}_{\alpha,a}\; 
m^l_a+ \hat\zeta^{(l)}_{\alpha,a} \; n^l_a \right)
 \\
 \left( \hat\zeta^{(l)}_{\alpha,a} \; n^l_a+\zeta^{(l)}_{\alpha,a} \; m^l_a \; , \; -\zeta^{(l)}_{\alpha,a}\; n^l_a +(\hat \zeta^{(l)}_{\alpha,a} -\zeta^{(l)}_{\alpha,a} ) \; m^l_a \right)
 \end{array}
 \\
 \hline
\end{array}
}{Z2Z6ExceptionalWrappingNumbers}{
The exceptional wrapping numbers of type I stem from a single contribution of a fixed point orbit, while those of type II result from a $\Z_3$ orbit contributing twice due to two different $\Z_2$ fixed points on $T^2_{(2)} \times T^2_{(3)}$. 
Details about the sign factor assignments $z_{\alpha, a}^{(1)}, \zeta^{(l)}_{\alpha,a},  \hat \zeta^{(l)}_{\alpha,a} \in \{\pm 1\}$ as well as $\hat z_{\alpha, a}^{(1)} \in \{0,\pm 2\}$ 
in terms of $\Z_2$ eigenvalues and discrete Wilson lines can be found in section 2.1.3. of~\cite{Ecker:2014hma}.
}
%%%%%%%%%%%%%%%%%%%%%%%%%%%%

The basic three-cycles used to decompose the fractional three-cycle $\Pi^{\text{frac}}_a$ in equation~(\ref{Eq:Z2Z6FractCycles}) do not correspond to $\OR$-even and $\OR$-odd three-cycles, which implies that also the bulk wrapping numbers $(P_a,Q_a,U_a,V_a)$ and exceptional wrapping numbers $(x^{(k)}_{\alpha, a}, y^{(k)}_{\alpha, a})$ transform under the orientifold projection. Their transformation can be deduced from the transformation properties of the basis three-cycles under the $\OR$-projection as summarised in table~\ref{tab:Z2Z6bulkexcept-Orient}. 
%%%%%%%%%%%%%%%%%%%%%%%%%%%%%%%%%%%%%%%%%%%%%%%%%
\mathtab{ {\footnotesize
\begin{array}{|c||c|c|c|c|}\hline
 \multicolumn{5}{|c|}{\text{\bf  Orientifold images of bulk and exceptional three-cycles } }\\
  \multicolumn{5}{|c|}{\text{\bf on } T^6/(\Z_2 \times \Z_6 \times \OR) \, \text{\bf with discrete torsion } \, (\eta = -1) }
\\\hline \hline 
 \multicolumn{5}{|c|}{\text{\bf Bulk 3-cycles }}\\
\hline \hline
{\rm three-cycle} &  \OR(\rho_1)= &  \OR(\rho_2)= &  \OR(\rho_3)= &  \OR(\rho_4)=
\\
{\bf a/bAA} & \rho_1-(2b)\rho_3 & \rho_1 - \rho_2-(2b)[\rho_3-\rho_4] & -\rho_3 & \rho_4 - \rho_3 
\\\hline \hline 
 \multicolumn{5}{|c|}{ \Z_2^{(1)} \text{\bf twisted sector}} \\
 \hline \hline 
 \text{three-cycle} & \OR(\varepsilon^{(1)}_{\alpha})= & \OR(\tilde{\varepsilon}^{(1)}_{\alpha})= & \alpha=\alpha' & \alpha \leftrightarrow \alpha'  \\
{\bf a/bAA} &  \eta_{(1)} \left( -\varepsilon^{(1)}_{\alpha'} + (2b) \tilde{\varepsilon}^{(1)}_{\alpha'} \right)
& \eta_{(1)} \, \tilde{\varepsilon}^{(1)}_{\alpha'}
& 0,1,2,3 & 4,5\\
\hline \hline
\multicolumn{5}{|c|}{ \Z_2^{(k)} \text{\bf twisted sector with}\; l = 2,3}\\
\hline \hline
 \text{three-cycle} & \OR(\varepsilon^{(l)}_{\alpha})= & \OR(\tilde{\varepsilon}^{(l)}_{\alpha})= & \alpha=\alpha' & \alpha \leftrightarrow \alpha' \\
{\bf a/bAA} & - \eta_{(l)} \, \varepsilon^{(l)}_{\alpha'} & \eta_{(l)} \left( \tilde{\varepsilon}^{(l)}_{\alpha'} -\varepsilon^{(l)}_{\alpha'} \right)
& 1,4 & 2+2b, 3-2b\\
\hline
%%%%%%%%%%%%%%%%%%%%%%%%%%%%%%%%%%%%%%%%%%%%%%%%%%%%%%%%5
\end{array}}
}{Z2Z6bulkexcept-Orient}{The orientifold projection acting on bulk and $\Z_2^{(k)}$ exceptional three-cycles on  $T^6/(\Z_2 \times \Z_6 \times \OR)$ with discrete torsion, depending on the choice of background lattice orientation and for $\Z_2^{(k)}$ twisted sectors also on the choice of exotic O6-plane orbit via the sign $\eta_{(k)} \equiv \eta_{\OR} \eta_{\OR\Z_2^{(k)}}$.}
%%%%%%%%%%%%%%%%%%%%%%%%%%%
To simplify the transformation rules for the $\Z_2^{(k)}$ twisted sectors, one introduces the sign factor $\eta_{(k)}$:
\begin{equation}\label{Eq:eta-for-Z2s}
\eta_{(k)} \equiv \eta_{\OR} \eta_{\OR \Z_2^{(k)}} \qquad \text{with constraint: } \quad \eta = \prod_{k=1}^3 \eta_{(k)},
\end{equation}
where the constraint is a simple rewriting of relation~(\ref{Eq:eta-for-ORZ2s}). Fractional three-cycles with their bulk part parallel to one of the four O6-plane orbits are characterised by $\OR$-invariant bulk wrapping numbers, as can be easily checked from table~\ref{tab:Z2Z6-Oplanes-torus}. 
%%%%%%%%%%%%%%%%%%%%%%%%%%%%%%%%%%%%
\begin{table}[h!]
\begin{minipage}[c]{0.75\textwidth}
\begin{equation*}\hspace{-30mm}
{\footnotesize
\begin{array}{|c|c||c|c|}\hline
\multicolumn{4}{|c|}{\text{\bf Torus and bulk wrapping numbers for the }}\\
\multicolumn{4}{|c|}{\text{\bf  four O6-plane orbits on  } T^6/(\Z_2 \times \Z_6 \times \OR)}
\\\hline\hline
\text{O6-plane} & \frac{\rm angle}{\pi} & \muc{2}{|c|}{\bf a/bAA \text{ lattice}} 
\\\hline
& & (n^i,m^i) & (P,Q,U,V)  
\\\hline\hline
\OR & (0,0,0) & (\frac{1}{1-b},\frac{-b}{1-b};1,0;1,0) & \frac{1}{1-b}(1,0,-b,0)
\\
\OR\Z_2^{(1)} & (0,\frac{1}{2},\frac{-1}{2}) &  (\frac{1}{1-b},\frac{-b}{1-b};-1,2;1,-2) & \frac{3}{1-b}(1,0,-b,0)
\\
\OR\Z_2^{(3)} & (\frac{1}{2},\frac{-1}{2},0) & (0,1;1,-2;1,0) & (0,0,1,-2)
\\
\OR\Z_2^{(2)}  &  (\frac{1}{2},0,\frac{-1}{2}) &  (0,1;1,0;1,-2) & (0,0,1,-2)
\\\hline
\end{array}}
\end{equation*}
\end{minipage}~\hspace{-0.6in}
\begin{minipage}[c]{0.40\textwidth}
\caption{The torus wrapping numbers $(n^i,m^i)_{i\in \{1,2,3\}}$ are given for the one representant of each O6-plane orbit on $T^6/(\Z_2 \times \Z_6 \times \OR)$, for which 
the angle w.r.t. the $\OR$-invariant  plane is listed in the second column. The bulk wrapping numbers $(P,Q,U,V)$ are independent of the choice of the representant. 
The number of identical O6-planes is $N_{O6}=2(1-b)$ with $b=0,1/2$ for the {\bf a}- and {\bf b}-type torus $T^2_{(1)}$, respectively.\label{tab:Z2Z6-Oplanes-torus}}
\end{minipage}
\end{table}
%%%%%%%%%%%%%%%%%%%%%%%%%%%
Note that a rectangular lattice configuration for $T_{(1)}^2$ also allows an O6-plane displaced by one-half of the lattice vector $\pi_1$ or $\pi_2$ (see figure~\ref{Fig:LatticesZ2Z6}), hence the factor $N_{O6} = 2 (1-b)$ to denote the number of  identical O6-plane orbits for the lattice configurations {\bf a/bAA}. The transformation of the  torus wrapping numbers under the $\OR$-projection, depending on the two-torus lattice orientation, is summarised as:
\begin{equation}\label{Eq:OR_on_n+m}
(n^1_{a'},m^1_{a'}) = \left\{\begin{array}{cc}
(n^1_a  , \, - m^1_a) & ({\bf a}) \\
(n^1_a , -n^1_a-m_a^1) & ({\bf b})
\end{array}\right. 
,
\qquad
(n^i_{a'},m^i_{a'})_{i=2,3} 
= \left\{\begin{array}{cc}
(n^i_a+m^i_a \, , \, - m^i_a) & ({\bf A}) \\
(m^i_a,n^i_a) & ({\bf B})
\end{array}\right.
.
\end{equation}

At the intersection points of two fractional three-cycles $\Pi_a^{\rm frac}$ and $\Pi_b^{\rm frac}$, chiral matter can arise in the bifundamental representation of the gauge groups supported by the corresponding D6-brane stacks. The amount of chiral matter is encoded in the net-chirality $\chi^{(\N_a, \ov \N_b)}$, which is computed as follows in the fractional three-cycle language reviewed above: 
\begin{equation}
\begin{aligned}\label{Eq:BifundChiral}
\chi^{(\N_a, \ov \N_b)} \equiv\Pi_a^{\text{frac}} \circ \Pi_b^{\text{frac}} &= \frac{1}{4} \Big( 2 \left( P_a U_b - P_b U_a + Q_a V_b - Q_b V_a \right) +  \left(  P_a V_b - P_b V_a +  Q_a U_b -  Q_b U_a \right)  \Big) \\
&- \frac{1}{4} \left( 3 \left[ x_{0,a}^{(1)} y_{0,b}^{(1)} -x_{0,b}^{(1)} y_{0,a}^{(1)} \right] + \sum_{\alpha=1}^5 \left[ x_{\alpha, a}^{(1)} y_{\alpha, b}^{(1)} - x_{\alpha, b}^{(1)} y_{\alpha, a}^{(1)} \right] \right)  \\
& - \frac{1}{4} \sum_{i=2}^3 \sum_{\alpha=1}^4 \left[ x_{\alpha,a}^{(i)}  y_{\alpha,b}^{(i)} - x_{\alpha,b}^{(i)}  y_{\alpha,a}^{(i)} \right].
\end{aligned}
\end{equation}
Intersections of a fractional three-cycle $\Pi_a^{\rm frac}$ with its orientifold image and with the O6-planes lead to chiral matter in the symmetric and/or antisymmetric representation, counted by the net-chiralities $\chi^{\Anti_a/\Sym_a}$:  
\begin{equation}\label{Eq:SymmAntiChiral}
\chi^{\Anti_a/\Sym_a} \equiv \frac{\Pi_a^{\text{frac}} \circ \Pi_{a'}^{\text{frac}} \pm \Pi_a^{\text{frac}} \circ \Pi_{O6}}{2}.
\end{equation}
In this expression, the fractional three-cycle $\Pi_{O6}$ only contains contributions from the bulk three-cycles of the O6-planes, 
\begin{equation}
\Pi_{O6} = \frac{N_{O6}}{4} \left( \eta_{\OR} \Pi_{\OR} + \sum_{k=1}^3 \eta_{\OR\Z_2^{(k)}} \Pi_{\OR\Z_2^{(k)}} \right),
\end{equation}
as they do not carry RR charges coming from twisted sectors. 

The formulae~\eqref{Eq:BifundChiral} and~\eqref{Eq:SymmAntiChiral}  compute the total net-chiralities for intersecting fractional D6-branes, but they do not offer a glance at the contributions to the net-chirality per sector $a(\omega^k \, b)_{k\in \{0,1,2\}}$. To obtain the net-chirality per sector, we can turn to the $\Z_2$ invariant toroidal intersection numbers as introduced in appendix A of~\cite{Gmeiner:2009fb} for $T^6/\Z_{2N}$ backgrounds and extended to $T^6/\Z_{2}\times\Z_{2M}$ orbifolds in~\cite{Forste:2010gw,Honecker:2011sm}. For example, the amount of chiral states in the  symmetric and antisymmetric representation per sector $(\omega^n a) (\omega^n a)' _{n=0,1,2}$ can be computed by the following formulae:
\begin{equation}\label{Eq:SymmAntiTorusExp}
\begin{aligned}
\chi^{\Anti_a/\Sym_a} 
&= - \sum_{n=0}^2 \frac{ \left(I_{(\omega^n a)(\omega^n a)'} + \sum_{k=1}^3 I_{(\omega^n a)(\omega^n a)'}^{\Z_2^{(k)}} \right)
\pm \left( \sum_{k=0}^3 \eta_{\OR\Z_2^{(k)}} \, \tilde{I}_{(\omega^n a)}^{\OR\Z_2^{(k)}} \right) }{8}\\
& \equiv \sum_{n=0}^2  \chi^{\Anti_a/\Sym_a}_{(\omega^n a)},
\end{aligned}
\end{equation}
where $\tilde{I}_{(\omega^n a)}^{\OR\Z_2^{(k)}} \equiv 2 (1-b) I_{(\omega^n a)}^{\OR\Z_2^{(k)}} $ represents the intersection number between an orbifold image three-cycle $(\omega^n a)$ and the O6-plane $\OR\Z_2^{(k)}$ on the underlying six-torus, with $2(1-b)$ the number of parallel O6-planes set by the shape of $T^2_{(1)}$. 

A complementary approach of calculating the total amount of matter per sector irrespective of its chirality
can be taken by computing the beta-function coefficients, as outlined in e.g.~table 7 of~\cite{Honecker:2011sm} or table 39 of~\cite{Honecker:2012qr}. 
This approach includes all vector-like states, in particular for D6-branes at some vanishing angle, for which net-chiralities vanish. 
For the systematic searches of MSSM-like models in section~\ref{S:MSSM-pheno} and left-right symmetric models in section~\ref{S:LRsymModels-pheno} of this article, we cross-checked multiplicities of states by combining the three options, namely net-chirality, net-chirality per sector and total counting irrespective of chirality by means of contributions to the beta function coefficients.

Last but not least, the stability and consistency conditions for intersecting D6-brane models rely on the supersymmetric nature of the D6-branes and the cancellation of the RR tadpoles along the internal directions of the string compactification. The supersymmetry conditions for a fractional D6-brane boil down to the requirements that its bulk three-cycle is {\it special Lagrangian}  on the underlying torus, while its $\Z_{2}^{(k)}$ twisted parts are completely specified by the eight independent parameters $(-)^{\tau_a^{\Z_2^{(k)}}},\sigma_a^i,\tau^j_a \in \{0,1\}$ detailed above. The geometric conditions on the bulk three-cycles consist of one necessary and one sufficient condition as displayed in the upper right corner of table~\ref{tab:Bulk-RR+SUSY-Z2Z6}. 
%%%%%%%%%%%%%%%%%%%%%%%%%%%%%%%%%%%%%%%%%%%%%%%%%%%%%%%%%
\mathtabfix{
\begin{array}{|c||c|c|}\hline
\multicolumn{3}{|c|}{\text{\bf Global  consistency conditions on } \; T^6/(\Z_2 \times \Z_6 \times \OR) \; \text{\bf with discrete torsion}\, (\eta = -1)}
\\\hline\hline
\text{\bf lattice} & \text{\bf Bulk RR tadpole cancellation} &  \text{\bf Bulk SUSY conditions}  
\\\hline\hline
{\bf a/bAA}& \begin{array}{l} 
\sum_a N_a \left( 2 \, P_a + Q_a \right) =   8 \, \left(\eta_{\OR} + 3 \,     \eta_{\OR\Z_2^{(1)}}\right)  \\
     - \sum_a N_a \frac{V_a + b \, Q_a}{1-b} =  8  \, \left(\eta_{\OR\Z_2^{(2)}} + \eta_{\OR\Z_2^{(3)}}\right)
    \end{array}
    & 
   \begin{array}{ll} 
   \text{\bf necessary:}& 3 Q_a + \varrho \, [ 2 U_a + V_a + b (2 P_a + Q_a)  ] =0 \\
    \text{\bf sufficient:}& 2 P_a + Q_a - \varrho \, [V_a + b Q_a] > 0 
    \end{array} \\
\hline 
\hline
\text{\bf lattice} & \text{\bf RR tcc in the $\Z_2^{(1)}$ twisted sector} &  \text{\bf RR tcc: $\Z_2^{(l)}$ twisted sector with } l = 2, 3 \\
\hline
\hline
{\bf a/bAA} & 
\begin{array}{l}
\sum_a N_a (1-\eta_{(1)}) x^{(1)}_{\alpha,a} = 0, \hspace{31mm} \alpha = 0,1,2,3 \\
\sum_a N_a [(1+ \eta_{(1)}) y^{(1)}_{\alpha,a} +
  \eta_{(1)} 2b \, x^{(1)}_{\alpha,a} ] = 0, \qquad \alpha = 0,1,2,3 \\
  \sum_a N_a [x^{(1)}_{4,a} - \eta_{(1)} x^{(1)}_{5,a}] = 0,\\
  \sum_a N_a [y^{(1)}_{4,a} + \eta_{(1)}y^{(1)}_{5,a} + b \, (x^{(1)}_{4,a} + \eta_{(1)}x^{(1)}_{5,a})] = 0,
\end{array}
& 
\begin{array}{l}
\sum_a N_a  
[(1-\eta_{(l)})x^{(l)}_{\alpha,a} -\eta_{(l)} y^{(l)}_{\alpha,a} ] =0, \qquad \alpha = 1,4 \\
\sum_a N_a  (1+\eta_{(l)})y^{(l)}_{\alpha,a} = 0, \hspace{28mm} \alpha = 1,4\\
 \sum_a N_a [x^{(l)}_{2,a} - \eta_{(l)} x^{(l)}_{2+2b,a}  - \eta_{(l)} y^{(l)}_{2+2b,a}]  = 0,\\
 \sum_a N_a [x^{(l)}_{3,a} - \eta_{(l)} x^{(l)}_{3-2b,a}  - \eta_{(l)} y^{(l)}_{3-2b,a}]  = 0,\\
 \sum_a N_a [y^{(l)}_{3,a} + \eta_{(l)} y^{(l)}_{3-2b,a}] = 0,\\
 \sum_a N_a [y^{(l)}_{2,a} + \eta_{(l)} y^{(l)}_{2+2b,a}] = 0,
\end{array}
\\
\hline
\end{array}
}{Bulk-RR+SUSY-Z2Z6}{Model building constraints for the bulk part and the exceptional parts of fractional D6-branes defined in equation~(\protect\ref{Eq:Z2Z6FractCycles}) for the {\bf a/bAA} lattice configurations, as first derived in~\cite{Forste:2010gw} with typos corrected in the $\Z_2^{(l),l \in \{2,3\}}$ twisted sectors for fixed point indices $\alpha=2,3$ in~\cite{Ecker:2014hma}. 
}
%%%%%%%%%%%%%%%%%%%%%%%%%%%%%%%%%%%%%%%%%%%%%%%%%%%%%%%
Both conditions depend on the complex structure modulus $\varrho$ and the lattice configuration of the two-torus $T_{(1)}^2$. 

The RR tadpole cancellation conditions on the other hand do contain both a bulk part and $\Z_2^{(k)}$ twisted parts, as listed in the upper left corner and the lower part of table~\ref{tab:Bulk-RR+SUSY-Z2Z6}. The bulk RR tadpole cancellation conditions express the need for the RR charges of the D6-branes to compensate the RR charges of the O6-planes, whereas the twisted RR charges of the various fractional D6-branes have to cancel among each other.

When confronting the supersymmetry conditions for the bulk three-cycles with the bulk RR tadpole cancellation conditions, one notices that some choices of exotic O6-plane configuration are
{\it a priori} ruled out for supersymmetric D6-brane model building, such as the specific choice $\eta_{\OR\Z_2^{(1)}}=-1$ or any combination of three exotic O6-planes. Note that {\it global} intersecting D6-brane models are not only characterised by vanishing RR tadpoles, but also satisfy the K-theory constraints, which will be elaborated on in section~\ref{Ss:KtheoryDiscrete} in the context of discrete $\Z_n$ remnants of massive Abelian gauge symmetries, since the K-theory constraints boil down to the existence of a specific $\Z_2$ symmetry.

%%%%%%%%%%%%%%%%%%%%%%%%%%%%%%%%%%%%%%%%%%%%%%%%%%%%%%%%%%%%%%%%%%%%%%%%%%%%%%%%%%%
%%%%%%%%%%%%%%%%%%%%%%%%%%%%%%%%%%%%%%%%%%%%%%%%%%%%%%%%%%%%%%%%%%%%%%%%%%%%%%%%%%%
\subsection{Elements of Intersecting D6-brane Model Building}\label{Ss:IntersectSummary}
A first step towards intersecting D6-brane model building on $T^6/(\Z_2\times \Z_6\times \OR)$ with discrete torsion consists in classifying the fractional three-cycles supporting enhanced $SO(2N)$ or $USp(2N)$ gauge groups, as the latter can be used to accommodate the $SU(2)_L$ left stack in the MSSM gauge factor and/or the $SU(2)_R$ right stack in left-right symmetric models. Furthermore, in section~\ref{Ss:KtheoryDiscrete} we will use this classification when discussing the derivation of the K-theory constraints by means of probe $USp(2)$ branes and when determining the conditions for the existence of discrete $\Z_n$ symmetries. In order for a fractional D6-brane $\Pi_a^{\rm frac}$ to support an enhanced $SO(2N)$ or $USp(2N)$ gauge group, its bulk three-cycle has to be parallel to one of the four O6-plane orbits and its discrete parameters $(\vec{\sigma}_a, \vec{\tau}_a)$ have to satisfy a set of topological conditions involving also the individual tiltedness of the two-tori~\cite{Forste:2010gw}. For the orientifold $T^6/(\Z_2\times \Z_6\times \OR)$ with discrete torsion the topological conditions are written out explicitly in the second column of table~\ref{Tab:Conditions-on_b+t+s-SOSp}. 
%%%%%%%%%%%%%%%%%%%%%%%%%%%%%%%%%%%%%%%%%%%%%%%%%
\begin{table}[ht!]
  \begin{center}
\begin{equation*}
\begin{array}{|c|c||c|c||c|c|}\hline
\multicolumn{6}{|c|}{\text{\bf Existence of $\OR$ invariant three-cycles on } T^6/(\Z_2 \times \Z_6 \times \OR)}
\\\hline\hline
\pp & (\eta_{(1)},\eta_{(2)},\eta_{(3)}) \stackrel{!}{=}
& \muc{2}{|c||}{(1,1,-1)} & \muc{2}{|c|}{(-1,-1,-1)}
\\
\text{O6} &  & b=0 & b=\frac{1}{2} & b=0 & b=\frac{1}{2} 
\\\hline\hline
\OR  & \hspace{-3mm}\left(\hspace{-3mm}\begin{array}{c} -(-1)^{\sigma^2\tau^2 + \sigma^3 \tau^3} \\ - (-1)^{2b\sigma^1\tau^1 + \sigma^3 \tau^3} \\  -(-1)^{2b\sigma^1\tau^1 +\sigma^2 \tau^2} \end{array}\hspace{-3mm}\right)\hspace{-3mm}
& \left(\!\!\!\!\begin{array}{c} \sigma^1;\tau^1\\ \underline{0;1} \\ 1;1\end{array}\!\!\!\! \right)
&   \left(\!\!\!\!\begin{array}{c} 1;1 \\ 1;1 \\ \underline{0;1} \end{array} \!\!\!\!\right)
& \left(\!\!\!\!\begin{array}{c} \sigma^1;\tau^1\\ \underline{0;1} \\ \underline{0;1} \end{array}\!\!\!\! \right)
&   \left(\!\!\!\!\begin{array}{c} 1;1 \\ 1;1 \\ 1;1 \end{array} \!\!\!\!\right)
\\\cline{3-6}
&  & USp(2N)& SO(2N) & USp(2N) & SO(2N)
\\
& & +1 \, \Anti & + 1 \, \Sym & + \emptyset & + \emptyset
\\\hline\hline
\!\!\OR\Z_2^{(1)}\!\!\!    &\hspace{-3mm} \left(\hspace{-3mm}\begin{array}{c} -(-1)^{\sigma^2\tau^2 + \sigma^3 \tau^3} \\ (-1)^{2b\sigma^1\tau^1 + \sigma^3 \tau^3} \\  (-1)^{2b\sigma^1\tau^1 + \sigma^2 \tau^2}\end{array} \hspace{-3mm}\right)\hspace{-3mm}
& \left(\!\!\!\!\begin{array}{c} \sigma^1;\tau^1\\ 1;1\\ \underline{0;1}\end{array}\!\!\!\! \right)
&   \left(\!\!\!\!\begin{array}{c} 1;1 \\  \underline{0;1} \\ 1;1 \end{array} \!\!\!\!\right)
& \left(\!\!\!\!\begin{array}{c} \sigma^1;\tau^1\\1;1 \\ 1;1 \end{array}\!\!\!\! \right)
&   \left(\!\!\!\!\begin{array}{c} 1;1 \\  \underline{0;1} \\  \underline{0;1} \end{array} \!\!\!\!\right)
\\\cline{3-6}
&  & USp(2N) & SO(2N) & SO(2N) & USp(2N) \hspace{-2mm}
\\
& & + 5 \, \Anti & + 5 \, \Sym & + 4 \, \Anti  & + 4 \, \Sym 
\\\hline\hline
\!\!\OR\Z_2^{(2)}\!\!\!    & \hspace{-3mm} \left(\hspace{-3mm}\begin{array}{c} (-1)^{\sigma^2\tau^2 + \sigma^3 \tau^3} \\ - (-1)^{2b\sigma^1\tau^1 + \sigma^3 \tau^3} \\  (-1)^{2b\sigma^1\tau^1 + \sigma^2 \tau^2}\end{array}\hspace{-3mm} \right)\hspace{-3mm}
& \left(\!\!\!\!\begin{array}{c} \sigma^1;\tau^1\\ 1;1 \\ 1;1\end{array}\!\!\!\! \right)
&   \left(\!\!\!\!\begin{array}{c} 1;1 \\  \underline{0;1} \\  \underline{0;1} \end{array} \!\!\!\!\right)
& \left(\!\!\!\!\begin{array}{c} \sigma^1;\tau^1\\ 1;1\\ \underline{0;1}\end{array}\!\!\!\! \right)
&   \left(\!\!\!\!\begin{array}{c} 1;1 \\   \underline{0;1} \\ 1;1 \end{array} \!\!\!\!\right)
\\\cline{3-6}
&  & SO(2N)& USp(2N) & USp(2N) & SO(2N)
\\
& & + 1\, \Anti & + 1 \, \Sym & + 2 \, \Anti & + 2 \, \Sym
\\\hline\hline
\!\!\OR\Z_2^{(3)}\!\!\!  & \hspace{-3mm} \left(\hspace{-3mm}\begin{array}{c} (-1)^{\sigma^2\tau^2 + \sigma^3 \tau^3} \\  (-1)^{2b\sigma^1\tau^1 + \sigma^3 \tau^3} \\  -(-1)^{2b\sigma^1\tau^1 + \sigma^2 \tau^2} \end{array}\hspace{-3mm}\right)\hspace{-3mm}
& \left(\!\!\!\!\begin{array}{c} \sigma^1;\tau^1\\  \underline{0;1} \\ \underline{0;1}\end{array}\!\!\!\! \right)
&   \left(\!\!\!\!\begin{array}{c} 1;1 \\ 1;1 \\ 1;1 \end{array} \!\!\!\!\right)
& \left(\!\!\!\!\begin{array}{c} \sigma^1;\tau^1\\ \underline{0;1} \\ 1;1 \end{array}\!\!\!\! \right)
&   \left(\!\!\!\!\begin{array}{c} 1;1 \\ 1;1 \\  \underline{0;1} \end{array} \!\!\!\!\right)
\\\cline{3-6}
&  & USp(2N) & SO(2N) & USp(2N) & SO(2N)
\\
& & +1 \, \Sym & + 1 \, \Anti & + 2\, \Anti & + 2 \, \Sym
\\\hline
\end{array}
\end{equation*}
\caption{Classification of $USp(2N)$ and $SO(2N)$ gauge groups and matter in the (anti)symmetric representation
 on $\OR$-invariant D6-branes. The configurations with $\eta_{\OR\Z_2^{(2)}}=-1$ can be obtained from the listed case $\eta_{\OR\Z_2^{(3)}}=-1$ by exchanging two-torus labels $2 \leftrightarrow 3$. The choice $\eta_{\OR\Z_2^{(1)}}=-1$ for the exotic O6-plane does not lead to {\it global} supersymmetric configurations due to the first bulk RR tadpole cancellation condition in table~\protect\ref{tab:Bulk-RR+SUSY-Z2Z6}. 
Underlining denotes three choices, e.g. $(\sigma^2;\tau^2) \in \{(0;0);(1;0), (0;1)\}$ since only $\sigma^2\tau^2=0$ is required. In case of underlining of both $(\sigma^2;\tau^2)$ and $(\sigma^3;\tau^3)$, the choices are independent - in other words, there are $3^2=9$ options. For $b=\frac{1}{2}$,  the cases with $\sigma^1\tau^1=0$ coincide with those listed for $b=0$, but those with $\sigma^1\tau^1=1$ differ and are listed explicitly here.\label{Tab:Conditions-on_b+t+s-SOSp}}
  \end{center}
\end{table}
%%%%%%%%%%%%%%%%%%%%%%%%%%%%%%%%%%%%%%%%%%%%%
The other columns in table~\ref{Tab:Conditions-on_b+t+s-SOSp} review for which combinations of discrete displacements $(\vec{\sigma}_a)$ and discrete Wilson lines $(\vec{\tau}_a)$ the $USp(2N)$ or $SO(2N)$ gauge group enhancement occurs in function of the choice of the exotic O6-plane. The table also indicates the full amount of  matter in the symmetric or antisymmetric representation under the respective gauge group arising in the three sectors $(\omega^k a)(\omega^k a)^{\prime} \simeq a(\omega^{-2k} a')_{k\in \{0,1,2\}}$. By counting the combinations of $\Z_2$ eigenvalues, displacements and Wilson lines we can determine the numbers $N_{USp}$ and $N_{SO}$ of configurations giving rise to $USp(2N)$ and $SO(2N)$ enhancement, respectively: for a rectangular $T_{(1)}^2$ ($b=0$) we have $N_{USp} = 240$ and $N_{SO} = 16$, whereas $N_{USp} = 216$ and $N_{SO} = 40$ for a tilted $T_{(1)}^2$ ($b=\frac{1}{2}$). For more details concerning gauge group enhancement on $T^6/(\Z_2\times \Z_6\times \OR)$ with discrete torsion, we refer to~\cite{Ecker:2014hma} where the classification was discussed for the first time.

In order to obtain phenomenologically appealing intersecting D6-brane models containing some MSSM-like or left-right symmetric model sector, the {\it QCD} stack and the $SU(2)_L$ left-stack ought to be constructed with {\it rigid} fractional three-cycles free of matter in the adjoint representation. This requirement prevents the corresponding gauge group to be spontaneously broken by (continuous) D-brane displacements or recombinations or Wilson lines. 
In~\cite{Ecker:2014hma} an exhaustive search for rigid fractional three-cycles without matter in the adjoint representation was presented, from which the following constraints on the fractional three-cycle for the {\bf a/bAA} lattice can be distilled:
\vspace{0.08in}\\ {\it
\hspace*{0.1in} (i) the bulk three-cycle is parallel to $\OR$ or to an orbit of the form $(n_a^1,m_a^1;1,0;1,-1)$,\\
\hspace*{0.1in} (ii) the discrete parameters $(\vec{\sigma}_a, \vec{\tau}_a)$ satisfy the relation: $\sigma^2_a \tau^2_a = \sigma_a^3 \tau_a^3 \in \{ 0, 1 \}$. \vspace{0.08in}\\}
All other discrete parameters, including the choice of the exotic O6-plane and $\Z_2$ eigenvalues, do not affect the amount of matter in the adjoint representation. The one-cycle wrapping numbers $(n_a^1,m_a^1)$ are related to the complex structure modulus of the two-torus $T_{(1)}^2$ through the necessary bulk supersymmetry condition in table~\ref{tab:Bulk-RR+SUSY-Z2Z6}:
\begin{equation}\label{Eq:RhoDependenceWrappingNumbers}
\varrho = 3 \frac{n_a^1}{m_a^1 + b\, n_a^1},
\end{equation}
and thus cannot be chosen at random. Note, however, that these conditions do not include those fractional three-cycles that are $\OR$-invariant and support an enhanced $USp(2)$ gauge group accompanied solely by matter in the antisymmetric representation, as listed in table~\ref{Tab:Conditions-on_b+t+s-SOSp}. This latter type of fractional three-cycles is well suited to support the $SU(2)_L$ and/or $SU(2)_R$ gauge group when constructing MSSM-like models or left-right symmetric models, respectively.

The absence of chiral matter in the symmetric representation under the {\it QCD} and the $U(2)_L$ gauge group requires us also to investigate the intersections between a fractional three-cycle and its orientifold image orbit. 
More explicitly, only fractional three-cycles with $\chi^{\Sym_a} = 0$ will be able to serve as candidate three-cycles to support the {\it QCD} stack or the $U(2)_L$ stack. In the case of the {\it QCD} stack, also the amount of chiral matter in the antisymmetric representation has to be constrained by the condition \mbox{$\left| \chi^{\Anti} \right| \leq 3$}. Otherwise, the {\it QCD} stack could be characterised by more than three generations of right handed $u_R$ (or $d_R$) quarks. 
We can now implement these extra conditions on the two types of bulk orbits, parallel to the $\OR$-plane or with representant $(n^1_a,m^1_a;1,0;1,-1)$, with $\sigma^2_a\tau^2_a=\sigma^3_a\tau^3_a$ specified above as sole configurations without matter in the adjoint representation:
\begin{itemize}
\item
The $\varrho$-independent configurations have their bulk orbit parallel to the $\OR$-plane, leading automatically to a vanishing intersection number with the O6-planes, \mbox{$\Pi_a \circ \Pi_{O6}=0$}, and thus to the requirement \mbox{$\chi^{\Anti_a} \equiv \chi^{\Sym_a} \stackrel{!}{=}0$}.
\item
For the $\varrho$-dependent configurations with bulk orbit $(n_a^1,m_a^1;1,0;1,-1)$, the constraint $\left| \chi^{\Anti_a} \right| \in \{0,1,2,3\}$ reduces the number of potential tuples $(n_a^1,m_a^1)$ significantly, as summarised in table~\ref{tab:RigidSymmFreeAntisymmetrics}. The bulk RR tadpole cancellation conditions in table~\ref{tab:Bulk-RR+SUSY-Z2Z6} can for $
\varrho$-dependent models only be satisfied for the choice $\eta_{\OR}=-1$ of exotic O6-plane.
For $\sigma_a^2 \tau_a^2 = \sigma_a^3 \tau_a^3  =1$, the list with six bulk orbits for the {\bf aAA} lattice and five bulk orbits for the {\bf bAA} lattice in table~\ref{tab:RigidSymmFreeAntisymmetrics} is exhaustive.
\\
For $\eta_{\OR}=-1$ and $\sigma_a^2 \tau_a^2 = \sigma_a^3 \tau_a^3 =0$, only the bulk orbit $(1,1;1,0;1,-1)$ on the {\bf aAA} lattice and the bulk orbit $(1,0;1,0;1,-1)$  on the {\bf bAA} lattice 
satisfy all constraints on adjoint and (anti)symmetric representations, cf. table~12 in~\cite{Ecker:2014hma}.\footnote{Also $\eta_{\OR\Z_2^{(2 \text{ or } 3)}}=-1$ and $\sigma_a^2 \tau_a^2 = \sigma_a^3 \tau_a^3 =0$ solve the constraints on the matter spectrum, but the second bulk RR tadpole cancellation condition in table~\ref{tab:Bulk-RR+SUSY-Z2Z6} cannot be satisfied in a supersymmetric way, see~\cite{Ecker:2014hma} for more details. For this last choice of discrete parameters, we note for completeness that also the bulk orbit $(1,1;1,0;1,-1)$ on the {\bf bAA} lattice satisfies all the above constraints on the matter spectrum, yet it does not offer opportunities for global supersymmetric model building as the bulk three-cycle suffers from the same obstruction -- deduced from the second bulk RR tadpole cancellation condition -- as the two other bulk orbits. }
\end{itemize}
 All these considerations provide us with a set of candidate fractional three-cycles to support the {\it QCD} stack, both for $\varrho$-independent as well as for $\varrho$-dependent configurations.    
%%%%%%%%%%%%%%%%%%%
\mathtab{{\footnotesize
\begin{array}{|c|c|c||c|c|c|}
\hline \multicolumn{6}{|c|}{\text{\bf Rigid fractional three-cycles with $\chi^{\Sym_a}=0$ and $\left| \chi^{\Anti_a} \right| \in \{0,1,2,3 \}$}} \\
\multicolumn{6}{|c|}{\text{\bf on $T^6/(\Z_2\times\Z_6\times\OR)$ with $\eta = \eta_{\OR}= -1$}} \\
\hline
\hline 
\multicolumn{3}{|c||}{\text{{\bf aAA} lattice}} & \multicolumn{3}{|c|}{\text{{\bf bAA} lattice}}\\
\hline\hline
(n_a^1, m_a^1)& \varrho & (\chi^{\Anti_a}_{(a)}, \chi^{\Anti_a}_{(\omega a)}, \chi^{\Anti_a}_{(\omega^2 a)} ) &  (n_a^1, m_a^1)& \varrho & (\chi^{\Anti_a}_{(a)}, \chi^{\Anti_a}_{(\omega a)}, \chi^{\Anti_a}_{(\omega^2 a)} )  \\
\hline
 (1,1) & 3 &  (0,2,0) & (1,0) &6 & (0,1,0)   \\
(1,2) & 3/2 &  (-1,3,-1) & (1,1) & 2& (-1,2,-1)   \\
(1,3) &1 &  (-2,4,-2) &  (1,2) & 6/5 & (-2,3,-2) \\
(1,4) &3/4  & (-3,5,-3)& (1,3) & 6/7 &(-3,4,-3) \\
(1,5) & 3/5  &(-4,6,-4) & (1,4) & 2/3& (-4,5,-4)  \\
(1,6) &1/2   &(-5,7,-5) & && \\
\hline
\end{array}}
}{RigidSymmFreeAntisymmetrics}{Overview of the net-chirality $(\chi^{\Anti_a}_{a}, \chi^{\Anti_a}_{(\omega a)}, \chi^{\Anti_a}_{(\omega^2 a)} )$ per sector $(\omega^k a)(\omega^k a)'$ for the rigid fractional three-cycles with bulk orbit $(n_a^1,m_a^1;1,0;1,-1)$ free from chiral states in the symmetric representation and with $\left| \chi^{\Anti_a} \right| \leq 3$ on the {\bf aAA} and {\bf bAA} lattice. 
The contributions $\chi^{\Anti_a/\Sym_a}_{(\omega^k a)}$ per sector 
$(\omega^k a)(\omega^k a)^{\prime}$ can be calculated~\cite{Ecker:2014hma} using formula (\ref{Eq:SymmAntiTorusExp}), and they scale with the one-cycle wrapping number $m_a^1$.
This list is valid and exhaustive regarding $(n_a^1,m_a^1)$, only if the $\OR$-plane is the exotic O6-plane and the discrete parameters along $T_{(2)}^2\times T_{(3)}^2$ satisfy the relations $\sigma_a^2 \tau_a^2 = \sigma_a^3 \tau_a^3  =1$,
ensuring the absence of matter in the adjoint representation. For the same exotic O6-plane choice and the relations $\sigma_a^2 \tau_a^2 = \sigma_a^3 \tau_a^3  =0$, the bulk orbits in the first row satisfy the constraints $\chi^{\Sym_a}=0$ and $|\chi^{\Anti_a}|\leq 3$ as well.  }
%%%%%%%%%%%%%%%%%%%%%%%%%

Once a fractional three-cycle for the {\it QCD} stack is identified, we have to determine an appropriate fractional three-cycle for the $SU(2)_L$ stack, such that the intersections between the two stacks give rise to three chiral generations of left handed quarks. Thus, another  indispensable  element of D6-brane model building consists in finding configurations of two fractional three-cycles $\Pi_a^{\rm frac}$ and $\Pi_b^{\rm frac}$ with $\chi^{ab} + \chi^{ab'} \stackrel{!}{=} \pm 3$. In case the {\it left} gauge group results from an enhanced $USp(2)_b$ gauge group on the $b$-stack, the condition on the total net-chirality reads $\chi^{ab} \equiv \chi^{ab'}=\pm3$ instead. In~\cite{Ecker:2014hma} an exhaustive search revealed a class of {\bf $\varrho$-independent} D6-brane configurations with three chiral generations, for which the {\it QCD}-stack is parallel to the $\OR$-plane and the $SU(2)_L$ stack is parallel to the $\OR\Z_{2}^{(1)}$-plane. More concretely, one can find 48 (36) combinations of {\it relative} $\Z_2^{(k)}$ eigenvalues, discrete Wilson lines and displacements for the {\it QCD}-stack and the {\it left} stack on the {\bf aAA} ({\bf bAA}) lattice yielding three chiral quark generations, provided that either the $\OR\Z_2^{(2)}$-plane or the $\OR\Z_2^{(3)}$-plane is the exotic O6-plane. The fractional three-cycles for the {\it QCD} stack are free from matter in the adjoint representation, as well as free from chiral matter in symmetric and antisymmetric representations, while the fractional three-cycles for the $SU(2)_L$ stack support enhanced $USp(2)$ gauge groups accompanied by five states in the antisymmetric ($\equiv$ singlet of $USp(2)$) representation (see table~\ref{Tab:Conditions-on_b+t+s-SOSp}).

In order to obtain a full MSSM-like or left-right symmetric spectrum, the ($\varrho$-independent) combinations of fractional three-cycles with three chiral left-handed quarks have to be completed with an appropriate $c$-and/or $d$-stack. For the D6$_c$- and D6$_d$-brane stacks, fractional three-cycles with a bulk orbit parallel to any $\OR\Z_2^{(k)}$-plane can serve as candidates, since these {\bf $\varrho$-independent} supersymmetric D6-brane stacks are allowed to be accompanied by matter in the adjoint representation. Regarding chiral matter in the (anti)symmetric representation, one can
easily verify that the constraint~\mbox{$\chi^{\Anti_a} \equiv \chi^{\Sym_a} \stackrel{!}{=}0$} is satisfied for all fractional three-cycles parallel to one of the $\OR\Z_2^{(k)}$-planes, independently of the choice of the exotic O6-plane, the $\Z_2^{(k)}$ eigenvalues, the discrete parameters $(\sigma_a^1,\tau_a^1)$ or the choice of the lattice orientation. 
An intensive search for {\bf $\varrho$-independent} global MSSM-like and left-right symmetric intersecting D6-brane models on the {\bf aAA} lattice will be presented in sections~\ref{Ss:MSSM-models} and~\ref{Ss:LR-models}, respectively. For these {\bf $\varrho$-independent} D6-brane configurations, the choice of the exotic O6-plane will be either the $\OR\Z_2^{(2)}$-plane or the $\OR\Z_2^{(3)}$-plane as dictated by the requirement of having three chiral generations of left-handed leptons.
Remember that in addition to the open string matter spectrum, the massless closed string spectrum on the {\bf aAA} lattice  for $\eta_{\OR\Z_2^{\text{(2 or 3)}}}=-1$ contains ${\cal N}=1$ supermultiplets with $h_{11}^+= 4_{\Z_6'}$ vectors, $h_{11}^-=3_{\text{bulk}} + 8_{\Z_3} + 4_{\Z_6'}$ K\"ahler moduli and  $h_{21}=1_{\text{bulk}} + 14_{\Z_2} + 2_{\Z_3} + 2_{\Z_6}$ complex structure moduli as first computed in~\cite{Forste:2010gw}. 

\clearpage
\subsubsection{Intermezzo: towards three generations in $\varrho$-dependent configurations}
The D6-brane configurations with the {\it QCD} stack wrapping a fractional three-cycle parallel to the $\OR$-plane and the {\it left} stack parallel to the $\OR\Z_2^{(1)}$-plane are the only $\varrho$-independent configurations which yield three chiral generations of left-handed quarks without exotic matter as specified above (i.e. no matter in the adjoint representation nor chiral matter in the (anti)symmetric representation). Indeed, considering a {\it left} stack parallel to one of the other O6-planes does not offer the right amount of chiral left-handed quarks, as shown in table 14 of~\cite{Ecker:2014hma}. This prompts 
us to consider the alternative roads of {\bf $\varrho$-dependent} D6-brane configurations consisting of two distinct choices, as expressed in tables 15 and 16 of~\cite{Ecker:2014hma}:
\begin{itemize}
\item[(1)]  \vspace{-0.2in} {\it Choice 1:}
\begin{itemize}
\item \vspace{-0.2in}
the {\it QCD} stack remains parallel to the $\OR$-plane, 
\item \vspace{-0.1in} but the $SU(2)_L$ stack has a bulk orbit $(1,m_b;1,0;1,-1)$ with $m_b \in \{1,3,5,7,9,11,13,15\}$ for the {\bf aAA} lattice and $m_b \in \{0,1,2,3,4,5,6,7\}$ for the {\bf bAA} lattice;
\end{itemize}
\item[(2)] {\it Choice 2:}
\begin{itemize}
\item \vspace{-0.2in} the {\it QCD} stack is characterised by a bulk orbit $(1,m_a;1,0;1,-1)$ with \mbox{$m_a\in \{1,3,4,5,6 \}$} for the {\bf aAA} lattice and \mbox{$m_a \in \{ 0,1,2,3,4\}$} for the {\bf bAA} lattice, 
\item \vspace{-0.1in} while the $SU(2)_L$ stack can be parallel to the $\OR$-plane or to a bulk orbit $(1,m_b;1,0;1,-1)$ with $m_b \in \{1,3\}$ for the {\bf aAA} lattice and $m_b \in \{0,1,4\}$ for the {\bf bAA} lattice,
see table~16 in~\cite{Ecker:2014hma} for the exact configurations. 
\end{itemize}
\end{itemize} 
Other choices of $(m^1_a,m^1_b)$ are excluded by requiring the existence of three chiral quark generations.

The $\varrho$-dependence of the D6-brane configurations, deducible from equation (\ref{Eq:RhoDependenceWrappingNumbers}), excludes the exotic O6-plane choices $\eta_{\OR\Z_2^{(2)}} = -1$ and $\eta_{\OR\Z_2^{(3)}} = -1$. That is to say, the bulk orbits preserving supersymmetry for specific $\varrho$-values are characterised by a bulk wrapping number $ V + b Q \neq 0$, implying that the bulk RR tadpole cancellation conditions in table~\ref{tab:Bulk-RR+SUSY-Z2Z6} are only satisfied when the $\OR$-plane plays the r\^ole of the exotic O6-plane. A more detailed account of the search for $\varrho$-dependent configurations yielding three chiral left-handed quark generations is offered in section 3.4.2 of~\cite{Ecker:2014hma}, accompanied by a precise counting of the number of consistent D6-brane configurations.
This exhaustive scan for three generations of left-handed quarks, however, still needs to be supplemented by the requirement of three right-handed quark generations and three lepton generations, after which global completions will have to be investigated. Such a systematic scan for global $\varrho$-dependent models is expected to be extremely time-consuming and cumbersome, and at this point we leave it for future work.

%%%%%%%%%%%%%%%%%%%%%%%%%%%%%%%%%%%%%%%%%%%%%%%%%%%%%%%%%%%%%%%%%%%%%%%%%%%%%%%%%%%
%%%%%%%%%%%%%%%%%%%%%%%%%%%%%%%%%%%%%%%%%%%%%%%%%%%%%%%%%%%%%%%%%%%%%%%%%%%%%%%%%%%
%%%%%%%%%%%%%%%%%%%%%%%%%%%%%%%%%%%%%%%%%%%%%%%%%%%%%%%%%%%%%%%%%%%%%%%%%%%%%%%%%%%
%%%%%%%%%%%%%%%%%%%%%%%%%%%%%%%%%%%%%%%%%%%%%%%%%%%%%%%%%%%%%%%%%%%%%%%%%%%%%%%%%%%
\section{D6-Brane Phenomenology on $T^6/(\Z_2 \times \Z_6 \times \OR)$}\label{S:D6branePheno}

%%%%%%%%%%%%%%%%%%%%%%%%%%%%%%%%%%%%%%%%%%%%%%%%%%%%%%%%%%%%%%%%%%%%%%%%%%%%%%%%%%%
%%%%%%%%%%%%%%%%%%%%%%%%%%%%%%%%%%%%%%%%%%%%%%%%%%%%%%%%%%%%%%%%%%%%%%%%%%%%%%%%%%%
\subsection{K-Theory Constraints and Discrete Symmetries}\label{Ss:KtheoryDiscrete}

%%%%%%%%%%%%%%%%%%%%%%%%%%%%%%%%%%%%%%%%%%%%%%%%%%%%%%%%%%%%%%%%%%%%%%%%%%%%%%%%%%%
\subsubsection{Basis of $\OR$-even three-cycles and K-theory constraints}

The RR charges of a D6-brane wrapping a fractional three-cycle as in expression (\ref{Eq:Z2Z6FractCycles}) are in first instance classified by (co-)homology theory, such that the required vanishing of RR charges on the orientifold $T^6/(\Z_2\times\Z_6\times\OR)$ with discrete torsion can be easily recast into the conditions of table~\ref{tab:Bulk-RR+SUSY-Z2Z6} in terms of homology,
$\sum_a N_a (\Pi_a + \Pi_a^{\prime}) = 4 \, \Pi_{O6}$. However, not all D6-brane charges are captured by homology, as D-branes carry additional $\Z_2$ valued K-theory charges~\cite{Witten:1998cd}. The presence of uncanceled K-theory charges in the compact internal space opens up the worrisome prospect of having an inconsistent compactification, even when the RR tadpoles vanish. We will call string vacua of this type {\it semi-local}.

In general, it is rather difficult to directly determine the conditions for vanishing K-theory charges on compact spaces, yet by using a probe brane argument~\cite{Uranga:2000xp} one can deduce necessary conditions for the vanishing of the K-theory charges:
\begin{equation}\label{Eq:KtheoryConstraints}
\Pi_{USp(2)_i}^{\text{frac}} \circ \left(\sum_a N_a \Pi_a^{\text{frac}}\right) \stackrel{!}{=} 0 \text{ mod } 2.
\end{equation}
This expression requires an even number of states in the fundamental representation of any $USp(2)$ probe brane, which is counted by the number of intersections of the set of D6-branes in the model weighted by the corresponding ranks. Violations of condition (\ref{Eq:KtheoryConstraints}) indicate the presence of a global field-theoretical anomaly in the $SU(2)\simeq USp(2)$ gauge theory on the probe brane~\cite{Witten:1982fp}. Thus, for a given D6-brane configuration with vanishing RR tadpoles, {\it global} consistency also requires that the K-theory constraints in~(\ref{Eq:KtheoryConstraints}) are satisfied.    

To assess the K-theory constraints, one requires the full classification of $\OR$-invariant fractional three-cycles supporting an enhanced $USp(2)$ gauge group as reviewed in table~\ref{Tab:Conditions-on_b+t+s-SOSp} of section~\ref{Ss:IntersectSummary}. Note, however, that not every $\OR$-invariant fractional three-cycle with $USp(2)$ gauge group leads to an independent constraint. In practice, we expect at most $b_3^{\text{bulk}+\Z_2}/2=h_{21}^{\text{bulk}+\Z_2}+1 = 16$ linearly independent conditions associated to the number of linearly independent $\OR$-even three-cycles on the orientifold $T^6/(\Z_2\times\Z_6\times \OR)$ with discrete torsion ($\eta=-1$). In order to reduce the number of independent constraints resulting from equation~(\ref{Eq:KtheoryConstraints}), we first determine all $\OR$-even and $\OR$-odd three-cycles, which are either purely of bulk or exceptional type, using table~\ref{tab:Z2Z6bulkexcept-Orient}. The result is displayed in table~\ref{tab:ORevenoddlattice} for the choice $\eta_{\OR\Z_2^{(3)}}=-1$ of exotic O6-plane on the {\bf aAA} lattice.\footnote{The constraints for the {\bf bAA} lattice can be obtained in an analogous manner by (i) replacing $m^1_a \to \tilde{m}^1_a=m^1_a+b \, n^1_a$ which amounts to $(U_a,V_a; y_{\alpha,a}^{(1)}) \to (\tilde{U}_a,\tilde{V}_a;\tilde{y}_{\alpha,a}^{(1)})$, (ii) permuting fixed point indices in the $\Z_2^{(2)}$ and $\Z_2^{(3)}$ sectors since probe branes parallel to the $\OR(\Z_2^{(1)})$-plane on $T^2_{(1)}$ now pass through fixed points
$\{1,4\}_{\sigma^1_a=0}$ and $\{2,3\}_{\sigma^1_a=0}$, see figure~\ref{Fig:LatticesZ2Z6}. However, we anticipate here that we only find three-generation models with cancelled RR tadpoles on the {\bf aAA} lattice as detailed in sections~\ref{S:MSSM-pheno} and~\ref{S:LRsymModels-pheno}.}
In the next step, we can express the $\OR$-invariant fractional three-cycles with $USp(2)$ gauge group in terms of the basis of purely bulk/exceptional $\OR$-even three-cycles and deduce which fractional three-cycles are truly linearly independent. More explicitly, by choosing different combinations of discrete parameters, one can easily show that various fractional three-cycles can be written as linear combinations of other fractional three-cycles, allowing us to reduce the 240 (216) $\OR$-invariant fractional three-cycles with $USp(2)$ gauge group on the {\bf aAA} ({\bf bAA}) lattice to only sixteen linearly independent combinations.
%%%%%%%%%%%%%%%%%%%%%%%%
\mathtab{\hspace{-0.4in}
\begin{array}{|c||c|c||c|}
\hline \multicolumn{4}{|c|}{\text{\bf $\OR$-even and -odd three-cycles on $T^6/(\Z_2 \times \Z_6 \times \OR)$ with $\eta = -1 = \eta_{\OR\Z_2^{(3)}}$ for {\bf aAA} lattice}} \\
\hline
\hline \text{\bf sector}& \OR-\text{\bf even} & \OR-\text{\bf odd} & \text{\bf Intersection Form } \\
\hline \hline 
\text{bulk} & \Pi_0^{\rm even} = \rho_1  & \Pi_0^{\rm odd} = \rho_3 &  \Pi_0^{\rm even} \cdot   \Pi_0^{\rm odd} = 8   \\
& \Pi_1^{\rm even} = \rho_3 - 2 \rho_4  & \Pi_1^{\rm odd} = \rho_1 - 2 \rho_2 & \Pi_1^{\rm even} \cdot  \Pi_1^{\rm odd} = -24   \\
\hline \hline
 & \Pi_{2 + \alpha}^{\rm even} =\tilde \varepsilon^{(1)}_{\alpha = 0,1,2,3} &  \Pi_{2 + \alpha}^{\rm odd} = \varepsilon^{(1)}_{\alpha = 0,1,2,3}  & \Pi^{\rm even}_2 \cdot \Pi^{\rm odd}_2  = 12,  \\
 \Z_{2}^{(1)} & && \Pi^{\rm even}_{a} \cdot \Pi^{\rm odd}_b  = 4 \delta_{ab }, \, a, b = 3,4,5   \\
 & \Pi_{6}^{\rm even} = \varepsilon_4^{(1)} -  \varepsilon_5^{(1)} &  \Pi_{6}^{\rm odd} =  \tilde \varepsilon_4^{(1)} - \tilde \varepsilon_5^{(1)} & \Pi_{6}^{\rm even} \cdot \Pi_{6}^{\rm odd} = -8 \\
 & \Pi_{7}^{\rm even} = \tilde  \varepsilon_4^{(1)} + \tilde  \varepsilon_5^{(1)} &  \Pi_{7}^{\rm odd} =   \varepsilon_4^{(1)} + \varepsilon_5^{(1)} & \Pi_{7}^{\rm even} \cdot \Pi_{7}^{\rm odd} = 8 \\
 \hline  \hline
  \Z_{2}^{(2)} & \Pi_{7 + \alpha}^{\rm even}  =  [\varepsilon^{(2)}_\alpha  - 2 \tilde \varepsilon^{(2)}_\alpha]_{\alpha = 1,2,3,4} &   \Pi_{7 + \alpha}^{\rm odd} = \varepsilon^{(2)}_{\alpha =1,2,3,4} & \Pi_{7 + \alpha}^{\rm even} \cdot \Pi_{7 + \beta}^{\rm odd}  = - 8 \delta_{\alpha \beta} \\
  \hline \hline
   \Z_{2}^{(3)} & \Pi_{11 + \alpha}^{\rm even}  = \varepsilon^{(3)}_{\alpha =1,2,3,4}  &   \Pi_{11 + \alpha}^{\rm odd} = [\varepsilon^{(3)}_\alpha  - 2 \tilde \varepsilon^{(3)}_\alpha]_{\alpha = 1,2,3,4} & \Pi_{11 + \alpha}^{\rm even} \cdot \Pi_{11 + \beta}^{\rm odd}  =  8 \delta_{\alpha \beta} \\
   \hline
\end{array}
}{ORevenoddlattice}{Overview of the $\OR$-even and $\OR$-odd pure bulk or exceptional three-cycles on \mbox{$T^6/(\Z_2 \times \Z_6 \times \OR)$} with discrete torsion $\eta = -1 = \eta_{\OR\Z_2^{(3)}}$ for the {\bf aAA} lattice. The right column lists all intersections number of the symplectic lattice (vanishing intersection numbers are omitted).}
%%%%%%%%%%%%%%%%%%%

As concrete example we consider the {\bf aAA} lattice with the choice of exotic O6-plane $\eta_{\OR\Z_2^{(3)}}=-1$ and 
the fractional three-cycles with bulk orbit parallel to the $\OR\Z_2^{(1)}$-plane and show how they can be written as linear combinations of the fractional three-cycles parallel to the $\OR$-plane. As indicated in table~\ref{Tab:Conditions-on_b+t+s-SOSp}, fractional three-cycles parallel to the $\OR\Z_2^{(1)}$-plane can support an enhanced $USp(2)$ gauge group for $\sigma^2_a \tau^2_a=1$. For the displacement parameters $(\vec{\sigma}_a)=(\sigma_a^1,1,1)$, the discrete Wilson lines have to be chosen as $(\vec{\tau}_a)=(\tau_a^1,1,0)$ in order to guarantee an enhanced $USp(2)$ gauge group. For this explicit choice of discrete parameters, the fractional three-cycles parallel to the $\OR\Z_2^{(1)}$-plane read as follows in terms of the basis of
purely bulk/exceptional $\OR$-even three-cycles:\footnote{Fractional three-cycles parallel to the $\OR\Z_2^{(1)}$-plane and with discrete displacements of the form $(\sigma_a^1,1,0)$ can also be written in this form upon a shift of the $\Z_2^{(k)}$ eigenvalues, i.e.~$\tau_a^{\Z_2^{(1)}} \rightarrow \tau_a^{\Z_2^{(1)}} +1$ and $\tau_a^{\Z_2^{(2)}} \rightarrow \tau_a^{\Z_2^{(2)}} +1$. For this choice of discrete displacements the Wilson line takes the values $\tau_a^3 \in \{ 0,1\}$.  
Note also that a different value for the discrete parameter $\sigma_a^3$ does not alter which $\OR$-even three-cycles are used in the $\Z_2^{(1)}$ and $\Z_2^{(2)}$ sectors to express the fractional three-cycles parallel to the $\OR\Z_2^{(1)}$-plane and invariant under the $\OR$-projection.
} 
\begin{equation}
\begin{array}{rcl}
\Pi^{{\rm frac},(\sigma_a^1,1,1)}_{a\pp \OR\Z_2^{(1)}}&=& \frac{3}{4} \Pi_0^{\rm even} -\frac{(-)^{\tau_a^{\Z_2^{(1)}}}}{4} \Pi_6^{\rm even}
+\frac{(-)^{\tau_a^{\Z_2^{(2)}}}}{4} \left[ \Pi^{\rm even}_{8+2 \sigma_a^1} + (-)^{\tau_a^1}  \Pi^{\rm even}_{9+2 \sigma_a^1} \right] \\
&&-3\frac{(-)^{\tau_a^{\Z_2^{(1)}}+\tau_a^{\Z_2^{(2)}}}}{4} \left[ \Pi^{\rm even}_{12+2 \sigma_a^1} + (-)^{\tau_a^1}  \Pi^{\rm even}_{13+2 \sigma_a^1} \right].
\end{array}
\end{equation}
By looking at the fractional three-cycles parallel to the $\OR$-plane with the following choice of discrete parameters $(\vec{\sigma}_a)=(\sigma_a^1,1,1)$ and  $(\vec{\tau}_a)=(\tau_a^1,0,1)$ allowing for an enhanced $USp(2N)$ gauge group (see table~\ref{Tab:Conditions-on_b+t+s-SOSp}): 
\begin{equation}
\begin{array}{rcl}
\Pi^{{\rm frac},(\sigma_a^1,1,1)}_{a\pp \OR}&=& \frac{1}{4} \Pi_0^{\rm even} -\frac{(-)^{\tau_a^{\Z_2^{(1)}}}}{4} \Pi_6^{\rm even}
+\frac{(-)^{\tau_a^{\Z_2^{(2)}}}}{4} \left[ \Pi^{\rm even}_{8+2 \sigma_a^1} + (-)^{\tau_a^1}  \Pi^{\rm even}_{9+2 \sigma_a^1} \right] \\
&&-\frac{(-)^{\tau_a^{\Z_2^{(1)}}+\tau_a^{\Z_2^{(2)}}}}{4} \left[ \Pi^{\rm even}_{12+2 \sigma_a^1} + (-)^{\tau_a^1}  \Pi^{\rm even}_{13+2 \sigma_a^1} \right],
\end{array}
\end{equation}
we can easily deduce the following relation among the fractional three-cycles:
\begin{equation}
\begin{array}{l}
 \Pi_{a\pp\OR}^{(\sigma^1_a,1,1)} [\tau_a^{\Z_2^{(1)}}, \tau_a^{\Z_2^{(2)}}, \tau_a^{\Z_2^{(3)}}]  +  \Pi_{a\pp\OR\Z_2^{(1)}}^{(\sigma^1_a,1,1)} [\tau_a^{\Z_2^{(1)}}+1, \tau_a^{\Z_2^{(2)}}+1, \tau_a^{\Z_2^{(3)}}] \\
 \qquad = \Pi_0^{\rm even} - (-)^{\tau_a^{\Z_2^{(1)}}+\tau_a^{\Z_2^{(2)}}} \left[ \Pi_{12+2 \sigma^1_a}^{\rm even} + (-)^{\tau^1_a} \Pi_{13+2 \sigma^1_a}^{\rm even} \right].
\end{array}
\end{equation}
This relation explicitly shows that the fractional three-cycles parallel to the $\OR\Z_2^{(1)}$-plane can be written as linear combinations of $\Pi_0^{\rm even}$, $\Pi_{12 + 2 \sigma_a^1}^{\rm even}$, $\Pi_{13+2 \sigma_a^1}^{\rm even}$, and  the fractional three-cycles parallel to the $\OR$-plane, by appropriately choosing the discrete Wilson lines $(\vec{\tau}_a)$ and the $\Z_2^{(k)}$ eigenvalues. Hence, the fractional three-cycles parallel to the $\OR\Z_2^{(1)}$-plane do not lead to independent K-theory constraints. Applying such reasonings we can reduce the number of K-theory constraints to the maximally possible number of 16 linearly independent constraints, associated to fractional three-cycles parallel to the $\OR$-plane or $\OR\Z_2^{(3)}$-plane and linear combinations thereof:
\begin{equation}\label{Eq:KTheoryZ2Z6}
\sum_a N_a 
\left(\begin{array}{c}
\frac{U_a + y_{5,a}^{(1)} - x_{1,a}^{(2)} -  x_{2,a}^{(2)} + x_{1,a}^{(3)} + x_{2,a}^{(3)} }{2}
\\  2 \, x_{1,a}^{(3)} 
\\- x_{1,a}^{(2)} + x_{1,a}^{(3)}
\\ 2\,x_{2,a}^{(3)}
\\ -x_{2,a}^{(2)} + x_{2,a}^{(3)}
\\y_{5,a}^{(1)} + x_{1,a}^{(3)} + x_{2,a}^{(3)}
\\\hline
\frac{U_a+ y_{5,a}^{(1)}  - x_{3,a}^{(2)} -  x_{4,a}^{(2)} + x_{3,a}^{(3)} + x_{4,a}^{(3)} }{2}
\\ 2\, x_{4,a}^{(3)} 
\\ -  x_{4,a}^{(2)} + x_{4,a}^{(3)} 
\\y_{5,a}^{(1)}+ x_{3,a}^{(3)} + x_{4,a}^{(3)}
\\\hline
24-\frac{3 P_a}{2}  + \frac{3 \,x_{0,a}^{(1)} + x_{1,a}^{(1)} + x_{2,a}^{(1)} + x_{3,a}^{(1)} }{4} - \frac{x_{2,a}^{(2)} +x_{3,a}^{(2)} }{2} + \frac{x_{2,a}^{(3)} +x_{3,a}^{(3)} }{2}\\
 \frac{3 \, x_{0,a}^{(1)} +  x_{2,a}^{(1)}}{2} \\
 \frac{x_{1,a}^{(1)} +  x_{2,a}^{(1)}}{2}\\
x_{3,a}^{(1)}\\
 \frac{x_{1,a}^{(1)} +  x_{3,a}^{(1)}}{2} + x_{2,a}^{(3)} +x_{3,a}^{(3)}\\
24-\frac{3 P_a}{2} + \frac{x_{1,a}^{(1)} +x_{5,a}^{(1)} }{2}- \frac{x_{2,a}^{(2)} +x_{3,a}^{(2)} }{2} - \frac{x_{2,a}^{(3)} +x_{3,a}^{(3)} }{2}
\end{array}\right) \stackrel{!}{=} 0 \text{ mod }2 . 
\end{equation}
Note that we have already used the RR tadpole cancellation conditions here to simplify the K-theory constraints, and that three conditions, namely in rows 2, 4 and 8, are now trivially satisfied.

%%%%%%%%%%%%%%%%%%%%%%%%%%%%%%%%
\subsubsection{Massless Abelian Symmetries and Discrete Gauge Symmetries}

Even though the RR tadpole cancellation conditions often do not suffice to guarantee the global consistency of Type IIA orientifold compactifications with intersecting D6-branes, they do guarantee the absence of non-Abelian gauge anomalies in the effective four-dimensional field theory. Mixed Abelian/non-Abelian as well as purely Abelian gauge anomalies vanish due to the generalised Green-Schwarz mechanism. In this process, some $U(1)$ gauge symmetry acquires a mass through St\"uckelberg couplings to closed string axions. More concretely, the dimensional reduction of the ten-dimensional RR-forms $C_{(3)}$ and $C_{(5)}$ along the basis of $\OR$-even and $\OR$-odd three-cycles,
\begin{equation}
\phi_i \equiv \frac{1}{\ell_s^3} \int_{\Pi_i^{\rm even}} C_{(3)}, \qquad  B_{(2)}^i \equiv  \frac{1}{\ell_s^5} \int_{\Pi_i^{\rm odd}} C_{(5)} 
\qquad 
\text{with} 
\quad
\ell_s \equiv 2 \pi \sqrt{\alpha'},
\end{equation}     
provides a set of closed string axions $\phi_i$ and their Hodge-dual two-forms $B_{(2)}^i$ in four dimensions, with $i \in \{0, 1, \ldots, h_{21}\}$. The reduction of the Chern-Simons action for the stack of D$6_a$-branes provides a set of St\"uckelberg couplings to the $U(1)_a \subset U(N_a)$ gauge group with field strength $F_a$ and a set of couplings
to the $SU(N_a) \times U(1)_a$ field strengths $G_a$ involving the closed string axions: 
\begin{equation}\label{Eq:CSD6Reduction}
{\cal S}_{D6-brane}^{CS} \supset \sum_a N_a \sum_{i=0}^{h_{21}} s_a^i \int_{\R^{1,3}} B_{(2)}^i \wedge F_a + \frac{1}{4\pi} \sum_a \sum_{i=0}^{h_{21}} r_a^i \int_{\R^{1,3}} \phi_i \Tr(G_a \wedge G_a).  
\end{equation} 
These two types of terms combined provide the Green-Schwarz couplings necessary to cancel the mixed gauge anomalies of the purely Abelian type $U(1)_a-U(1)_b^2$ and the Abelian/non-Abelian type $U(1)_a-SU(N_b)^2$. The (rational) wrapping numbers $r_a^i$ and $s_a^i$ follow from decomposing the fractional three-cycle $\Pi_a$ with respect to the basis of $\OR$-even and $\OR$-odd three-cycles:
\begin{equation}\label{Eq:Frac3CycleExpansion}
\Pi_a^{\rm frac} = \sum_{i=0}^{h_{21}} \left( r_a^i \Pi_i^{\rm even} + s_a^i \Pi_i^{\rm odd}  \right).
\end{equation}
A linear combination of $U(1)$'s, say $U(1)_X = \sum_a q_a U(1)_a$ with $q_a \in 
\Q $, remains as massless anomaly-free $U(1)$ gauge symmetry if all its associated St\"uckelberg couplings in equation (\ref{Eq:CSD6Reduction}) vanish. The vanishing of the St\"uckelberg couplings can be rewritten in terms of the following set of topological conditions: 
\begin{equation}\label{Eq:MasslessU(1)}
\Pi_i^{\rm even} \circ \left( \sum_a q_a N_a \Pi_a \right) = 0 \qquad \forall i \in \{0,\ldots, h_{21}\}.
\end{equation} 
Massive Abelian $U(1)$ symmetries obviously do not satisfy this condition. Instead they couple to (some linear combination of) a closed string axion $\phi_i$ and acquire mass through the St\"uckelberg mechanism. At energies below the St\"uckelberg mass scale, these $U(1)$ symmetries behave as perturbative global symmetries that are broken further to discrete $\Z_n$ symmetries~\cite{BerasaluceGonzalez:2011wy} by non-perturbative corrections. The existence conditions for discrete $\Z_n$ symmetries can also be written through a set of topological conditions:
\begin{equation}\label{Eq:DiscreteSymmetriesGeneral}
\Pi_i^{\rm even} \circ \left( \sum_a k_a N_a \Pi_a \right) = 0 \, \text{ mod } n \quad \forall i \quad \text{ with } \Z_n \subset \sum_a k_a U(1)_a.
\end{equation}
In order to unambiguously determine the correct value of $n$, the coefficients $k_a \in \Z$ are chosen such that they lie within the interval $0\leq k_a < n$ and satisfy the condition gcd$(k_a, k_b, \ldots, n)=1$. In case all the coefficients satisfy $k_a \equiv 1\, (\forall a)$, we reproduce the K-theory constraint equations in (\ref{Eq:KtheoryConstraints}), which in turn imply the existence of a discrete $\Z_2$ symmetry.  Note that the interpretation of the K-theory constraint as $\Z_2$ symmetry is only valid if the full lattice of $\OR$-even three-cycles can be spanned by cycles, which support $USp(2)$ gauge factors (and not $SO(2)$ gauge groups) as in the present situation.

In order to clarify the conditions (\ref{Eq:DiscreteSymmetriesGeneral}) on the existence of discrete $\Z_n$ symmetries in the low-energy effective field theory, we work them out explicitly for the orientifold \mbox{$T^6/(\Z_2\times \Z_6 \times \OR)$} with discrete torsion $\eta=-1$. Anticipating the results of our search for global MSSM and left-right symmetric models, we restrict our discussion to the {\bf aAA} lattice configuration with the $\OR\Z_2^{(3)}$-plane as the exotic O6-plane.  For this configuration, the basis of $\OR$-even and $\OR$-odd three-cycles of pure bulk/exceptional type
in table~\ref{tab:ORevenoddlattice} does not form a uni-modular lattice, given that it satisfies the relations:
\begin{equation} \label{Eq:ORevenoddIntersectionForm}
\Pi_i^{\rm even} \circ \Pi_j^{\rm odd} = c_i \delta_{ij}, \qquad \text{ with }\, c_i = \left\{\begin{array}{cc} 8 & i=0,7,12,13,14,15,\\
-8 & i = 6,8,9,10,11, \\
-24 & i = 1, \\
12& i=2, \\
4 & i=3,4,5. \end{array}\right. 
\end{equation}
As an immediate consequence, the wrapping numbers $r_a^i$ and $s_a^i$ for the fractional three-cycles (\ref{Eq:Frac3CycleExpansion}) on this lattice are therefore rational numbers taking value in $\frac{1}{8} \Z$ in agreement with the general form of the expansion displayed in equation~\eqref{Eq:Z2Z6FractCycles}.
 It also implies that the discrete $\Z_n$ symmetry conditions obtained from (\ref{Eq:DiscreteSymmetriesGeneral}) by using the basic purely bulk or exceptional three-cycles $\Pi_i^{\rm even}$ 
 in table~\ref{tab:ORevenoddlattice} do not provide for all constraints, but rather only provide for a set of sixteen {\bf necessary} conditions~\cite{Honecker:2013hda,Honecker:2013kda,Honecker:2015ela}. 
 Said differently, the cycles $\Pi_i^{\rm even}$ of table~\ref{tab:ORevenoddlattice} only form a sublattice of the full lattice of $\OR$-even three-cycles, as clearly suggested by the structure of their intersection form in (\ref{Eq:ORevenoddIntersectionForm}). The {\bf necessary} conditions on the existence of discrete $\Z_n$ symmetries can be written out in terms of bulk and exceptional wrapping numbers as follows:
\begin{equation}\label{Eq:Zn-condition-nec}
\sum_a k_a \, N_a \, \left(\begin{array}{c} 2U_a + V_a \\ 3 \, Q_a \\\hline 3 x_{0,a}^{(1)} \\ x_{1,a}^{(1)} \\ x_{2,a}^{(1)} \\ x_{3,a}^{(1)} \\ x^{(1)}_{4,a} + x^{(1)}_{5,a}
\\ - [y^{(1)}_{4,a} - y^{(1)}_{5,a}]
\end{array}\right)
\stackrel{!}{=} 0 \text{ mod } n \stackrel{!}{=}
\sum_a k_a \, N_a \, \left(\begin{array}{c}  - ( 2 \, x^{(2)}_{1, a} +  y ^{(2)}_{1, a}) \\  - ( 2 \, x^{(2)}_{2, a} +  y ^{(2)}_{2, a}) \\ 
 - ( 2 \, x^{(2)}_{3, a} +  y ^{(2)}_{3, a}) \\  - ( 2 \, x^{(2)}_{4, a} +  y ^{(2)}_{4, a}) \\\hline
 -y ^{(3)}_{1, a}  \\  -y ^{(3)}_{2, a} \\  -y ^{(3)}_{3, a} \\ -y ^{(3)}_{4, a}  
\end{array}\right).
\end{equation}
These conditions have to be supplemented with a set of {\bf sufficient} conditions which derive from (\ref{Eq:DiscreteSymmetriesGeneral}) by taking the set of linearly independent $\OR$-even fractional three-cycles supporting an enhanced $USp(2N)$ or $SO(2N)$ gauge group. 
For the {\bf aAA} lattice configuration with $\eta_{\OR\Z_2^{(3)}}=-1$, fractional three-cycles parallel to the $\OR\Z_2^{(2)}$-plane support an enhanced $SO(2N)$ gauge group for the choice of the discrete displacements $(\vec{\sigma}_a)=(\sigma_a^1,1,1)$ and the discrete Wilson lines $(\vec{\tau}_a)=(\tau_a^1,1,1)$, see table~\ref{Tab:Conditions-on_b+t+s-SOSp}. The corresponding three-cycles can be written down in terms of the $\OR$-even basis of table~\ref{tab:ORevenoddlattice}:
 \begin{equation}
 \begin{array}{rcl}
\Pi^{{\rm frac},(\sigma_a^1,1,1)}_{a\pp \OR\Z_2^{(2)}}&=& \frac{1}{4} \Pi_1^{\rm even} +\frac{(-)^{\tau_a^{\Z_2^{(1)}}}}{4} \left[ -2 \Pi_5^{\rm even} + \Pi_7^{\rm even}  \right]
+\frac{(-)^{\tau_a^{\Z_2^{(2)}}}}{4} \left[ \Pi^{\rm even}_{8+\sigma_a^1} + (-)^{\tau_a^1}  \Pi^{\rm even}_{11-  \sigma_a^1} \right] \\
&&-3\frac{(-)^{\tau_a^{\Z_2^{(1)}}+\tau_a^{\Z_2^{(2)}}}}{4} \left[ \Pi^{\rm even}_{12+ \sigma_a^1} + (-)^{\tau_a^1}  \Pi^{\rm even}_{15- \sigma_a^1} \right]  .
\end{array}
 \end{equation}
 Writing out the fractional three-cycles parallel to the $\OR\Z_3^{(3)}$-plane and supporting an enhanced $USp(2N)$ gauge group for the choice of discrete parameters $(\vec{\sigma}_a)=(\sigma_a^1,1,1)$ and $(\vec{\tau_a})=(\tau_a^1,0,0)$:
   \begin{equation}
 \begin{array}{rcl}
\Pi^{{\rm frac},(\sigma_a^1,1,1)}_{a\pp \OR\Z_2^{(3)}}&=& \frac{1}{4} \Pi_1^{\rm even} +\frac{(-)^{\tau_a^{\Z_2^{(1)}}}}{4} \left[ -2 \Pi_5^{\rm even} + \Pi_7^{\rm even}  \right]
-\frac{(-)^{\tau_a^{\Z_2^{(2)}}}}{4} \left[ \Pi^{\rm even}_{8+\sigma_a^1} + (-)^{\tau_a^1}  \Pi^{\rm even}_{11-  \sigma_a^1} \right] \\
&&-\frac{(-)^{\tau_a^{\Z_2^{(1)}}+\tau_a^{\Z_2^{(2)}}}}{4} \left[ \Pi^{\rm even}_{12+ \sigma_a^1} + (-)^{\tau_a^1}  \Pi^{\rm even}_{15- \sigma_a^1} \right],
\end{array}
 \end{equation}
 allows us to deduce the following relation among the fractional three-cycles:
 \begin{equation}
 \begin{array}{l}
\Pi^{{\rm frac},(\sigma_a^1,1,1)}_{a\pp \OR\Z_2^{(2)}} [\tau_a^{\Z_2^{(1)}},\tau_a^{\Z_2^{(2)}}+1,\tau_a^{\Z_2^{(3)}}+1] \\= \Pi^{{\rm frac},(\sigma_a^1,1,1)}_{a\pp \OR\Z_2^{(3)}} [\tau_a^{\Z_2^{(1)}},\tau_a^{\Z_2^{(2)}},\tau_a^{\Z_2^{(3)}}]  +  (-)^{\tau_a^{\Z_2^{(1)}}+\tau_a^{\Z_2^{(2)}}}  \left[ \Pi^{\rm even}_{12+ \sigma_a^1} + (-)^{\tau_a^1}  \Pi^{\rm even}_{15- \sigma_a^1} \right].   
\end{array}
  \end{equation}
This relation implies that the fractional three-cycles with an enhanced $SO(2N)$ gauge group do not provide for additional conditions, and the sixteen linearly independent fractional three-cycles found when deriving the K-theory constraints suffice - as expected - to derive the {\bf sufficient} conditions on the existence of some  $\Z_n$ gauge  symmetry:
\begin{equation}\label{Eq:Zn-condition-suf}
\sum_a N_a \, k_a \left(
\hspace{-3mm}
\begin{array}{c}
\frac{2U_a + V_a - [y^{(1)}_{4,a} - y^{(1)}_{5,a}] -( 2 \, x^{(2)}_{1, a} +  y ^{(2)}_{1, a}) -( 2 \, x^{(2)}_{2, a} +  y ^{(2)}_{2, a})  -y_{1,a}^{(3)}-y_{2,a}^{(3)}}{4} 
\\
-y_{1,a}^{(3)}\\
- \frac{( 2 \, x^{(2)}_{1, a} +  y ^{(2)}_{1, a}) + y_{1,a}^{(3)}}{2} \\
-y_{2,a}^{(3)}\\
- \frac{( 2 \, x^{(2)}_{2, a} +  y ^{(2)}_{2, a}) +y_{2,a}^{(3)}}{2} \\
- \frac{[y^{(1)}_{4,a} - y^{(1)}_{5,a}]  + y_{1,a}^{(3)} + y_{2,a}^{(3)}}{2} \\\hline 
\frac{2U_a + V_a - [y^{(1)}_{4,a} - y^{(1)}_{5,a}] -( 2 \, x^{(2)}_{3, a} +  y ^{(2)}_{3, a}) -( 2 \, x^{(2)}_{4, a} +  y ^{(2)}_{4, a}) -y_{3,a}^{(3)} - y_{4,a}^{(3)}}{4} \\
- y_{4,a}^{(3)}\\ 
- \frac{( 2 \, x^{(2)}_{4, a} +  y ^{(2)}_{4, a}) +  y_{4,a}^{(3)} }{2} \\ 
- \frac{[y^{(1)}_{4,a} - y^{(1)}_{5,a}] + y_{3,a}^{(3)} + y_{4,a}^{(3)}}{2} \\\hline 
\frac{3 Q_a + [3 x^{(1)}_{0,a} + x^{(1)}_{1,a} + x^{(1)}_{2,a} + x^{(1)}_{3,a}]  -( 2 \, x^{(2)}_{2, a} +  y ^{(2)}_{2, a})  -( 2 \, x^{(2)}_{3, a} +  y ^{(2)}_{3, a})-y_{2,a}^{(3)} - y_{3,a}^{(3)}}{4}\\ 
\frac{3 x^{(1)}_{0,a} + x^{(1)}_{2,a}}{2} \\
\frac{x^{(1)}_{1,a} + x^{(1)}_{2,a}}{2} \\
x_{3,a}^{(1)}\\
 \frac{x^{(1)}_{1,a} + x^{(1)}_{3,a} -y_{2,a}^{(3)} - y_{3,a}^{(3)}}{2}\\ 
\frac{3 Q_a + 2 x^{(1)}_{1,a} + [x^{(1)}_{4,a}+x^{(1)}_{5,a}]  - ( 2 \, x^{(2)}_{2, a} +  y ^{(2)}_{2, a})  -( 2 \, x^{(2)}_{3, a} +  y ^{(2)}_{3, a}) + y_{2,a}^{(3)} + y_{3,a}^{(3)} }{4}
\end{array}
\hspace{-3mm}
\right) \stackrel{!}{=} 0 \text{ mod } n .
\end{equation}
Each entry in the necessary and sufficient conditions corresponds to an intersection number with some $\OR$-even three-cycle and is thus integer-valued. 
Ultimately, of course only $h_{21}^{\text{bulk}+\Z_2}+1=16$ conditions on the existence of $\Z_n$ symmetries are independent, but for practical purposes it is usually convenient to first check the simpler set of {\bf necessary} conditions and then refine the search by verifying which of the candidate $n$ also obey the {\bf sufficient} conditions.

Massless Abelian gauge symmetries correspond to those choices of $(k_a,k_b,\ldots)$, for which the entries in each line of equations~\eqref{Eq:Zn-condition-nec} and~\eqref{Eq:Zn-condition-suf} add up to exactly zero (without `mod $n$'). We will clarify these considerations through the explicit examples in sections~\ref{Ss:MSSMDiscrete} and~\ref{Ss:LRSDiscrete}. 

As shown in~\cite{BerasaluceGonzalez:2011wy,Honecker:2013hda}, discrete $\Z_n$ symmetries are left unbroken by the non-perturbative effects inherent to string theory, such as Euclidean D-brane instantons. In this respect, gauged $\Z_n$ symmetries constrain (also) the form of the non-perturbative part of the four-dimensional superpotential, whereas the massive Abelian $U(1)$ symmetries only constrain the perturbative superportential. This observation matches nicely the field theoretic motivation for the existence of discrete symmetries to explain the absence of dangerous lepton/baryon-number violating operators in supersymmetric field theories~\cite{Ibanez:1991pr,Dreiner:2005rd}. Any appropriate discrete symmetry should allow for the presence of the traditional Yukawa couplings, such that a generic discrete $\Z_n$ symmetry in the MSSM with generator $g_n$ can be decomposed~\cite{Ibanez:1991pr} in terms of three independent generators ${\cal R}_n=e^{i 2 \pi {\cal R}/n}$, $ \mathfrak{L}_n = e^{i 2 \pi \mathfrak{L}/n}$ and ${\cal A}_n = e^{i 2 \pi {\cal A}/n}$:
\begin{equation}\label{Eq:GenRALMSSM}
g_n = {\cal R}_n^m \cdot {\cal A}_n^k \cdot {\mathfrak{L}_n^p} , \hspace{0.4in}  m, k, p = 0, 1, \ldots, n-1, 
\end{equation}
under which the MSSM states are charged as follows,
\begin{eqnarray}
\begin{array}{l@{\hspace{0.4in}}l@{\hspace{0.4in}}l}
\alpha_{Q_L} = 0, & \alpha_{u_R} = -m, & \alpha_{d_R} = m-k,\\
\alpha_L = -k-p, & \alpha_{e_R} = m + p & \alpha_{\nu_R} = -m + k + p, \\
\alpha_{H_u} = m, & \alpha_{H_d}= - m + k. &
\end{array} \label{Eq:MSSMcharges}
\end{eqnarray}
The charges of the MSSM fields are chosen such that the standard Yukawa couplings are allowed by the discrete $\Z_n$ symmetry generated by $g_n$. In section~\ref{Ss:MSSMDiscrete} we will investigate the discrete symmetries in a global five-stack intersecting D6-brane model with a MSSM-like gauge group and spectrum, and compare the discrete symmetries to the decomposition in~(\ref{Eq:GenRALMSSM}).
Taking into account the anomaly constraints concerning the discrete $\Z_n$ symmetries eliminates all but three discrete symmetries compatible with the MSSM: matter parity ${\cal R}_2$, baryon-triality ${\cal B}_3 \equiv {\cal R}_3 {\cal L}_3$ and proton-hexality ${\cal P}_6\equiv {\cal R}_6^5 {\cal L}_6^2$.

%%%%%%%%%%%%%%%%%%%%%%%%%%%%%%%%%%%%%%%%%%%%%%%%%%%%%%%%%%%%%%%%%%%%%%%%%%%%%%%%%%%
%%%%%%%%%%%%%%%%%%%%%%%%%%%%%%%%%%%%%%%%%%%%%%%%%%%%%%%%%%%%%%%%%%%%%%%%%%%%%%%%%%%
\subsection{Yukawa and Other Cubic Couplings}\label{Ss:Yukawa}
A correct identification of the massless open string states as left/right-handed quarks or leptons requires in the first place that the considered open string state transforms in the correct representation under the MSSM or left-right symmetric gauge group. Nonetheless, there exist situations in which the identification remains ambiguous for massless open string states arising from different sectors but with the same quantum numbers under the visible gauge group. A recurring example of two states whose identification is not always straightforward is the candidate left-handed leptons $L$ versus the candidate down-type Higgsinos $\tilde{H}_d$ in MSSM-like D6-brane models. Furthermore, intersecting D6-brane models also come with various massless singlet fermions under the visible gauge factor, which at first sight are all able to serve as candidate right-handed neutrinos $\nu_R$. To identify the matter on massless open string states unambiguously, we have to determine the Yukawa and other three-point couplings involving the left-handed leptons, the Higgses $H_d$ and/or the right-handed neutrinos $\nu_R$. Apart from a correct identification of the chiral spectrum, the computation of the Yukawa couplings also forms an essential litmus test to assess how close a consistent string theory model comes to real-world physics. The Yukawa couplings arising from a string compactification should for instance be able to exhibit the mass hierarchies among the different quark and lepton generations. 
 
Generally, determining the allowed Yukawa and three-point couplings consists of two steps. First of all, a three-point coupling is allowed whenever it satisfies the {\bf charge selection rule}: a set of three massless open string states $\phi_{ab}^x \in \Pi_{a}^{\rm frac} \cap \Pi_b^{\rm frac}$, $\phi_{bc}^y \in \Pi_{b}^{\rm frac} \cap \Pi_c^{\rm frac}$ and $\phi_{ca}^z \in \Pi_{c}^{\rm frac} \cap \Pi_a^{\rm frac}$ combines into a three-point coupling in the perturbative part of the superpotential ${\cal W}$,
\begin{equation}\label{Eq:3pointCoupling}
{\cal W}_{\rm per} \ni W_{xyz} \phi_{ab}^x \phi_{bc}^y \phi_{ca}^z,
\end{equation}  
provided that the total three-point coupling forms a singlet representation under the full gauge factor (including hidden gauge groups). In this expression the subscripts $a$, $b$ and $c$ refer to the fractional three-cycles $\Pi_a^{\rm frac}$, $\Pi_b^{\rm frac}$ and $\Pi_c^{\rm frac}$ of the corresponding D6-branes whose intersections provide for the massless states, while the superscripts $x, y, z$ are related to the multiplicity or generation of the respective massless state. Invariance under the full gauge group also implies invariance under massless Abelian gauge symmetries and gauged discrete $\Z_n$ symmetries. In this respect, non-trivial discrete $\Z_n$ symmetries, which are not homomorphic to 
the centre of some non-Abelian gauge factor, are able to rule out non-perturbative $m$-point couplings, analogously to their field theoretic ``raison d'\^etre'' discussed at the end of the previous section. 
 An explicit example of a non-trivial discrete $\Z_3$ symmetry is presented below in section~\ref{Ss:MSSMDiscrete} for a prototype global five-stack MSSM-like D6-brane model, which is characterised by an abundant collection of up-type Higgses $(H_u, \tilde H_u)$ and down-type Higgses $(H_d, \tilde H_d)$. The Higgs doublets $\tilde H_u$ and $\tilde H_d$ carry different $\Z_3$-charges from their untilted counterparts, from which one can immediately deduce that the Yukawa coupling $Q_L\cdot \tilde H_u d_R$ is allowed while the coupling $Q_L\cdot H_u d_R$ is forbidden according to the $\Z_3$ selection rule. Other examples of $\Z_3$-forbidden and -allowed couplings will be discussed in section~\ref{Ss:MSSMDiscrete}.

A second criterium for the existence of the three-point coupling in (\ref{Eq:3pointCoupling}) relies on the microscopic intersecting D6-brane realisation and goes under the name of {\bf stringy selection rule}: the bulk three-cycles $\Pi_a^{\rm bulk}$, $\Pi_b^{\rm bulk}$ and $\Pi_c^{\rm bulk}$ of the intersecting D6-branes  have to form a closed triangle sequence $[a,b,c]=[a,b][b,c][c,a]$ on each two-torus $T_{(i)}^2$, whose apexes correspond to the D6-brane intersections at which the massless states $\phi_{ab}^x$, $\phi_{bc}^y$ and $\phi_{ca}^z$ are located. In a more formal language~\cite{Kachru:2000ih,Cremades:2003qj}, the one-cycles of the factorisable bulk three-cycles $\Pi_a^{\rm bulk}$, $\Pi_b^{\rm bulk}$ and $\Pi_c^{\rm bulk}$ conspire to enclose a worldsheet instanton with a planar triangular shape 
ending on $\Pi_a^{\rm bulk}\cup \Pi_b^{\rm bulk} \cup \Pi_c^{\rm bulk}$ and connecting the intersections $\Pi_a^{\rm bulk}\cap \Pi_b^{\rm bulk}$, $\Pi_b^{\rm bulk}\cap \Pi_c^{\rm bulk} $ and $\Pi_a^{\rm bulk}\cap \Pi_c^{\rm bulk} $  on each two-torus $T_{(i)}^2$.
The coefficient $W_{xyz}$ is then related to the area sum of the triangles enclosed by the three intersecting bulk three-cycles $\Pi_a^{\rm bulk}$, $\Pi_b^{\rm bulk}$ and $\Pi_c^{\rm bulk}$:
\begin{equation}\label{Eq:AmplitudeT6}
{W}_{xyz} \simeq e^{- \sum_{i=1}^3{\cal A}_{xyz}^{(i)}/(2\pi \alpha')},
\end{equation}
where ${\cal A}_{xyz}^{(i)}$ represents the area of the closed triangle $[a,b,c]$ on the two-torus $T_{(i)}^2$. In case the three-cycles $a$, $b$ and $c$ intersect in a single point on a two-torus, the corresponding contribution to the amplitude ${W}_{xyz}$ is of the order ${\cal O}(1)$. When the three-cycles $a$, $b$ and  $c$ do not form a closed sequence (on at least one of the three two-tori), the coefficient ${W}_{xyz}$ vanishes. 
Notice that while charge selection and stringy selection coincide on the mere six-torus, for orbifolds the stringy selection rule plays a vital rule due to the existence of orbifold image cycles $(\omega^k a)$ on the underlying torus.
In all cases with a closed triangle of non-vanishing size, the amplitude ${W}_{xyz}$ is exponentially suppressed by the area ${\cal A}_{xyz}^{(i)}$ which scales with the K\"ahler modulus $v_i$ measuring the area of the two-torus $T_{(i)}^2$. The expression in (\ref{Eq:AmplitudeT6}) corresponds to the worldsheet instanton at leading order, 
and an infinite set of copies with larger areas will refine the size of the coupling~\cite{Cremades:2003qj}.

A first consideration to take into account is that the form of the amplitude in (\ref{Eq:AmplitudeT6}) is valid for the ambient space $T^6$, 
 thus neglecting a potential overall numerical factor $1/(2\cdot 6)$ accounting for the $\Z_2 \times \Z_6$ orbifold geometry.
In this respect, expression (\ref{Eq:AmplitudeT6}) should be considered as a reasonable order of magnitude for the three-point coupling, such that it allows for instance to identify hierarchies among the Yukawa couplings in intersecting D6-brane models on $T^6/(\Z_2\times \Z_6 \times \OR)$. In the absence of cubic couplings, one can conceive perturbative non-renormalisable higher $m$-point couplings which are string mass scale suppressed with the appropriate power $M_{\rm string}^{3-m}$ and where the worldsheet instanton takes the shape of an $m$-polygon, in the same spirit as the construction for the cubic couplings outlined above.
Next, we also point out that the expressions in (\ref{Eq:3pointCoupling}) and (\ref{Eq:AmplitudeT6}) only contain the classical part of the coupling. The quantum contribution to the Yukawa coupling takes into account the proper normalisation of the matter fields $\phi_{ab}^x$, $\phi_{bc}^y$ and $\phi_{ca}^z$ and can be deduced by computing four-point scattering amplitudes involving the matter fields as external legs~\cite{Abel:2003yx,Cvetic:2003ch,Lust:2004cx}. The normalisation of a matter field is in principle proportional to its K\"ahler metric upon dimensional reduction to four dimensions, and the K\"ahler metrics can contribute to establishing mass hierarchies among the different particle generations~\cite{Honecker:2012jd,Honecker:2012fn}. 
The K\"ahler metrics for the matter fields on the orbifold $T^6/(\Z_2\times \Z_6 \times \OR)$ can also be deduced from the one-loop computation of the running gauge couplings~\cite{Blumenhagen:2007ip,Honecker:2011sm,Honecker:2011hm}, offering an alternative (and often simpler) method to obtain the proper normalisation of the matter fields. We end our list of considerations with a specific feature of the toroidal orientifold $T^6/(\Z_2\times \Z_6\times \OR)$ regarding various three-point couplings: The invariance of the first two-torus $T_{(1)}^2$ under the $\Z_6$ orbifold action in equation (\ref{Eq:1-cycle-orbits}) indicates that a bulk three-cycle $a$ will have orbifold images $(\omega a)$ and $(\omega^2 a)$ parallel to the original bulk three-cycle on $T_{(1)}^2$. This immediately implies that three bulk three-cycles $a$, $b$ and $c$ with identical torus wrapping numbers $(n^1,m^1)$ along $T_{(1)}^2$ will have a vanishing three-point coupling on the ambient space $T^6$, which might possibly be subsequently generalised to three-point couplings involving their orbifold images. For such cases we will nevertheless compute the (classical) contributions to the amplitude associated to the ambient space $T^2_{(2)} \times T^2_{(3)}$. 
This approach is motivated by the fact that the vanishing of the Yukawa coupling on just the six-torus is related to the extended ${\cal N}=2$ supersymmetry if one angle vanishes, while on $T^6/(\Z_2 \times \Z_{2M} \times \OR)$ only ${\cal N}=1$ supersymmetry is preserved due to the $\Z_2$ symmetries, and $m$-point couplings on such orbifolds containing $\Z_2$ symmetries have to our best knowledge not been computed so far - in particular the option to have a non-vanishing classical contribution remains.
To clarify some of the points discussed in this section, we will compute various Yukawa and other cubic couplings for the global five-stack MSSM-like D6-brane model in section~\ref{Ss:MSSMYukawa} and for the global six-stack left-right symmetric  D6-brane models in section~\ref{Ss:LRSMYukawa}.

%%%%%%%%%%%%%%%%%%%%%%%%%%%%%%%%%%%%%%%%%%%%%%%%%%%%%%%%%%%%%%%%%%%%%%%%%%%%%%%%%%%
%%%%%%%%%%%%%%%%%%%%%%%%%%%%%%%%%%%%%%%%%%%%%%%%%%%%%%%%%%%%%%%%%%%%%%%%%%%%%%%%%%%
%%%%%%%%%%%%%%%%%%%%%%%%%%%%%%%%%%%%%%%%%%%%%%%%%%%%%%%%%%%%%%%%%%%%%%%%%%%%%%%%%%%
%%%%%%%%%%%%%%%%%%%%%%%%%%%%%%%%%%%%%%%%%%%%%%%%%%%%%%%%%%%%%%%%%%%%%%%%%%%%%%%%%%%
\section{Phenomenology of Global MSSM-like Models}\label{S:MSSM-pheno}

%%%%%%%%%%%%%%%%%%%%%%%%%%%%%%%%%%%%%%%%%%%%%%%%%%%%%%%%%%%%%%%%%%%%%%%%%%%%%%%%%%%
%%%%%%%%%%%%%%%%%%%%%%%%%%%%%%%%%%%%%%%%%%%%%%%%%%%%%%%%%%%%%%%%%%%%%%%%%%%%%%%%%%%
\subsection{Searching for MSSM-like D6-Brane Models}\label{Ss:MSSM-models}

%%%%%%%%%%%%%%%%%%%%%%%%%%%%%%%%%%%%%%%%%%%%%%%%%%%%%%%%%%%%%%%%%%%%%%%%%%%%%%%%%%%
\subsubsection{Global $\varrho$-independent configurations}
As shown in~\cite{Ecker:2014hma} and reviewed in section~\ref{Ss:IntersectSummary}, $\varrho$-independent D6-brane configurations yielding three left-handed quark generations 
without excessive exotic matter can only be realised for the following bulk three-cycles:
\begin{equation}\label{Eq:D6ConfQCDLeftMSSM}
\begin{array}{llll}
QCD:& a \pp &\OR: (\frac{1}{1-b},\frac{-b}{1-b};1,0;1,0)& N_a = 3 \text{ without enhancement},\\
SU(2)_L:& b \pp &\OR\Z_{2}^{(1)}: (\frac{1}{1-b},\frac{-b}{1-b};-1,2;1,-2) & N_b = 1 \text{ with $USp(2)$ enhancement},
\end{array}
\end{equation}
provided that either the $\OR\Z_2^{(2)}$- or $\OR\Z_2^{(3)}$-plane plays the r\^ole of the exotic O6-plane ($\eta_{\OR\Z_{2}^{(2 \text{ or } 3)}}=-1$). In a next step, we complete the MSSM gauge groups and chiral spectrum by embedding additional $U(1)$ gauge factors on fractional three-cycles that are supersymmetric for all values of the complex structure modulus $\varrho$.\footnote{It was argued in~\cite{Ecker:2014hma}, based on the bulk RR tadpole conditions in table~\ref{tab:Bulk-RR+SUSY-Z2Z6} with $\eta_{\OR\Z_2^{(2 \text{ or } 3)}}=-1$, that the supersymmetric D6-brane configurations in (\ref{Eq:D6ConfQCDLeftMSSM}) can only be completed consistently using fractional D6-branes with bulk wrapping number $V + b Q = 0$. From the classification of supersymmetric three-cycles in appendix A of~\cite{Ecker:2014hma}, one can then deduce the four candidate bulk orbits listed in table~\ref{tab:BulkCyclesV=0abAA}, which happen to be supersymmetric for all values of the complex structure modulus $\varrho$. 
These considerations thus exclude {\it ab initio} the possibility to use $\varrho$-dependent supersymmetric fractional three-cycles to account for missing $U(1)$ gauge factors when completing the MSSM gauge group. Also any potential hidden sector will consist of $\varrho$-independent D6-branes.}
The three generations of right-handed quarks and left-handed leptons then ought to be realised at the intersections between these $U(1)$ D6-brane stacks, the {\it QCD} stack and the $SU(2)_L$ stack, according to table~\ref{tab:OverTopologIntersections34Stack} for three-stack and four-stack D6-brane models. 
%%%%%%%%%%%%%%%%
\mathtab{
\begin{array}{|c||c|c||c|c|}
\hline \multicolumn{5}{|c|}{\text{\bf Overview of topological intersection \# for chiral MSSM spectrum}} \\
\hline
\hline
&\multicolumn{2}{|c||}{U(3)_a\times USp(2)_b\times U(1)_c} & \multicolumn{2}{|c|}{U(3)_a\times USp(2)_b\times U(1)_c\times U(1)_d} \\
\hline
\text{ state }  & \text{sector} & \text{chirality} &  \text{sector} & \text{chirality}  \\
\hline
Q_L & ab & \chi^{ab} = \chi^{ab'} = \pm 3   & ab & \chi^{ab} = \chi^{ab'} = \pm 3    \\
d_R & ac & \chi^{ac}=\mp3 &  ac + ad & \chi^{ac} + \chi^{ad} =\mp 3 \\
u_R & ac' & \chi^{ac'}=\mp3 & ac' + ad' & \chi^{ac'} + \chi^{ad'} = \mp 3\\
L & bc& \chi^{bc} =\pm 3 & bc + bd &  \chi^{bc} +  \chi^{bd}  =\pm 3 \\
e_R & cc' &  \chi^{\Sym_c} = \pm 3 & cc'+ dd' + cd'  &  \chi^{\Sym_c} +  \chi^{\Sym_d}+  \chi^{cd'}  = \pm  3 \\
\hline
\end{array}
}{OverTopologIntersections34Stack}{Topological intersection numbers of a three generation chiral MSSM-like spectrum in compliance with the D6-brane configuration in eq. (\ref{Eq:D6ConfQCDLeftMSSM}) for the {\it QCD} and $SU(2)_L$ stacks and with the hypercharge prescription $x_c=x_d=1$ in eq. (\ref{Eq:HyperChargePrescr34Stacks}). 
The upper signs in the net-chiralities correspond to the convention where a positive net-chirality $\chi^{ab}>0$ gives rise to  chiral states in the bifundamental representation $(N_a, \ov N_b)$, the lower signs correspond to the opposite convention. 
In principle, right-handed $d_R$ quarks can also be realised as chiral states in the antisymmetric representation of the $U(3)_a$ gauge factor. However, for the particular choice in eq.~(\ref{Eq:D6ConfQCDLeftMSSM}) with respect to the {\it QCD} stack such chiral states in the antisymmetric representation are not present. Right-handed neutrinos $\nu_R$ can be realised through singlet states under the MSSM gauge group, provided that the singlet states allow for the existence of the appropriate Yukawa coupling.}
%%%%%%%%%%%%%%
The explicit construction of the chiral MSSM-like spectrum with three-stack and four-stack D6-brane models is further constrained~\cite{Gmeiner:2005vz,Cvetic:2010mm} by the realisation of the $U(1)_Y$ hypercharge as a linear combination of the $U(1)$ gauge groups: 
\begin{equation}\label{Eq:HyperChargePrescr34Stacks}
\text{3-stack: } Q_Y = \frac{1}{6} Q_a +  \frac{x_c}{2} Q_c, \qquad \text{4-stack: } Q_Y = \frac{1}{6} Q_a +  \frac{x_c}{2} Q_c +  \frac{x_d}{2} Q_d,
\end{equation} 
where $x_c, x_d \in \{\pm1\}$. 
In first instance, one notices that - if at all - only the relative sign between the charges  $Q_c$ and $Q_d$ might provide distinguishable physical situations under the assumed D6-brane set-up in equation~(\ref{Eq:D6ConfQCDLeftMSSM}), for which none of the right handed $d_R$ quarks are realised as chiral states in the antisymmetric representation of the {\it QCD} gauge group.\footnote{More exotic expressions~\cite{Cvetic:2010mm} for the hypercharge, such as $Q_Y = \frac{1}{6} Q_a +  \frac{1}{2} Q_c \pm \frac{3}{2} Q_d $, are also excluded based on the consideration that here the $d_R$ quarks cannot be realised through chiral states in the antisymmetric representation located in the $aa'$ sector of the {\it QCD} stack.} For the three-stack models we can pick $x_c=1$, as the other sign choice reproduces the same chiral spectrum upon exchanging $c\leftrightarrow c'$. Hence, by including the orientifold images of all fractional three-cycles in the set of candidate $c$-stacks to complete the three-stack MSSM-like model, we cover both choices for $x_c$. 
Similarly, the choices $(x_c,x_d)= - (1,\pm1)$ are equivalent to the choices $(x_c,x_d)=(1,\pm1)$ upon exchanging $(c,d)\leftrightarrow (c',d')$ in the four-stack set-up, such that there remain at most two distinguishable situations to consider: $(x_c,x_d)=(1,1)$ and $(x_c,x_d)=(1,-1)$. 
Then again, flipping relative sign among the $U(1)_c$ and $U(1)_d$ factors in (\ref{Eq:HyperChargePrescr34Stacks}), i.e.~$(Q_c + Q_d) \leftrightarrow (Q_c - Q_d)$, boils down to exchanging
the multiplicities $\chi^{xd} \leftrightarrow \chi^{xd'}$ for $x \in \{a,b,c\}$ in table~\ref{tab:OverTopologIntersections34Stack}.
Hence, by including the orientifold images of all fractional three-cycles supporting a $U(1)$ gauge factor in the set of candidate $c$-stacks and $d$-stacks, all possible `standard' realisations of the $U(1)_Y$ hypercharge within the initial gauge group  $U(3)_a\times USp(2)_b\times U(1)_c\times U(1)_d$ are automatically taken into account. 
The only independent choice in~(\ref{Eq:D6ConfQCDLeftMSSM}) is $(x_c,x_d)=(1,1)$, for which the phenomenological constraints for MSSM-like spectra
on the topological intersection numbers are listed in table~\ref{tab:OverTopologIntersections34Stack}. 
When identifying suitable bulk orbits, first intuition is provided by the bulk RR tadpole cancellation conditions for $\eta_{\OR\Z_2^{(2 \text{ or } 3)}}=-1$,
\begin{equation}\label{Eq:BulkRRRhoIndMSSM}
\sum_{x\in\{a,b,c,d\}} N_x (2 P_x + Q_x) \leq 32, \qquad \sum_{x\in\{a,b,c,d\}} N_x (V_x + b\, Q_x)  = 0,
\end{equation} 
which help us to exclude various options. More precisely, in order not to overshoot the bulk RR tadpole cancellation conditions, the bulk wrapping numbers have to satisfy $2P_x~+~Q_x~\leq~20\, (8)$ and $V_x + b\, Q_x = 0$ for $x\in \{c,d\}$ on the {\bf aAA} ({\bf bAA}) lattice.\footnote{Observe that the bulk wrapping numbers of supersymmetric D6-branes always satisfy the conditions $2P_x + Q_x \geqslant 0$ and $-(V_x + b Q_x) \geqslant 0$, resulting from the bulk supersymmetry conditions in table~\ref{tab:Bulk-RR+SUSY-Z2Z6} upon using the expansionss in one-cycle wrapping numbers in eq.~\eqref{Eq:Def-PQUV-via-nm}.} The bulk orbits that are supersymmetric irrespective of the $\varrho$-value and satisfy the latter constraint are listed in table~\ref{tab:BulkCyclesV=0abAA} for both lattices {\bf a/bAA}.  
%%%%%%%%%%%%%%%%%%%%
\mathtab{\hspace*{-5mm}
\begin{array}{|c|c||c|c|}
\hline \multicolumn{4}{|c|}{\text{\bf Overview of SUSY bulk three-cycles in compliance with eq.~}(\ref{Eq:BulkRRRhoIndMSSM})\,\, \forall\, \varrho }\\
\hline
\hline
\multicolumn{2}{|c||}{\text{{\bf aAA} lattice}} &\multicolumn{2}{|c|}{\text{{\bf bAA} lattice}} 
\\
\hline
\text{bulk wrapping numbers} & (2P +Q , V) & \text{bulk wrapping numbers} & (2P +Q , V+\frac{1}{2} Q)\\
\hline
\OR: (1,0;1,0;1,0) & (2,0) & \OR: (2,-1;1,0;1,0) & (4,0)\\
\OR\Z_{2}^{(1)}:(1,0;-1,2;1,-2) & (6,0) & \OR\Z_{2}^{(1)}:(2,-1;-1,2;1,-2) & (12,0)\\
(1,0;2,1;3,-1)  & (14,0)  & (2,-1;2,1;3,-1) & (28,0) \\
(1,0;4,-1;3,1)& (26,0) & (2,-1;4,-1;3,1)  & (52,0) \\
\hline
\end{array}
}{BulkCyclesV=0abAA}{The bulk wrapping numbers of $\varrho$-independent supersymmetric three-cycles on the {\bf a/bAA} lattices with $V+bQ = 0$. The last bulk orbit on the {\bf bAA} lattice does not play a r\^ole in supersymmetric model building as it overshoots the first bulk RR tadpole cancellation condition in equation~(\ref{Eq:BulkRRRhoIndMSSM}).}
%%%%%%%%%%%%%%%%%%%%%

Based on the list in table~\ref{tab:BulkCyclesV=0abAA} we can speculate which combinations of bulk orbits for the $c$-stack and the $d$-stack would allow for favourable MSSM-like configurations. We have to make sure that the choice of the bulk orbits does not lead to a violation of the first bulk RR tadpole condition in (\ref{Eq:BulkRRRhoIndMSSM}). Thus, given the implied constraint $2P_x+Q_x < 8$ for $x\in \{c,d\}$ on the {\bf bAA} lattice, this boils down to considering the three- and four-stack configurations as listed in table~\ref{tab:4stackMSSMrhoIndepbAA}. Hence, the $c$-stack and the $d$-stack can only have bulk orbits parallel to the $\OR$-plane. 
%%%%%%%%%%%%%%%%%%%%%%%%%%
\mathtab{
\begin{array}{|c|c|c||c|c||c|c|}
\hline \multicolumn{7}{|c|}{\text{\bf 3- or 4-stack combinations with gauge group } U(3)_a\times USp(2)_b \times U(1)_c \left(\times U(1)_d\right)} \\
\hline
\hline & \text{$c$-stack} & \text{$d$-stack}  & \text{RR tadpoles: } \sum_{x\in \{a,b,c,d\}} N_x (2 P_x + Q_x) & \leq 32& 3\, q_R & 3\, L  \\ 
\hline
3-\text{stack}&\OR & & 4N_a + 12 N_b + 4N_c = 28  & {\color{mygr} \checkmark} & {\color{red}\lightning} &   {\color{mygr} \checkmark}_{36}   \\
4-\text{stack}&\OR & \OR & 4N_a + 12 N_b + 4N_c + 4N_d = 32  & {\color{mygr} \checkmark} & {\color{red}\lightning} &  {\color{mygr} \checkmark}_{19008} \\
\hline
\end{array}
}{4stackMSSMrhoIndepbAA}{Combinations of supersymmetric bulk orbits for three-stack and four-stack models aiming at $\varrho$-independent configurations of the MSSM spectrum on the {\bf bAA} lattice for \mbox{$T^6/(\Z_2\times \Z_6\times \OR)$} with discrete torsion ($\eta=-1$) and exotic O6-plane $\eta_{\OR\Z_2^{(2 \text{ or } 3)}}=-1$. The second and third column indicate the bulk orbit for the $c$-stack and $d$-stack, respectively, the fourth and fifth column test whether the bulk RR tadpole cancellation conditions (\ref{Eq:BulkRRRhoIndMSSM}) are not violated, the second-to-last column verifies if three right-handed quark generations can be realised as prescribed by table~\ref{tab:OverTopologIntersections34Stack}, and the last column does the same for three left-handed lepton generations, with the subscript indicating the number of combinatorial possibilities of $(\vec{\sigma}_x)$, $(\vec{\tau}_x)$ and {\it relative} $(-)^{\Delta\tau^{\Z_2^{(k)}}_{xy}}$ for $x,y\in \{a,b,c,d\}$  for one choice $\eta_{\OR\Z_2^{(2 \text{ or } 3)}}=-1$ (both equivalent upon permutation of two-torus indices). 
}
%%%%%%%%%%%%%%%%%%%%%%

Once the bulk RR tadpole cancellation conditions are verified, we also have to check whether the intersections between the $c$-stack (and $d$-stack) and the {\it QCD} stack in the set-up of equation~(\ref{Eq:D6ConfQCDLeftMSSM}) can provide for three chiral generations of right-handed quarks $d_R$ and $u_R$, i.e. $|\chi^{ac}(+ \chi^{ad})|=3=|\chi^{ac'} (+ \chi^{ad'})|$ where the sign of the net-chirality has to be chosen opposite to the one of the net-chirality $\chi^{ab} \stackrel{USp(2)_b}{\equiv} \chi^{ab'}$, as indicated in table~\ref{tab:OverTopologIntersections34Stack}. As the corresponding fractional three-cycles for the $c$-stack (and $d$-stack) should at this point not support an enhanced $USp(2)$ gauge group, the discrete displacement parameters $(\vec{\sigma})$ and discrete Wilson lines $(\vec{\tau})$ have to be chosen accordingly for the respective fractional three-cycles. And by verifying the topological intersection numbers for all candidate fractional three-cycles on the {\bf bAA} lattice, we end up with the results in table~\ref{tab:4stackMSSMrhoIndepbAA}, from which we conclude that three-stack (and four-stack) configurations with three chiral right-handed quark generations cannot be found for the cases where the $c$-stack (and the $d$-stack) is (are) parallel to the $\OR$-plane. In summary, three-stack and four-stack intersecting D6-brane models on the {\bf bAA} lattice do not allow for $\varrho$-independent global MSSM-like models.

Next, we turn our attention to the {\bf aAA} lattice and repeat the same reasoning as above. Upon identifying which bulk orbits do not overshoot the first bulk RR tadpole cancellation condition in equation~(\ref{Eq:BulkRRRhoIndMSSM}), we can list all potential combinations of bulk orbits for the $c$-stack and the $d$-stack in table~\ref{tab:4stackMSSMrhoIndepaAA} to identify potential three-stack and four-stack configurations, with the definition of the hypercharge given in equation (\ref{Eq:HyperChargePrescr34Stacks}). 
%%%%%%%%%%%%%%%%%%%%
\mathtab{\hspace*{-14mm}
\begin{array}{|c|c|c||c|c||c|c|}
\hline \multicolumn{7}{|c|}{\text{\bf Three-stack combinations with gauge group } U(3)_a\times USp(2)_b \times U(1)_c} \\
\hline
\hline & \multicolumn{2}{|c||}{\text{$c$-stack}} & \text{RR tadpoles: } \sum_{x\in \{a,b,c\}} N_x (2 P_x + Q_x) & \leq 32 & 3\, q_R & 3\, L   \\ 
\hline
1& \multicolumn{2}{|c||}{\OR}  & 2N_a + 6 N_b + 2N_c + = 14  & {\color{mygr} \checkmark} &{\color{red}\lightning}& {\color{mygr} \checkmark}_{48}\\
2& \multicolumn{2}{|c||}{\OR\Z_2^{(1)}}  & 2N_a + 6 N_b + 6N_c  = 18  & {\color{mygr} \checkmark} &{\color{red}\lightning}& {\color{mygr} \checkmark}_{960}\\
3& \multicolumn{2}{|c||}{(1,0;2,1;3,-1)}  & 2N_a + 6 N_b + 14N_c = 26  & {\color{mygr} \checkmark} &{\color{red}\lightning}& {\color{mygr} \checkmark}_{48}\\
\hline
\hline \multicolumn{7}{|c|}{\text{\bf Four-stack combinations with gauge group } U(3)_a\times USp(2)_b \times U(1)_c \times U(1)_d} \\
\hline
\hline & \text{$c$-stack} & \text{$d$-stack}  & \text{RR tadpoles: } \sum_{x\in \{a,b,c,d\}} N_x (2 P_x + Q_x) & \leq 32 & 3\, q_R & 3\, L   \\ 
\hline
1&\OR & \OR & 2N_a + 6 N_b + 2N_c + 2N_d = 16  & {\color{mygr} \checkmark} &{\color{red}\lightning}& {\color{mygr} \checkmark}_{25344}\\
2&\OR & \OR\Z_{2}^{(1)} & 2N_a + 6 N_b + 2N_c + 6N_d = 20  &{\color{mygr} \checkmark} & {\color{red}\lightning}& {\color{mygr} \checkmark}_{157824}\\
3&\OR  &(1,0;2,1;3,-1)  & 2N_a + 6 N_b + 2 N_c + 14 N_d = 28 &{\color{mygr} \checkmark} & {\color{red}\lightning}& {\color{mygr} \checkmark}_{23760} \\
4&\OR\Z_{2}^{(1)}& \OR& 2N_a + 6 N_b + 6 N_c + 2N_d =  20 & {\color{mygr} \checkmark}&{\color{red}\lightning}& {\color{mygr} \checkmark}_{157824} \\
5&\OR\Z_{2}^{(1)}& \OR\Z_{2}^{(1)} &2N_a + 6 N_b + 6 N_c + 6 N_d = 24 &{\color{mygr} \checkmark} & {\color{mygr} \checkmark}_{1152} & {\color{mygr} \checkmark}_{\hspace{-1.5mm} 316800} (  {\color{mygr} \checkmark}_{\hspace{-1.5mm} 576}) \\
6&\OR\Z_{2}^{(1)} &(1,0;2,1;3,-1)  & 2N_a + 6 N_b + 6 N_c + 14 N_d = 32  & {\color{mygr} \checkmark}& {\color{mygr} \checkmark}_{576} &  {\color{mygr} \checkmark}_{\hspace{-1.5mm} 201024} ({\color{mygr} \checkmark}_{\hspace{-1.5mm} 144} )  \\
7&(1,0;2,1;3,-1) &\OR &2N_a + 6 N_b + 14 N_c + 2N_d= 28   &{\color{mygr} \checkmark}& {\color{red}\lightning}&  {\color{mygr} \checkmark}_{23760} \\
8&(1,0;2,1;3,-1)  & \OR\Z_{2}^{(1)}& 2N_a + 6 N_b + 14 N_c + 6 N_d = 32&{\color{mygr} \checkmark} & {\color{mygr} \checkmark}_{576} & {\color{mygr} \checkmark}_{\hspace{-1.5mm} 201024} ({\color{mygr} \checkmark}_{\hspace{-1.5mm} 144} )  \\ 
9&(1,0;2,1;3,-1)& (1,0;2,1;3,-1) &2N_a + 6 N_b + 14 N_c + 14 N_d = 40  & {\color{red}\lightning} & {\color{red}\lightning} &{\color{mygr} \checkmark}_{24768} \\
\hline
\end{array}
}{4stackMSSMrhoIndepaAA}{Combinations of supersymmetric bulk orbits for the $c$- and $d$-stacks aiming at $\varrho$-independent three-stack and four-stack D6-brane configurations of the MSSM on the {\bf aAA} lattice for $T^6/(\Z_2\times \Z_6\times \OR)$ with discrete torsion ($\eta=-1$) and exotic O6-plane $\eta_{\OR\Z_2^{(2 \text{ or } 3)}}=-1$. The second and third column indicate the bulk orbit for the $c$-stack and $d$-stack, respectively, the fourth and fifth column test whether the bulk RR tadpole conditions (\ref{Eq:BulkRRRhoIndMSSM}) are not violated, the second-to-last column verifies if three right-handed quark generations can be realised through $|\chi^{ac}(+ \chi^{ad})|=3=|\chi^{ac'}(+ \chi^{ad'})|$, and the last column does the same for three left-handed lepton generations with $|\chi^{bc}(+\chi^{bd})|=3$, 
 with the proper relative sign among the net-chiralities as dictated by table~\ref{tab:OverTopologIntersections34Stack}. 
For combinations (5,6,8), three chiral generations of both $q_R$ and  $L$ can be realised, and the compatibility between the two constraints is indicated by the symbol in parenthesis in the last column.  
The subscript indicates the number of combinatorial possibilities of $(\vec{\sigma}_x)$, $(\vec{\tau}_x)$ and {\it relative} $(-)^{\Delta\tau^{\Z_2^{(k)}}_{xy}}$ for $x,y\in \{a,b,c,d\}$ and one given choice of exotic O6-plane $\eta_{\OR\Z_2^{(2 \text{ or }3)}}=-1$. 
}
%%%%%%%%%%%%%%%%%%%%%%%%%%%%%%%%%%%
 All but one of the nine combinations comply with the bulk RR tadpole cancellation conditions, but only three combinations of four-stack D6-brane models give rise to three chiral generations of right-handed quarks $u_R$ and $d_R$. In the last column of table~\ref{tab:4stackMSSMrhoIndepaAA} we also indicate whether the three-stack and four-stack configurations yield three chiral generations of  left-handed leptons. In this way we end up with the combinations (5,6,8) for which three generation intersecting D6-brane models with chiral quarks and left-handed leptons can be constructed.

Looking further into these three combinations of table~\ref{tab:4stackMSSMrhoIndepaAA}, we find that the combinations 6 and 8 allow for D6-brane configurations with three generations of right-handed quarks and/or three generations of left-handed leptons,
and the bulk RR tadpoles are saturated by just the four stacks required to engineer the MSSM gauge group.
Note that only a fraction of the fractional D6-brane configurations represented by the combinations 6 and 8 allow for three generations of right-handed quarks and  left-handed leptons simultaneously. More explicitly, for both combinations 6 and 8 we found 576 D6-brane configurations with three generations of right-handed quarks and 201024 configurations with three generations of left-handed leptons. Yet there exist only 144 configurations where the three generations of right-handed quarks $u_R$ and $d_R$ are compatible with the three generations of left-handed leptons.\footnote{The numbers are given here for the exotic O6-plane choice $\eta_{\OR\Z_2^{(3)}} = -1$, but the same numbers are also valid for the choice $\eta_{\OR\Z_2^{(2)}} = -1$, as expected from the permutation symmetry $T^2_{(2)} \leftrightarrow T^2_{(3)}$ for the {\bf aAA} lattice.} The identical counting of models in configurations 6 and 8 agrees with the expectation that they yield physically equivalent models upon exchanging $(c,d) \leftrightarrow (d,c)$. An insurmountable obstruction to completing these D6-brane configurations into global intersecting D6-brane models, however, is the observation that bulk RR tadpoles for these cases
are always saturated, while the twisted RR tadpoles are never cancelled, regardless of the specific fractional D6-brane configuration under consideration. 
We remark that the configuration counting reflects different combinatorial possibilities for the $\Z_2^{(i)}$ eigenvalues $(-)^{\tau^{\Z_2^{(i)}}_x}$, the discrete displacements $(\vec{\sigma}_x)$ and discrete Wilson lines $(\vec{\tau}_x)$ with $x\in \{a,b,c,d\}$. Combinations of four fractional three-cycles $(a,b,c,d)$ with identical {\it relative} $\Z_2^{(i)}$ eigenvalues $\Delta \vec{\tau}^{\Z_2^{(i)}}_x$, but identical {\it absolute} displacements $(\vec{\sigma}_x)$ and Wilson lines $(\vec{\tau}_x)$ have been counted as one independent configuration only, as they all provide identical chiral and non-chiral massless spectra and field theoretical results at the current state-of-the-art, i.e. gauge couplings and K\"ahler metrics with formal expressions collected in~\cite{Honecker:2011sm}.
 Further identifications might exist for combinations 6 and 8, but due to the {\it local} character of these models, we will not pursue this issue here but only explore it further in the context of {\it global} models, where all RR tapdoles are cancelled and the K-theory constraints are satisfied.

Combination n$^\circ$ 5 on the other hand leads to a class of {\it global} five-stack MSSM-like models with initial gauge group $U(3)_a\times USp(2)_b\times U(1)_c\times U(1)_d \times U(4)_h$, with 
288 distinguishable five-stack fractional D6-brane configurations
 for the choice of the exotic O6-plane $\eta_{\OR\Z_2^{(2)}} = -1$, and as required by the permutation symmetry of  $T^2_{(2)} \leftrightarrow T^2_{(3)}$ the same number for  the choice $\eta_{\OR\Z_2^{(3)}} = -1$, as we checked explicitly.
Out of the 576 {\it local} four-stack models in table~\ref{tab:4stackMSSMrhoIndepaAA}, only 288 can account for three generations of right-handed electrons $e_R$ and satisfy all RR tadpoles for the maximal hidden gauge group $U(4)_h$. The K-theory constraints are then automatically satisfied, as we explicitly checked. 
Again, the number 288 counts fractional D6-brane configurations with different combinations of {\it relative} $\Z_2^{(i)}$ eigenvalues and {\it absolute} discrete displacements and Wilson lines. 
Thus, the 288 D6-brane configurations correspond to the maximal set of physically inequivalent D6-brane configurations. One can nevertheless show that the chiral and non-chiral massless spectra for the 288 D6-brane configurations are all identical (upon a potential exchange of $c\leftrightarrow d$ and $h\leftrightarrow h'$), suggesting a further reduction to a smaller set (maybe even a unique version) of physically inequivalent D6-brane configurations by virtue of to date unknown additional maps between non-identical relative discrete parameters.

An explicit sample of fractional D6-branes providing such a global five-stack MSSM-like model is given in table~\ref{tab:5stackMSSMaAAPrototypeI} for $\eta_{\OR\Z_2^{(3)}} = -1$, and the resulting massless spectrum is summarised in table~\ref{tab:5stackMSSMaAAPrototypeISpectrum}. In the next section we will determine the massless $U(1)$ symmetries and the discrete $\Z_n$ symmetries for this model, yet the charges under the massless hypercharge $U(1)_Y$ and the discrete $\Z_3$ symmetry are already indicated in table~\ref{tab:5stackMSSMaAAPrototypeISpectrum} for all massless states.  For later reference, we also list the charges under the massive Peccei-Quinn symmetry, $Q_{PQ} \equiv Q_c-Q_d$.
Note that the absence of a massless $U(1)_{B-L}$ symmetry slightly complicates the proper identification of the chiral MSSM states as it prevents an unambiguous distinction between the chiral states corresponding to the left-handed lepton multiplets $L$ and those corresponding to the down-type Higgs multiplets $H_d$ and $\tilde H_d$.

%%%%%%%%%%%%%%%%%%%%%%%%%%%%%%%%%%%%%%%%%%%%%%%%%%%%%%%%%%%%%%%%%%%%%%%%%%%
\mathtabfix{
\begin{array}{|c||c|c||c|c|c||c|}\hline 
\muc{7}{|c|}{\text{\bf D6-brane configuration of a global 5-stack MSSM configuration on the {aAA} lattice}}
\\\hline \hline
&\text{\bf wrapping numbers} &\frac{\rm Angle}{\pi}&\text{\bf $\Z_2^{(i)}$ eigenvalues}  & (\vec \tau) & (\vec \sigma)& \text{\bf gauge group}\\
\hline \hline
 a&(1,0;1,0;1,0)&(0,0,0)&(--+)&(0,1,1) & (0,1,1)& U(3)\\
 b&(1,0;-1,2;1,-2)&(0,\frac{1}{2},-\frac{1}{2})&(+++)&(0,1,0) & (0,1,0)&USp(2)\\
 c&(1,0;-1,2;1,-2)&(0, \frac{1}{2},-\frac{1}{2})&(+--)&(0,1,1) & (0,1,1)&U(1)\\ 
  d&(1,0;-1,2;1,-2)&(0, \frac{1}{2},-\frac{1}{2})&(-+-)&(0,0,1) & (0,0,1)& U(1)\\
  \hline
    h&(1,0;1,0;1,0)&(0,0,0)&(+++)&(0,1,1) & (0,1,1)& U(4)\\
 \hline
\end{array}
}{5stackMSSMaAAPrototypeI}{D6-brane configuration of a global 5-stack MSSM-like configuration with initial gauge group 
$U(3)_a\times USp(2)_b \times U(1)_c\times U(1)_d \times U(4)_h$ on the {\bf aAA} lattice of the orientifold \mbox{$T^6/(\Z_2 \times \Z_6 \times \OR)$} with discrete torsion ($\eta=-1$) and the $\OR\Z_2^{(3)}$-plane as the exotic O6-plane ($\eta_{\OR\Z_2^{(3)}}=-1$).}
%%%%%%%%%%%%%%%%%%%%%%%%%%%%%%%%%%%%%%%%%%%%%%%%%%%%%%%%%%%%%%%%%%%%%%%%%%%

Furthermore, a closer look at the chiral spectrum shows an abundance of right-handed down-quarks $d_R$ from the $ac$ sector and left-handed leptons $L$ from the $bd$ sector. The proper identification of the first follows by looking at possible Yukawa couplings among the quarks. More precisely, charge conservation only allows the following types of Yukawa couplings:
\begin{equation}\label{Eq:PreYukMSSM}
Q_L^{ab} \cdot \tilde H_u^{bd'} u_R^{ad'} \qquad \text{and} \qquad Q_L^{ab} \cdot H_d^{bc} d_R^{ac}
,
\end{equation}
where we indicated explicitly the $xy$ sectors from which the states emerge as a superscript. With respect to the MSSM gauge group $(SU(3)_a\times USp(2)_b)_{U(1)_Y}$ the three chiral states $\ov{d_R}$ from the $ad$ sector form hermitian conjugates to the right handed down-quarks from the $ac$ sector. In order for these states to be heavy, we consider cubic couplings involving three of the six down-quarks $d_R$ from the $ac$ sector and some Standard Model singlet states $\Sigma^{cd}$:
\begin{equation}\label{Eq:d-Sigma-dbar-coupling}
 d_R^{ac}\,  \Sigma^{cd}\, \ov d_R^{ad},
\end{equation}
where $\Sigma^{cd}$ can receive a non-vanishing vacuum expectation value.
A similar consideration is valid for the Higgses from the $bc$ sector and the three surplus left-handed leptons from the $bd$ sector, which can be combined into cubic couplings of the form:
\begin{equation}\label{Eq:Hu-L-Sigma-coupling}
H_u^{bc} \cdot L^{bd}\,  \Sigma^{cd}.
\end{equation}
Using the argument of charge conservation, we can schematically write down cubic couplings which are expected to lift the abundant $d_R$-quarks, $H_u$ Higgses and three of the six leptons $L$ upon giving a {\it vev} to the Standard Model singlet states $\Sigma^{cd}$. A more in-depth analysis 
 involving the stringy selection rules will be performed in section~\ref{Ss:MSSMYukawa}, where we will verify explicitly whether such mechanisms can be invoked to give masses to the abundant vector-like pairs of matter states in table~\ref{tab:5stackMSSMaAAPrototypeISpectrum} and effectively obtain a three-generation MSSM-like model with continuous gauge group $SU(3)_a\times USp(2)_b\times U(1)_Y \times SU(4)_h$.
%%%%%%%%%%%%%%%%%%%%%%%%%%%%%%%%%%%%%%%%%%%%%%%%%%%%%%%%%%%%%%%%%%%%%%%%%%%
\mathtab{
\hspace{-0.5in}
\begin{array}{|c||c|c||c|c||c|c|}
\hline \multicolumn{7}{|c|}{\text{\bf Overview of the massless matter spectrum for global 5-stack MSSM on the {aAA} lattice}}\\
\hline \hline
\text{sector} & \text{state} &(SU(3)_a \times USp(2)_b \times SU(4)_h)_{U(1)_a \times U(1)_c\times U(1)_d \times U(1)_h}& Q_Y & Q_{PQ}& \Z_3 & \Z_6\\
\hline \hline
ab \equiv ab'& Q_L &3 \times  (\3, \2, \1)_{(1,0,0,0)} & 1/6 & 0& 0 & 0 \\
 ac & d_R&6 \times (\ov \3, \1,\1)_{(-1,1,0,0)}  & 1/3 &1 &1 & 2\\
 ad &\ov{d_R} & 3 \times (\3, \1, \1)_{(1,0,-1,0)}&-1/3& 1 & 1 & 2 \\
 ad' &u_R& 3 \times  (\ov \3, \1, \1)_{(-1,0,-1,0)} & -2/3& 1 &1& 2 \\
 bc \equiv b'c & H_u &3 \times (\1, \2,\1)_{(0,1,0,0)} &1/2 & 1 & 1 & 2 \\
  bc \equiv b'c & H_u + H_d &3 \times \left[ (\1, \2,\1)_{(0,1,0,0)} + h.c. \right] & \pm 1/2 & \pm1 &1||2 &2||4 \\
  bd \equiv b'd & L  &6 \times (\1,\2,\1)_{(0,0,-1,0)}  & -1/2& 1 & 1 & 2 \\
  bd \equiv b'd & \tilde H_u  + \tilde H_d &2 \times \left[ (\1,\2,\1)_{(0,0,1,0)}  + h.c. \right] &\pm 1/2& \mp1 & 2||1 & 4||2 \\
  cd & \nu_R &3 \times (\1,\1,\1)_{(0,-1,1,0)} & 0 & -2 & 1 & 2\\
  cd & \Sigma^{cd} +  \tilde\Sigma^{cd} &3 \times \left[ (\1,\1,\1)_{(0,-1,1,0)} + h.c. \right]& 0 & \mp 2 &1||2 & 2||4 \\
    cd' &e_R &3 \times (\1,\1,\1)_{(0,1,1,0)} & 1 & 0 &0 & 0 \\
  cd' & X^{cd'} + \tilde X^{cd'} &3 \times \left[ (\1,\1,\1)_{(0,1,1,0)} + h.c. \right]& \pm 1& 0 &0 & 0 \\
 \hline \hline
 ah& &2 \times \left[ (\3,\1,\ov\4)_{(1,0,0,-1)}  +h.c.\right] & \pm1/6& 0  & 1||2 & 5||1\\
 ah' & &(\3,\1,\4)_{(1,0,0,1)}  +h.c. & \pm1/6 & 0 & 2 ||1 & 1||5 \\
 bh \equiv b'h & &3 \times (\1,\2,\4)_{(0,0,0,1)} &0 & 0 & 2 &1\\
 ch' &&6 \times (\1,\1,\ov\4)_{(0,-1,0,-1)}  & -1/2 & -1 & 0&3\\
 dh & &3 \times (\1,\1,\ov\4)_{(0,0,1,-1)} & 1/2& -1 & 0&3 \\
 dh'&&3 \times (\1,\1,\4)_{(0,0,1,1)}  & 1/2& -1 &1  &5\\
 \hline \hline
 aa'& & 2\times[ ({\bf 3_{A}},\1,\1)_{(2,0,0,0)} + h.c.] & \pm1/3 & 0 &0&0\\
  bb'\equiv bb & \Anti_b^{(i)} &  5 \times  (\1,\1_{\bf A},\1)_{(0,0,0,0)}& 0 & 0 &0&0\\
   cc& & 4 \times (\1,\1,\1)_{(0,0,0,0)}& 0 & 0 &0 &0\\
    dd& \Adj_d^{(i)} & 5 \times (\1,\1,\1)_{(0,0,0,0)} & 0 & 0 &0& 0\\
        dd'& \Sym_d+ \ov\Sym_d  & (\1,\1,\1)_{(0,0,2,0)} + (\1,\1,\1)_{(0,0,-2,0)}   & \pm 1 & \mp 2 &1||2& 2||4 \\
    hh'& &2 \times [ (\1,\1,{\bf 6_{A}})_{(0,0,0,2)} + h.c.] & 0 & 0 & 1||2& 2||4\\
 \hline
\end{array}
}{5stackMSSMaAAPrototypeISpectrum}{Chiral and non-chiral massless matter spectrum for the five-stack D6-brane model with initial gauge group $U(3)_a\times USp(2)_b \times U(1)_c\times U(1)_d \times U(4)_h$ corresponding to the configuration from table~\ref{tab:5stackMSSMaAAPrototypeI}. For vector-like states different charges under the discrete $\Z_3$ ($\Z_6$) symmetry are denoted using the logic symbol~$||$.
The closed string sector for this model contains $(h_{11}^+,h_{11}^-,h_{21})=(4,15,19)$ vectors, K\"ahler and complex structure moduli, respectively.
}
%%%%%%%%%%%%%%%%%%%%%%%%%%%%%%%%%%%%%%%%%%%%%%%%%%%%%%%%%%%%%%%%%%%%%%%%%%%

\clearpage
%%%%%%%%%%%%%%%%%%%%%%%%%%%%%%%%%%%%%%%%%%%%%%%%%%%%%%%%%%%%%%%%%%%%%%%%%%%%%%%%%%%
%%%%%%%%%%%%%%%%%%%%%%%%%%%%%%%%%%%%%%%%%%%%%%%%%%%%%%%%%%%%%%%%%%%%%%%%%%%%%%%%%%%
\subsection{Discrete Symmetries}\label{Ss:MSSMDiscrete}
Next, we focus on the phenomenological aspects of the global five-stack MSSM-like model presented in table~\ref{tab:5stackMSSMaAAPrototypeI} starting with revealing the presence of discrete symmetries. The main motivation to discuss discrete symmetries for this model consists in potentially prohibiting undesired cubic and/or baryon/lepton-number violating couplings and in reinforcing the interpretation of the chiral spectrum presented in table~\ref{tab:5stackMSSMaAAPrototypeISpectrum} by virtue of non-trivially acting gauged $\Z_n$ symmetries. To this end, we first write down the necessary existence conditions (\ref{Eq:Zn-condition-nec}) for the D6-brane configuration given in table~\ref{tab:5stackMSSMaAAPrototypeI}:
\begin{align}\hspace{-25mm}
&\left\{k_a \left(\begin{array}{c} 0 \\ 0 \\\hline 0 \\ 0 \\ 0 \\ 6 \\ -6 \\ 0
\end{array}\right)
+ k_c \left(\begin{array}{c} 0 \\  0 \\\hline 0 \\ 0 \\ 0 \\ -2 \\ 2 \\ 0
\end{array}\right)
+ k_d \left(\begin{array}{c} 0 \\  0 \\\hline 0 \\ 0 \\ 0 \\ 0 \\ 0 \\ 0 
\end{array}\right)
+ k_h \left(\begin{array}{c} 0 \\  0 \\\hline 0 \\ 0 \\ 0 \\ -8 \\ 8 \\ 0 
\end{array}\right) \right\}
\stackrel{!}{=} 0 \text{ mod } n, \\
&\left\{ k_a \left(\begin{array}{c} 0 \\ 0 \\ 0 \\ 0 \\\hline 6 \\ 6 \\ 0 \\ 0
\end{array}\right)
+ k_c \left(\begin{array}{c} 6 \\ 6 \\ 0 \\ 0 \\\hline 0 \\ 0 \\ 0 \\ 0
\end{array}\right)
+ k_d \left(\begin{array}{c} -6 \\ -6 \\ 0 \\ 0 \\\hline -2 \\ -2 \\ 0 \\ 0
\end{array}\right)
+ k_h \left(\begin{array}{c}  0 \\ 0 \\ 0 \\ 0 \\\hline 8 \\ 8 \\ 0 \\ 0
\end{array}\right) \right\}\stackrel{!}{=} 0 \text{ mod } n. 
\end{align}
A row-by-row comparison clearly shows that ten of the conditions are trivially satisfied, and that the remaining six conditions correspond to only three independent conditions:
\begin{equation}\label{Eq:NecCon5StackMSSM}
\begin{array}{rcl}
-6k_a+2k_c+8k_h&\stackrel{!}{=}& 0 \text{ mod } n  ,\\
6k_c-6k_d&\stackrel{!}{=}& 0 \text{ mod } n  ,\\
6k_a-2k_d+8k_h&\stackrel{!}{=}& 0 \text{ mod } n  .
\end{array}
\end{equation}
These relations have to be supplemented with the sufficient existence conditions (\ref{Eq:Zn-condition-suf}):
\begin{align}\label{Eq:SufCon5StackMSSMRaw}
\left\{k_a\left(\begin{array}{c}
3\\
6\\
3\\
6\\
3\\
6\\
\hline
0\\
0\\
0\\
0\\
\hline
3\\
0\\
0\\
6\\
6\\
-3
\end{array}\right)+k_c\left(\begin{array}{c}
3\\
0\\
3\\
0\\
3\\
0\\
\hline
0\\
0\\
0\\
0\\
\hline
1\\
0\\
0\\
-2\\
-1\\
2
\end{array}\right)+k_d\left(\begin{array}{c}
-4\\
-2\\
-4\\
-2\\
-4\\
-2\\
\hline
0\\
0\\
0\\
0\\
\hline
-2\\
0\\
0\\
0\\
-1\\
-1
\end{array}\right)+k_h\left(\begin{array}{c}
4\\
8\\
4\\
8\\
4\\
8\\
\hline
0\\
0\\
0\\
0\\
\hline
0\\
0\\
0\\
-8\\
0\\
0
\end{array}\right) \right\}=0\text{  mod  }n,
\end{align}
where various rows turn out to be linearly dependent of each other, and some of the rows (i.e.~rows 2, 4, 6 and 14) yield the same conditions as the necessary conditions in~\eqref{Eq:NecCon5StackMSSM}. Moreover, the last sufficient condition in (\ref{Eq:SufCon5StackMSSMRaw}) is a linear combination of the first and fifth sufficient condition in the third block, such that the sufficient conditions only provide three linearly independent constraints:
\begin{equation} \label{Eq:SufCon5StackMSSM}
\begin{array}{rcl}
3 k_a + 3 k_c - 4 k_d + 4k_h &\stackrel{!}{=}& 0 \text{ mod } n ,\\
3 k_a+ k_c -2 k_d&\stackrel{!}{=}& 0 \text{ mod } n  ,\\
6k_a -k_c - k_d&\stackrel{!}{=}& 0 \text{ mod } n .
\end{array}
\end{equation}
The sufficient conditions allow to further reduce the number of linearly independent necessary conditions: more explicitly, subtracting twice the second condition in (\ref{Eq:SufCon5StackMSSM}) from the third condition in (\ref{Eq:SufCon5StackMSSM}) corresponds to the second constraint in (\ref{Eq:NecCon5StackMSSM}). Adding the first condition in (\ref{Eq:NecCon5StackMSSM}) to two times the third condition in (\ref{Eq:SufCon5StackMSSM}) reproduces the third constraint in (\ref{Eq:NecCon5StackMSSM}). Hence, there are effectively four linearly independent constraints, i.e. the first condition in (\ref{Eq:NecCon5StackMSSM}) and the three conditions in (\ref{Eq:SufCon5StackMSSM}), which agrees with the existence of four Abelian gauge factors $U(1)_a \times U(1)_c \times U(1)_d \times U(1)_h$ as starting point. In order to identify the Abelian massless and the massive $\Z_n$ symmetries for the global five-stack MSSM-like model, the four linearly independent conditions from (\ref{Eq:NecCon5StackMSSM}) and (\ref{Eq:SufCon5StackMSSM}) have to be satisfied simultaneously.  
A first observation is that the linear combination $Q_Y = \frac{1}{6} Q_a + \frac{1}{2} Q_c + \frac{1}{2} Q_d$ satisfies the constraints exactly, for any value of $n$, implying that this linear combination of $U(1)$'s corresponds to the massless hypercharge, in line with the discussion surrounding equation (\ref{Eq:HyperChargePrescr34Stacks}). In our search for discrete $\Z_n$ symmetries, this massless hypercharge can be used to set the $\Z_n$ charges of the left-handed quarks to zero.

The full set of solutions to the constraints (\ref{Eq:NecCon5StackMSSM}) and (\ref{Eq:SufCon5StackMSSM}) can then be summarised as:
\begin{itemize}
\item The configuration $(k_a,k_c,k_d,k_h) = (1,0,0,0)$ is a discrete $\Z_3$ symmetry homomorphic to the centre of the $SU(3)_a$ gauge symmetry, playing the r\^ole of a baryon-like discrete symmetry. Upon a massless hypercharge rotation this discrete symmetry acts trivially on the visible and hidden sector.
\item The configuration $(k_a,k_c,k_d,k_h) = (0,0,0,1)$ corresponds to the discrete $\Z_4$ symmetry homomorphic to the centre of the `hidden' gauge group $SU(4)_h$ and acts only non-trivially on exotic states charged under the hidden gauge group, reproducing the same charge selection rule as the non-Abelian `hidden' $SU(4)_h$.
\item The linear combination $(k_a,k_c,k_d,k_h) = (1,1,1,1)$ corresponds to the discrete $\Z_2$ symmetry guaranteed by the K-theory constraints, which acts trivially on the massless spectrum upon a rotation over the massless hypercharge. 
Note that this $\Z_2$ symmetry corresponds to a linear combination of the $\Z_2$ symmetry hiding within the massless hypercharge and the $\Z_2$ symmetry within the former $\Z_4$ symmetry. As such, the discrete $\Z_2$ symmetry associated to $(k_a,k_c,k_d,k_h) = (1,1,1,1)$ should not be considered as an independent discrete symmetry.
\item Finally, we also find a discrete $\Z_6$ symmetry for the combination $(k_a,k_c,k_d,k_h) = (0,2,4,1)$, for which the charges of the massless open string states are listed in the last column of table~\ref{tab:5stackMSSMaAAPrototypeISpectrum}. Nonetheless, the order 6 does not correspond to a viable discrete symmetry in the low-energy effective field theory,
as this $\Z_6$ can be reduced to a discrete $\Z_3$ symmetry. More explicitly, in order to identify the discrete $\Z_n$ symmetry acting independently from the centres of the non-Abelian gauge factors, we have to mod out those centres from the independent discrete $\Z_n$ symmetries found above, being the discrete $\Z_3$, $\Z_4$ and $\Z_6$ symmetry. Thus, when we consider the quotient group $(\Z_3 \times \Z_4 \times \Z_6)/(\Z_3\times \Z_2\times \Z_4) \simeq \Z_3$, we notice that it is homomorphic to the discrete $\Z_3$ gauge symmetry arising from $(k_a,k_c,k_d,k_h) = (0,1,2,2)$ and acting non-trivially on the massless spectrum as indicated in the second-to-last column of table~\ref{tab:5stackMSSMaAAPrototypeISpectrum}. The $\Z_6$ charges are mapped to the $\Z_3$ charges as follows:
\begin{equation}\label{Eq:Z6RedZ3ChargeAssign}
\begin{array}{ccc}
\Z_6 \text{ charges}& \longrightarrow &\Z_3 \text{ charges}\\
\hline \hline 0,3 & \longrightarrow & 0\\
2, 5 & \longrightarrow & 1\\
1,4 & \longrightarrow & 2\\
\hline
\end{array}
\end{equation}
Hence, this global five-stack MSSM-like model contains a non-trivial discrete $\Z_3$ symmetry, which can however not be decomposed according to (\ref{Eq:GenRALMSSM}) when comparing the $\Z_3$ charges in table~\ref{tab:5stackMSSMaAAPrototypeISpectrum} to the generic expressions for the charges in~(\ref{Eq:MSSMcharges}). This conundrum can be traced back to the appearance of two up-type Higgses $H_u^{bc}$ and $\tilde H_u^{bd}$, where the first ones are required to compose the Yukawa couplings for the right-handed neutrinos and the latter ones to compose the Yukawa couplings for the right-handed down-quarks $d_R$. If we relax the required existence of the Yukawa couplings for the right-handed neutrinos, then they do not have to be identified with the singlet states from the $cd$ sector and we could identify the right-handed neutrinos with the singlet states in the $bb$, $cc$ or $dd$ sectors. Under these assumptions, the discrete $\Z_3$ symmetry can be reinterpreted as the $\Z_3$ symmetry ${\cal R}_3^2 {\cal A}_3 {\cal L}_3 {\cal }$. This simple example of a discrete $\Z_3$ symmetry exhibits the intimate r\^ole between the assumed existence of Yukawa couplings and the classification of a discrete symmetry. At the same time, it also shows that (global) intersecting D-brane models can realise discrete $\Z_n$ symmetries which do not appear in the purely field theoretic set-up of the MSSM, due to the presence of extended Higgs sectors in the massless spectrum of intersecting D-brane models.
For the extended Higgs sector listed in table~\ref{tab:5stackMSSMaAAPrototypeISpectrum} one can clearly see that the up-type Higgses $H_u$ and $\tilde{H}_u$ have different charges under the $\Z_3$ symmetry, which forbids Yukawa couplings to $H_u$ for the up-quarks. A similar consideration for the down-type Higgses $H_d$ and $\tilde{H}_d$ teaches that the discrete $\Z_3$ symmetry also forbids Yukawa couplings to $\tilde H_d$ for the down-quarks, consistent with the observations surrounding equation (\ref{Eq:PreYukMSSM}).

In the same way, one can use the discrete $\Z_3$ symmetry to verify that the up-type Higgs ${H}_u$ allows for Yukawa-type couplings~\eqref{Eq:Hu-L-Sigma-coupling}
 involving the left-handed leptons $L$ and the neutral states located in the chiral $cd$ sector. Other neutral states under the Standard Model gauge group, as listed in table~\ref{tab:5stackMSSMaAAPrototypeISpectrum} for the five-stack MSSM-like model, require the other up-type Higgses $\tilde{H}_u$ to participate in the respective three-point couplings. In the next section, we will investigate  in more detail which three-point couplings are allowed from the stringy selection rules of closed polygons. This will allow us to verify which Yukawa couplings are present in the perturbative superpotential, and to justify our identification of the chiral states in the $cd$ sector as the right-handed neutrinos. At this point, we point out that the neutral states in the non-chiral $cd$ sector seem to be suitable candidates to construct supersymmetric versions of the DFSZ axion model, through the $\Z_3$ preserving couplings of the form $H_u \cdot \tilde{H}_d\, \Sigma^{cd}$ and $\tilde{H}_u \cdot H_d\, \tilde \Sigma^{cd}$ and with the Peccei-Quinn symmetry identified as one of the two natural options, $Q_{PQ}= Q_c - Q_d$, for open string axion models~\cite{Honecker:2013mya,Honecker:2015ela}. In the next section, we will devote more attention to this consideration and derive the associated scalar Higgs-axion potential in full detail.

\end{itemize} 

In summary, the full gauge group for the five-stack MSSM-like model is  given by $SU(3)_a\times USp(2)_b\times U(1)_Y \times SU(4)_h \times \Z_3$ below the string mass scale.

%%%%%%%%%%%%%%%%%%%%%%%%%%%%%%%%%%%%%%%%%%%%%%%%%%%%%%%%%%%%%%%%%%%%%%%%%%%%%%%%%%%
%%%%%%%%%%%%%%%%%%%%%%%%%%%%%%%%%%%%%%%%%%%%%%%%%%%%%%%%%%%%%%%%%%%%%%%%%%%%%%%%%%%
\subsection{Yukawa Couplings and Higgs-Axion Potential}\label{Ss:MSSMYukawa}

Focusing on the spectrum associated to the visible sector in table~\ref{tab:5stackMSSMaAAPrototypeISpectrum}, we can easily identify three generations of quarks and leptons, but we are also confronted with an extended Higgs-sector and various vector-like matter pairs. To probe the phenomenological viability of this global five-stack MSSM-like model, we have to determine the Yukawa couplings and justify why the vector-like states acquire larger masses than the quarks and leptons. As reviewed in section~\ref{Ss:Yukawa}, the first selection rule for a cubic coupling (composed of three massless open string states) consists in verifying whether it forms a singlet under all gauge symmetries of the model, including the discrete $\Z_3$ gauge symmetry identified in the previous section. Cubic couplings generated through worldsheet instantons also have to form singlets under the global $U(1)_{PQ}$ symmetry from table~\ref{tab:5stackMSSMaAAPrototypeISpectrum}. The massive linear combination $U(1)_{PQ} \equiv U(1)_c - U(1)_d$ forms an orthogonal direction to the massless hypercharge and acquires its mass through the St\"uckelberg mechanism (involving a closed string axion). Our interpretation of this massive linear combination of $U(1)$ gauge factors as a Peccei-Quinn $U(1)_{PQ}$ symmetry follows from the charge assignment of the quarks, leptons and Higgses under $U(1)_{PQ}$, following similar reasoning as the one presented in~\cite{Honecker:2013mya,Honecker:2015ela}. The second part of our argument is based on the form of the perturbative superpotential, which contains the following three contributions: 
\begin{equation}\label{Eq:SumDFSZ}
{\cal W}_{\rm per} \supset {\cal W}_{DFSZ} + {\cal W}_{MSSM} + {\cal W}_{extra},
\end{equation}
and where the three contributions can be written (schematically) as:
\begin{subequations}
\begin{align}
{\cal W}_{DFSZ}&= \mu\, H_u \cdot \tilde H_d \Sigma^{cd} + \tilde \mu\, \tilde H_u \cdot H_d \tilde \Sigma^{cd}, \label{Eq:WDFSZ}\\
{\cal W}_{MSSM} &= y_u\, Q_L \cdot \tilde H_u u_R + y_d\, Q_L\cdot H_d d_R + y_e\, L\cdot H_d e_R + y_\nu\, L\cdot H_u \nu_R, \label{Eq:WMSSM}\\
{\cal W}_{extra} &= \kappa\,  \ov d_R \Sigma^{cd} d_R + \tilde \kappa\, L \cdot H_u \Sigma^{cd}. \label{Eq:Wextra}
\end{align}
\end{subequations}
The superpotential contribution~(\ref{Eq:WDFSZ}) forms the straightforward supersymmetrised version of the DFSZ axion model as proposed in~\cite{Rajagopal:1990yx,Honecker:2013mya}. Note that the Standard Model singlets $\Sigma^{cd}$ and $\tilde \Sigma^{cd}$ couple linearly to the Higgs doublets, which should be contrasted to the quadratic coupling proposed in~\cite{Kim:1983dt} as a means to solve the $\mu$-problem and the strong CP-problem simultaneously. Since the singlet fields $\Sigma^{cd}$ and $\tilde \Sigma^{cd}$ are charged under a Peccei-Quinn symmetry containing the $U(1)_c$ factor, this model forms an alternative realisation of the supersymmetric DFSZ axion model within Type IIA string theory with intersecting D6-branes compared to the example discussed in detail in~\cite{Honecker:2013mya,Honecker:2015ela}. The superpotential contribution (\ref{Eq:WMSSM}) contains the Yukawa couplings for the quarks and leptons, but differs slightly from the usual Yukawa superpotential of the MSSM: the up-type Higgs $H_u$ responsible for the Yukawa couplings involving the right-handed neutrinos $\nu_R$ is here not the same as the up-type Higgs $\tilde H_u$ appearing in the Yukawa couplings for the right-handed quarks $u_R$. The last renormalisable contribution~\eqref{Eq:Wextra} to the perturbative superpotential contains cubic couplings for the abundant right-handed quarks and left-handed leptons. These couplings form the key elements to generate the supersymmetric masses for three out of the six right-handed quarks and left-handed leptons by giving a {\it vev} to the singlet states $\Sigma^{cd}$, such that the model in table~\ref{tab:5stackMSSMaAAPrototypeISpectrum} effectively becomes the three-generation MSSM 
(possibly up to additional  MSSM-singlet states) at low energies, as suggested at the end of section~\ref{Ss:MSSM-models}.    

%%%%%%%%%%%%%%%%%%%%%%%%%%%%%%%%%%%%%%%%%%%%%%%%%%%%%%%%%%%%%%%%%%%%%%%%%%%
\begin{sidewaystable}[h]
\begin{minipage}{21cm}
\begin{center}$\begin{array}{|c||c|c|c|c|c|}
\hline  \multicolumn{6}{|c|}{\text{\bf Total amount of matter per sector for a 5-stack MSSM model on the {aAA} lattice}}\\
\hline \hline (\chi^{x y},\chi^{x (\omega y)}, \chi^{x (\omega^2 y)})& y=a & y=b& y=c& y=d & y=h \\  
\hline
x= a & (0,0,0) & (2,1,0) & (-4,-1,-1) & (2,1,0)  & (|2|,1,-1) \\
x= b & &(0,``\frac{3 +|2|}{2}'',``\frac{-3+|2|}{2}'')  & (0,-3 + |2|,|4|) & (0,3+|2|,3+|2|) &(-2,-1,0) \\
x= c & & &(0,\frac{|4|}{2},\frac{|4|}{2})  & (0,|4|,-3+|2|) & (0,0,0)\\
x= d & & & & (0,``\frac{-3 +|2|}{2}'',``\frac{3+|2|}{2}'')  & (2,1,0) \\
x= h & & & & & (0,0,0)\\
\hline
\end{array}$
\caption{Overview of the total amount of chiral and non-chiral massless matter per sector $x(\omega^k y)$ for the global five-stack MSSM-like model with fractional D6-brane configuration given in table~\ref{tab:5stackMSSMaAAPrototypeI}. If the net-chirality $|\chi^{x(\omega^ky)}| < \varphi^{x(\omega^ky)}$, the sector $x(\omega^ky)$ comes with a set of non-chiral pairs of matter states, whose multiplicity corresponds to $n^{x(\omega^ky)}_{NC} \equiv\varphi^{x(\omega^ky)} - |\chi^{x(\omega^ky)}|$, e.g. $|n^{a(\omega^0h)}_{NC}|=|2|$ denotes one non-chiral pair of bifundamentals in sector $a(\omega^0h)$. 
Such non-chiral pairs are indicated as $|n^{x(\omega^ky)}_{NC}|$. 
The diagonal entries $\varphi^{x(\omega x)} = \varphi^{x(\omega^2 x)} = \frac{\varphi^{\Adj_x}}{2}$
count the number of states in the adjoint representation for the $x$-stack, e.g. $(0,``\frac{3+|2|}{2}'',``\frac{-3+|2|}{2}'')$ for the $b$-stack denotes five multiplets in the adjoint representation with half of the d.o.f. localised in the $b(\theta b)$ sector and the other half in the $b(\theta^2 b) \equiv (\theta b)b$ sector.
\label{tab:Z2Z65stackMSSMTotalSpectrumI}}\end{center}
\end{minipage}
\\
\\ \\ \\
\begin{minipage}{21cm}\begin{center}
$\begin{array}{|c||c|c|c|c|}
\hline  \multicolumn{5}{|c|}{\text{\bf Total amount of matter per sector for a 5-stack MSSM model}}\\
\hline \hline (\chi^{x y'},\chi^{x (\omega y)'}, \chi^{x (\omega^2 y)'})& y=a & y=c& y=d & y=h\\  
\hline
x= a & \tarh{(|2|,-1,1)}{(0,0,0)}  & (0,0,0) & (-2,0,-1)& (|2|,0,0)  \\
x= c &  & \tarh{(|2|,3+|2|,-3+|2|)}{(0,0,0)}  &(0,3+|2|,|4|) & (-4,-1,-1)   \\
x= d &  & & \tarh{(0,3+|2|,-3+|2|)}{(|2|,0,0)}  & (2,1,0)  \\
x= h &  & & & \tarh{(|2|,-1,1)}{(0,0,0)}   \\
\hline
\end{array}
$
\caption{Overview of the total amount of chiral and non-chiral massless matter per sector $x(\omega^k y)'$ for the global five-stack MSSM-like model with the fractional D6-brane configuration given in table~\ref{tab:5stackMSSMaAAPrototypeI}. 
The notation for the counting of bifundamental states is identical to table~\protect\ref{tab:Z2Z65stackMSSMTotalSpectrumI}. For $x(\theta^k x)'$ sectors, the upper entries count the numbers of antisymmetric representations and the lower entries the symmetric ones. The $\OR$-invariance of the $b$-stack implies $b(\theta^k x)'=b(\theta^k x)$, which are already listed in table~\protect\ref{tab:Z2Z65stackMSSMTotalSpectrumI}. 
For D6-brane stacks $c$ and $d$ supporting $U(1)$ gauge groups, the states in the antisymmetric representation do not exist, but are included for completeness and for consistency when checking the anomaly cancellation conditions.
\label{tab:Z2Z65stackMSSMTotalSpectrumII}}\end{center}
\end{minipage}
\end{sidewaystable}
%%%%%%%%%%%%%%%%%%%%%%%%%%%%%%%%%%%%%%%%%%%%%%%%%%%%%%%%%%%%%%%%%%%%%%%%%%%

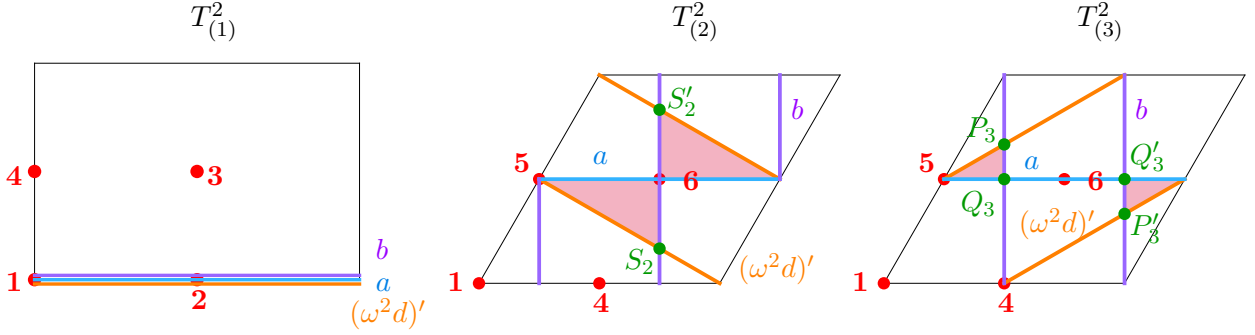
\begin{figure}[h]
\begin{center}
\vspace{0.2in}
\begin{tabular}{c@{\hspace{0.5in}}c@{\hspace{0.1in}}c}
\includegraphics[width=4.5cm]{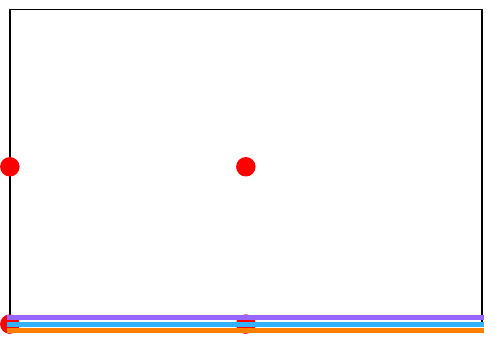} \begin{picture}(0,0)\put(-70,100){$T_{(1)}^2$}\put(-140,0){\bf \color{red} \bf 1} \put(-140,40){\bf \color{red} \bf 4} \put(-70,-7){\bf \color{red} \bf 2} \put(-64,40){\bf \color{red} \bf 3} \put(0,0){\color{myblue} $a$} \put(-10,-12){\color{myorange} $(\omega^2 d)'$}  \put(0,12){\color{mypurple} $b$}\end{picture}
& \includegraphics[width=5cm]{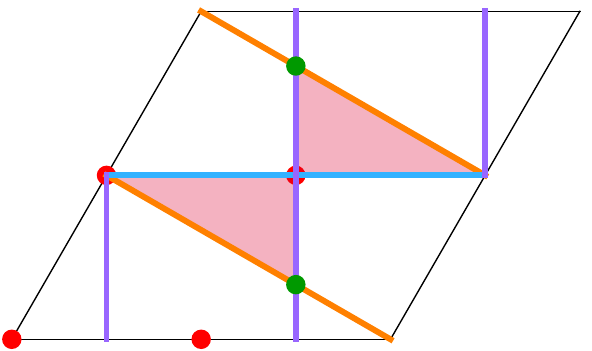} \begin{picture}(0,0) \put(-70,100){$T_{(2)}^2$} \put(-155,0){\color{red} \bf 1}  \put(-100,-8){\color{red} \bf 4}  \put(-130,45){\color{red} \bf 5}  \put(-66,38){\color{red} \bf 6} \put(-88,8){\color{mygr} \bf $S_2$} \put(-72,69){\color{mygr} \bf $S_2'$} \put(-100,48){\color{myblue} $a$} \put(-45,5){\color{myorange} $(\omega^2 d)'$}  \put(-25,65){\color{mypurple} $b$} \end{picture} 
&  \includegraphics[width=5cm]{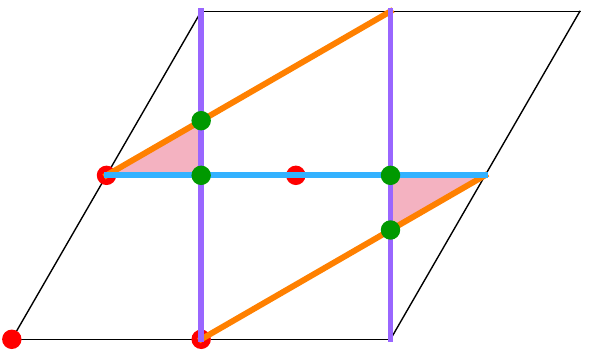} \begin{picture}(0,0) \put(-70,100){$T_{(3)}^2$} \put(-155,0){\color{red} \bf 1}  \put(-100,-8){\color{red} \bf 4}  \put(-130,45){\color{red} \bf 5}  \put(-66,38){\color{red} \bf 6}  \put(-114,30){\color{mygr} \bf $Q_3$} \put(-50,48){\color{mygr} \bf $Q_3'$}  \put(-112,58){\color{mygr} \bf $P_3$} \put(-50,21){\color{mygr} \bf $P_3'$}  \put(-90,46){\color{myblue} $a$} \put(-92,22){\color{myorange} $(\omega^2 d)'$}  \put(-48,65){\color{mypurple} $b$}  \end{picture}
\end{tabular}
\vspace{0.1in}
\caption{Pictorial view of the D6-brane configuration involving the closed sequence of the bulk three-cycles $[{\color{myblue}a}, {\color{mypurple}b}, {\color{myorange}(\omega^2 d)'}]$
of the five-stack MSSM, compatible with the perturbatively allowed Yukawa couplings $Q_L^{(2)} \cdot \tilde H_u^{(2)} u_R^{(1)}$ and $Q_L^{(1)} \cdot \tilde H_u^{(2)} u_R^{(1)}$. The points $(S_2,S_2')$, $(P_3,P_3')$ and $(Q_3,Q_3')$ correspond each to $\Z_2^{(i)}$-invariant pairs of intersection points per two-torus. \label{Fig:ExampleCubicCoupling}}
\end{center}
\end{figure}

In order for the parameters of the cubic couplings to be non-vanishing, the corresponding couplings also have to satisfy the stringy selection rule as explained in section~\ref{Ss:Yukawa}. The first step in determining the closed triangle sequences for the three-point couplings consists in indicating from which sectors $x(\omega^ky)_{k=0,1,2}$ the matter states arise. A full overview of the matter states per sector is given in tables~\ref{tab:Z2Z65stackMSSMTotalSpectrumI} and \ref{tab:Z2Z65stackMSSMTotalSpectrumII} for the global five-stack MSSM-like D6-brane configuration from  table~\ref{tab:5stackMSSMaAAPrototypeI}. 
Furthermore, to determine the shapes and sizes of the triangles enclosed by the intersecting one-cycles on $T_{(2)}^2$ and $T_{(3)}^2$ we also have to pin down at which intersection points the matter states are located. 
A comprehensive description of the technical precedure using Chan-Paton labels for fractional D-branes is provided in appendix~\ref{A:ChanPatonMethod}, where we also clarify the subtlety of discrete Wilson lines.
For chiral matter states, one can uniquely identify a $\Z_2^{(i)}$-invariant point at which an ${\cal N}=1$ supersymmetric chiral multiplet is located. Even when the intersection points correspond to points $R$ and $R'$ that are not $\Z_2^{(i)}$ fixed points, one can always form a $\Z_2$ invariant orbit $(R,R')$ at which the chiral multiplet is located.  
An explicit example of a closed sequence is presented in figure~\ref{Fig:ExampleCubicCoupling} using the bulk three-cycles $a$, $b$ and $(\omega^2 d)'$. Note that the cycles are all coincident along $T_{(1)}^2$, as a consequence of the invariance of the first two-torus under the $\Z_6$ orbifold action, such that we only focus on the intersections along the remaining ambient space $T_{(2)}^2\times T_{(3)}^2$, as anticipated in section~\ref{Ss:Yukawa}.
To the intersections on $T_{(2)}^2\times T_{(3)}^2$ between $a$ and $b$ we can allocate two left-handed quarks: $Q_L^{(2)}$ at $\{5,(Q_3,Q_3')\}$ and $Q_L^{(1)}$ at $\{6,(Q_3,Q_3')\}$, where $(Q_3,Q_3')$ is an example of a $\Z_2^{(i)}$-invariant pair of intersection points on the two-torus $T_{(3)}^2$. The right-handed quark $u_R^{(1)}$ arising from the intersections between $a$ and $(\omega^2 d)'$ is located at the point $\{5,5\}$ on $T_{(2)}^2\times T_{(3)}^2$. The intersecting three-cycles $b$ and $(\omega^2 d)'$ give rise to three chiral left-handed leptons $L$ ($L^{(2)}$ at the points $\{5,4\}$, $L^{(3)}$ at the pair of points $\{5, (P_3,P_3')\}$ and $L^{(1)}$ at the pair of points $\{(S_2,S_2'),4\}$) and one non-chiral pair of states interpreted as the Higgs doublets $\tilde H_u + \tilde H_d$ at the quadruplet $\{(S_2,S_2'), (P_3,P_3')\}$
under  $\Z_2 \times \Z_2$.
Taking these allocations of the open string matter states into account, we find the two allowed cubic couplings $y_u^{(221)} \,Q_L^{(2)} \cdot \tilde H_u^{(2)} u_R^{(1)}$ and $y_u^{(121)} \,Q_L^{(1)} \cdot \tilde H_u^{(2)} u_R^{(1)}$, for which the parameters scale at leading order as follows: 
\begin{equation}
y_u^{(221)} \sim {\cal O}\left(e^{- \frac{16v_2 + v_3}{48}}\right), \qquad y_u^{(121)} \sim {\cal O}\left(e^{- \frac{4v_2+v_3}{48} }\right),
\end{equation}
where $v_i$ corresponds to the area  (i.e.~the real part of the bulk K\"ahler modulus) of the two-torus $T_{(i)}^2$ in units of $\alpha'$. The superscripts labelling the generation of the left-handed and right-handed quarks have been chosen in such a way that a realistic pattern of the Yukawa couplings among the different generations can be inferred when taking into account the other allowed cubic couplings listed in tables~\ref{tab:YukaCouplingMSSMpartI} and~\ref{tab:YukaCouplingMSSMpartII} as well. The subscripts for the Higgs doublets are chosen with the convention that the non-chiral pairs $H_u^{(1)} + H_d^{(1)}$ and $\tilde H_u^{(1)} + \tilde H_d^{(1)}$ are situated in the $b(\omega c)$ and $b(\omega d)$ sector, respectively, while the non-chiral pairs $H_u^{(2,3)} + H_d^{(2,3)}$ and $\tilde H_u^{(2)} + \tilde H_d^{(2)}$ emerge from the $b(\omega^2 c)$ and $b(\omega^2 d)$ sector, respectively.
%%%%%%%%%%%%%%%%%%%%%%%%%%
%%%%%%%%%%%%%%%%%%%%%%

 A first look at tables~\ref{tab:YukaCouplingMSSMpartI}--\ref{tab:YukaCouplingMSSMpartVII} reveals that the cubic couplings in the superpotentials (\ref{Eq:WDFSZ}), (\ref{Eq:WMSSM}) and (\ref{Eq:Wextra}) should be seen as a schematic representation of the types of couplings to expect, since the exact structure of the cubic couplings turns out to be more involved due to the presence of generation-mixing and of the extended Higgs-sector, expressed through the various superscripts on the parameters $y, \kappa, \tilde{\kappa}, \mu, \tilde{\mu}$  in the last column of tables~\ref{tab:YukaCouplingMSSMpartI},~\ref{tab:YukaCouplingMSSMpartII}, \ref{tab:YukaCouplingMSSMpartIII}, \ref{tab:YukaCouplingMSSMpartIV}, and~\ref{tab:YukaCouplingMSSMpartV}.  
 Focusing on the quark sector, we notice the absence of a diagonal Yukawa coupling for the up-quark $u_R^{(2)}$, yet the latter does enter in a non-diagonal Yukawa coupling with the left-handed quark $Q_L^{(3)}$. A second observation concerns the strengths of the Yukawa couplings and the possible realisation of hierarchies among distinct generations: the Yukawa parameter $y_u^{(313)}$ is larger than the parameter $y_u^{(121)}$ for $T^6/(\Z_2\times \Z_6 \times \OR)$ backgrounds with $v_3 < v_2$, while the off-diagonal up-type Yukawa couplings $y_u^{(221)}$ and $y_u^{(321)}$ are more suppressed than the diagonal terms $y_u^{(121)}$ and $y_u^{(313)}$ when assuming $v_2< 5 \, v_3$. For the Yukawa couplings involving the down-quarks, we also notice a suppression of the off-diagonal terms $y_d^{112}$, $y_d^{231}$ and $y_d^{33i \in \{4,5,6\}}$ with respect to the diagonal couplings $y_d^{(212)}$, $y_d^{(131)}$ and $y_d^{(333)}$, respectively. The Yukawa couplings involving the down-type Higgs $H_d^{(2)}$, however, do not show this pattern, as the off-diagonal term $y_d^{221}$ is for instance larger than the diagonal coupling $y_d^{121}$. Table~\ref{tab:YukaCouplingMSSMpartII} consists of the cubic couplings suited to make the three abundant generations of quarks $d_R^{(4,5,6)}$ and $\ov{d_R}^{(1,2,3)}$ sufficiently massive. 
Unfortunately, the cubic couplings also involve the down-quarks $d_R^{(1,2,3)}$ which appear in the Yukawa couplings, such that a more elaborate reasoning involving the {\it vev}s of the singlets $\Sigma^{cd(i)}$ has to be developed in order to argue why the masses for the right-handed down-quarks $d_R^{(4,5,6)}$ are lifted and only the right-handed quarks $d_R^{(1,2,3)}$ appear effectively at low energies.

Before doing so, we consider the cubic couplings (\ref{Eq:WDFSZ})-(\ref{Eq:Wextra}) involving those states that are uncharged under the strong gauge group  and investigate in detail how these couplings can be realised through worldsheet instantons on the ambient space $T^{2}_{(2)}\times T_{(3)}^2$. Recalling that the leptons, Higgses and singlet states $\Sigma^{cd(i)}$ and $\tilde\Sigma^{cd(i)}$ arise from intersections between the D6-brane stacks $b$, $c$ and $d$ (including their orbifold and orientifold images), and that these D6-brane stacks are characterised by the same bulk wrapping numbers as listed in table~\ref{tab:5stackMSSMaAAPrototypeI}, matter states are only expected to arise from the intersections  $b(\omega^k c)_{k=1,2}$, $b(\omega^k d)_{k=1,2}$, $c(\omega^k d)_{k=1,2}$ and  $c(\omega^k d)'_{k=1,2}$, which has been verified explicitly in tables~\ref{tab:Z2Z65stackMSSMTotalSpectrumI} and \ref{tab:Z2Z65stackMSSMTotalSpectrumII}. Another characteristic of this D6-brane configuration, which has not been encountered in previous studies of Yukawa couplings for fractional intersecting D6-branes~\cite{Honecker:2012jd,Honecker:2012qr,Honecker:2013mya}, is the potential appearance of both chiral and non-chiral matter in bifundamental representations from the same sector.\footnote{Note that we explicitly exclude the self-intersections between a D6-brane and its orientifold images in this statement, as those sectors are known to potentially give rise to vector-like matter pairs besides chiral matter states in the (anti)symmetric representation, see for instance~\cite{Gmeiner:2008xq,Honecker:2013mya}.
Since the distinction of chiral states versus non-chiral pairs stems from $\Z_2$-invariant intersection points versus pairs of points under the same $\Z_2$, the two types of states do not simultaneously exist on the six-torus~\cite{Aldazabal:2000cn,Aldazabal:2000dg,Ibanez:2001nd,Cremades:2003qj,MarchesanoBuznego:2003hp} or its $T^6/(\Z_2 \times \Z_2 \times \OR)$ orbifold without discrete torsion, see e.g.~\cite{Cvetic:2001tj,Cvetic:2001nr,Cvetic:2002wh,Blumenhagen:2005mu,Gmeiner:2005vz}.} 
Following reasonings similar to the ones presented in~\cite{Gmeiner:2008xq} and in appendix B.1 of~\cite{Forste:2010gw}, one can verify that one non-chiral pair of matter states in the bifundamental representation is located at the $\Z_2 \times \Z_2$-invariant quadruplet $\{(S_2,S_2'),(P_3,P_3')\}$, while the chiral matter states -- if present -- can be allocated to some $\Z_2^{(i)}$-invariant intersection point (with $i \in \{2,3\}$ for $\varrho$-independent models). 
As we demonstrate in appendix~\ref{A:ChanPatonMethod} in detail for the $c(\omega \, d)$ sector, further non-chiral pairs of matter states can arise from combining two different $\Z_2 \times \Z_2$-invariant doublets of intersection points. Taking into account the correct allocations of the matter states following the logic of appendix~\ref{A:ChanPatonMethod}, one can compute the non-vanishing Higgs-axion couplings as in table~\ref{tab:YukaCouplingMSSMpartIII}, the Yukawa couplings for the leptons as in table~\ref{tab:YukaCouplingMSSMpartIV} and the cubic $L\cdot H_u \Sigma$ couplings as in table~\ref{tab:YukaCouplingMSSMpartV}. Zooming in on the Yukawa couplings involving the leptons, we notice that only the third generation is characterised by diagonal Yukawa terms, while the other two generations only appear in non-diagonal couplings in combination with the third generation leptons $L^{(3)}$, $e_R^{(3)}$ or $\nu_R^{(3)}$. The generation-label for the leptons has been chosen such that the Yukawa parameters for the second generation are larger than the ones for the first generation: $y_e^{(322)} \approx y_e^{(223)} > y_e^{(133)} \approx  y_e^{(331)}$ and $y_\nu^{(622)} \approx  y_\nu^{(523)} > y_\nu^{(433)} \approx  y_\nu^{(631)}$.  An interesting observation is that the Yukawa couplings for the right-handed charged leptons $e_R$ and right-handed neutrinos $\nu_R$ involve left-handed leptons from different sectors, namely $L^{(1,2,3)}$ for the righth-handed leptons $e_R$ and $L^{(4,5,6)}$ for the right-handed neutrinos $\nu_R$. This consideration has non-trivial consequences for the argument establishing three effective generations of left-handed leptons, since the couplings of the right-handed neutrinos to $L^{(i=4,5,6)}$  undermine the provisioned mechanism to make the left-handed leptons $L^{(i=4,5,6)}$ heavier than $L^{(i=1,2,3)}$ by cranking up the {\it vev} for the scalar field in the multiplet $\Sigma^{cd(3)}$ appearing in the couplings $L^{(i=4,5,6)} \cdot H_u^{(j=2,3)} \Sigma^{cd (3)}$. Considering a large {\it vev} for $\Sigma^{cd (3)}$ would also imply a large supersymmetric mass for the right-handed down-quarks $d_R^{(1)}$ and $d_R^{(2)}$, which is phenomenologically unacceptable. More explicitly, the right-handed quarks $d_R^{(1)}$ and $d_R^{(2)}$ appear in the Yukawa couplings in table~\ref{tab:YukaCouplingMSSMpartI}, where they fulfil the r\^ole of the down-quark and the strange-quark, respectively, and whose mass cannot be made parametrically large.  This last reflection suggests that a proper reasoning arguing for three effective generations of left-handed leptons is intimately connected to the argument for three effective right-handed down-quarks. Moreover, table~\ref{tab:YukaCouplingMSSMpartV} teaches us that also the other left-handed leptons $L^{(1,2,3)}$ appear in cubic couplings of the form $L\cdot H_u \Sigma^{cd(1,2)}$. This implies that also the {\it vevs} of the singlets $\Sigma^{cd(1,2)}$ cannot be taken randomly large, as this would suggest a large supersymmetric mass for the left-handed leptons $L^{(1,2,3)}$.

%%%%%%%%%%%%%%%%%%%%%%%%%%%
\begin{figure}[h]
\begin{center}
\vspace{0.2in}
\begin{tabular}{c@{\hspace{0.5in}}c@{\hspace{0.1in}}c}
\includegraphics[width=4.5cm]{T1Coupling1} \begin{picture}(0,0)\put(-70,100){$T_{(1)}^2$}\put(-140,0){\bf \color{red} \bf 1} \put(-140,40){\bf \color{red} \bf 4} \put(-70,-7){\bf \color{red} \bf 2} \put(-64,40){\bf \color{red} \bf 3}  \put(0,0){\color{myblue} $b$} \put(-10,-12){\color{myorange} $(\omega^2 d)'$}  \put(-8,12){\color{mypurple} $(\omega^2c)$} \end{picture}
& \includegraphics[width=5cm]{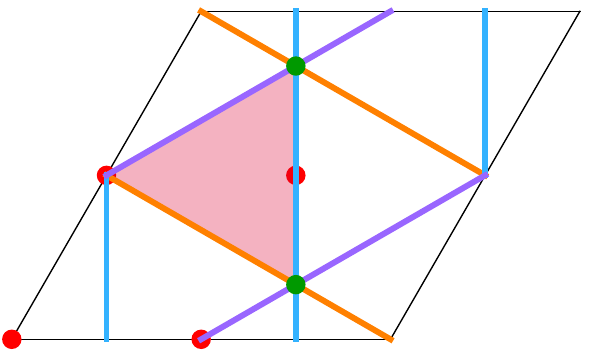} \begin{picture}(0,0) \put(-70,100){$T_{(2)}^2$} \put(-155,0){\color{red} \bf 1}  \put(-100,-8){\color{red} \bf 4}  \put(-130,45){\color{red} \bf 5}  \put(-66,38){\color{red} \bf 6} \put(-68,12){\color{mygr} \bf $S_2$} \put(-68,65){\color{mygr} \bf $S_2'$} \put(-120,62){\color{mypurple} $(\omega^2c)$} \put(-45,5){\color{myorange} $(\omega^2 d)'$}  \put(-25,65){\color{myblue} $b$}\end{picture} 
&  \includegraphics[width=5cm]{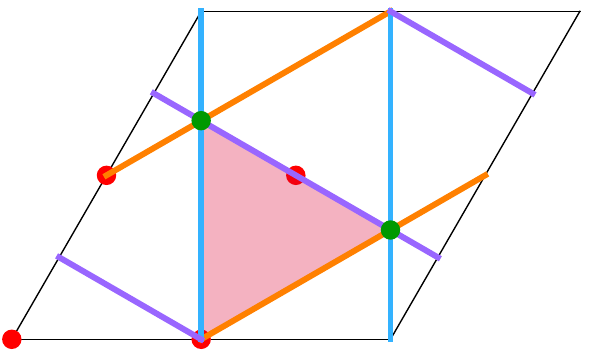} \begin{picture}(0,0) \put(-70,100){$T_{(3)}^2$} \put(-155,0){\color{red} \bf 1}  \put(-100,-8){\color{red} \bf 4}  \put(-130,45){\color{red} \bf 5}  \put(-66,38){\color{red} \bf 6}   \put(-98,64){\color{mygr} \bf $P_3$} \put(-67,13){\color{mygr} \bf $P_3'$}   \put(-48,65){\color{myblue} $b$} \put(-32,32){\color{myorange} $(\omega^2 d)'$}  \put(-158,20){\color{mypurple} $(\omega^2 c)$}   \end{picture}
\end{tabular}
\caption{Pictorial view of the D6-brane configuration involving the closed sequence of the bulk three-cycles $[{\color{myblue}b}, {\color{mypurple}(\omega^2 c)}, {\color{myorange}(\omega^2 d)'}]$ for the five-stack MSSM, compatible with the non-vanishing, perturbatively allowed Yukawa couplings $L\cdot H_d \, e_R$ appearing in the first half of table~\ref{tab:YukaCouplingMSSMpartIV}. The points $(S_2,S_2')$ and $(P_3,P_3')$ correspond to $\Z_2^{(i)}$-invariant pairs of intersection points per two-torus, and there exist four $\Z_2^{(i)}$-invariant orbits on $T_{(2)}^2 \times T_{(3)}^2$: $\{5,4\}$, $\{5,(P_3,P_3')\}$, $\{(S_2,S_2'),4\}$ and $\{(S_2,S_2'),(P_3,P_3')\}$. \label{Fig:ExampleCubicCoupling2}}
\end{center}
\end{figure}
%%%%%%%%%%%%%%%%%%%%%%%%%%%%%%%
The indispensable couplings for a supersymmetric DFSZ axion model are the cubic couplings in (\ref{Eq:WDFSZ}) of the form $H_u \cdot \tilde H_d \Sigma^{cd}$ and $\tilde H_u \cdot H_d \tilde \Sigma^{cd}$, through which the Higgses are forced to be charged under the $U(1)_{PQ}$ symmetry. For  the explicit MSSM-like model of this section, the perturbatively allowed cubic couplings between the Higgses and the axion multiplets are listed in table~\ref{tab:YukaCouplingMSSMpartIII} and involve the non-chiral states from the $b(\omega^{k=1,2} c)$, $b(\omega^{k=2,1})d$ and $c(\omega^{k=1,2}d)$ sectors. Given the specific allocation of these non-chiral states at $\Z_2\times \Z_2$-invariant doublets of intersection points on $T_{(2)}^2\times T_{(3)}^2$, it can occur that three states are located at points which cannot serve as the apexes of a closed triangle. In this case, the area of the triangle is infinity, and the respective cubic coupling vanishes. We give two explicit examples of such a situation in the first and third row of the table~\ref{tab:YukaCouplingMSSMpartIII}. When we take into account the structure of the Yukawa couplings for the quarks and leptons, we observe that the down-type Higgses $\tilde H_d^{(1)}$ and  $\tilde H_d^{(2)}$ do not enter at all in the discussion as a consequence of $U(1)_{PQ}$ invariance of the Yukawa interactions. Hence, we can anticipate that the most relevant Higgs-axion couplings to consider are the ones on rows 9, 10, 13 and 14 of table~\ref{tab:YukaCouplingMSSMpartIII}, as they are the ones that require the $U(1)_{PQ}$ charged nature of the Higgses appearing in the Yukawa coupling. Due to the appearance of the up-type Higgses $H_u^{(2,3)}$ in the Higgs-axion couplings on rows 11 and 12, we should also take these two couplings into account. The other Higgs-axion couplings involve Higgses which do not appear in the Yukawa couplings and given that they are slightly more suppressed, we are able to neglect them to simplify the discussion. The resulting scalar Higgs-axion potential is now expected to have the same structure as the one derived for the $T^6/\Z_6^{(\prime)}$ models in~\cite{Honecker:2013mya,Honecker:2015ela}, namely consisting of four separate contributions: F-term contributions set by the superpotential (\ref{Eq:SumDFSZ}), D-term contributions associated to the $USp(2)_b$ gauge symmetry, D-term contributions associated to the $U(1)_{PQ}$ symmetry (which acted as a local symmetry before the St\"uckelberg mechanism) and soft terms added ``by hands'' at this point, possibly arising from a gaugino condensate in the hidden sector.

\begin{figure}[h]
\begin{center}
\vspace{0.2in}
\begin{tabular}{c@{\hspace{0.5in}}c@{\hspace{0.1in}}c}
\includegraphics[width=4.5cm]{T1Coupling1} \begin{picture}(0,0)\put(-70,100){$T_{(1)}^2$}\put(-140,0){\bf \color{red} \bf 1} \put(-140,40){\bf \color{red} \bf 4} \put(-70,-7){\bf \color{red} \bf 2} \put(-64,40){\bf \color{red} \bf 3}  \put(0,0){\color{myblue} $b$} \put(-10,-12){\color{myorange} $(\omega^2 d)$}  \put(-6,12){\color{mypurple} $(\omega d)$} \end{picture}
& \includegraphics[width=5cm]{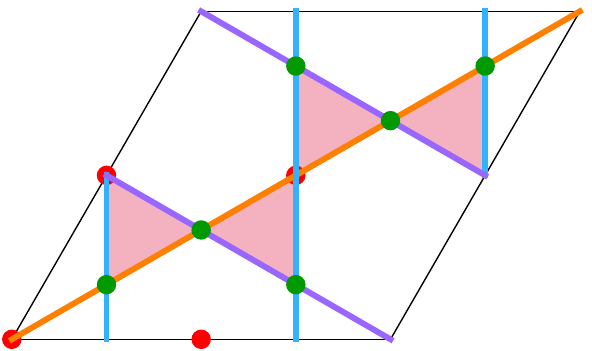} \begin{picture}(0,0) \put(-70,100){$T_{(2)}^2$} \put(-155,0){\color{red} \bf 1}  \put(-100,-8){\color{red} \bf 4}  \put(-130,45){\color{red} \bf 5}  \put(-66,38){\color{red} \bf 6} \put(-138,17){\color{mygr} \bf $R_2$} \put(-27,62){\color{mygr} \bf $R_2'$} \put(-102,35){\color{mygr} \bf $2$} \put(-57,62){\color{mygr} \bf $3$} \put(-72,17){\color{mygr} \bf $S_2$} \put(-90,60){\color{mygr} \bf $S_2'$} \put(-10,87){\color{myorange} $(\omega^2 d)$} \put(-45,5){\color{mypurple} $(\omega d)$}  \put(-33,85){\color{myblue} $b$}\end{picture} 
&  \includegraphics[width=5cm]{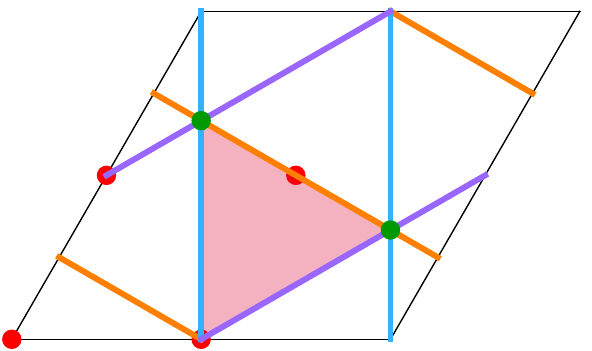} \begin{picture}(0,0) \put(-70,100){$T_{(3)}^2$} \put(-155,0){\color{red} \bf 1}  \put(-100,-8){\color{red} \bf 4}  \put(-130,45){\color{red} \bf 5}  \put(-66,38){\color{red} \bf 6}  
\put(-66,12){\color{mygr} \bf $P_3'$} \put(-98,64){\color{mygr} \bf $P_3$}   \put(-53,85){\color{myblue} $b$} \put(-32,32){\color{mypurple} $(\omega d)$}  \put(-158,20){\color{myorange} $(\omega^2 d)$}   \end{picture}
\end{tabular}
\caption{Pictorial representation of the D6-brane configuration involving the closed sequence of the bulk three-cycles $[{\color{myblue}b}, {\color{mypurple}(\omega d)}, {\color{myorange}(\omega^2 d)}]$ for the five-stack MSSM, compatible with the perturbatively allowed Yukawa couplings $L^{(i=1,2,3)}\cdot \tilde H_u^{(1)} \Adj_d^{(k=1,2,3,4,5)}$. The intersection points $(R_2,R_2')$, $(S_2,S_2')$, $(2,3)$ and $(P_3,P_3')$ correspond to $\Z_2^{(i)}$-invariant pairs of intersection points per two-torus. 
\label{Fig:ExampleCubicCoupling3}}
\end{center}
\end{figure}

Another aspect, which we should turn our attention to, is the presence of additional singlets under the Standard Model gauge group, which can serve as candidate right-handed neutrinos, namely the five $ \Anti_b \equiv(\1,\1_{\bf A},\1)_{(0,0,0,0)}$ states in the antisymmetric representation of $USp(2)_b$, the four multiplets in the adjoint representation of $U(1)_c$ and the five multiplets in the adjoint representation of $U(1)_d$. One might even consider the superpartners of the geometric moduli, but here we focus on open string states. Focusing first on the cubic couplings of the form $L\cdot \tilde H_u^{(i=1,2)} \Adj_c$, we notice that these couplings are perfectly allowed from the field theory side based on charge conservation arguments. Nevertheless, from the stringy side, we notice that the cubic couplings involving the multiplets $\Adj_c$ in the adjoint representation of $U(1)_c$ are not allowed based on the violation of the stringy selection rule. More explicitly, as both the left-handed leptons and the up-type Higgses $\tilde H_u^{(i=1,2)}$ arise from the $b(\omega^k d)_{k=1,2}$ sectors, the four multiplets in the adjoint representation of $U(1)_c$ located in the $c(\omega^k c)_{k=1,2}$ sectors do not allow for closed sequences. The other singlet states do allow for closed sequences, as listed in tables~\ref{tab:YukaCouplingMSSMpartVI} and~\ref{tab:YukaCouplingMSSMpartVII}, such that we have to include the following perturbative cubic couplings in the superpotential:
\begin{equation}\label{Eq:CubicCoupAntiAdj}
\begin{array}{rl}
{\cal W}_{per} \supset & \mathfrak{B}^{(i1k)} L^{(i)}\cdot \tilde H_u^{(1)} \Anti_b^{(k)} + \mathfrak{ \tilde B}^{(j1k)} L^{(j)}\cdot \tilde H_u^{(2)} \Anti_b^{(k)} \\
&+  \mathfrak{A}^{(i1k)} L^{(i)}\cdot \tilde H_u^{(1)} \Adj_d^{(k)}+ \mathfrak{\tilde  A}^{(j1k)} L^{(j)}\cdot \tilde H_u^{(2)} \Adj_d^{(k)},
\end{array}
\end{equation}
with $i\in \{1,2,3 \}$, $j \in \{4,5,6\}$ and $k\in \{1,2,3,4,5\}$. 
The explicit form of the closed sequences as well as the leading order behaviour of the non-vanishing coupling constants are elaborated in tables~\ref{tab:YukaCouplingMSSMpartVI} and~\ref{tab:YukaCouplingMSSMpartVII}, while a pictorial representation of the worldsheet instantons is presented in figure~\ref{Fig:ExampleCubicCoupling3} for the third kind of couplings in (\ref{Eq:CubicCoupAntiAdj}). 
%%%%%%%%%%%%%%%%%%%%%%%%%%%%%%%%
%%%%%%%%%%%%%%%%%%%%%%%%%%%%%%%%
Apart from representing alternative Yukawa couplings when the r\^ole of the right-handed neutrinos is played by the states $\Anti_b^{(k)}$ or $\Adj_d^{(k)}$, the cubic couplings in (\ref{Eq:CubicCoupAntiAdj}) can in principle also be useful to lift the masses for the leptons $L^{(4,5,6)}$ with respect to the other three leptons $L^{(1,2,3)}$ by cranking up the {\it vev}s for a selected number of matter states in the antisymmetric or adjoint representation. However, as the up-type Higgs $\tilde H_u^{(2)}$ appears in the Yukawa couplings involving the up-quarks, one cannot give it randomly a large supersymmetric mass to argue for three effective leptons generations without giving a large supersymmetric mass to the left-handed quarks $Q_R^{(1,2)}$ and right-handed quark $u_R^{(1)}$. Moreover, the symmetry between the cubic couplings in tables~\ref{tab:YukaCouplingMSSMpartVI} or~\ref{tab:YukaCouplingMSSMpartVII} indicate that the leptons $L^{(1,2,3)}$ will acquire a large supersymmetric mass as well, when turning on a large {\it vev} for the states $\Anti_b^{(k)}$ or $\Adj_d^{(k)}$.

%%%%%%%%%%%%%%%%%%%%%%%%%%%%%%%%
\begin{sidewaystable}[h]
\begin{center}
\begin{tabular}{|c|c|c|c|c|}
\hline \multicolumn{5}{|c|}{\bf Cubic couplings for the superpotential (\ref{Eq:SumDFSZ}) of a global 5-stack MSSM (part I)} \\
\hline
\hline {\bf Coupling} & {\bf Sequence} & {\bf Triangles on $T_{(2)}^2 \times T_{(3)}^2$} & {\bf Enclosed Area} & {\bf Parameter}\\
\hline 
\hline
$Q_L^{(2)} \cdot \tilde H_u^{(2)} u_R^{(1)}$& \multirow{2}{*}{$[a,b,(\omega^2 d)']$} & $\left\{ [5,(S_2,S_2'),5], [(Q_3,Q_3'), (P_3,P_3'),5] \right\}$& $\frac{v_2}{3}+\frac{v_3}{48} $ & $y_u^{(221)} \sim {\cal O}\left(e^{- \frac{16 v_2 + v_3}{48} }\right)$ \\
$Q_L^{(1)} \cdot \tilde H_u^{(2)} u_R^{(1)}$& & $\left\{ [6,(S_2,S_2'),5], [(Q_3,Q_3'), (P_3,P_3'),5] \right\}$& $ \frac{v_2}{12} + \frac{v_3}{48}$ & $y_u^{(121)} \sim {\cal O}\left(e^{- \frac{4v_2 + v_3}{48} }\right)$  \\
\hline
$Q_L^{(3)} \cdot \tilde H_u^{(1)} u_R^{(2)}$ & \multirow{2}{*}{$[a,(\omega b),d']$}  & $\left\{ [6,(P_2,P_2'),(Q_2,Q_2')], [6,(R_3,R_3'),6] \right\}$ & $\frac{v_2}{48} +  \frac{v_3}{3} $ &$y_u^{(312)} \sim {\cal O}\left(e^{- \frac{v_2+16 v_3}{48} }\right)$  \\
$Q_L^{(3)} \cdot \tilde H_u^{(1)} u_R^{(3)}$ & &$\left\{ [6,(P_2,P_2'),(Q_2,Q_2')], [6,(R_3,R_3'),5] \right\}$ & $ \frac{v_2}{48}  +  \frac{v_3}{12}$  & $y_u^{(313)} \sim {\cal O}\left(e^{ - \frac{v_2 + 4 v_3}{48}}\right)$ \\
\hline
$Q_L^{(2)} \cdot H_d^{(1)} d_R^{(2)}$ & \multirow{2}{*}{$[a, b,(\omega c)]$}  &$\left\{ [5,(R_2,R_2'),6], [(Q_3,Q_3'), (P_3,P_3'),5] \right\}$  & $ \frac{v_2 }{12} + \frac{v_3}{48}$ &  $y_d^{(212)} \sim {\cal O}\left(e^{ - \frac{4v_2 +v_3}{48}}\right)$  \\
$Q_L^{(1)} \cdot H_d^{(1)}  d_R^{(2)}$& & $\left\{ [6,(R_2,R_2'),6], [(Q_3,Q_3'), (P_3,P_3'),5] \right\}$ & $\frac{v_2 }{3} +\frac{v_3}{48} $ & $y_d^{(112)} \sim {\cal O}\left(e^{- \frac{16v_2 + v_3}{48}}\right)$ \\
\hline
$Q_L^{(2)} \cdot H_d^{(2)}  d_R^{(1)}$ & \multirow{4}{*}{$[a,b,(\omega^2 c)]$}  & $\left\{ [5], [(Q_3,Q_3'), (P_3,P_3'),6] \right\}$ & $ \frac{v_3}{48} $ & $y_d^{(221)} \sim {\cal O}\left(e^{- \frac{ v_3}{48} }\right)$ \\
$Q_L^{(2)} \cdot H_d^{(3)}  d_R^{(1)}$ &  & $\left\{ [5,(S_2,S_2'),5], [(Q_3,Q_3'), (P_3,P_3'),6]\right\}$ & $\frac{v_2 }{3} +\frac{v_3}{48} $ & $y_d^{(231)} \sim {\cal O}\left(e^{- \frac{16 v_2 + v_3}{48} }\right)$ \\
$Q_L^{(1)} \cdot H_d^{(2)}  d_R^{(1)}$& & $\left\{ [6,5,5],[(Q_3,Q_3'), (P_3,P_3'),6] \right\}$ & $ \frac{3v_2}{4} + \frac{v_3}{48} $ &  $y_d^{(121)} \sim {\cal O}\left(e^{- \frac{36v_2 +v_3}{48}}\right)$  \\
$Q_L^{(1)} \cdot H_d^{(3)}  d_R^{(1)}$& & $\left\{ [6,(S_2,S_2'),5], [(Q_3,Q_3'), (P_3,P_3'),6]\right\}$ & $ \frac{v_2}{12} + \frac{v_3}{48} $ &  $y_d^{(131)} \sim {\cal O}\left(e^{- \frac{4v_2 +v_3}{48}}\right)$  \\
\hline
$Q_L^{(3)} \cdot H_d^{(2)} d_R^{(3)}$ &  \multirow{8}{*}{$[a,(\omega b),c]$}  & $\left\{ [6,6,5], [6,(R_3,R_3'),5] \right\}$   & $\frac{3v_2}{4} + \frac{v_3}{12}$ &  $y_d^{(323)} \sim {\cal O}\left(e^{-\frac{9v_2+ v_3}{12}}\right)$ \\
$Q_L^{(3)} \cdot H_d^{(3)} d_R^{(3)}$ &   & $\left\{ [6,(R_2,R_2'),5], [6,(R_3,R_3'),5] \right\}$   & $\frac{v_2}{12} + \frac{v_3}{12}$ &  $y_d^{(333)} \sim {\cal O}\left(e^{-\frac{v_2+ v_3}{12}}\right)$ \\
$Q_L^{(3)} \cdot H_d^{(2)} d_R^{(4)}$ &  & $\left\{ [6,6,5], [6,(R_3,R_3'),6] \right\}$ & $\frac{3v_2}{4} + \frac{v_3}{3}$ & $y_d^{(324)}  \sim {\cal O}\left( e^{- \frac{9v_2+4 v_3}{12} }  \right)$ \\
$Q_L^{(3)} \cdot H_d^{(3)} d_R^{(4)}$ &  & $\left\{ [6,(R_2,R_2'),5], [6,(R_3,R_3'),6] \right\}$ & $\frac{v_2}{12} + \frac{v_3}{3}$ & $y_d^{(334)}  \sim {\cal O}\left( e^{- \frac{v_2+4 v_3}{12} }  \right)$ \\
$Q_L^{(3)} \cdot H_d^{(2)} d_R^{(5)}$ &  & $\left\{ [6], [6,(R_3,R_3'),5] \right\}$ & $ \frac{v_3}{12}$ &  $y_d^{(325)}  \sim {\cal O}\left( e^{- \frac{v_3}{12} }  \right)$ \\
$Q_L^{(3)} \cdot H_d^{(3)} d_R^{(5)}$ &  & $\left\{ [6,(R_2,R_2'),6], [6,(R_3,R_3'),5] \right\}$ & $\frac{v_2}{3} + \frac{v_3}{12}$ &  $y_d^{(335)}  \sim {\cal O}\left( e^{- \frac{4v_2+v_3}{12} }  \right)$ \\
$Q_L^{(3)} \cdot H_d^{(2)} d_R^{(6)}$ &  & $\left\{ [6], [6,(R_3,R_3'),6] \right\}$ & $  \frac{v_3}{3}$ &  $y_d^{(326)}  \sim {\cal O}\left( e^{-\frac{ v_3}{3}  }  \right)$ \\
$Q_L^{(3)} \cdot H_d^{(3)} d_R^{(6)}$ &  & $\left\{ [6,(R_2,R_2'),6], [6,(R_3,R_3'),6] \right\}$ & $\frac{v_2}{3} + \frac{v_3}{3}$ &  $y_d^{(336)}  \sim {\cal O}\left( e^{-\frac{v_2+ v_3}{3}  }  \right)$ \\
\hline
\end{tabular}
\caption{Overview of the Yukawa couplings involving the left-handed and right-handed quarks. The third column lists the (triangular) worldsheet instantons $[x]$ or $[x,y,z]$ spanned by the indicated apexes (intersection points) $x,y,z$ on $T_{(i=2,3)}^2$ for the respective cubic couplings, see figures~\protect\ref{Fig:ExampleCubicCoupling},~\protect\ref{Fig:ExampleCubicCoupling2} and~\protect\ref{Fig:ExampleCubicCoupling3} for details. 
The fourth column presents the corresponding area for the worldsheet instantons expressed in terms of the areas $v_i$ of the two-tori $T_{(i=2,3)}^2$, and the last column shows the scaling of the coupling constant corresponding to the considered cubic coupling.   \label{tab:YukaCouplingMSSMpartI}}
\end{center}
\end{sidewaystable}
%%%%%%%%%%%%%%%%%%%%%%%%%%%%%%%%

%%%%%%%%%%%%%%%%%%%%%%%%%%%%%%%%
\begin{sidewaystable}[h]
\begin{center}
\hspace*{-0.3in}
\begin{tabular}{|c|c|c|c|c|}
\hline \multicolumn{5}{|c|}{\bf Cubic couplings for the superpotential (\ref{Eq:SumDFSZ}) of a global 5-stack MSSM (part II)} \\
\hline
\hline {\bf Coupling} & {\bf Sequence} & {\bf Triangles on $T_{(2)}^2 \times T_{(3)}^2$} & {\bf Enclosed Area} & {\bf Parameter}\\
\hline 
\hline
$\ov{d_R}^{(3)}\Sigma^{cd(1)} d_R^{(3)}$ & \multirow{8}{*}{$[a,c, (\omega d)]$} & $\left\{ [5], [5,(S_3,S_3'),5] \right\}$   &  $\frac{v_3}{3}$ & $\kappa^{(313)} \sim {\cal O}\left( e^{-\frac{v_3}{3}  }\right)$ \\
$\ov{d_R}^{(3)}\Sigma^{cd(2)} d_R^{(3)}$ & & $\left\{ [5,(S_2,S_2'),5], [5,(S_3,S_3'),5] \right\}$   &  $\frac{v_2}{3}+\frac{v_3}{3}$ & $\kappa^{(323)} \sim {\cal O}\left( e^{-\frac{v_2 + v_3}{3}  }\right)$ \\
$\ov{d_R}^{(3)}\Sigma^{cd(1)} d_R^{(4)}$ & & $\left\{ [5], [6,(S_3,S_3'),5]\right\}$ & $\frac{v_3}{12}$& $\kappa^{(314)} \sim {\cal O}\left( e^{-\frac{ v_3}{12}  } \right)$  \\
$\ov{d_R}^{(3)}\Sigma^{cd(2)} d_R^{(4)}$ & & $\left\{ [5,(S_2,S_2'),5], [6,(S_3,S_3'),5]\right\}$ & $\frac{v_2}{3}+\frac{v_3}{12}$& $\kappa^{(324)} \sim {\cal O}\left( e^{-\frac{4v_2 + v_3}{12}  } \right)$  \\
$\ov{d_R}^{(3)}\Sigma^{cd(1)} d_R^{(5)}$ & &$\left\{ [6,5,5], [5,(S_3,S_3'),5] \right\}$ &$ \frac{3v_2}{4}+\frac{v_3}{3}$& $\kappa^{(315)} \sim {\cal O}\left(  e^{-\frac{9v_2 + 4v_3}{12}  } \right)$  \\
$\ov{d_R}^{(3)}\Sigma^{cd(2)} d_R^{(5)}$ & &$\left\{ [6,(S_2,S_2'),5], [5,(S_3,S_3'),5] \right\}$ &$ \frac{v_2}{12}+\frac{v_3}{3}$& $\kappa^{(325)} \sim {\cal O}\left(  e^{-\frac{v_2 + 4v_3}{12}  } \right)$  \\
$\ov{d_R}^{(3)}\Sigma^{cd(1)} d_R^{(6)}$ & & $\left\{ [6,5,5], [6,(S_3,S_3'),5] \right\}$ & $\frac{3v_2}{4} + \frac{v_3}{12}$ & $\kappa^{(316)}   \sim {\cal O}\left( e^{-\frac{9v_2 + v_3}{12}  }  \right)$ \\
$\ov{d_R}^{(3)}\Sigma^{cd(2)} d_R^{(6)}$ & & $\left\{ [6,(S_2,S_2'),5], [6,(S_3,S_3'),5] \right\}$ & $\frac{v_2}{12} + \frac{v_3}{12}$ & $\kappa^{(326)}   \sim {\cal O}\left( e^{-\frac{v_2 + v_3}{12}  }  \right)$ \\
\hline
$\ov{d_R}^{(1)}\Sigma^{cd(3)} d_R^{(1)}$ & $[a,(\omega^2 c), (\omega d)]$ & $\left\{ [5,(S_2,S_2'),5], [6,(P_3,P_3'),5]\right\}$ & $\frac{v_2}{6}+\frac{v_3}{24}$ & $\kappa^{(131)} \sim {\cal O}\left( e^{- \frac{4v_2 +v_3}{24} }  \right)$ \\
\hline
$\ov{d_R}^{(2)}\Sigma^{cd(3)} d_R^{(2)}$ & \multirow{2}{*}{$[a,(\omega c), d]$} & $\left\{ [6,(P_2,P_2'),(Q_2,Q_2')], [5,(S_3,S_3'),5] \right\}$ & $\frac{v_2}{48}+\frac{v_3}{3}$  & $\kappa^{(232)}  \sim {\cal O}\left(e^{- \frac{v_2+16 v_3}{48} }\right)$  \\ 
$\ov{d_R}^{(1)}\Sigma^{cd(3)} d_R^{(2)}$& & $\left\{ [6,(P_2,P_2'),(Q_2,Q_2')], [5,(S_3,S_3'),6] \right\}$& $\frac{v_2}{48} + \frac{v_3}{12}$ & $\kappa^{(132)}  \sim {\cal O}\left(e^{- \frac{v_2+4v_3 }{48} }\right)$ \\
\hline
$\ov{d_R}^{(2)}\Sigma^{cd(1)} d_R^{(1)}$ & \multirow{4}{*}{$[a,(\omega^2 c), d]$} & $\left\{ [5,4,(Q_2,Q_2')] , [6,(R_3,R_3'),5] \right\}$ &  $\frac{3v_2}{16}   + \frac{v_3}{12}$ & $\kappa^{(211)}  \sim {\cal O}\left(e^{- \frac{12v_2 + 4 v_3}{48} }\right)$ \\ 
$\ov{d_R}^{(2)}\Sigma^{cd(2)} d_R^{(1)}$ & & $\left\{  [5,(P_2,P_2'),(Q_2,Q_2')], [6,(R_3,R_3'),5] \right\}$ &  $ \frac{v_2}{48} + \frac{v_3}{12}$ & $\kappa^{(221)}  \sim {\cal O}\left(e^{- \frac{v_2 + 4 v_3}{48} }\right)$ \\ 
$\ov{d_R}^{(1)}\Sigma^{cd(1)} d_R^{(1)}$& &$\left\{ [5,4,(Q_2,Q_2')], [6,(R_3,R_3'),6] \right\}$ & $ \frac{3v_2}{16} +\frac{v_3}{3} $  & $\kappa^{(111)}  \sim {\cal O}\left(e^{- \frac{9v_2 + 16 v_3}{48} }\right)$  \\
$\ov{d_R}^{(1)}\Sigma^{cd(2)} d_R^{(1)}$& &$\left\{ [5,(P_2,P_2'),(Q_2,Q_2')], [6,(R_3,R_3'),6] \right\}$ & $\frac{v_2}{48} +\frac{v_3}{3} $  & $\kappa^{(121)}  \sim {\cal O}\left(e^{- \frac{v_2 + 16 v_3}{48} }\right)$  \\
\hline
\end{tabular}
\caption{Overview of the cubic couplings $\ov{d_R} \Sigma^{cd} d_R$ in (\ref{Eq:Wextra}) involving the right-handed down-type quarks. The third column lists the (triangular) worldsheet instantons $[x]$ or $[x,y,z]$ bounded by three branes pairwise intersecting in the  indicated apexes $x,y,z$ on $T_{(i=2,3)}^2$ for the respective cubic couplings. The fourth column presents the corresponding area for the worldsheet instantons expressed in terms of the areas $v_i$ of the two-tori $T_{(i=2,3)}^2$, and the last column shows the scaling of the coupling constant corresponding to the considered cubic coupling. \label{tab:YukaCouplingMSSMpartII}}
\end{center}
\end{sidewaystable}
%%%%%%%%%%%%%%%%%%%%%%%%%%%%%%%%

%%%%%%%%%%%%%%%%%%%%%%%%%%%%%%%%
\begin{sidewaystable}[h]
\begin{center}
\hspace*{-0.3in}
\begin{tabular}{|c|c|c|c|c|}
\hline \multicolumn{5}{|c|}{\bf Cubic couplings for the superpotential (\ref{Eq:SumDFSZ}) of a global 5-stack MSSM (part III)} \\
\hline
\hline {\bf Coupling} & {\bf Sequence} & {\bf Triangles on $T_{(2)}^2 \times T_{(3)}^2$} & {\bf Enclosed Area} & {\bf Parameter}\\
\hline \hline
$H_u^{(4)} \cdot \tilde H_d^{(2)} \Sigma^{cd (1)}$ &  \multirow{10}{*}{$[b,(\omega c), (\omega^2 d)]$} & $    \{[6,(R_2,R_2'),6],[4,(P_3,P_3'),(P_3,P_3')]\}$& $\infty + \frac{v_3}{6}$ & $ \mu^{(421)} \sim {\cal O} (0)$\\
$H_u^{(4)} \cdot \tilde H_d^{(2)} \Sigma^{cd (2)}$ & & $    \{[6,(R_2,R_2'),(R_2,R_2')],[4,(P_3,P_3'),(P_3,P_3')]\}$& $\frac{v_2}{6} + \frac{v_3}{6}$ & $ \mu^{(422)} \sim {\cal O} \left(e^{-\frac{v_2+v_3}{6}}\right)$\\
$H_u^{(5)} \cdot \tilde H_d^{(2)} \Sigma^{cd (1)}$ & & $    \{[6,(R_2,R_2'),6],[(P_3,P_3')]\}$& $\infty $ & $ \mu^{(521)} \sim {\cal O} (0)$\\
$H_u^{(5)} \cdot \tilde H_d^{(2)} \Sigma^{cd (2)}$ & & $    \{[6,(R_2,R_2'),(R_2,R_2')],[(P_3,P_3')]\}$& $\frac{v_2}{6}$ & $ \mu^{(522)} \sim {\cal O} \left(e^{-\frac{v_2}{6}}\right)$\\
$H_u^{(6)} \cdot \tilde H_d^{(2)} \Sigma^{cd (1)}$ &  & $    \{[(R_2,R_2'),(R_2,R_2'),6],[4,(P_3,P_3'),(P_3,P_3')]\}$&  $\frac{v_2}{6} + \frac{v_3}{6}$ & $ \mu^{(621)} \sim {\cal O} \left(e^{-\frac{v_2+v_3}{6}}\right)$\\
$H_u^{(6)} \cdot \tilde H_d^{(2)} \Sigma^{cd (2)}$ & & $    \{[(R_2,R_2')],[4,(P_3,P_3'),(P_3,P_3')]\}$&  $\frac{v_3}{6}$ & $ \mu^{(622)} \sim {\cal O} \left(e^{-\frac{v_3}{6}}\right)$\\
$H_u^{(1)} \cdot \tilde H_d^{(2)} \Sigma^{cd (1)}$ & & $    \{[(R_2,R_2'),(R_2,R_2'),6],[(P_3,P_3')]\}$&  $\frac{v_2}{6} $ & $ \mu^{(121)} \sim {\cal O} \left(e^{-\frac{v_2}{6}}\right)$\\
$H_u^{(1)} \cdot \tilde H_d^{(2)} \Sigma^{cd (2)}$ & & $    \{[(R_2,R_2')],[(P_3,P_3')]\}$& 0 & $ \mu^{(122)} \sim {\cal O} (1)$\\
$\tilde H_u^{(2)}  \cdot  H_d^{(1)} \tilde \Sigma^{cd (1)} $&  & $    \{[(R_2,R_2')],[(P_3,P_3'),(P_3,P_3'),4]\}$&  $\frac{v_3}{6}$   & $   \tilde \mu^{(211)}  \sim {\cal O} \left(e^{-\frac{v_3}{6}}\right)$\\
$\tilde H_u^{(2)}  \cdot  H_d^{(1)} \tilde \Sigma^{cd (2)} $& & $    \{[(R_2,R_2')],[(P_3,P_3')]\}$  & 0 & $   \tilde \mu^{(212)} \sim {\cal O} (1)$\\
\hline
$H_u^{(2)} \cdot \tilde H_d^{(1)} \Sigma^{cd (3)}$ &  \multirow{4}{*}{$[b,(\omega^2 c), (\omega d)]$} &$    \{[(S_2,S_2')],[4,(P_3,P_3'),(P_3,P_3')]\}$
& $\frac{v_3}{6}$    & $  \mu^{(213)} \sim {\cal O} \left(e^{-\frac{v_3}{6}}\right)$\\
$H_u^{(3)} \cdot \tilde H_d^{(1)} \Sigma^{cd (3)}$ &  & $    \{[(S_2,S_2')],[(P_3,P_3')]\}$
&  0 & $  \mu^{(313)} \sim {\cal O} (1)$\\
$ \tilde H_u^{(1)}   \cdot H_d^{(2)} \tilde \Sigma^{cd (3)} $&  & $\{ [(S_2,S_2'),5,(S_2,S_2')],[(P_3,P_3')] \}$ & $\frac{v_2}{6}$ & $\tilde \mu^{(123)} \sim {\cal O} \left(e^{-\frac{v_2}{6}}\right)$ \\
$ \tilde H_u^{(1)}   \cdot H_d^{(3)} \tilde \Sigma^{cd (3)} $&  & $\{ [(S_2,S_2')],[(P_3,P_3')] \}$& 0 & $\tilde \mu^{(133)} \sim {\cal O} (1)$\\
\hline
\end{tabular}
\caption{Overview of the Higgs-axion couplings in (\ref{Eq:WDFSZ}). The third column lists the (triangular) worldsheet instantons $[x]$ or $[x,y,z]$ bounded by three branes pairwise intersecting in the  indicated apexes $x,y,z$ on $T_{(i=2,3)}^2$ for the respective cubic couplings. The fourth column presents the corresponding area for the worldsheet instantons expressed in terms of the areas $v_i$ of the two-tori $T_{(i=2,3)}^2$, and the last column shows the scaling of the coupling constant corresponding to the considered cubic coupling. \label{tab:YukaCouplingMSSMpartIII}}
\end{center}
\end{sidewaystable}
%%%%%%%%%%%%%%%%%%%%%%%%%%%%%%%%

%%%%%%%%%%%%%%%%%%%%%%%%%%%%%%%%
\begin{sidewaystable}[h]
\begin{center}
\hspace*{-0.3in}
\begin{tabular}{|c|c|c|c|c|}
\hline \multicolumn{5}{|c|}{\bf Cubic couplings for the superpotential (\ref{Eq:SumDFSZ}) of a global 5-stack MSSM (part IV)} \\
\hline 
 \hline {\bf Coupling} & {\bf Sequence} & {\bf Triangles on $T_{(2)}^2 \times T_{(3)}^2$} & {\bf Enclosed Area} & {\bf Parameter}\\
\hline \hline
$ L^{(3)} \cdot H_d^{(2)}  e_R^{(2)}$ & \multirow{5}{*}{$[b,(\omega^2 c),(\omega^2 d)']$}  & $ \{[5],[(P_3,P_3'),(P_3,P_3'),4]\}$ & $\frac{v_3}{6}$  & $y_e^{(322)} \sim {\cal O} \left(e^{- \frac{ v_3}{6}  }\right)$\\
$ L^{(2)} \cdot H_d^{(2)}  e_R^{(3)}$ & & $ \{[5],[4,(P_3,P_3'),(P_3,P_3')]\}$ & $\frac{v_3}{6}$  & $y_e^{(223)} \sim {\cal O} \left(e^{- \frac{ v_3}{6}  }\right)$\\
$ L^{(3)} \cdot H_d^{(2)}  e_R^{(3)}$ &  &$ \{[5],[(P_3,P_3')]\}$  & 0 & $y_e^{(323)} \sim {\cal O} (1)$\\
$ L^{(1)} \cdot H_d^{(3)}  e_R^{(3)}$ &   &$ \{[(S_2,S_2'),(S_2,S_2'),5],[4,(P_3,P_3'),(P_3,P_3')]\}$& $\frac{v_2+v_3}{6}$  & $y_e^{(133)} \sim {\cal O} (e^{- \frac{v_2 + v_3}{6}  })$\\
$ L^{(3)} \cdot H_d^{(3)}  e_R^{(1)}$ &   &$ \{[5,(S_2,S_2'),(S_2,S_2')],[(P_3,P_3'),(P_3,P_3'),4]\}$& $\frac{v_2+v_3}{6}$  & $y_e^{(331)} \sim {\cal O} (e^{- \frac{v_2 + v_3}{6}  })$\\
\hline
$ L^{(6)} \cdot H_u^{(2)}  \nu_R^{(2)}$ & \multirow{5}{*}{$[b,(\omega^2 c),(\omega d)]$} &$ \{[(S_2,S_2'),(S_2,S_2'),5],[4]\}$ & $\frac{v_2}{6}$  & $y_\nu^{(622)} \sim {\cal O} (e^{- \frac{v_2 }{6}  })$\\
$ L^{(6)} \cdot H_u^{(3)}  \nu_R^{(1)}$ & &$ \{[(S_2,S_2'),(S_2,S_2'),5],[4,(P_3,P_3'),(P_3,P_3')]\}$ & $\frac{v_2+v_3}{6}$  & $y_\nu^{(631)} \sim {\cal O} (e^{- \frac{v_2 + v_3}{6}  })$\\
$ L^{(5)} \cdot H_u^{(2)}  \nu_R^{(3)}$ & &$ \{[5,(S_2,S_2'),(S_2,S_2')],[4]\}$ & $\frac{v_2}{6}$  & $y_\nu^{(523)} \sim {\cal O} (e^{- \frac{v_2}{6}  })$\\
$ L^{(4)} \cdot H_u^{(3)}  \nu_R^{(3)}$ & &$ \{[5,(S_2,S_2'),(S_2,S_2')],[(P_3,P_3'),(P_3,P_3'),4]\}$ & $\frac{v_2+v_3}{6}$  & $y_\nu^{(433)} \sim {\cal O} (e^{- \frac{v_2 + v_3}{6}  })$\\
$ L^{(6)} \cdot H_u^{(2)}  \nu_R^{(3)}$ & &$ \{[(S_2,S_2')],[4]\}$ & 0  & $y_\nu^{(623)} \sim {\cal O} (1)$\\
\hline
\end{tabular}
\caption{Overview of the non-vanishing Yukawa couplings involving the left-handed and right-handed leptons. The third column lists the (triangular) worldsheet instantons $[x]$ or $[x,y,z]$ bounded by three branes pairwise intersecting in the  indicated apexes $x,y,z$ on $T_{(i=2,3)}^2$ for the respective cubic couplings. The fourth column presents the corresponding area for the worldsheet instantons expressed in terms of the areas $v_i$ of the two-tori $T_{(i=2,3)}^2$, and the last column shows the scaling of the coupling constant corresponding to the considered cubic coupling. \label{tab:YukaCouplingMSSMpartIV}}
\end{center}
\end{sidewaystable}
%%%%%%%%%%%%%%%%%%%%%%%%%%%%%%%%

%%%%%%%%%%%%%%%%%%%%%%%%%%%%%%%%
\begin{sidewaystable}[h]
\begin{center}
\hspace*{-0.3in}
\begin{tabular}{|c|c|c|c|c|}
\hline \multicolumn{5}{|c|}{\bf Cubic couplings for the superpotential (\ref{Eq:SumDFSZ}) of a global 5-stack MSSM (part V)} \\
\hline 
 \hline {\bf Coupling} & {\bf Sequence} & {\bf Triangles on $T_{(2)}^2 \times T_{(3)}^2$} & {\bf Enclosed Area} & {\bf Parameter}\\
\hline \hline
$L^{(5)} \cdot H_u^{(3)} \Sigma^{cd (3)}$ &  \multirow{4}{*}{$[b, (\omega^2 c), (\omega d)]$}& $\{[5,(S_2,S_2'),(S_2,S_2')],[4,(P_3,P_3'),(P_3,P_3')]\}$ & $\frac{v_2+v_3}{6}$ 
& $\tilde \kappa^{(533)}  \sim {\cal O} (e^{- \frac{v_2 + v_3}{6}  }) $ \\
$L^{(4)} \cdot H_u^{(2)} \Sigma^{cd (3)}$& & $\{[5,(S_2,S_2'),(S_2,S_2')],[(P_3,P_3'),4,(P_3,P_3')]\}$ & $\frac{v_2+v_3}{6}$ 
& $\tilde \kappa^{(423)}  \sim {\cal O} (e^{- \frac{v_2 + v_3}{6}  }) $ \\
$L^{(4)} \cdot H_u^{(3)} \Sigma^{cd (3)}$ && $\{[5,(S_2,S_2'),(S_2,S_2')],[(P_3,P_3')]\}$ & $\frac{v_2}{6}$ 
& $\tilde \kappa^{(433)}  \sim {\cal O} (e^{- \frac{v_2 }{6}  }) $ \\
$L^{(6)} \cdot H_u^{(3)} \Sigma^{cd (3)}$ & & $\{[(S_2,S_2')],[4,(P_3,P_3'),(P_3,P_3')]\}$ & $\frac{v_3}{6}$ 
& $\tilde \kappa^{(633)}  \sim {\cal O} (e^{- \frac{ v_3}{6}  }) $ \\
\hline
$L^{(3)} \cdot H_u^{(4)} \Sigma^{cd (1)}$ & \multirow{9}{*}{$[b, (\omega c), (\omega^2 d)]$} & $\{[6],[(P_3,P_3'),4,(P_3,P_3')]\}$ & $\frac{v_3}{6}$ 
& $\tilde \kappa^{(341)}  \sim {\cal O} (e^{- \frac{v_3}{6}  }) $ \\
$L^{(2)} \cdot H_u^{(5)} \Sigma^{cd (1)}$ && $\{[6],[4,(P_3,P_3'),(P_3,P_3')]\}$ & $\frac{v_3}{6}$ 
& $\tilde \kappa^{(251)}  \sim {\cal O} (e^{- \frac{ v_3}{6}  }) $ \\
$L^{(3)} \cdot H_u^{(5)} \Sigma^{cd (1)}$ && $\{[6],[(P_3,P_3')]\}$ & 0
& $\tilde \kappa^{(351)}  \sim {\cal O} (1) $ \\
$L^{(1)} \cdot H_u^{(5)} \Sigma^{cd (2)}$ && $\{[(R_2,R_2'),6,(R_2,R_2')],[4,(P_3,P_3'),(P_3,P_3')]\}$ & $\frac{v_2+v_3}{6}$ 
& $\tilde \kappa^{(152)}  \sim {\cal O} (e^{- \frac{ v_2+ v_3}{6}  }) $ \\
$L^{(3)} \cdot H_u^{(6)} \Sigma^{cd (2)}$ && $\{[6,(R_2,R_2'),(R_2,R_2')],[(P_3,P_3'),4,(P_3,P_3')]\}$ & $\frac{v_2+v_3}{6}$ 
& $\tilde \kappa^{(362)}  \sim {\cal O} (e^{- \frac{ v_2+ v_3}{6}  }) $ \\
$L^{(2)} \cdot H_u^{(1)} \Sigma^{cd (2)}$ && $\{[6,(R_2,R_2'),(R_2,R_2')],[4,(P_3,P_3'),(P_3,P_3')]\}$ & $\frac{v_2+v_3}{6}$ 
& $\tilde \kappa^{(212)}  \sim {\cal O} (e^{- \frac{ v_2+ v_3}{6}  }) $ \\
$L^{(3)} \cdot H_u^{(1)} \Sigma^{cd (2)}$ && $\{[6,(R_2,R_2'),(R_2,R_2')],[(P_3,P_3')]\}$ & $\frac{v_2}{6}$ 
& $\tilde \kappa^{(312)}  \sim {\cal O} (e^{- \frac{ v_2}{6}  }) $ \\
$L^{(1)} \cdot H_u^{(1)} \Sigma^{cd (1)}$ && $\{[(R_2,R_2'),(R_2,R_2'),6],[4,(P_3,P_3'),(P_3,P_3')]\}$ & $\frac{v_2+v_3}{6}$ 
& $\tilde \kappa^{(111)}  \sim {\cal O} (e^{- \frac{ v_2+ v_3}{6}  }) $ \\
$L^{(1)} \cdot H_u^{(1)} \Sigma^{cd (2)}$ && $\{[(R_2,R_2')],[4,(P_3,P_3'),(P_3,P_3')]\}$ & $\frac{v_3}{6}$ 
& $\tilde \kappa^{(112)}  \sim {\cal O} (e^{- \frac{ v_3}{6}  }) $ \\
\hline
\end{tabular}
\caption{Overview of the non-vanishing cubic couplings $L\cdot H_u \Sigma^{cd}$ in (\ref{Eq:Wextra}). The third column lists the (triangular) worldsheet instantons $[x]$ or $[x,y,z]$ bounded by three branes pairwise intersecting in the  indicated apexes $x,y,z$ on $T_{(i=2,3)}^2$ for the respective cubic couplings. The fourth column presents the corresponding area for the worldsheet instantons expressed in terms of the areas $v_i$ of the two-tori $T_{(i=2,3)}^2$, and the last column shows the scaling of the coupling constant corresponding to the considered cubic coupling. \label{tab:YukaCouplingMSSMpartV}}
\end{center}
\end{sidewaystable}
%%%%%%%%%%%%%%%%%%%%%%%%%%%%%%%%

%%%%%%%%%%%%%%%%%%%%%%%%%%%%%%%%
\begin{sidewaystable}[h]
\begin{center}
\hspace*{-0.6in}
\begin{tabular}{|c|c|c|c|c|}
\hline \multicolumn{5}{|c|}{\bf Cubic couplings for the superpotential (\ref{Eq:SumDFSZ}) of a global 5-stack MSSM (part VI)} \\
\hline
\hline {\bf Coupling} & {\bf Sequence} & {\bf Triangles on $T_{(2)}^2 \times T_{(3)}^2$} & {\bf Enclosed Area} & {\bf Parameter}\\
\hline 
\hline
$L^{(2)}\cdot \tilde H_u^{(1)} \Anti_b^{(4)}$ & \multirow{8}{*}{$[b,(\omega^2 d),(\omega^2 b)']$} & $\{[6,(R_2,R_2'),(R_2,R_2')],[4,(R_3,R_3'),(2,3)]\}$    & $\frac{v_2}{6}+\frac{v_3}{24}$& $ \mathfrak{B}^{(214)}  \sim {\cal O} (e^{- \frac{4 v_2 + v_3}{24}})  $  \\
$L^{(2)}\cdot \tilde H_u^{(1)} \Anti_b^{(5)}$ &  &  $\{[6,(R_2,R_2'),(R_2,R_2')],[4,(R_3,R_3'),(2,3)]\}$    & $\frac{v_2}{6}+\frac{v_3}{24}$& $ \mathfrak{B}^{(215)}  \sim {\cal O} (e^{- \frac{4 v_2 + v_3}{24}})  $  \\

$L^{(3)}\cdot \tilde H_u^{(1)} \Anti_b^{(2)}$ &  &  $\{[6,(R_2,R_2'),(R_2,R_2')],[(P_3,P_3'),(R_3,R_3'),1)]\}$    & $\frac{v_2}{6}+\frac{v_3}{24}$& $ \mathfrak{B}^{(312)}  \sim {\cal O} (e^{- \frac{4 v_2 + v_3}{24}})  $  \\
$L^{(3)}\cdot \tilde H_u^{(1)} \Anti_b^{(4)}$ &  &  $\{[6,(R_2,R_2'),(R_2,R_2')],[(P_3,P_3'),(R_3,R_3'),(2,3)]\}$    & $\frac{v_2}{6}+\frac{3v_3}{8}$& $ \mathfrak{B}^{(314)}  \sim {\cal O} (e^{- \frac{4 v_2 + 9 v_3}{24}})  $  \\
$L^{(3)}\cdot \tilde H_u^{(1)} \Anti_b^{(5)}$ &  &  $\{[6,(R_2,R_2'),(R_2,R_2')],[(P_3,P_3'),(R_3,R_3'),(2,3)]\}$    & $\frac{v_2}{6}+\frac{3v_3}{8}$& $ \mathfrak{B}^{(315)}  \sim {\cal O} (e^{- \frac{4v_2+9 v_3}{24}})  $  \\

$L^{(1)}\cdot \tilde H_u^{(1)} \Anti_b^{(3)}$ & & $\{[(R_2,R_2'),(R_2,R_2'),6],[4,(R_3,R_3'),(2,3)]\}$    & $\frac{v_2}{6}+\frac{v_3}{24}$& $ \mathfrak{B}^{(113)}  \sim {\cal O} (e^{- \frac{4v_2+ v_3}{24}})  $  \\
$L^{(1)}\cdot \tilde H_u^{(1)} \Anti_b^{(4)}$ &  &  $\{[(R_2,R_2'),[4,(R_3,R_3'),(2,3)]\}$    & $\frac{v_3}{24}$& $ \mathfrak{B}^{(114)}  \sim {\cal O} (e^{- \frac{ v_3}{24}})  $  \\
$L^{(1)}\cdot \tilde H_u^{(1)} \Anti_b^{(5)}$ & & $\{[(R_2,R_2')],[4,(R_3,R_3'),(2,3)]\}$    & $\frac{v_3}{24}$& $ \mathfrak{B}^{(115)}  \sim {\cal O} (e^{- \frac{ v_3}{24}})  $  \\
\hline

$L^{(5)}\cdot \tilde H_u^{(2)} \Anti_b^{(4)}$ & \multirow{8}{*}{$[b,(\omega d),(\omega b)']$} & $\{[5,(S_2,S_2'),(S_2,S_2')],[4,(S_3,S_3'),(2,3)]\}$    & $\frac{v_2}{6}+\frac{v_3}{24}$& $ \mathfrak{B}^{(524)}  \sim {\cal O} (e^{- \frac{4 v_2 + v_3}{24}})  $  \\
$L^{(5)}\cdot \tilde H_u^{(2)} \Anti_b^{(5)}$ &  &  $\{[5,(S_2,S_2'),(S_2,S_2')],[4,(S_3,S_3'),(2,3)]\}$    & $\frac{v_2}{6}+\frac{v_3}{24}$& $ \mathfrak{B}^{(525)}  \sim {\cal O} (e^{- \frac{4 v_2 + v_3}{24}})  $  \\

$L^{(4)}\cdot \tilde H_u^{(2)} \Anti_b^{(2)}$ &  &  $\{[5,(S_2,S_2'),(S_2,S_2')],[(P_3,P_3'),(S_3,S_3'),1)]\}$    & $\frac{v_2}{6}+\frac{v_3}{24}$& $ \mathfrak{B}^{(422)}  \sim {\cal O} (e^{- \frac{4 v_2 + v_3}{24}})  $  \\
$L^{(4)}\cdot \tilde H_u^{(2)} \Anti_b^{(4)}$ &  &  $\{[5,(S_2,S_2'),(S_2,S_2')],[(P_3,P_3'),(S_3,S_3'),(2,3)]\}$    & $\frac{v_2}{6}+\frac{3v_3}{8}$& $ \mathfrak{B}^{(424)}  \sim {\cal O} (e^{- \frac{4 v_2 + 9 v_3}{24}})  $  \\
$L^{(4)}\cdot \tilde H_u^{(2)} \Anti_b^{(5)}$ &  &  $\{[5,(S_2,S_2'),(S_2,S_2')],[(P_3,P_3'),(S_3,S_3'),(2,3)]\}$    & $\frac{v_2}{6}+\frac{3v_3}{8}$& $ \mathfrak{B}^{(425)}  \sim {\cal O} (e^{- \frac{4v_2+9 v_3}{24}})  $  \\

$L^{(6)}\cdot \tilde H_u^{(2)} \Anti_b^{(3)}$ & & $\{[(S_2,S_2'),(S_2,S_2'),5],[4,(S_3,S_3'),(2,3)]\}$    & $\frac{v_2}{6}+\frac{v_3}{24}$& $ \mathfrak{B}^{(623)}  \sim {\cal O} (e^{- \frac{4v_2+ v_3}{24}})  $  \\
$L^{(6)}\cdot \tilde H_u^{(2)} \Anti_b^{(4)}$ &  &  $\{[(S_2,S_2'),[4,(S_3,S_3'),(2,3)]\}$    & $\frac{v_3}{24}$& $ \mathfrak{B}^{(624)}  \sim {\cal O} (e^{- \frac{ v_3}{24}})  $  \\
$L^{(6)}\cdot \tilde H_u^{(2)} \Anti_b^{(5)}$ & & $\{[(S_2,S_2')],[4,(S_3,S_3'),(2,3)]\}$    & $\frac{v_3}{24}$& $ \mathfrak{B}^{(625)}  \sim {\cal O} (e^{- \frac{ v_3}{24}})  $  \\
\hline
\end{tabular}
\caption{Overview of the allowed cubic couplings involving the alternative candidates for the right-handed neutrinos. The third column lists the (triangular) worldsheet instantons $[x]$ or $[x,y,z]$ with indicated apexes $x,y,z$ on $T_{(i=2,3)}^2$ for the respective cubic couplings. The fourth column presents the corresponding area for the worldsheet instantons expressed in terms of the areas $v_i$ of the two-tori $T_{(i=2,3)}^2$, and the last column shows the scaling of the coupling constant corresponding to the considered cubic coupling. \label{tab:YukaCouplingMSSMpartVI}}
\end{center}
\end{sidewaystable}
%%%%%%%%%%%%%%%%%%%%%%%%%%%%%%%%

%%%%%%%%%%%%%%%%%%%%%%%%%%%%%%%%
\begin{sidewaystable}[h]
\begin{center}
\hspace*{-0.6in}
\begin{tabular}{|c|c|c|c|c|}
\hline \multicolumn{5}{|c|}{\bf Cubic couplings for the superpotential (\ref{Eq:SumDFSZ}) of a global 5-stack MSSM (part VII)} \\
\hline
\hline {\bf Coupling} & {\bf Sequence} & {\bf Triangles on $T_{(2)}^2 \times T_{(3)}^2$} & {\bf Enclosed Area} & {\bf Parameter}\\
\hline 
\hline
$L^{(5)}\cdot \tilde H_u^{(2)} \Adj_d^{(4)}$ &  \multirow{8}{*}{$[b,(\omega d),(\omega^2 d)]$} & $\{ [5,(R_2,R_2'),(2,3)], [4,(P_3,P_3'),(P_3,P_3')] \}$ & $\frac{v_2}{24} + \frac{v_3}{6}$ & $  \mathfrak{A}^{(524)} \sim {\cal O}\left(e^{- \frac{v_2 + 4v_3}{24} }\right)$\\
$L^{(5)}\cdot \tilde H_u^{(2)} \Adj_d^{(5)}$ && $\{ [5,(R_2,R_2'),(2,3)], [4,(P_3,P_3'),(P_3,P_3')] \}$ & $\frac{v_2}{24} + \frac{v_3}{6}$ & $  \mathfrak{A}^{(525)} \sim {\cal O}\left(e^{- \frac{v_2 + 4v_3}{24} }\right)$\\

$L^{(4)}\cdot \tilde H_u^{(2)} \Adj_d^{(2)}$ && $\{ [5,(R_2,R_2'),(2,3)], [4,(P_3,P_3'),(P_3,P_3')] \}$ & $\frac{v_2}{24} + \frac{v_3}{6}$ & $  \mathfrak{A}^{(422)} \sim {\cal O}\left(e^{- \frac{v_2 + 4v_3}{24} }\right)$\\
$L^{(4)}\cdot \tilde H_u^{(2)} \Adj_d^{(4)}$ && $\{ [5,(R_2,R_2'),(2,3)], [(P_3,P_3')] \}$ & $\frac{v_2}{24} $ & $  \mathfrak{A}^{(424)} \sim {\cal O}\left(e^{- \frac{v_2 }{24} }\right)$\\
$L^{(4)}\cdot \tilde H_u^{(2)} \Adj_d^{(5)}$ && $\{ [5,(R_2,R_2'),(2,3)], [(P_3,P_3')] \}$ & $\frac{v_2}{24} $ & $  \mathfrak{A}^{(425)} \sim {\cal O}\left(e^{- \frac{v_2 }{24} }\right)$\\

$L^{(6)}\cdot \tilde H_u^{(2)} \Adj_d^{(3)}$ && $\{ [(S_2,S_2'),(R_2,R_2'),1], [4,(P_3,P_3'),(P_3,P_3')] \}$ & $\frac{v_2}{24} + \frac{v_3}{6}$ & $  \mathfrak{A}^{(623)} \sim {\cal O}\left(e^{- \frac{v_2 + 4v_3}{24} }\right)$\\
$L^{(6)}\cdot \tilde H_u^{(2)} \Adj_d^{(4)}$ && $\{ [(S_2,S_2'),(R_2,R_2'),(2,3)],  [4,(P_3,P_3'),(P_3,P_3')] \}$ & $\frac{3v_2}{8} + \frac{v_3}{6} $ & $  \mathfrak{A}^{(624)} \sim {\cal O}\left(e^{- \frac{9v_2 + 4v_3}{24} }\right)$\\
$L^{(6)}\cdot \tilde H_u^{(2)} \Adj_d^{(5)}$ && $\{ [(S_2,S_2'),(R_2,R_2'),(2,3)],  [4,(P_3,P_3'),(P_3,P_3')]  \}$ & $\frac{3v_2}{8} + \frac{v_3}{6} $ & $  \mathfrak{A}^{(625)} \sim {\cal O}\left(e^{- \frac{9v_2 + 4v_3 }{24} }\right)$\\
\hline
$L^{(2)}\cdot \tilde H_u^{(1)} \Adj_d^{(4)}$ &  \multirow{8}{*}{$[b,(\omega^2 d),(\omega d)]$} & $\{ [6,(S_2,S_2'),(2,3)], [4,(P_3,P_3'),(P_3,P_3')] \}$ & $\frac{v_2}{24} + \frac{v_3}{6}$ & $  \mathfrak{A}^{(214)} \sim {\cal O}\left(e^{- \frac{v_2 + 4v_3}{24} }\right)$\\
$L^{(2)}\cdot \tilde H_u^{(1)} \Adj_d^{(5)}$ && $\{ [6,(S_2,S_2'),(2,3)], [4,(P_3,P_3'),(P_3,P_3')] \}$ & $\frac{v_2}{24} + \frac{v_3}{6}$ & $  \mathfrak{A}^{(215)} \sim {\cal O}\left(e^{- \frac{v_2 + 4v_3}{24} }\right)$\\

$L^{(3)}\cdot \tilde H_u^{(1)} \Adj_d^{(2)}$ && $\{ [6,(S_2,S_2'),(2,3)], [(P_3,P_3'),(P_3,P_3'),4] \}$ & $\frac{v_2}{24} + \frac{v_3}{6}$ & $  \mathfrak{A}^{(312)} \sim {\cal O}\left(e^{- \frac{v_2 + 4v_3}{24} }\right)$\\
$L^{(3)}\cdot \tilde H_u^{(1)} \Adj_d^{(4)}$ && $\{ [6,(S_2,S_2'),(2,3)], [(P_3,P_3')] \}$ & $\frac{v_2}{24} $ & $  \mathfrak{A}^{(314)} \sim {\cal O}\left(e^{- \frac{v_2 }{24} }\right)$\\
$L^{(3)}\cdot \tilde H_u^{(1)} \Adj_d^{(5)}$ && $\{ [6,(S_2,S_2'),(2,3)], [(P_3,P_3')] \}$ & $\frac{v_2}{24} $ & $  \mathfrak{A}^{(315)} \sim {\cal O}\left(e^{- \frac{v_2 }{24} }\right)$\\

$L^{(1)}\cdot \tilde H_u^{(1)} \Adj_d^{(3)}$ && $\{ [(S_2,S_2'),(R_2,R_2'),1], [4,(P_3,P_3'),(P_3,P_3')] \}$ & $\frac{v_2}{24} + \frac{v_3}{6}$ & $  \mathfrak{A}^{(113)} \sim {\cal O}\left(e^{- \frac{v_2 + 4v_3}{24} }\right)$\\
$L^{(1)}\cdot \tilde H_u^{(1)} \Adj_d^{(4)}$ && $\{ [(S_2,S_2'),(R_2,R_2'),(2,3)],  [4,(P_3,P_3'),(P_3,P_3')] \}$ & $\frac{3v_2}{8} + \frac{v_3}{6} $ & $  \mathfrak{A}^{(114)} \sim {\cal O}\left(e^{- \frac{9v_2 + 4v_3}{24} }\right)$\\
$L^{(1)}\cdot \tilde H_u^{(1)} \Adj_d^{(5)}$ && $\{ [(S_2,S_2'),(R_2,R_2'),(2,3)],  [4,(P_3,P_3'),(P_3,P_3')]  \}$ & $\frac{3v_2}{8} + \frac{v_3}{6} $ & $  \mathfrak{A}^{(115)} \sim {\cal O}\left(e^{- \frac{9v_2 + 4v_3 }{24} }\right)$\\\hline
\end{tabular}
\caption{Overview of the allowed cubic couplings involving the alternative candidates for the right-handed neutrinos. The third column lists the (triangular) worldsheet instantons $[x]$ or $[x,y,z]$ with indicated apexes $x,y,z$ on $T_{(i=2,3)}^2$ for the respective cubic couplings. The fourth column presents the corresponding area for the worldsheet instantons expressed in terms of the areas $v_i$ of the two-tori $T_{(i=2,3)}^2$, and the last column shows the scaling of the coupling constant corresponding to the considered cubic coupling. \label{tab:YukaCouplingMSSMpartVII}}
\end{center}
\end{sidewaystable}
%%%%%%%%%%%%%%%%%%%%%%%%%%%%%%%%

\clearpage
%%%%%%%%%%%%%%%%%%%%%%%%%%%%%%%%%%%%%%%%%%%%%%%%%%%%%%%%%%%%%%%%%%%%%%%%%%%%%%%%%%%
%%%%%%%%%%%%%%%%%%%%%%%%%%%%%%%%%%%%%%%%%%%%%%%%%%%%%%%%%%%%%%%%%%%%%%%%%%%%%%%%%%%
%%%%%%%%%%%%%%%%%%%%%%%%%%%%%%%%%%%%%%%%%%%%%%%%%%%%%%%%%%%%%%%%%%%%%%%%%%%%%%%%%%%
%%%%%%%%%%%%%%%%%%%%%%%%%%%%%%%%%%%%%%%%%%%%%%%%%%%%%%%%%%%%%%%%%%%%%%%%%%%%%%%%%%%
\section{Phenomenology of Global L-R Symmetric Models}\label{S:LRsymModels-pheno}

%%%%%%%%%%%%%%%%%%%%%%%%%%%%%%%%%%%%%%%%%%%%%%%%%%%%%%%%%%%%%%%%%%%%%%%%%%%%%%%%%%%
%%%%%%%%%%%%%%%%%%%%%%%%%%%%%%%%%%%%%%%%%%%%%%%%%%%%%%%%%%%%%%%%%%%%%%%%%%%%%%%%%%%
\subsection{Searching for Left-Right Symmetric D6-Brane Models}\label{Ss:LR-models}

The starting point for our search of {\it global} left-right symmetric models is the same as the one for the MSSM-like models, as formulated in~(\ref{Eq:D6ConfQCDLeftMSSM}). 
 More explicitly, $\varrho$-independent D6-brane configurations with three chiral left-handed quarks are only realisable when the {\it QCD} stack is parallel to the $\OR$-plane, while the $SU(2)_L$ stack wraps a fractional three-cycle parallel to the $\OR\Z_2^{(1)}$-plane and supports an enhanced $USp(2)$ gauge group. It is also understood that either the $\OR\Z_2^{(2)}$-plane or the $\OR\Z_2^{(3)}$-plane takes on the r\^ole of the exotic O6-plane. 
To obtain left-right symmetric models, the gauge group has to be completed with a right-symmetric $SU(2)_c$ and an Abelian $U(1)_d$ gauge group with wrapping numbers as classified in
 table~\ref{tab:BulkCyclesV=0abAA}, and without overshooting the bulk RR tadpole cancellation conditions~(\ref{Eq:BulkRRRhoIndMSSM}). Analogously to the last row  in table~\ref{tab:4stackMSSMrhoIndepbAA}, the only possible configuration on the {\bf bAA} lattice not violating~(\ref{Eq:BulkRRRhoIndMSSM}) consists in taking the $c$-stack and $d$-stack both parallel to the $\OR$-plane. The left-right symmetry of the gauge group requires the $c$-stack to support a $U(2)$ or $USp(2)$ non-Abelian gauge group, while intersecting with the {\it QCD} stack to yield three chiral generations of right-handed quark doublets $Q_R\equiv(u_R, d_R)$. In case the $c$-stack supports a $U(2)$ gauge group, the three chiral generations of right-handed quarks arise for the net-chirality $\left| \chi^{ac} + \chi^{ac'} \right| = 3$, where the sign has to be opposite to the sign of the net-chirality $\chi^{ab} \equiv \chi^{ab'}$. For an enhanced $USp(2)_c$ gauge group, the net-chirality associated to the right handed quarks has to satisfy $|\chi^{ac}| \equiv |\chi^{ac'}|=3$ with $\sgn(\chi^{ac}) = - \sgn(\chi^{ab}) $
 instead. Recall from the discussion in section~\ref{Ss:IntersectSummary} that D6-brane configurations, where the $c$-stack is parallel to the $\OR$-plane, do not give rise to three chiral generations of right-handed quarks -- neither for a $U(2)_c$ nor a $USp(2)_c$ group,
 implying that the {\bf bAA} lattice does not allow for any $\varrho$-independent {\it global} left-right symmetric models. 
 Note that the geometric conditions on the fractional three-cycles associated to candidate $SU(2)_R$ branes are less stringent in comparison to the ones for the {\it left} stack. That is to say, the {\it right} stack can be accompanied by (chiral) matter in the symmetric and/or adjoint representation, and the only requirement we impose for the $SU(2)_R$-stack is the existence of three chiral generations of right-handed quarks.

Turning to the {\bf aAA} lattice configuration, one notices from table~\ref{tab:4stackLRSMrhoIndepaAA} that the potential combinations of bulk orbits for the $c$-stack and $d$-stack are considerably more numerous than for the {\bf bAA} lattice. But also here, the requirement to have three chiral generations of right-handed quarks $(u_R, d_R)$ eliminates most combinations. 
On the other hand, the condition to obtain three chiral generations of left-handed leptons, i.e.~$\left|\chi^{bd}\right| = 3$ given that the $b$-stack has to be $\OR$-invariant in $\varrho$-independent configurations without adjoint/symmetric representations of $SU(2)_L$ in the spectrum,  
 does not constrain any of the bulk three-cycle combinations in table~\ref{tab:4stackLRSMrhoIndepaAA}. Combining all three requirements (bulk RR tadpoles and three chiral generations of right-handed quarks and left-handed leptons) leaves us with only the combinations (8,10,12) of bulk three-cycles to realise $\varrho$-independent left-right symmetric models on the {\bf aAA} lattice.
%%%%%%%%%%%%%%%%%%%%%%
\mathtab{\hspace*{-24mm}
\begin{array}{|c|c|c||c|c||c|c|}
\hline \multicolumn{7}{|c|}{\text{\bf Four-stack combinations with gauge group $U(3)_a\times USp(2)_b \times U(2)_c||USp(2)_c \times U(1)_d$ on aAA}} \\
\hline
\hline & \text{$c$-stack} & \text{$d$-stack}  & \text{RR tadpoles: } \sum_{x\in \{a,b,c,d\}} N_x (2 P_x + Q_x) & \leq 32 & 3\, q_R & 3\, L  \\ 
\hline
1&\OR & \OR & 2N_a + 6 N_b + 2N_c + 2N_d = 18  & {\color{mygr} \checkmark} & {\color{red}\lightning}& {\color{mygr} \checkmark}_{9840} \\
2&\OR \text{ with } USp(2)_c & \OR & 2N_a + 6 N_b + 2N_c + 2N_d = 16  & {\color{mygr} \checkmark} & {\color{red}\lightning} & {\color{mygr} \checkmark}_{2304}   \\
3&\OR & \OR\Z_{2}^{(1)} & 2N_a + 6 N_b + 2N_c + 6N_d = 22  &{\color{mygr} \checkmark}& {\color{red}\lightning}&{\color{mygr} \checkmark}_{198720} \\
4&\OR \text{ with } USp(2)_c  & \OR\Z_{2}^{(1)} & 2N_a + 6 N_b + 2N_c + 6N_d = 20  &{\color{mygr} \checkmark} & {\color{red}\lightning} & {\color{mygr} \checkmark}_{46080}\\
5&\OR  &(1,0;2,1;3,-1)  & 2N_a + 6 N_b + 2 N_c + 14 N_d = 30 &{\color{mygr} \checkmark} &  {\color{red}\lightning} & {\color{mygr} \checkmark}_{9936} \\
6&\OR  \text{ with } USp(2)_c  &(1,0;2,1;3,-1)  & 2N_a + 6 N_b + 2 N_c + 14 N_d = 28 &{\color{mygr} \checkmark} & {\color{red}\lightning}&{\color{mygr} \checkmark}_{2304}  \\
7&\OR\Z_{2}^{(1)}& \OR& 2N_a + 6 N_b + 6 N_c + 2N_d =  26 & {\color{mygr} \checkmark} &{\color{red}\lightning} &{\color{mygr} \checkmark}_{9984} \\
8&\OR\Z_{2}^{(1)} \text{ with } USp(2)_c   & \OR& 2N_a + 6 N_b + 6 N_c + 2N_d =  20 & {\color{mygr} \checkmark}  &   {\color{mygr} \checkmark}_{59616} & {\color{mygr} \checkmark}_{2304} ( {\color{mygr} \checkmark}_{288} ) \\
9&\OR\Z_{2}^{(1)}& \OR\Z_{2}^{(1)} &2N_a + 6 N_b + 6 N_c + 6 N_d = 30 &{\color{mygr} \checkmark} & {\color{red}\lightning}  & {\color{mygr} \checkmark}_{197760} \\
10&\OR\Z_{2}^{(1)} \text{ with } USp(2)_c & \OR\Z_{2}^{(1)} &2N_a + 6 N_b + 6 N_c + 6 N_d = 24 &{\color{mygr} \checkmark} & {\color{mygr} \checkmark}_{59904}&  {\color{mygr} \checkmark}_{46080}  ({\color{mygr} \checkmark}_{5760} )  \\
11&\OR\Z_{2}^{(1)} &(1,0;2,1;3,-1)  & 2N_a + 6 N_b + 6 N_c + 14 N_d = 38  & {\color{red}\lightning}  &  {\color{red}\lightning}  & {\color{mygr} \checkmark}_{9984}  \\
12&\OR\Z_{2}^{(1)} \text{ with } USp(2)_c &(1,0;2,1;3,-1)  & 2N_a + 6 N_b + 6 N_c + 14 N_d = 32  & {\color{mygr} \checkmark} &{\color{mygr} \checkmark}_{73728}  &{\color{mygr} \checkmark}_{2304} ( {\color{mygr} \checkmark}_{288} )  \\
13&(1,0;2,1;3,-1) &\OR &2N_a + 6 N_b + 14 N_c + 2N_d= 42   & {\color{red}\lightning} & {\color{red}\lightning}& {\color{mygr} \checkmark}_{12288} \\
14&(1,0;2,1;3,-1)  & \OR\Z_{2}^{(1)}& 2N_a + 6 N_b + 14 N_c + 6 N_d = 46& {\color{red}\lightning}& {\color{red}\lightning} &{\color{mygr} \checkmark}_{245760}  \\ 
15&(1,0;2,1;3,-1)& (1,0;2,1;3,-1) &2N_a + 6 N_b + 14 N_c + 14 N_d = 54  & {\color{red}\lightning} & {\color{red}\lightning}& {\color{mygr} \checkmark}_{12240} \\
\hline
\end{array}
}{4stackLRSMrhoIndepaAA}{Combinations of supersymmetric bulk orbits for $c$- and $d$-stacks aiming at $\varrho$-independent configurations of left-right symmetric models on the {\bf aAA} lattice for $T^6/(\Z_2\times \Z_6\times \OR)$ with discrete torsion, $\eta=-1$, and exotic O6-plane, $\eta_{\OR\Z_2^{(2 \text{ or } 3)}}=-1$. The second and third column indicate the bulk orbit of the $c$-stack and $d$-stack, respectively, the fourth and fifth column test whether the bulk RR tadpole cancellation conditions (\ref{Eq:BulkRRRhoIndMSSM}) are not over-shot, the second-to-last column verifies if three right-handed quark generations can be realised through 
 $|\chi^{ac}| \equiv |\chi^{ac'}| = 3$ for $USp(2)_c$ or $|\chi^{ac} + \chi^{ac'}| = 3$ for $U(2)_c$, 
and the last column does the same for three left-handed lepton generations with  $|\chi^{bd}| \equiv   |\chi^{bd'}| = 3$, in all cases with consistent relative sign choices. 
For the three bulk orbit combinations (8,10,12) allowing for three chiral generations of right-handed quarks {\it and} left-handed leptons simultaneously, we note that the constraints are mutually compatible as indicated in parenthesis in the last column. The subscript indicates the number of combinatorial possibilities of $(\vec{\sigma}_x)$, $(\vec{\tau}_x)$ and {\it relative} $(-)^{\tau^{\Z_2^{(k)}}_{xy}}$ for $x,y\in \{a,b,c,d\}$ and with exotic O6-plane $\eta_{\OR\Z_2^{(3)}}=-1$. Equivalent results with the same number of combinatorial possibilities are valid for the exotic O6-plane $\eta_{\OR\Z_2^{(2)}}=-1$ upon permutation of two-torus indices $2 \leftrightarrow 3$.
}
%%%%%%%%%%%%%%%%%%%%%%%%%%%%

As the bulk RR tadpoles are saturated for the choice of bulk three-cycles in any combination of type n$^\circ$ 12, there is no room left to add `hidden' fractional D6-branes in order to compensate the twisted RR charges coming from the four D6-brane stacks with initial gauge group $U(3)_a\times USp(2)_b \times USp(2)_c \times U(1)_d$. Hence, the resulting D6-brane models associated to combination n$^\circ$ 12 only provide for {\it local} left-right symmetric models, given that none of the 
288 four-stack fractional D6-brane configurations is characterised by vanishing twisted RR tadpoles.\footnote{If we also take into consideration the constraint for right-handed leptons, $|\chi^{cd}|=3$, the number of {\it local} four-stack fractional D6-brane configurations reduces by a factor two to 144 {\it local} models.} 
Also here the number of independent fractional D6-brane configurations has been reduced by taking into account that configurations with identical {\it relative} $\Z_2^{(i)}$ eigenvalues, but identical {\it absolute} discrete displacements and Wilson lines give rise to the same chiral and non-chiral massless spectrum and low-energy effective field theory at the current state-of-the-art.

Combination n$^\circ$ 8 on the other hand allows for the construction of two prototypes of global left-right symmetric models, with the hidden sector gauge group as the defining difference between the prototypes. The hidden D6-brane stacks of the first prototype consist of two stacks of D6-branes wrapping fractional three-cycles parallel to the $\OR$-plane supporting the gauge factors $U(3)_{h_1}\times U(3)_{h_2}$. An explicit D6-brane configuration for prototype I is given in table~\ref{tab:6stackLRSaAAPrototypeI}, with the corresponding massless matter spectrum listed in table~\ref{tab:6stackLRSMaAAPrototypeISpectrum}. 
%%%%%%%%%%%%%%%%%%%%%%%%%%%%%%%%%%%%%%%%%%%%%%%%%%%%%%%%%%%%%%%%%%%%%%%%%%%
\mathtabfix{
\begin{array}{|c||c|c||c|c|c||c|}\hline 
\muc{7}{|c|}{\text{\bf D6-brane configuration for a 6-stack LRS model (prototype I)  on the {aAA} lattice}}
\\\hline \hline
&\text{\bf wrapping numbers} &\frac{\rm Angle}{\pi}&\text{\bf $\Z_2^{(i)}$ eigenvalues}  & (\vec \tau) & (\vec \sigma)& \text{\bf gauge group}\\
\hline \hline
 a&(1,0;1,0;1,0)&(0,0,0)&(+++)&(0,1,1) & (0,1,1)& U(3)\\
 b&(1,0;-1,2;1,-2)&(0,\frac{1}{2},-\frac{1}{2})&(+++)&(0,1,0) & (0,1,0)&USp(2)\\
 c&(1,0;-1,2;1,-2)&(0, \frac{1}{2},-\frac{1}{2})&(-+-)&(0,1,0) & (0,1,0)&USp(2)\\ 
  d&(1,0;1,0;1,0)&(0, 0,0)&(+--)&(0,1,1) & (0,1,1)& U(1)\\
  \hline
    h_1&(1,0;1,0;1,0)&(0,0,0)&(+++)&(0,0,0) & (0,0,0)& U(3)\\
        h_2&(1,0;1,0;1,0)&(0,0,0)&(+--)&(0,0,0) & (0,0,0)& U(3)\\
 \hline
\end{array}
}{6stackLRSaAAPrototypeI}{D6-brane configuration for a six-stack Left-Right Symmetric model (prototype I) with initial gauge group 
$SU(3)_a\times USp(2)_b \times USp(2)_c\times SU(3)_{h_1}\times SU(3)_{h_2}\times  U(1)_a \times U(1)_d \times U(1)_{h_1} \times U(1)_{h_2}$ on the {\bf aAA} lattice of the orientifold $T^6/(\Z_2 \times \Z_6 \times \OR)$ with discrete torsion ($\eta=-1$) and the $\OR\Z_2^{(3)}$-plane as the exotic O6-plane ($\eta_{\OR\Z_2^{(3)}}=-1$).}
%%%%%%%%%%%%%%%%%%%%%%%%%%%%%%%%%%%%%%%%%%%%%%%%%%%%%%%%%%%%%%%%%%%%%%%%%%%
%%%%%%%%%%%%%%%%%%%%%%%%%%%
\mathtab{\hspace*{-22mm}
\begin{array}{|c||c|c||c|c|c|}
\hline \multicolumn{6}{|c|}{\text{\bf Overview of the spectrum for prototype I LRS model on the {aAA} lattice}}\\
\hline \hline
\text{sector} & \text{state} & (SU(3)_a \times USp(2)_b \times USp(2)_c \times SU(3)_{h_1}\times SU(3)_{h_2} )_{U(1)_a \times U(1)_d \times U(1)_{h_1} \times U(1)_{h_2}}& \Z_2 & \Z_3 & \Z_6\\
\hline 
ab \equiv ab'& Q_L &3 \times  (\3,\2,\1,\1,\1)_{(1,0,0,0)} & 1& 1 &1 \\
ac \equiv ac'& Q_R &3 \times  (\ov \3,\1,\2,\1,\1)_{(-1,0,0,0)} &1&2 &5\\
ad&  & (\3,\1,\1,\1,\1)_{(1,-1,0,0)} + h.c.  &0& 1||2 &4||2\\ 
ad'& &2 \times \left[ (\3,\1,\1,\1,\1)_{(1,1,0,0)} + h.c.\right] &0&1||2&4||2\\
bc \equiv bc'& \!\!\!(H_u, H_d)\!\!\! &10 \times (\1,\2,\2,\1,\1)_{(0,0,0,0)} &0&0&0\\
bd \equiv b'd& L &3 \times (\1,\2,\1,\1,\1)_{(0,-1,0,0)} &1&0&3 \\
cd \equiv c'd & R &3 \times (\1,\1,\2,\1,\1)_{(0,1,0,0)} &1&0&3 \\
\hline \hline
ah_1& &2 \times  (\ov\3,\1,\1,\3,\1)_{(-1,0,1,0)}  &0&0&0\\
ah_2& &2 \times (\3,\1,\1,\1,\ov\3)_{(1,0,0,-1)}  &0&0&0\\
bh_1 \equiv b'h_1& &  (\1,\2,\1,\ov\3,\1)_{(0,0,-1,0)} &1 &2&5\\
bh_1 \equiv b'h_1& &  (\1,\2,\1,\ov\3,\1)_{(0,0,-1,0)} +h.c. & 1&2||1&5||1\\
bh_2 \equiv b'h_2& &  (\1,\2,\1,\1,\3)_{(0,0,0,1)} &1&1 &1\\
bh_2 \equiv b'h_2& &   (\1,\2,\1,\1,\3)_{(0,0,0,1)} + h.c. &1& 1||2&1||5 \\
ch_1\equiv  c'h_1& & (\1,\1,\2,\ov\3,\1)_{(0,0,-1,0)} &1&2 &5\\
ch_1 \equiv c'h_1 & & (\1,\1,\2,\ov\3,\1)_{(0,0,-1,0)} +h.c. &1& 2||1&5||1 \\
ch_2 \equiv c'h_2& &(\1,\1,\2,\1,\3)_{(0,0,0,1)} &1&1&1 \\
ch_2 \equiv  c'h_2& & (\1,\1,\2,\1,\3)_{(0,0,0,1)} + h.c.&1&1||2&1||5 \\
dh_1& & 2 \times (\1,\1,\1,\ov\3,\1)_{(0,1,-1,0)} & 0&2&2 \\
dh_2& &2 \times(\1,\1,\1,\1,\3)_{(0,-1,0,1)}  &0&1&4 \\
h_1 h_2 &&(\1,\1,\1,\3,\ov\3)_{(0,0,1,-1)} + h.c.  &0&0 &0\\
h_1 h_2'&& 2\times\left[ (\1,\1,\1,\3,\3)_{(0,0,1,1)} + h.c.  \right] &0&2||1 &2||4 \\
\hline \hline
 aa'& & 2\times[ ({\bf 3_{A}},\1,\1,\1,\1)_{(2,0,0,0)} + h.c.] &0&2||1&2||4 \\
  bb' \equiv bb & &  5 \times  (\1,\1_{\bf A},\1,\1,\1)_{(0,0,0,0)}&0&0 &0  \\
   cc' \equiv cc & &  5 \times  (\1,\1,\1_{\bf A},\1,\1)_{(0,0,0,0)}& 0&0 &0\\
    h_1 h_1'& & 2\times[ (\1,\1,\1,{\bf 3_{A}},\1)_{(0,0,2,0)} + h.c.] &0&2||1 &2||4 \\
     h_2 h_2'& & 2\times[ (\1,\1,\1,\1,{\bf 3_{A}})_{(0,0,0,2)} + h.c.] &0&2||1 &2||4 \\
\hline
\end{array}
}{6stackLRSMaAAPrototypeISpectrum}{Chiral and non-chiral massless matter spectrum for the global six-stack D6-brane model (prototype I) with initial gauge group $U(3)_a\times USp(2)_b \times USp(2)_c \times U(1)_d \times U(3)_{h_1} \times U(3)_{h_2}$ corresponding to the configuration in table~\ref{tab:6stackLRSaAAPrototypeI}. The last three columns list the charges of the massless open string states under the discrete $\Z_n$ symmetries identified in section~\ref{Ss:LRSDiscrete}.
}
%%%%%%%%%%%%%%%%%%%%%%%%%%

\clearpage

To obtain prototype II left-right symmetric models, we have to add two D6-branes wrapping fractional three-cycles parallel to the $\OR\Z_2^{(1)}$-plane with suitable $\Z_2^{(i)}$-eigenvalues $(-1)^{\vec{\tau}^{\Z_2^{(k)}}_x}$, discrete displacements $(\vec{\sigma}_x)$ and discrete Wilson lines $(\vec{\tau}_x)$ with $x\in \{h_1, h_2\}$. 
In the prototype II models, the hidden D6-brane stacks support the Abelian gauge group $U(1)_{h_1} \times U(1)_{h_2}$, as indicated in the explicit example in table~\ref{tab:6stackLRSaAAPrototypeII} with the corresponding massless open string spectrum summarised in table~\ref{tab:6stackLRSMaAAPrototypeIISpectrum}. 
%%%%%%%%%%%%%%%%%%%%%%%%%%%%%%%%%%%%%%%%%%%%%%%%%%%%%%%%%%%%%%%%%%%%%%%%%%%
\mathtabfix{
\begin{array}{|c||c|c||c|c|c||c|}\hline 
\muc{7}{|c|}{\text{\bf D6-brane configuration for a 6-stack LRS model (prototype II)  on the {aAA} lattice}}
\\\hline \hline
&\text{\bf wrapping numbers} &\frac{\rm Angle}{\pi}&\text{\bf $\Z_2^{(i)}$ eigenvalues}  & (\vec \tau) & (\vec \sigma)& \text{\bf gauge group}\\
\hline \hline
 a&(1,0;1,0;1,0)&(0,0,0)&(+++)&(0,1,1) & (0,1,1)& U(3)\\
 b&(1,0;-1,2;1,-2)&(0,\frac{1}{2},-\frac{1}{2})&(+++)&(0,1,0) & (0,1,0)&USp(2)\\
 c&(1,0;-1,2;1,-2)&(0, \frac{1}{2},-\frac{1}{2})&(-+-)&(0,1,0) & (0,1,0)&USp(2)\\ 
  d&(1,0;1,0;1,0)&(0, 0,0)&(+--)&(0,1,1) & (0,1,1)& U(1)\\
  \hline
    h_1&(1,0;-1,2;1,-2)&(0,\frac{1}{2},-\frac{1}{2})&(+++)&(0,0,0) & (0,0,0)& U(1)\\
        h_2&(1,0;-1,2;1,-2)&(0,\frac{1}{2},-\frac{1}{2})&(+--)&(0,0,0) & (0,0,0)& U(1)\\
 \hline
\end{array}
}{6stackLRSaAAPrototypeII}{D6-brane configuration for a six-stack Left-Right Symmetric model (prototype II) with initial gauge group 
$SU(3)_a\times USp(2)_b \times USp(2)_c\times  U(1)_a \times U(1)_d \times U(1)_{h_1} \times U(1)_{h_2}$ on the {\bf aAA} lattice of the orientifold $T^6/(\Z_2 \times \Z_6 \times \OR)$ with discrete torsion ($\eta=-1$) and the $\OR\Z_2^{(3)}$-plane as the exotic O6-plane ($\eta_{\OR\Z_2^{(3)}}=-1$).}
%%%%%%%%%%%%%%%%%%%%%%%%%%%%%%%%%%%%%%%%%%%%%%%%%%%%%%%%%%%%%%%%%%%%%%%%%%%
%
%%%%%%%%%%%%%%%%%%%%%%%%%%%
\mathtab{
\hspace*{-0.5in}\begin{array}{|c||c|c||c|c|}
\hline \multicolumn{5}{|c|}{\text{\bf Overview of the spectrum for prototype II LRS model on the {aAA} lattice}}\\
\hline \hline
\text{sector} & \text{state} & (SU(3)_a \times USp(2)_b \times USp(2)_c )_{U(1)_a \times U(1)_d \times U(1)_{h_1} \times U(1)_{h_2}}& \widetilde{U(1)}_{B-L} &\Z_6\\
\hline 
ab \equiv ab'& Q_L &3 \times  (\3,\2,\1)_{(1,0,0,0)} & 1/3&1  \\
ac \equiv ac'& Q_R &3 \times  (\ov \3,\1,\2)_{(-1,0,0,0)}& -1/3&5 \\
ad& X^{ad} + \tilde X^{ad}  &(\3,\1,\1)_{(1,-1,0,0)} + h.c.  &\pm4/3& 4||2 \\ 
ad'& X^{ad'(i)} + \tilde X^{ad'(i)} &2 \times \left[ (\3,\1,\1)_{(1,1,0,0)} + h.c.\right] & \mp 2/3 &4||2\\
bc \equiv bc'& (H_u, H_d) &10 \times (\1,\2,\2)_{(0,0,0,0)} & 0& 0   \\
bd \equiv b'd& L &3 \times (\1,\2,\1)_{(0,-1,0,0)} & 1 & 3 \\
cd \equiv c'd & R &3 \times (\1,\1,\2)_{(0,1,0,0)} & -1 &3\\
\hline \hline
ah_1& &2 \times \left[ (\3,\1,\1)_{(1,0,-1,0)} + h.c. \right] & \pm 4/3&4||2 \\
ah_1'& &(\3,\1,\1)_{(1,0,1,0)} + h.c.& \mp 2/3 &4||2 \\
ah_2& &2 \times \left[ (\3,\1,\1)_{(1,0,0,-1)} + h.c. \right] & \mp 2/3 &4||2\\
ah_2'& &(\3,\1,\1)_{(1,0,0,1)} + h.c.& \pm 4/3&4||2\\
bh_1 \equiv b'h_1& &3 \times  (\1,\2,\1)_{(0,0,1,0)} & -1 & 3 \\
bh_1 \equiv b'h_1& &3 \times \left[ (\1,\2,\1)_{(0,0,-1,0)} +h.c. \right]& \pm 1& 3  \\
bh_2 \equiv b'h_2& &3 \times  (\1,\2,\1)_{(0,0,0,-1)} & -1 & 3 \\
bh_2 \equiv b'h_2& &3 \left[ \times  (\1,\2,\1)_{(0,0,0,1)} + h.c. \right] & \pm 1 & 3\\
ch_1\equiv  c'h_1& &3 \times (\1,\1,\2)_{(0,0,-1,0)} & 1 & 3 \\
ch_1 \equiv c'h_1 & &3 \times \left[  (\1,\1,\2)_{(0,0,1,0)} +h.c. \right]& \mp 1  & 3\\
ch_2 \equiv c'h_2& &3 \times (\1,\1,\2)_{(0,0,0,1)} & 1& 3 \\
ch_2 \equiv  c'h_2& &3 \left[ \times (\1,\1,\2)_{(0,0,0,1)} + h.c. \right]& \pm 1 & 3 \\
dh_1& X^{dh_1(i)} + \tilde X^{dh_1(i)}  &2 \times \left[ (\1,\1,\1)_{(0,1,-1,0)} + h.c. \right]  & 0 &0\\
dh_1'& X^{dh_1'} + \tilde X^{dh_1'} & (\1,\1,\1)_{(0,1,1,0)} + h.c.  & \mp 2&0 \\
d h_2 &X^{dh_2(i)} + \tilde X^{dh_2(i)}& 2\times \left[(\1,\1,\1)_{(0,1,0,-1)} + h.c. \right]  & \mp 2&0 \\
d h_2'&X^{dh_2'} + \tilde X^{dh_2'}& (\1,\1,\1)_{(0,1,0,1)} + h.c.  & 0&0\\
h_1 h_2& &5 \times \left[ (\1,\1,\1)_{(0,0,1,-1)} + h.c. \right]   & \mp 2&0 \\
h_1 h_2'& &6\times \left[ (\1,\1,\1)_{(0,0,1,1)} + h.c. \right] &0&0\\
\hline \hline
 aa'& & 2\times[ ({\bf 3_{A}},\1,\1)_{(2,0,0,0)} + h.c.] & \pm 2/3& 2||4 \\
  bb' \equiv bb & &  5 \times  (\1,\1_{\bf A},\1)_{(0,0,0,0)}& 0&0  \\
   cc' \equiv cc & &  5 \times  (\1,\1,\1_{\bf A})_{(0,0,0,0)}&0 &0  \\
    h_1h_1& & 4 \times (\1,\1,\1)_{(0,0,0,0)} &0 &0  \\
    h_2h_2& &4 \times (\1,\1,\1)_{(0,0,0,0)}& 0 &0  \\
\hline
\end{array}
}{6stackLRSMaAAPrototypeIISpectrum}{Chiral and non-chiral massless matter spectrum for the global six-stack D6-brane model with initial gauge group $U(3)_a\times USp(2)_b \times USp(2)_c \times U(1)_d \times U(1)_{h_1} \times U(1)_{h_2}$ corresponding to the configuration from table~\ref{tab:6stackLRSaAAPrototypeII}, with 
the massless $\widetilde{U(1)}_{B-L}$ symmetry listed in the third column. 
%and the charges under the discrete $\Z_6$ symmetry identified in section~\ref{Ss:LRSDiscrete} in the last column.
}
%%%%%%%%%%%%%%%%%%%%%%%%%%%%%

Note that the twisted RR tadpole cancellation conditions prevent the hidden D6-branes from supporting enhanced gauge groups of the $USp$ or $SO$ type.

 \clearpage

The D6-brane combination n$^\circ$~10 can give rise to two types of left-right symmetric models: five-stack models with the hidden D6-brane stack parallel to the $\OR$-plane and supporting a $U(4)_{h}$ hidden gauge group, or six-stack models with the two hidden D6-brane stacks parallel to the $\OR$-plane and the $\OR\Z_2^{(1)}$-plane, respectively, and each supporting a $U(1)_{h_i}$ hidden gauge group. 
A superficial analysis of the chiral and non-chiral massless open string spectrum reveals that various of these global six-stack models correspond to prototype II models 
as in table~\ref{tab:6stackLRSaAAPrototypeII} or the variants IIb and IIc in tables~\ref{tab:6stackLRSaAAPrototypeIIb} and~\ref{tab:6stackLRSaAAPrototypeIIc} of appendix~\ref{A:ProtoIIModels},
where now one of the hidden stacks $h_i$ that was parallel to the $\OR\Z_2^{(1)}$-plane has been permuted with the $d$-stack which before was parallel to the $\OR$-plane. 
This consideration suggests that the global six-stack models arising from combination n$^\circ$~10 with hidden gauge group $U(1)_{h_1}\times U(1)_{h_2}$ might form a subset of the  prototype II models identified from combination n$^\circ$~8. In order to verify this speculative statement, a more thorough analysis of the massless spectra of all 20736 global left-right symmetric six-stack models associated to combination n$^\circ$~10 has to be performed, which we postpone for future work.

Apart from the hidden gauge factors, there happens to be another appreciable difference between the two prototypes of left-right symmetric models: the absence of a massless $U(1)_{B-L}$ symmetry for prototype I and the presence of a generalized massless $\widetilde{U(1)}_{B-L}$ symmetry for prototype II, as we will show in the next section. This observation implies a different approach when identifying the left-handed and right-handed quarks and leptons. For the prototype~I model, the absence of a massless $U(1)_{B-L}$ symmetry might entice us to exchange the r\^ole of one of the hidden stacks $h_i$ with the {\it QCD}-$U(3)_a$-stack. Indeed, the chiral state in the $bh_2$ sector can equally be interpreted as a left-handed quark based on its quantum numbers.  Nevertheless, the lack of three generations prevents us to exchange the r\^oles of the $U(3)$-stacks and provides a solid argument for the identification of the chiral states presented in table~\ref{tab:6stackLRSMaAAPrototypeISpectrum}. In this interpretation, the chiral and non-chiral massless states in the $bh_1$, $bh_2$, $ch_1$ and $ch_2$ sectors form a portal between the visible sector and a dark sector, instead of being inherent to the visible sector.

An argument against exchanging the r\^oles of the $U(1)$ stacks in the prototype~II left-right symmetric models can be made based on the generalised massless $B-L$ symmetry defined in equation (\ref{Eq:GenerBLSymm}) below. More precisely, due to the presence of this $\widetilde{U(1)}_{B-L}$ symmetry the visible sector branes are uniquely determined, and the identification of the chiral states charged under the visible gauge group corresponds  unequivocally to the one presented in table~\ref{tab:6stackLRSMaAAPrototypeIISpectrum}.
The key observation leading to this conclusion results from considering the $d$-stack along the $\OR\Z_2^{(1)}$-plane instead of the set-up of table~\ref{tab:6stackLRSaAAPrototypeII}, which also requires exchanging the r\^ole of the $b$-stack and $c$-stack in order to correctly identify the left-handed and right-handed leptons. However, the combined exchange of the  r\^oles of the $b$-stack and $c$-stack conflicts with the desired representations for the left- and right-handed quarks under the $\widetilde{U(1)}_{B-L}$ gauge group, and thereby excludes a potential liberty to place the $d$-stack along the $\OR\Z_2^{(1)}$-plane.

The five-stack models with hidden gauge group $U(4)_h$ are new and form an entirely independent prototype for which bulk and twisted RR tadpole cancellation conditions are satisfied. As explained in appendix~\ref{A:5StackLRS} through an explicit example, the K-theory constraints for this prototype of five-stack models are, however, not fulfilled and the models can therefore not be considered as globally consistent models, but rather as {\it semi-local} models. An example of a five-stack model can be found in appendix~\ref{A:5StackLRS}, more explicitly in table~\ref{tab:5stackLRSaAAPrototypeI} with corresponding massless open string spectrum in table~\ref{tab:5stackLRSMaAAPrototypeISpectrum}.  In total we can identify 1296 {\it semi-local} fractional D6-brane configurations for the five bulk orbits presented in table~\ref{tab:5stackLRSaAAPrototypeI} and with different values of the $\Z_2^{(i)}$ eigenvalues $(-)^{\tau_x^{\Z_2^{(i)}}}$ and the discrete parameters $(\vec{\sigma}_x)$ and $(\vec{\tau}_x)$ with $x\in \{a,b,c,d,h\}$. Five-stack configurations with the same {\it relative} $\Z_2^{(i)}$ eigenvalues $(-)^{\Delta\tau^{\Z_2^{(k)}}_{xy}}$ and identical {\it absolute} discrete parameters $(\vec{\sigma}_x)$ and $(\vec{\tau}_x)$ with $x,y \in \{a,b,c,d,h\}$ are counted as a single configuration.

%%%%%%%%%%%%%%%%%%%%%
\begin{table}[h]
\begin{center}\hspace*{-15mm}
\begin{tabular}{|c||c|c|c|}
\hline \multicolumn{4}{|c|}{\bf Summary of $\varrho$-independent Left-Right Symmetric Models on the {\bf aAA} lattice} \\
\hline
\hline
{\bf Combination} & {\bf Type} & {\bf Hidden Gauge factor} &{\bf Number of Configurations} \\
\hline \hline 
n$^\circ$8 & 6-stack prototype I & $U(3)_{h_1} \times U(3)_{h_2} $ & 165888 (\text{\it semi-local}) \& 105408 (\text{\it global}) \\
n$^\circ$8  & 6-stack prototype II & $U(1)_{h_1} \times U(1)_{h_2} $ & 165888 (\text{\it semi-local}) \& 105984 (\text{\it global}) \\
 n$^\circ$10 & 6-stack prototype II & $U(1)_{h_1} \times U(1)_{h_2} $ & 912384 (\text{\it semi-local}) \& 20736 (\text{\it global}) \\
n$^\circ$10 & 5-stack & $U(4)_{h}$ & 1296 (\text{\it semi-local}) \\
 \hline
\end{tabular}
\caption{Overview of the $\varrho$-independent left-right symmetric models with vanishing bulk+twisted RR tadpoles on the {\bf aAA} lattice configuration for $T^6/(\Z_2\times\Z_6\times\OR)$ with discrete torsion ($\eta=-1$). The amount of configurations is given for $\eta_{\OR\Z_2^{(3)}}=-1$, yet the same numbers of configurations are valid for the other choice of exotic O6-plane, $\eta_{\OR\Z_2^{(2)}}=-1$, upon permutation of two-torus indices. 
The number of configurations corresponds to the number of independent combinatorial possibilities for $(\vec{\sigma}_x)$, $(\vec{\tau}_x)$ and $(-)^{\tau^{\Z_2^{(k)}}_x}$ with $x\in \{a,b,c,d,h_1,h_2\}$ for the six-stack models and $x\in \{a,b,c,d,h\}$ for the five-stack models, as explained in the main text.
\label{tab:SummaryLRSModelsaAA}}
\end{center}
\end{table}
%%%%%%%%%%%%%%%%%%%%%%%%%%%%%%%%%%%%%%%%
To finish this section, in table~\ref{tab:SummaryLRSModelsaAA} we give a summary of the various $\varrho$-independent left-right symmetric models that can be constructed on the {\bf aAA} lattice. We list the numbers for one particular choice of the exotic O6-plane, namely for the $\OR\Z_2^{(3)}$-plane, but remark that we cross-checked that the same summary is valid in case the $\OR\Z_2^{(2)}$-plane plays the r\^ole of the exotic O6-plane, as expected by the permutation symmetry of the two-tori $T^2_{(2)} \times T^2_{(3)}$.
A first observation concerns the prototype I and II left-right symmetric models, for which the K-theory constraints are not satisfied for all fractional D6-brane configurations with vanishing bulk and twisted RR tadpoles.
From the full set of fractional D6-brane configurations with vanishing RR tadpoles, 39\% (39\%) of prototype I (II) left-right symmetric models are fully global, as indicated in table~\ref{tab:SummaryLRSModelsaAA}. The numbers listed in that table count the different combinatorial possibilities for the $\Z_2^{(i)}$ eigenvalues, discrete parameters $(\vec{\sigma}_x)$ and $(\vec{\tau}_x)$, while identical {\it relative} $\Z_2^{(i)}$ eigenvalues $(-)^{\Delta\tau^{\Z_2^{(k)}}_{xy}}$  are counted as a single fractional D6-brane configuration for $x,y \in \{a,b,c,d,h_1,h_2\}$.

 Secondly, one can verify that the {\it semi-local} five-stack and six-stack models do not allow for a massless (generalised) $B-L$ symmetry. This observation implies a subtle difference between {\it semi-local} and {\it global} prototype II models, whose gauge group and spectrum are fully equivalent, and requires us to define ``prototype" more precisely: 
 the term  ``prototype" captures all fractional D6-brane models with six D6-brane stacks, whose bulk three-cycles are identical to the ones in table~\ref{tab:6stackLRSaAAPrototypeI} (prototype I) or in table~\ref{tab:6stackLRSaAAPrototypeII} (prototype II), and with the same left-right symmetric gauge structure and massless open string spectrum in the purely visible sector of table~\ref{tab:6stackLRSMaAAPrototypeISpectrum} or table~\ref{tab:6stackLRSMaAAPrototypeIISpectrum}, respectively. Within the prototypes, one can find subclasses of global left-right symmetric models whose chiral and non-chiral spectrum in the hidden sector slightly differs. 
 How many physically distinguishable subclasses\footnote{By physically distinguishable subclasses, we refer to two six-stack intersecting D6-brane models whose spectrum cannot be related to each other by a mere exchange $b\leftrightarrow c$, $h_1 \leftrightarrow h_2$. 
As usual, we also treat models as identical if they merely differ in the choice of orbifold/orientifold representant, e.g. $c \leftrightarrow c'$ or $h_{i=1,2} \leftrightarrow h_{i=1,2}'$.
 } 
 there exist requires a full comparison of the massless spectrum for all global D6-brane models, which is postponed for future research. In appendix~\ref{A:ProtoIIModels} we provide two other examples of six-stack intersecting D6-brane models fitting within the prototype II models.        

Apart from the six-stack left-right symmetric models presented above, we also searched for six-stack D6-brane models associated to combination n$^\circ$8 with hidden gauge groups $U(3)_{h_1}\times U(1)_{h_2}$, where the $h_1$-stack is parallel to the $\OR$-plane and the $h_2$-stack parallel to the $\OR\Z_2^{(1)}$-plane, and for six-stack D6-brane models associated to  combination n$^\circ$10 with hidden gauge groups $U(2)\times U(2)$, $USp(4)\times U(2)$ or $USp(4)\times USp(4)$, where the hidden $h_1$ and $h_2$-stack are both parallel to the $\OR$-plane. For all those combinations of hidden D6-brane stacks we observed that the resulting six-stack left-right symmetric models are able to satisfy the RR tadpole cancellation conditions, yet always violate some of the K-theory constraints, such that no global six-stack models can be found.

%\clearpage
%%%%%%%%%%%%%%%%%%%%%%%%%%%%%%%%%%%%%%%%%%%%%%%%%%%%%%%%%%%%%%%%%%%%%%%%%%%%%%%%%%%
%%%%%%%%%%%%%%%%%%%%%%%%%%%%%%%%%%%%%%%%%%%%%%%%%%%%%%%%%%%%%%%%%%%%%%%%%%%%%%%%%%%
\subsection{Discrete Symmetries}\label{Ss:LRSDiscrete}
Also for left-right symmetric models, a classification of discrete $\Z_n$ symmetries can be useful to constrain the cubic couplings among the massless open string states. 
At the same time, this computation will determine if the commonly required massless $U(1)_{B-L}$ symmetry exists.
That is why this section will be devoted to the search for discrete $\Z_n$ symmetries for the two prototype I and II examples of global six-stack left-right symmetric models presented in table~\ref{tab:6stackLRSaAAPrototypeI} and~\ref{tab:6stackLRSaAAPrototypeII}, respectively. 
We will briefly comment on the differences of the prototype IIb and IIc examples in tables~\ref{tab:6stackLRSaAAPrototypeIIb} and~\ref{tab:6stackLRSaAAPrototypeIIc} compared to the prototype II example of~\ref{tab:6stackLRSaAAPrototypeII} in appendix~\ref{A:ProtoIIModels}. The discrete symmetries for the {\it semi-local} five-stack left-right symmetric model will be discussed in appendix~\ref{A:5StackLRS}.

%%%%%%%%%%%%%%%%%%%%%%%%%
{\bf Prototype I Left-Right Symmetric Model}\\
Zooming in on the first prototype left-right symmetric model with hidden gauge group $U(3)_{h_1}\times U(3)_{h_2}$, we write down the necessary conditions (\ref{Eq:Zn-condition-nec}) 
on the existence of some $\Z_n$ gauge symmetry for the D6-brane configuration listed in table~\ref{tab:6stackLRSaAAPrototypeI}:
\begin{align}\hspace{-25mm}
&\left\{k_a \left(\begin{array}{c} 0 \\ 0 \\\hline 0 \\ 0 \\ 0 \\ -6 \\ 6 \\ 0
\end{array}\right)
+ k_d \left(\begin{array}{c} 0 \\  0 \\\hline 0 \\ 0 \\ 0 \\ -2 \\ 2 \\ 0
\end{array}\right)
+ k_{h_1} \left(\begin{array}{c} 0 \\  0 \\\hline 9 \\ 3 \\ 3 \\ 3 \\ 0 \\ 0 
\end{array}\right)
+ k_{h_2} \left(\begin{array}{c} 0 \\  0 \\\hline 9 \\ 3 \\ 3 \\ 3 \\ 0 \\ 0 
\end{array}\right) \right\}
\stackrel{!}{=} 0 \text{ mod } n, \\
&\left\{ k_a \left(\begin{array}{c} 0 \\ 0 \\ 0 \\ 0 \\\hline 6 \\ 6 \\ 0 \\ 0
\end{array}\right)
+ k_d \left(\begin{array}{c} 0 \\ 0 \\ 0 \\ 0 \\\hline -2 \\ -2 \\ 0 \\ 0
\end{array}\right)
+ k_{h_1} \left(\begin{array}{c} -6 \\ -6 \\ 0 \\ 0 \\\hline 0 \\ 0 \\ 0 \\ 0
\end{array}\right)
+ k_{h_2} \left(\begin{array}{c}  6 \\ 6 \\ 0 \\ 0 \\\hline 0 \\ 0 \\ 0 \\ 0
\end{array}\right) \right\}\stackrel{!}{=} 0 \text{ mod } n, 
\end{align}
which can be reduced to four linearly independent constraints, since various rows are trivially satisfied or can be related to each other:
\begin{equation}\label{Eq:NecConLRSProtoI}
\begin{array}{rcl}
3k_{h_1} + 3 k_{h_2}&\stackrel{!}{=}& 0 \text{ mod } n  ,\\
6k_a + 2 k_d&\stackrel{!}{=}&0 \text{ mod } n  ,\\
-6k_{h_1} + 6 k_{h_2}&\stackrel{!}{=}& 0 \text{ mod } n ,\\
6k_a - 2 k_d&\stackrel{!}{=}&0 \text{ mod } n .
\end{array}
\end{equation}
These four constraints have to be completed with the linearly independent constraints coming from the sufficient conditions (\ref{Eq:Zn-condition-suf}), which read for the D6-brane configuration in table~\ref{tab:6stackLRSaAAPrototypeI}:
\begin{align}\label{Eq:SuffConLRSprotoIRaw}
\left\{k_a\left(\begin{array}{c}
3\\
6\\
3\\
6\\
3\\
6\\
\hline
0\\
0\\
0\\
0\\
\hline
0\\
0\\
0\\
-6\\
0\\
0
\end{array}\right)+k_d\left(\begin{array}{c}
-1\\
-2\\
-1\\
-2\\
-1\\
-2\\
\hline
0\\
0\\
0\\
0\\
\hline
-1\\
0\\
0\\
-2\\
-2\\
1
\end{array}\right)+k_{h_1}\left(\begin{array}{c}
-3\\
0\\
-3\\
0\\
-3\\
0\\
\hline
0\\
0\\
0\\
0\\
\hline
3\\
6\\
3\\
3\\
3\\
0
\end{array}\right)+k_{h_2}\left(\begin{array}{c}
3\\
0\\
3\\
0\\
3\\
0\\
\hline
0\\
0\\
0\\
0\\
\hline
6\\
6\\
3\\
3\\
3\\
3
\end{array}\right)\right\}=0\text{  mod  }n.
\end{align}
A closer inspection of the sufficient conditions shows that the first block only leads to two independent conditions (row 1 and row 2), where the second constraint already appeared as one of the necessary conditions. The second block does not impose any additional constraint, as all conditions are trivially satisfied. The third block yields five independent conditions (row 11, 13, 14, 15, 16), for which two conditions already appeared before. Note also that the last sufficient condition in (\ref{Eq:SuffConLRSprotoIRaw}) forms a linear combination of the first and fifth row of the third block. Hence, there are at most three independent constraints coming from the sufficient conditions:
\begin{equation}\label{Eq:SufConLRSProtoI}
\begin{array}{rcl}
3k_a - k_d - 3 k_{h_1} + 3k_{h_2}&\stackrel{!}{=}& 0 \text{ mod } n,\\
-k_d +3 k_{h_1} + 6 k_{h_2}&\stackrel{!}{=}& 0 \text{ mod } n,\\
-2k_d +3 k_{h_1} + 3 k_{h_2}&\stackrel{!}{=}& 0 \text{ mod } n.
\end{array}
\end{equation}
Notice, however, that the last {\it sufficient} condition in equation~(\ref{Eq:SufConLRSProtoI}) for example can be reduced to \mbox{$2k_d \stackrel{!}{=} 0 \text{ mod } n$} upon inserting the first 
{\it necessary} condition from equation~\eqref{Eq:NecConLRSProtoI}, which in turn renders the second and fourth {\it necessary} condition identical to \mbox{$6 k_a \stackrel{!}{=} 0 \text{ mod } n$}.
Continuing along these lines, the set of truly independent conditions can be reduced to match the number $U(1)$ factors in the model.
Combining the  four independent necessary conditions (\ref{Eq:NecConLRSProtoI}) and three independent sufficient conditions (\ref{Eq:SufConLRSProtoI}),
one can easily notice that no non-trivial combination $(k_a,k_d,k_{h_1}, k_{h_2})$ can solve them simultaneously for all $n$, implying that this left-right symmetric model does not come with a massless $U(1)_{B-L}$ symmetry or possible extensions thereof involving the hidden $U(1)$ factors.

Let us thus continue with the classification of discrete $\Z_n$ symmetries for the prototype I left-right symmetric model:
\begin{itemize}
\item The combination $(k_a,k_d,k_{h_1}, k_{h_2})=(1,1,1,1)$ gives rise to the discrete $\Z_2$ symmetry guaranteed by the K-theory constraints. In first instance, we might feel the urge to see the symmetry as a remnant of a massive $B-L$ like symmetry, based on the charges of the visible sector under this discrete symmetry. The fact, that also exotic matter charged under the hidden gauge groups carries discrete $\Z_2$ charges, indicates a more general form than a massive $B-L$ like symmetry.   From the low-energy viewpoint this discrete symmetry acts trivially on the massless open string spectrum: since only open string states transforming in the fundamental representation ${\bf 2}$ of $USp(2)_b$ or $USp(2)_c$ carry a non-trivial charge (cf. table~\ref{tab:6stackLRSMaAAPrototypeISpectrum}), this discrete $\Z_2$ symmetry provides the same selection rules as the ones coming from the centres of the non-Abelian gauge factors $USp(2)_b$ and $USp(2)_c$.  
\item There exists a set of three discrete $\Z_3$ symmetries, corresponding to the combinations $(k_a,k_d,k_{h_1}, k_{h_2})=(1,0,0,0)$, $(0,0,1,0)$, and $(0,0,0,1)$, such that each $\Z_3$ symmetry is homomorphic to the centre of a $SU(3)_{x\in \{a, h_1, h_2\}}$ gauge symmetry. Hence, these discrete symmetries do not offer any other selection rules for the $m$-point couplings beyond the ones associated to the non-Abelian gauge symmetries.
\item The vector $(k_a,k_d,k_{h_1}, k_{h_2})=(1,3,1,1)$ corresponds to a discrete $\Z_6$ symmetry, with the charges for the massless open string states given in the last column of table~\ref{tab:6stackLRSMaAAPrototypeISpectrum}. Note that the K-theory $\Z_2$ symmetry is a subgroup of this discrete symmetry, suggesting that the truly independent discrete symmetry is only a $\Z_3$ symmetry. This $\Z_3$ symmetry also pops up when we mod out the centres of the overall non-Abelian gauge group from the discrete symmetries found as solutions to the necessary and sufficient conditions~\eqref{Eq:NecConLRSProtoI} and~\eqref{Eq:SufConLRSProtoI}, 
i.e.~we find that the quotient group $(\Z_2\times\Z_3^3 \times \Z_6)/(\Z_3^3 \times\Z_2^2)$ 
is homomorphic to a $\Z_3$ symmetry with charge assignments listed in table~\ref{tab:6stackLRSMaAAPrototypeISpectrum}, after reduction of $\Z_6 \to \Z_3$.
In this particular case, the $\Z_3$ symmetry can also be associated to the combination $(k_a,k_d,k_{h_1}, k_{h_2})=(1,0,1,1)$, such that the $\Z_3$ symmetry acts effectively as a linear combination of the three discrete $\Z_3$ symmetries identified above. As such, this $\Z_3$ symmetry should not be considered as an independent discrete symmetry and is not expected to yield additional selection rules apart from those associated to the centres of the non-Abelian gauge groups.

\end{itemize}
In conclusion, the gauge group encountered for prototype I below the string mass scale corresponds to $SU(3)_a\times USp(2)_b\times USp(2)_c \times SU(3)_{h_1}\times SU(3)_{h_2}\times \Z_3$
with the $\Z_3$ acting trivially on massless matter states.

{\bf Prototype II Left-Right Symmetric Model}\\
Next, we discuss the discrete symmetries arising in the second prototype left-right symmetric model 
through the example presented in table~\ref{tab:6stackLRSaAAPrototypeII} following the same line of thought as for prototype I. Writing down the necessary conditions for the existence of discrete $\Z_n$ gauge symmetries with respect to the example in table~\ref{tab:6stackLRSaAAPrototypeII}:
\begin{align}\hspace{-25mm}
&\left\{k_a \left(\begin{array}{c} 0 \\ 0 \\\hline 0 \\ 0 \\ 0 \\ -6 \\ 6 \\ 0
\end{array}\right)
+ k_d \left(\begin{array}{c} 0 \\  0 \\\hline 0 \\ 0 \\ 0 \\ -2 \\ 2 \\ 0
\end{array}\right)
+ k_{h_1} \left(\begin{array}{c} 0 \\  0 \\\hline 3 \\ 1 \\ 1 \\ 1 \\ 0 \\ 0 
\end{array}\right)
+ k_{h_2} \left(\begin{array}{c} 0 \\  0 \\\hline 3 \\ 1 \\ 1 \\ 1 \\ 0 \\ 0 
\end{array}\right) \right\}
\stackrel{!}{=} 0 \text{ mod } n, \label{Eq:NecFullLRSProtoIIa} \\
&\left\{ k_a \left(\begin{array}{c} 0 \\ 0 \\ 0 \\ 0 \\\hline 6 \\ 6 \\ 0 \\ 0
\end{array}\right)
+ k_d \left(\begin{array}{c} 0 \\ 0 \\ 0 \\ 0 \\\hline -2 \\ -2 \\ 0 \\ 0
\end{array}\right)
+ k_{h_1} \left(\begin{array}{c} 0 \\ 0 \\ 0 \\ 0 \\\hline 2 \\ 2 \\ 0 \\ 0
\end{array}\right)
+ k_{h_2} \left(\begin{array}{c}  0 \\ 0 \\ 0 \\ 0 \\\hline -2 \\ -2 \\ 0 \\ 0
\end{array}\right) \right\}\stackrel{!}{=} 0 \text{ mod } n,  \label{Eq:NecFullLRSProtoIIb}
\end{align}
we deduce three linearly independent constraints:
\begin{equation}\label{Eq:NecConLRSProtoII}
\begin{array}{rcl}
k_{h_1} +  k_{h_2}&\stackrel{!}{=}& 0 \text{ mod } n  ,\\
6k_a + 2 k_d&\stackrel{!}{=}&0 \text{ mod } n  ,\\
6k_a - 2 k_d +2k_{h_1} -2 k_{h_2}&\stackrel{!}{=}& 0 \text{ mod } n ,
\end{array}
\end{equation}
from those rows in (\ref{Eq:NecFullLRSProtoIIa}) and (\ref{Eq:NecFullLRSProtoIIb}) that are not trivially satisfied. To these four necessary constraints we have to add the
subset of the linearly independent constraints coming from the sufficient conditions~(\ref{Eq:Zn-condition-suf}) written out for the D6-brane configuration in table~\ref{tab:6stackLRSaAAPrototypeII}:
\begin{align} \label{Eq:SufFullLRSProtoII}
\left\{k_a\left(\begin{array}{c}
3\\ 6\\ 3\\ 6\\3\\ 6\\
\hline
0\\ 0\\ 0\\ 0\\
\hline
0\\ 0\\ 0\\ -6\\ 0\\0
\end{array}\right)+k_d\left(\begin{array}{c}
-1\\ -2\\ -1\\ -2\\ -1\\ -2\\
\hline
0\\0\\0\\0\\
\hline
-1\\0\\0\\-2\\-2\\1
\end{array}\right)
+k_{h_1}\left(\begin{array}{c}
1\\2\\1\\2\\1\\2\\
\hline
0\\0\\0\\0\\
\hline
2\\2\\1\\1\\2\\0
\end{array}\right)
+k_{h_2}\left(\begin{array}{c}
-1\\-2\\-1\\-2\\-1\\-2\\
\hline
0\\0\\0\\0\\
\hline
1\\2\\1\\1\\0\\1
\end{array}\right)\right\}=0\text{  mod  }n.
\end{align}
Clearly, the first block of the sufficient conditions provides one linearly independent constraint, while the second block contains only trivially satisfied conditions. The third block gives three conditions (rows 11, 15, and 16) which have not appeared yet before in the necessary conditions (\ref{Eq:NecConLRSProtoI}). 
Since row 11 is the sum of rows 15 and 16, we are naively left with three new and linearly independent constraints:
\begin{equation}\label{Eq:SufConLRSProtoII}
\begin{array}{rcl}
3k_a - k_d + k_{h_1} - k_{h_2}&\stackrel{!}{=}& 0 \text{ mod } n,\\
-2 k_{d} +2 k_{h_1} &\stackrel{!}{=}& 0 \text{ mod } n,\\
k_d+k_{h_2}&\stackrel{!}{=}& 0 \text{ mod } n .  
\end{array}
\end{equation}
The last {\it necessary} condition in equation~\eqref{Eq:NecConLRSProtoII} turns out to equal twice the first {\it sufficient} condition in equation~\eqref{Eq:SufConLRSProtoII}.
Since also the second {\it sufficient} condition can be expressed as twice the linear combination of the  first {\it necessary} condition minus the last {\it sufficient} condition, 
only four conditions are truly independent, as expected from the four initial $U(1)$ factors in the model. A closer inspection of the three necessary constraints (\ref{Eq:NecConLRSProtoII}) and the four sufficient constraints (\ref{Eq:SufConLRSProtoII}) teaches us that the non-trivial combination $(k_a,k_d,k_{h_1}, k_{h_2}) = (1,-3,-3,3)$ satisfies all seven constraints irrespective of the value of $n$. This combination points towards the presence of a massless linear combination of $U(1)$'s:
\begin{equation}\label{Eq:GenerBLSymm}
\widetilde{U(1)}_{B-L} = \frac{1}{3} U(1)_a - U(1)_d - U(1)_{h_1} + U(1)_{h_2},
\end{equation}
which plays the r\^ole of a (generalised) $B-L$ symmetry. 
Turning our attention to the discrete $\Z_n$ gauge symmetries allowed by the constraints (\ref{Eq:NecConLRSProtoII}) and (\ref{Eq:SufConLRSProtoII}), we obtain the following classification:
\begin{itemize}
\item Also for this prototype we encounter the discrete $\Z_2$ symmetry guaranteed by the K-theory constraints for the combination $(k_a,k_d,k_{h_1}, k_{h_2})=(1,1,1,1)$,  but in this model the $\Z_2$ symmetry is a discrete subgroup of the massless $\widetilde{U(1)}_{B-L}$ gauge symmetry. This can be seen explicitly by shifting the charges of the massless open string states under the $\Z_2$ symmetry by virtue of the massless $\widetilde{U(1)}_{B-L}$ symmetry, after which all charges are set to zero (modulo 2) simultaneously.
\item We only encounter one discrete $\Z_3$ symmetry, namely for $(k_a,k_d,k_{h_1}, k_{h_2})=(1,0,0,0)$, which is homomorphic to the centre of the non-Abelian $SU(3)_a$ gauge group. This discrete symmetry will thus not provide any new selection rules for cubic and higher order couplings.  Also here we can perform a rotation over the massless $\widetilde{U(1)}_{B-L}$ symmetry, setting the charges for all open string states to zero (modulo 3), to verify that the action of the $\Z_3$ symmetry is trivial from the effective low-energy perspective. 
\item Finally, we also encounter a discrete $\Z_6$ symmetry corresponding to the linear combination $(k_a,k_d,k_{h_1}, k_{h_2})=(1,3,3,3)$, with the charges of the massless open string states given in the last column of table~\ref{tab:6stackLRSMaAAPrototypeIISpectrum}. 
We must ask ourselves again whether this discrete symmetry should not be reduced to a discrete $\Z_3$ symmetry, given the quotient group $(\Z_2\times \Z_3\times \Z_6)/(\Z_3\times \Z_2\times \Z_2) \simeq \Z_3$, where the subgroup $\Z_3\times \Z_2\times \Z_2$ corresponds to the centres of the non-Abelian gauge factors for the prototype II models. Indeed, since the K-theory $\Z_2$ symmetry forms a subgroup of the discrete $\Z_6$ symmetry, the truly independent discrete symmetry is rather the $\Z_3$ symmetry associated with $(k_a,k_d,k_{h_1}, k_{h_2})=(1,0,0,0)$ discussed in the previous bullet point. Alternatively, to argue for the triviality of the $\Z_6$-action we also point out that the $\Z_6$ symmetry forms a discrete subgroup of the massless $\widetilde{U(1)}_{B-L}$ gauge symmetry and that we can set the charges of the open string states to zero by virtue of a shift over $\widetilde{U(1)}_{B-L}$. 
\end{itemize}
Hence, the full gauge group for the prototype II left-right symmetric model, exemplified by the D-brane configuration in table~\ref{tab:6stackLRSaAAPrototypeII},
 is given by $SU(3)_a \times USp(2)_b \times USp(2)_c \times \widetilde{U(1)}_{B-L}$ below the string scale.

%%%%%%%%%%%%%%%%%%%%%%%%%%%%%%%%%%%%%%%%%%%%%%%%%%%%%%%%%%%%%%%%%%%%%%%%%%%%%%%%%%%
%%%%%%%%%%%%%%%%%%%%%%%%%%%%%%%%%%%%%%%%%%%%%%%%%%%%%%%%%%%%%%%%%%%%%%%%%%%%%%%%%%%
\subsection{Yukawa and other Cubic Couplings}\label{Ss:LRSMYukawa}
Comparing the fractional D6-brane configurations in tables~\ref{tab:6stackLRSaAAPrototypeI},~\ref{tab:6stackLRSaAAPrototypeII},~\ref{tab:6stackLRSaAAPrototypeIIb} and~\ref{tab:6stackLRSaAAPrototypeIIc} for the explicit examples representing the prototype I, II, IIb and IIc models, respectively, one notices that the bulk orbits of the visible D-brane stacks $a$, $b$, $c$ and $d$ are identical and that 
their displacement  $\sigma_{x}^{i=2,3}$
and Wilson line $\tau_x^{i=2,3}$ parameters, with $x,y\in\{a,b,c,d\}$, along $T^2_{(2)} \times T^2_{(3)}$ are also equal for all examples. 
Moreover, within each example the parameters $\sigma^1_x$ and $\tau^1_x$ for $x\in \{a,b,c,d\}$ are respectively equal along the two-torus $T^2_{(1)}$, where all D-branes are positioned at vanising angle w.r.t.~the $\OR$-invariant plane.
Therefore, all the prototype examples are characterised by the same massless visible open string spectrum, which consists of states  charged only under the visible gauge group $U(3)_a\times USp(2)_b\times USp(2)_c\times U(1)_d$. As discussed in appendix~\ref{A:ChanPatonMethod}, the correct localisation of massless open string states at intersection points also depends on the relative $\Z_2^{(i)}$ eigenvalues 
$(-1)^{\tau_{xy}^{\Z_2^{(i)}}}$ between the stacks $x,y \in \{a,b,c,d \}$, which are all the same for the explicit D6-brane configurations given in tables~\ref{tab:6stackLRSaAAPrototypeI},~\ref{tab:6stackLRSaAAPrototypeII},~\ref{tab:6stackLRSaAAPrototypeIIb} and~\ref{tab:6stackLRSaAAPrototypeIIc}. Hence, it suffices to discuss the Yukawa couplings for one prototype model to obtain the Yukawa couplings for the other prototype models as well.

\begin{figure}[h]
\begin{center}
\vspace{0.2in}
\begin{tabular}{c@{\hspace{0.5in}}c@{\hspace{0.5in}}c}
\includegraphics[width=4.5cm]{T1Coupling1} \begin{picture}(0,0)\put(-70,100){$T_{(1)}^2$}\put(-140,0){\bf \color{red} \bf 1} \put(-140,40){\bf \color{red} \bf 4} \put(-70,-7){\bf \color{red} \bf 2} \put(-64,40){\bf \color{red} \bf 3} \put(0,0){\color{myblue} $a$} \put(0,-12){\color{myorange} $c$}  \put(-6,12){\color{mypurple} $(\omega^2 b)$}\end{picture}& \includegraphics[width=5cm]{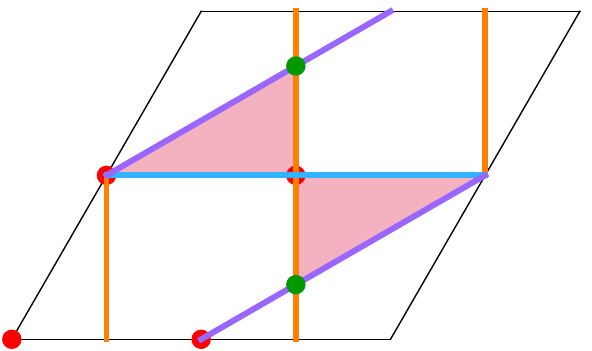} \begin{picture}(0,0) \put(-70,100){$T_{(2)}^2$} \put(-155,0){\color{red} \bf 1}  \put(-100,-8){\color{red} \bf 4}  \put(-130,45){\color{red} \bf 5}  \put(-66,38){\color{red} \bf 6} \put(-90,18){\color{mygr}  $S_2$} \put(-70,62){\color{mygr} $S_2'$}  \put(-23,39){\color{myblue} $a$} \put(-125,60){\color{mypurple} $(\omega^2 b)$}  \put(-31,85){\color{myorange} $c$} \end{picture} &  \includegraphics[width=5cm]{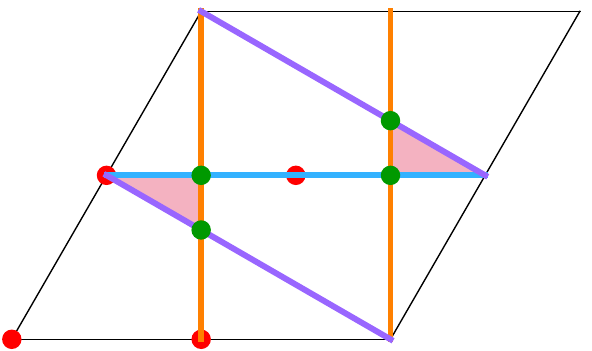} \begin{picture}(0,0) \put(-70,100){$T_{(3)}^2$} \put(-155,0){\color{red} \bf 1}  \put(-100,-8){\color{red} \bf 4}  \put(-130,45){\color{red} \bf 5}  \put(-66,38){\color{red} \bf 6}  \put(-114,48){\color{mygr} $Q_3$} \put(-50,30){\color{mygr} $ Q_3'$}  \put(-107,20){\color{mygr} $2$} \put(-48,57){\color{mygr} $3$} \put(-55,87){\color{myorange} $c$} \put(-45,0){\color{mypurple} $(\omega^2 b)$}   \put(-23,39){\color{myblue} $a$}  \end{picture}
\end{tabular}
\caption{Pictorial representation of the D6-brane configuration involving the closed sequence of the bulk three-cycles $[{\color{myblue}a}, {\color{mypurple}(\omega^2 b)}, {\color{myorange}c}]$ for the six-stack left-right symmetric prototype II model, compatible with the perturbatively allowed Yukawa couplings $Q_L^{(2)}\cdot  (H_u,H_d)^{(3,4,5,6)} Q_R^{(1,2)}$. The intersection points $(S_2,S_2')$, $(Q_3,Q_3')$ and $(2,3)$ correspond to $\Z_2^{(i)}$-invariant pairs of intersection points per two-torus. 
\label{Fig:ExampleCubicCouplingLRS}}
\end{center}
\end{figure}

Let us thus, for instance, consider the Yukawa couplings for the prototype II model with D6-brane configuration in table~\ref{tab:6stackLRSaAAPrototypeII}. The first step consists in determining the cubic couplings that are allowed by charge conservation: 
\begin{equation}\label{Eq:YukaCouplingLRS}
{\cal W}_{\rm Yuk} = y_Q Q_L (H_u, H_d) Q_R + y_L L (H_u, H_d) R, 
\end{equation}
where we wrote down the Yukawa couplings in a schematic way involving the quarks, leptons and Higgses appearing in the first block of table~\ref{tab:6stackLRSMaAAPrototypeIISpectrum}. 
 In order to assess which Yukawa couplings are non-vanishing, we first have to allocate the massless open string states unambiguously to $\Z_2\times \Z_2$ invariant intersection points 
or orbits and then verify that also the stringy selection rules are satisfied. These steps require  us to determine from which sectors $x (\omega^k y)_{k=0,1,2}$ and $x (\omega^k y)'_{k=0,1,2}$ with $x,y \in \{a,b,c,d\}$ the massless states arise, as listed in tables~\ref{tab:Z2Z65stackLRSTotalSpectrumI} and~\ref{tab:Z2Z65stackLRSTotalSpectrumII}, after which we can use the techniques involving the Chan-Paton labels from appendix~\ref{A:ChanPatonMethod} to allocate the states explicitly. As a last step, we investigate the area of the closed triangle sequences on $T_{(2)}^2\times T_{(3)}^2$ with the allocated massless states at their apexes, as explained in section~\ref{Ss:Yukawa}.   
 An overview of the non-vanishing Yukawa couplings involving the quark sector is given in table~\ref{tab:YukaCouplingLRSpartI}, while the non-vanishing leptonic Yukawa couplings are listed in table~\ref{tab:YukaCouplingLRSpartII}. A quick comparison between tables~\ref{tab:YukaCouplingLRSpartI} and~\ref{tab:YukaCouplingLRSpartII} reveals a subtle symmetry among the Yukawa couplings involving the quarks and leptons: upon exchanging $Q_L^{(i)} \leftrightarrow L^{(i)}$ and  $Q_R^{(i)} \leftrightarrow R^{(i)}$ we find the same order of magnitude for the corresponding coupling constants. This allows us to discuss solely the quark Yukawa couplings and deduce the same conclusions for the leptonic sector. The numbering of the Higgses $(H_u,H_d)^{(i)}$ emerging from the $b(\omega^k c)_{k=0,1,2}$ sector follows a normal ordering with $i\in \{1,2\}$ for $k=0$, $i\in \{3,4,5,6\}$ for $k=1$ and $i\in \{7,8,9,10\}$ for $k=2$. Figure~\ref{Fig:ExampleCubicCouplingLRS} provides a pictorial representation of the perturbatively allowed Yukawa couplings $Q_L^{(2)}\cdot  (H_u,H_d)^{(3,4,5,6)} Q_R^{(1,2)}$ associated to the closed sequence $[a,(\omega b), c]$. 

By taking a closer look at the Yukawa couplings for the quarks, we notice that all Yukawa couplings are exponentially suppressed. The diagonal Yukawa couplings for the second and third generation only occur for the third Higgs doublet $(H_u, H_d)^{(3)}$ and are accidentally equal to each other. We also notice the absence of the diagonal Yukawa coupling for the first generation $Q_L^{(1)}$ and $Q_R^{(1)}$, yet both chiral states appear in non-diagonal Yukawa couplings to the third and second generation, respectively. For the Yukawa couplings involving the third Higgs doublet $(H_u, H_d)^{(3)}$ we observe that the off-diagonal terms are more suppressed than the diagonal Yukawa couplings. For the Yukawa couplings involving the  Higgs doublets $(H_u, H_d)^{(4,5,6)}$ we notice the opposite pattern, which complicates a clear microscopic explanation of the hierarchies within the CKM matrix. Notice that the off-diagonal Yukawa couplings between the second and third generation proceed according to a separate Higgs-sector from the other off-diagonal Yukawa couplings, which might be a useful observation to explain some hierarchical structure in the CKM matrix entries based on a hierarchy among the {\it vevs} for different Higgs sectors. 
The Higgses $(H_u,H_d)^{(1,2)}$ attributed to the $bc$-sector are somewhat special as they cannot be unambiguously assigned to $\Z_2\times\Z_2$ invariant intersection points. This feature can be traced back to the fact that both bulk orbits are fully parallel to each other on all three two-tori. In this respect they give the impression of being a remnant (local) ${\cal N}=2$ supersymmetric multiplet, and it is not entirely clear if the presence of the $\Z_2\times \Z_2$ symmetries, which lead to manifestly only ${\cal N}=1$ supersymmetry, will change the existence of cubic couplings involving these states. 
For that reason, we have not treated Yukawa couplings involving the Higgses $(H_u,H_d)^{(1,2)}$ and hope to address this conundrum in future work.

To enhance the phenomenological appeal of the massless open string spectrum in table \ref{tab:6stackLRSMaAAPrototypeIISpectrum}, we should also discuss mechanisms to lift the masses of the non-chiral matter pairs from the $ad$ and $ad'$ sectors. First of all, observe that cubic couplings involving the non-chiral matter pairs from the $ad$ or $ad'$ sectors combined with some Standard Model singlet
cannot occur due to the absence of massless states in the adjoint or symmetric representation under the $U(1)_d$ gauge group. Hence, we have to look at quartic couplings involving the states from the $ad$ and $ad'$ sectors. 
By virtue of the matter states in the $dh_1$, $dh_1'$, $dh_2$ and $dh_2$ sector, we are able to write down the gauge-invariant quartic couplings:  
\begin{equation}\label{Eq:QuarticCouplingLRS}
\begin{array}{rcl}
{\cal W}_{\rm extra} &=& \frac{\mu^{ij}_{h_1}}{M_{\rm string}} X^{ad} \tilde X^{ad' (i)} X^{dh_1(j)} X^{dh_1' } +  \frac{\mu^{ij}_{h_2}}{M_{\rm string}} X^{ad} \tilde X^{ad' (i)} X^{dh_2(j)} X^{dh_2' }  \\
&& + \frac{\tilde \mu^{ij}_{h_1}}{M_{\rm string}} \tilde X^{ad}  X^{ad' (i)} \tilde X^{dh_1(j)} \tilde X^{dh_1' } +  \frac{\tilde \mu^{ij}_{h2}}{M_{\rm string}} \tilde X^{ad} X^{ad' (i)} \tilde X^{dh_2(j)} \tilde X^{dh_2' },  
\end{array}
\end{equation}
with $i,j \in \{1,2 \}$ and $M_{\rm string}$ the string mass scale. Note, however, that this discussion can only be pursued for the prototype II and IIc left-right symmetric models, as the existence of the quartic couplings is tied to the existence of `messenger' states in the $dh_1$, $dh_1'$, $dh_2$ and $dh_2$ sectors. The next step then comprises the compution of the non-vanishing perturbatively allowed quartic couplings, using the same techniques as explained in section~\ref{Ss:Yukawa} - including localisations analogous to those in appendix~\ref{A:ChanPatonMethod} -  generalised to quartic couplings and their associated quadrilateral worldsheet instantons. Looking carefully at the $ad$ sector, we observe that the non-chiral matter pair $X^{ad} + \tilde X^{ad}$ arises solely from two D6-brane stacks whose bulk orbits are completely parallel to each other on all three two-tori.  In this regard, the non-chiral pair cannot be localised at $\Z_2\times \Z_2$ invariant intersection points from a geometric perspective, giving the impression that the non-chiral pair is a remnant (local) ${\cal N}=2$ supermultiplet. Analogously to the Higgses from the $bc$-sector, it is not entirely clear how the $\Z_2\times \Z_2$ symmetries act on the quartic couplings involving the non-chiral pair $X^{ad} + \tilde X^{ad}$. In order to asses whether the quartic couplings are non-vanishing, a better understanding of the CFT computations for $m$-point couplings  on orbifolds with $\Z_2$ factors is required.

%%%%%%%%%%%%%%%%%%%%%%%%%%%%%%%%%%%%%%%%%%%%%%%%%%%%%%%%%%%%%%%%%%%%%%%%%%%
\begin{sidewaystable}[h]
\begin{minipage}{21cm}
\begin{center}
\hspace*{-14mm}
$\begin{array}{|c||c|c|c|c|c|c|}
\hline  \multicolumn{7}{|c|}{\text{\bf Total amount of matter per sector for a 6-stack Left-Right Symmetric model on the {aAA} lattice}}\\
\hline \hline (\chi^{x y},\chi^{x (\omega y)}, \chi^{x (\omega^2 y)})& y=a & y=b& y=c& y=d & y=h_1& y=h_2\\  
\hline
x= a & (0,0,0) & (2,0,1) & (-2,-1,0) &  (|2|,0,0)    & (|2|,-1,1)  & (|2|,1,-1)  \\
x= b & &(0,\frac{3 +|2|}{2},\frac{-3+|2|}{2})  & (|2|,|4|, |4|) &  (2,1,0)   &(0,|4|,-3+|2|)& (0,|4|,3+|2|) \\
x= c & & &(0,\frac{3 +|2|}{2},\frac{-3+|2|}{2}) &  (-2,0,-1)  & (0,3+|2|,|4|) &(0,-3+|2|,|4|)  \\
x= d & & & & (0,0,0)& (|2|,1,-1) & (|2|,-1,1) \\
x= h_1 & & & & & (0,\frac{|4|}{2},\frac{|4|}{2})  & (|2|,|4|,|4|) \\
x= h_2 & & & & & &(0,\frac{|4|}{2},\frac{|4|}{2}) \\
\hline
\end{array}
$
\caption{Overview of the total amount of chiral and non-chiral matter per sector $x(\omega^k y)$ for the six-stack prototype II LRS model with fractional D6-brane configuration given in table~\ref{tab:6stackLRSaAAPrototypeII}. If the net-chirality $|\chi^{x(\omega^ky)}| < \varphi^{x(\omega^ky)}$, the sector $x(\omega^ky)$ comes with a set of non-chiral pairs of matter states,whose multiplicity corresponds to $n^{x(\omega^ky)}_{NC} \equiv\varphi^{x(\omega^ky)} - |\chi^{x(\omega^ky)}|$, , e.g. $|n^{a(\omega^0h)}_{NC}|=|2|$ denotes one non-chiral pair of bifundamentals in the sector $a(\omega^0h)$.  
Such non-chiral pairs are indicated as $|n^{x(\omega^ky)}_{NC}|$. 
The diagonal entries $\varphi^{x(\omega x)} = \varphi^{x(\omega^2 x)} = \frac{\varphi^{\Adj_x}}{2}$
count the number of states in the adjoint representation for the $x$-stack, e.g. $(0,\frac{3+|2|}{2},\frac{-3+|2|}{2})$ for the $b$-stack denotes five multiplets in the adjoint representation with half of the d.o.f. localised in the $b(\theta b)$ sector and the other half in the $b(\theta^2 b) \equiv (\theta b)b$ sector.
\label{tab:Z2Z65stackLRSTotalSpectrumI}}
\end{center}
\end{minipage}
\\
\\ \\ \\
\begin{minipage}{21cm}\begin{center}
$\begin{array}{|c||c|c|c|c|}
\hline  \multicolumn{5}{|c|}{\text{\bf Total amount of matter per sector for a 6-stack Left-Right Symmetric model}}\\
\hline \hline (\chi^{x y'},\chi^{x (\omega y)'}, \chi^{x (\omega^2 y)'})& y=a & y=d & y=h_1& y=h_2\\  
\hline
x= a &  \tarh{(|2|,-1,1)}{(0,0,0)} & (|2|,-1,1)&(|2|,0,0)  & (|2|,0,0) \\
x= d & &   \tarh{(|2|,-1,1)}{(0,0,0)} & (|2|,0,0) & (|2|,0,0) \\
x= h_1 & & &   \tarh{(|2|,3+|2|,-3+|2|)}{(0,0,0)} &(|2|,3+|2|,-3+|2|) \\
x= h_2 & & & & \tarh{(|2|,3+|2|,-3+|2|)}{(0,0,0)} \\
\hline
\end{array}
$
\caption{Overview of the total amount of chiral and non-chiral matter per sector $x(\omega^k y)'$ for the six-stack prototype II LRS model with fractional D6-brane configuration given in table~\ref{tab:6stackLRSaAAPrototypeII}. The notation for the counting of bifundamental states is identical to table~\protect\ref{tab:Z2Z65stackLRSTotalSpectrumI}. For $x(\theta^k x)'$ sectors, the upper entries count the numbers of antisymmetric representations and the lower entries the symmetric ones. The $\OR$-invariance of the $b$-stack and $c$-stack implies $b(\theta^k x)'=b(\theta^{3-k} x)$ and $c(\theta^k x)'=c(\theta^{3-k} x)$, respectively. For D6-brane stacks $d$, $h_1$ and $h_2$ supporting $U(1)$ gauge groups, the states in the antisymmetric representation do not exist, but their would-be multiplicities are included for completeness and for consistency when checking the anomaly cancellation conditions.
\label{tab:Z2Z65stackLRSTotalSpectrumII}}
\end{center}
\end{minipage}
\end{sidewaystable}
%%%%%%%%%%%%%%%%%%%%%%%%%%%%%%%%%%%%%%%%%%%%%%%%%%%%%%%%%%%%%%%%%%%%%%%%%%%

\begin{sidewaystable}[h]
\begin{center}
\begin{tabular}{|c|c|c|c|c|}
\hline \multicolumn{5}{|c|}{\bf Yukawa couplings (\ref{Eq:YukaCouplingLRS}) for prototype II Left-Right Symmetric Model (part I) } \\
\hline
\hline {\bf Coupling} & {\bf Sequence} & {\bf Triangles on $T_{(2)}^2 \times T_{(3)}^2$} & {\bf Enclosed Area} & {\bf Parameter}\\
\hline 
\hline

\hline
$Q_L^{(1)} (H_u, H_d)^{(3)} Q_R^{(3)}$ & \multirow{6}{*}{$[a,b,(\omega c)]$}  & $\left\{ [5,6,6],   [(Q_3,Q_3'),(2,3),6]  \right\}$ & $  \frac{3v_2}{4}+ \frac{v_3}{48} $ &$y_Q^{(113)} \sim  {\cal O}(e^{-\frac{36 v_2+v_3}{48}})$  \\
 $Q_L^{(1)} (H_u, H_d)^{(4)} Q_R^{(3)}$ & &$\left\{[5,(R_2,R_2'),6],   [(Q_3,Q_3'),1,6] \right\}$ & $ \frac{v_2}{12} +  \frac{3v_3}{16} $  & $y_Q^{(143)} \sim {\cal O}\left(e^{-\frac{4 v_2 + 9 v_3}{48}} \right)$ \\
 $Q_L^{(1)} (H_u, H_d)^{(i=5,6)} Q_R^{(3)}$ & & $\left\{[5,(R_2,R_2'),6],   [(Q_3,Q_3'),(2,3),6] \right\}$ & $ \frac{v_2}{12}+\frac{v_3}{48} $ &$y_Q^{(1i3)} \sim  {\cal O}(e^{-\frac{4v_2+v_3}{48}})$  \\
 
 $Q_L^{(3)} (H_u, H_d)^{(3)} Q_R^{(3)}$ &   & $\left\{ [6],   [(Q_3,Q_3'),(2,3),6]  \right\}$ & $ \frac{v_3}{48} $ &$y_Q^{(333)} \sim  {\cal O}(e^{-\frac{v_3}{48}})$  \\
 $Q_L^{(3)} (H_u, H_d)^{(4)} Q_R^{(3)}$ & &$\left\{[6,(R_2,R_2'),6],   [(Q_3,Q_3'),1,6] \right\}$ & $ \frac{v_2}{3} +  \frac{3v_3}{16} $  & $y_Q^{(343)} \sim {\cal O}\left(e^{-\frac{16 v_2 + 9 v_3}{48}} \right)$ \\
 $Q_L^{(3)} (H_u, H_d)^{(i=5,6)} Q_R^{(3)}$ & & $\left\{[6,(R_2,R_2'),6],   [(Q_3,Q_3'),(2,3),6] \right\}$ & $ \frac{v_2}{3}+\frac{v_3}{48} $ &$y_Q^{(3i3)} \sim  {\cal O}(e^{-\frac{16v_2+v_3}{48}})$  \\
 
\hline

 $Q_L^{(2)} (H_u, H_d)^{(7)} Q_R^{(3)}$ & \multirow{3}{*}{ $[a,(\omega^2 b),(\omega c)]$}  &$\left\{  [5,4,6], [5,(2,3),6]\right\}$  & $  \frac{3v_2}{8} + \frac{v_3}{24}$ &  $y_Q^{(273)} \sim {\cal O}\left( e^{-\frac{9v_2 + v_3}{24}} \right)$  \\
  $Q_L^{(2)} (H_u, H_d)^{(8)} Q_R^{(3)}$ & &$\left\{  [5,(P_2,P_2'),6], [5,1,6]\right\}$  & $ \frac{v_2}{24} +  \frac{3v_3}{8}$ &  $y_Q^{(283)} \sim {\cal O}\left( e^{-\frac{v_2 + 9 v_3}{24}} \right)$  \\
   $Q_L^{(2)} (H_u, H_d)^{(i=9,10)} Q_R^{(3)}$ &  &$\left\{  [5,(P_2,P_2'),6], [5,(2,3),6]\right\}$  & $ \frac{v_2 + v_3}{24}$ &  $y_Q^{(2i3)} \sim {\cal O}\left( e^{-\frac{v_2 + v_3}{24}} \right)$  \\
   
\hline
$Q_L^{(2)} (H_u, H_d)^{(3)} Q_R^{(2)} $ & \multirow{6}{*}{$[a,(\omega^2 b),c]$}  & $\left\{ [5], [5,(2,3), (Q_3,Q_3')] \right\}$ & $ \frac{v_3}{48} $ & $y_Q^{(232)} \sim {\cal O}\left(e^{-\frac{ v_3}{48}}\right)$ \\
$Q_L^{(2)} (H_u, H_d)^{(4)} Q_R^{(2)} $ &   & $\left\{ [5,(S_2,S_2'),5], [5,1, (Q_3,Q_3')] \right\}$ & $ \frac{v_2}{3}+\frac{3v_3}{16} $ & $y_Q^{(242)} \sim {\cal O}\left(e^{-\frac{ 16 v_2 +9v_3}{48}}\right)$ \\
$Q_L^{(2)} (H_u, H_d)^{(i=5,6)} Q_R^{(2)} $ &   & $\left\{ [5,(S_2,S_2'),5], [5,(2,3), (Q_3,Q_3')] \right\}$ & $ \frac{v_2}{3}+\frac{v_3}{48} $ & $y_Q^{(2i2)} \sim {\cal O}\left(e^{-\frac{ 16 v_2 +v_3}{48}}\right)$ \\

$Q_L^{(2)} (H_u, H_d)^{(3)} Q_R^{(1)} $ & & $\left\{ [5,5,6], [5,(2,3), (Q_3,Q_3')] \right\}$ & $\frac{3v_2}{4} + \frac{v_3}{48} $ & $y_Q^{(231)} \sim {\cal O}\left(e^{-\frac{ 36v_2+ v_3}{48}}\right)$ \\
$Q_L^{(2)} (H_u, H_d)^{(4)} Q_R^{(1)} $ &   & $\left\{ [5,(S_2,S_2'),6], [5,1, (Q_3,Q_3')] \right\}$ & $ \frac{v_2}{12}+\frac{3v_3}{16} $ & $y_Q^{(241)} \sim {\cal O}\left(e^{-\frac{ 4 v_2 +9v_3}{48}}\right)$ \\
$Q_L^{(2)} (H_u, H_d)^{(i=5,6)} Q_R^{(1)} $ &   & $\left\{ [5,(S_2,S_2'),6], [5,(2,3), (Q_3,Q_3')] \right\}$ & $ \frac{v_2}{12}+\frac{v_3}{48} $ & $y_Q^{(2i1)} \sim {\cal O}\left(e^{-\frac{ 4 v_2 +v_3}{48}}\right)$ \\

\hline
\end{tabular}
\caption{Overview of the Yukawa couplings for the global six-stack left-right symmetric model of prototype II with D6-brane configuration of table~\ref{tab:6stackLRSaAAPrototypeII}. The third column lists the (triangular) worldsheet instantons $[x]$ or $[x,y,z]$ spanned by the indicated apexes $x,y,z$ on $T_{(i=2,3)}^2$ for the respective cubic couplings, where $x,y,z$ correspond to $\Z_2^{(i)}$-invariant orbits of intersection points. The fourth column presents the corresponding areas of the worldsheet instantons expressed in terms of the areas $v_i$ of the two-tori $T_{(i=2,3)}^2$, and the last column shows the scaling of the coupling constant corresponding to the considered cubic coupling.   \label{tab:YukaCouplingLRSpartI}}
\end{center}
\end{sidewaystable}

\begin{sidewaystable}[h]
\begin{center}
\begin{tabular}{|c|c|c|c|c|}
\hline \multicolumn{5}{|c|}{\bf Yukawa couplings (\ref{Eq:YukaCouplingLRS}) for prototype II Left-Right Symmetric Model (part II)  } \\
\hline
\hline {\bf Coupling} & {\bf Sequence} & {\bf Triangles on $T_{(2)}^2 \times T_{(3)}^2$} & {\bf Enclosed Area} & {\bf Parameter}\\
\hline 
\hline
$L^{(1)} (H_u, H_d)^{(3)} R^{(3)}$ & \multirow{6}{*}{$[b,(\omega c), d]$}  & $\left\{ [5,6,6],   [(Q_3,Q_3'),(2,3),6]  \right\}$ & $  \frac{3v_2}{4}+ \frac{v_3}{48} $ &$y_L^{(133)} \sim  {\cal O}(e^{-\frac{36 v_2+v_3}{48}})$  \\
 $L^{(1)} (H_u, H_d)^{(4)} R^{(3)}$ & &$\left\{[5,(R_2,R_2'),6],   [(Q_3,Q_3'),1,6] \right\}$ & $ \frac{v_2}{12} +  \frac{3v_3}{16} $  & $y_L^{(143)} \sim {\cal O}\left(e^{-\frac{4 v_2 + 9 v_3}{48}} \right)$ \\
 $L^{(1)} (H_u, H_d)^{(i=5,6)} R^{(3)}$ & & $\left\{[5,(R_2,R_2'),6],   [(Q_3,Q_3'),(2,3),6] \right\}$ & $ \frac{v_2}{12}+\frac{v_3}{48} $ &$y_L^{(1i3)} \sim  {\cal O}(e^{-\frac{4v_2+v_3}{48}})$  \\
 
 $L^{(3)} (H_u, H_d)^{(3)} R^{(3)}$ &   & $\left\{ [6],   [(Q_3,Q_3'),(2,3),6]  \right\}$ & $ \frac{v_3}{48} $ &$y_L^{(333)} \sim  {\cal O}(e^{-\frac{v_3}{48}})$  \\
 $L^{(3)} (H_u, H_d)^{(4)} R^{(3)}$ & &$\left\{[6,(R_2,R_2'),6],   [(Q_3,Q_3'),1,6] \right\}$ & $ \frac{v_2}{3} +  \frac{3v_3}{16} $  & $y_L^{(343)} \sim {\cal O}\left(e^{-\frac{16 v_2 + 9 v_3}{48}} \right)$ \\
 $L^{(3)} (H_u, H_d)^{(i=5,6)} R^{(3)}$ & & $\left\{[6,(R_2,R_2'),6],   [(Q_3,Q_3'),(2,3),6] \right\}$ & $ \frac{v_2}{3}+\frac{v_3}{48} $ &$y_L^{(3i3)} \sim  {\cal O}(e^{-\frac{16v_2+v_3}{48}})$  \\
 
\hline

 $L^{(2)} (H_u, H_d)^{(7)} R^{(3)}$ & \multirow{3}{*}{ $[b,(\omega^2 c),(\omega d)]$}  &$\left\{  [6,5,4], [4,(2,3),5]\right\}$  & $  \frac{3v_2}{8} + \frac{v_3}{24}$ &  $y_L^{(273)} \sim {\cal O}\left( e^{-\frac{9v_2 + v_3}{24}} \right)$  \\
  $L^{(2)} (H_u, H_d)^{(8)} R^{(3)}$ & &$\left\{  [6,(S_2,S_2'),4], [4,1,5]\right\}$  & $ \frac{v_2}{24} +  \frac{3v_3}{8}$ &  $y_L^{(283)} \sim {\cal O}\left( e^{-\frac{v_2 + 9 v_3}{24}} \right)$  \\
   $L^{(2)} (H_u, H_d)^{(i=9,10)} R^{(3)}$ &  &$\left\{  [6,(S_2,S_2'),4], [4,(2,3),5]\right\}$  & $ \frac{v_2 + v_3}{24}$ &  $y_L^{(2i3)} \sim {\cal O}\left( e^{-\frac{v_2 + v_3}{24}} \right)$  \\
   
\hline
$L^{(2)} (H_u, H_d)^{(3)} R^{(2)} $ & \multirow{6}{*}{$[b,(\omega c),(\omega d)]$}  & $\left\{ [6], [4,(2,3), (R_3,R_3')] \right\}$ & $ \frac{v_3}{48} $ & $y_L^{(232)} \sim {\cal O}\left(e^{-\frac{ v_3}{48}}\right)$ \\
$L^{(2)} (H_u, H_d)^{(4)} R^{(2)} $ &   & $\left\{ [6,(R_2,R_2'),6], [4,1, (R_3,R_3')] \right\}$ & $ \frac{v_2}{3}+\frac{3v_3}{16} $ & $y_L^{(242)} \sim {\cal O}\left(e^{-\frac{ 16 v_2 +9v_3}{48}}\right)$ \\
$L^{(2)} (H_u, H_d)^{(i=5,6)} R^{(2)} $ &   & $\left\{ [6,(R_2,R_2'),6], [4,(2,3), (R_3,R_3')] \right\}$ & $ \frac{v_2}{3}+\frac{v_3}{48} $ & $y_L^{(2i2)} \sim {\cal O}\left(e^{-\frac{ 16 v_2 +v_3}{48}}\right)$ \\

$L^{(2)} (H_u, H_d)^{(3)} R^{(1)} $ & & $\left\{ [6,6,4], [4,(2,3), (R_3,R_3')] \right\}$ & $\frac{3v_2}{4} + \frac{v_3}{48} $ & $y_L^{(231)} \sim {\cal O}\left(e^{-\frac{ 36v_2+ v_3}{48}}\right)$ \\
$L^{(2)} (H_u, H_d)^{(4)} R^{(1)} $ &   & $\left\{ [6,(R_2,R_2'),4], [4,1, (R_3,R_3')] \right\}$ & $ \frac{v_2}{12}+\frac{3v_3}{16} $ & $y_L^{(241)} \sim {\cal O}\left(e^{-\frac{ 4 v_2 +9v_3}{48}}\right)$ \\
$L^{(2)} (H_u, H_d)^{(i=5,6)} R^{(1)} $ &   & $\left\{ [6,(R_2,R_2'),4], [4,(2,3), (R_3,R_3')] \right\}$ & $ \frac{v_2}{12}+\frac{v_3}{48} $ & $y_L^{(2i1)} \sim {\cal O}\left(e^{-\frac{ 4 v_2 +v_3}{48}}\right)$ \\

\hline
\end{tabular}
\caption{Overview of the Yukawa couplings for the global six-stack left-right symmetric model of prototype II with D6-brane configuration of table~\ref{tab:6stackLRSaAAPrototypeII}. The third column lists the (triangular) worldsheet instantons $[x]$ or $[x,y,z]$ spanned by the indicated apexes $x,y,z$ on $T_{(i=2,3)}^2$ for the respective cubic couplings, where $x,y,z$ correspond to $\Z_2^{(i)}$-invariant orbits of intersection points. The fourth column presents the corresponding areas of the worldsheet instantons expressed in terms of the areas $v_i$ of the two-tori $T_{(i=2,3)}^2$, and the last column shows the scaling of the coupling constant corresponding to the considered cubic coupling.   \label{tab:YukaCouplingLRSpartII}}
\end{center}
\end{sidewaystable}

\clearpage

%%%%%%%%%%%%%%%%%%%%%%%%%%%%%%%%%%%%%%%%%%%%%%%%%%%%%%%%%%%%%%%%%%%%%%%%%%%%%%%%%%%
%%%%%%%%%%%%%%%%%%%%%%%%%%%%%%%%%%%%%%%%%%%%%%%%%%%%%%%%%%%%%%%%%%%%%%%%%%%%%%%%%%%
%%%%%%%%%%%%%%%%%%%%%%%%%%%%%%%%%%%%%%%%%%%%%%%%%%%%%%%%%%%%%%%%%%%%%%%%%%%%%%%%%%%
%%%%%%%%%%%%%%%%%%%%%%%%%%%%%%%%%%%%%%%%%%%%%%%%%%%%%%%%%%%%%%%%%%%%%%%%%%%%%%%%%%%
\section{Conclusions and Outlook}\label{S:conclu}
This article proceeds with the study of intersecting D6-brane model building on the fertile \linebreak \mbox{$T^6/(\Z_2 \times \Z_6 \times \OR)$} background with discrete torsion, which was initiated in a previous article by the same authors. The emphasis in this article lies on systematic scans for MSSM-like and left-right symmetric models on the considered toroidal orbifold background, which form consistent Type IIA/$\OR$ string vacua where the gauge degrees of freedom are attributed to D6-branes wrapping fractional, ideally rigid, three-cycles stuck at $\Z_2 \times \Z_2$ orbifold singularities. The scans presented in this article are exhaustive for D-brane configurations that are supersymmetric irrespective of the choice of the complex structure parameter $\varrho$ on the two-torus that is invariant under the $\Z_6$ and only feels the $\Z_2$ orbifold action.

As starting point for the systematic scans, we considered the requirement that the D6-brane stacks supporting the {\it QCD}  and the $SU(2)_L$ gauge groups are not accompanied by matter states in the adjoint representation. From a physical perspective, this requirement is motivated by ensuring that neither of the two gauge groups can be continuously broken by a non-vanishing {\it vev} of a matter state in the adjoint representation under the respective gauge group. A summary of all fractional three-cycles satisfying this constraint on the orbifold \mbox{$T^6/(\Z_2 \times \Z_6 \times \OR)$} is offered in section~\ref{Ss:IntersectSummary}. In our systematic search for MSSM-like and left-right symmetric models, this requirement has to be supplemented by additional constraints reflecting the correct massless open string spectra with respect to the gauge group configuration under consideration. These latter constraints can be decomposed into two separate requirements: the required absence of chiral matter states in the symmetric representation under the {\it QCD} or $SU(2)_L$ gauge group on the one hand, and the presence of three chiral generations of quarks and leptons on the other hand. An important observation following from these requirements is the fact that $\varrho$-independent models are only able to satisfy all of the aforementioned restrictions provided that the $SU(2)_L$ gauge group is realised as an enhanced $USp(2)$ gauge group, and the exotic O6-plane is chosen to be the $\OR\Z_2^{(2 \text{ or } 3)}$-plane. Moreover, by virtue of all these requirements, we can exclude the existence of $\varrho$-independent {\it local} MSSM-like and left-right symmetric models on the {\bf bAA} lattice with three right-handed quark generations and a $SU(2)_L$-stack realised as an enhanced $USp(2)$ gauge group. For that reason, our systematic search focused on the only remaining independent {\bf aAA} lattice configuration of the orientifold $T^6/(\Z_2 \times \Z_6 \times \OR)$ with discrete torsion, for which we have more room to manoeuvre with respect to the number of D6-brane stacks without overshooting the bulk RR tadpole cancellation conditions. 

Regarding MSSM-like D6-brane model searches on the {\bf aAA} lattice, we noticed the absence of {\it local} three-stack MSSM-like configurations, confirmed the existence of {\it local} four-stack MSSM-like models and identified a class of {\it global} five-stack MSSM-like D-brane configurations. When counting the number of four- and five-stack D6-brane configurations, we took into account the obvious symmetries among the models due to identical {\it relative} $\Z_2^{(i)}$ eigenvalues and identical {\it absolute} discrete Wilson lines and displacements characterising the fractional three-cycles. Nonetheless, identical massless spectra among different D6-brane configurations might suggest the potential existence of more intricate pairwise symmetries among non-identical {\it relative} discrete parameters, such that the number of physically inequivalent models might even be further reduced.

With respect to left-right symmetric D6-brane model searches on the {\bf aAA} lattice, we confirmed the existence of {\it local} four-stack left-right symmetric D-brane configurations, stumbled upon the existence of {\it semi-local} five-stack left-right symmetric models and identified two prototypes of six-stack left-right symmetric models based on the ranks of the hidden gauge groups. Both prototype models contain D6-brane configurations yielding {\it semi-local} models and D6-brane configurations giving rise to {\it global} models. Within the prototype II models, we were also able to identify subclasses IIb and IIc based on the massless open string states in the `messenger' and `hidden' sectors. Subclass IIb represents examples of global six-stack left-right symmetric models with both hidden gauge groups completely decoupled from the visible sector, while subclass IIc captures examples of global six-stack left-right symmetric models with one of the hidden gauge groups completely decoupled from the visible sector. Another subtle difference between the various classes of models is the absence of a massless (generalised) $B-L$ symmetry for the prototype I, IIb and IIc models, whereas the other prototype II models -- not belonging to IIb or IIc -- do come with a massless generalised $B-L$ symmetry. This observation begs the question whether it is possible to identify a massless $U(1)_Y$ hypercharge at all upon spontaneous breaking of the $SU(2)_R$ gauge group for the prototype models I, IIb and IIc. 

Apart from the systematic search for MSSM-like and left-right symmetric models, we also studied various phenomenological properties of the global models, such as the presence of discrete $\Z_n$ gauge symmetries and the existence of non-vanishing Yukawa and other cubic couplings, for one of the representants per prototype of the global models. In particular for the global five-stack MSSM-like model, the discrete symmetries and the explicit form of the superpotential play an indispensable r\^ole in identifying correctly the right-handed quarks and distinguishing the left-handed lepton multiplets from the down-type Higgs multiplets. The matter state assignment of the chiral spectrum in table~\ref{tab:5stackMSSMaAAPrototypeISpectrum} is the only interpretation, for which $d_R$ quark candidates enter in standard Yukawa couplings that are compatible with the discrete $\Z_3$ symmetry. This $\Z_3$ symmetry also acts non-trivially on the extended Higgs sector and the left-handed leptons, such that Yukawa couplings involving right-handed electrons or right-handed neutrinos can only occur for the considered identification of left-handed leptons and down-type Higgses in table~\ref{tab:5stackMSSMaAAPrototypeISpectrum}. These considerations also constrain the origin of the three right-handed Standard Model neutrinos.
Looking further at the visible part of the massless spectrum, one notices that the discrete $\Z_3$ symmetry provides the same selection rules as the global, anomalous linear combination $U(1)_{c} - U(1)_d$ that acts as  Peccei-Quinn symmetry on the MSSM-like models. In this respect, the global five-stack MSSM-like D6-brane configurations found here present explicit supersymmetric realisations of the DFSZ axion model, as we confirmed through the existence of non-vanishing perturbatively allowed Higgs-axion couplings. In this type of models, the $QCD$ axion is located in the ${\cal N}=1$ supermultiplet of a massless open string state
similar to the scenario suggested in~\cite{Berenstein:2012eg}, which should be contrasted to other type II superstring scenarios where the r\^ole of the $QCD$ axion is played by a closed string axion~\cite{Conlon:2006tq,Svrcek:2006yi,Cicoli:2012sz,Belhaj:2015zra}.  

In the case of the global six-stack left-right symmetric models, we did not find any discrete $\Z_n$ symmetry acting non-trivially on the massless open string spectrum that could have provided selection rules beyond the ones associated to the non-Abelian gauge factors. We point out that the Yukawa couplings present a form of universality, due to the fact that the prototype models have an identical visible sector and only differ in the choice of hidden D-branes.

In order to study the related low-energy effective Yukawa and higher order couplings more in-depth, it will be necessary to perform reliable CFT computations for $m$-point couplings on orbifolds containing $\Z_2$ factors, since  the argument of vanishing couplings for some vanishing angle~\cite{Lust:2004cx} is based on extended ${\cal N}=2$ supersymmetry on the six-torus, which, however is broken here to ${\cal N}=1$ by the $\Z_2 \times \Z_2$ symmetries. 

Other phenomenological aspects to be studied in the future include possible deformations of the exceptional three-cycles away from the singular point in moduli space in analogy to~\cite{Blaszczyk:2014xla,Blaszczyk:2015oia}, which will usually lead to a splitting of previously identical gauge couplings at tree level for some deformations and stabilisation of other twisted moduli at the orbifold point. When also taking one-loop corrections to the gauge couplings into account  in analogy to section~5 of~\cite{Honecker:2012qr}, it will be interesting to see how low values of the string scale are compatible with the measured strengths of the strong and electro-weak gauge couplings, and if our global models 
fit into the analysis of  low string scale scenarios at the LHC as discussed e.g. in~\cite{Accomando:1999sj,Cullen:2000ef,Burikham:2004su,Lust:2008qc,Chemtob:2008cb,Anchordoqui:2009mm,Kitazawa:2010gh,Anchordoqui:2011eg,Anchordoqui:2012wt,Berenstein:2014wva}.

All models presented here preserve ${\cal N}=1$ supersymmetry at the string scale. Another pressing question thus consists in identifying possible supersymmetry breaking scenarios.
While we expect that non-supersymmetric deformations away from the singular orbifold point will predominantly stabilise moduli at the singularity as argued in~\cite{Blaszczyk:2014xla,Blaszczyk:2015oia}, 
it remains to be seen if the maximal non-Abelian hidden gauge groups $SU(4)$ or $SU(3) \times SU(3)$ in the present D6-brane configurations
 are suitable to generate a gaugino condensate, which breaks supersymmetry, and if so study gauge mediation versus gravity mediation scenarios. 

\vspace{5mm}

\noindent
{\bf Acknowledgements:} 
W.S. would like to thank Fernando Marchesano for useful discussions. This work is partially supported by the {\it Cluster of Excellence `Precision Physics, Fundamental Interactions and Structure of Matter' (PRISMA)} DGF no. EXC 1098,
the DFG research grant HO 4166/2-1 and the DFG Research Training Group {\it `Symmetry Breaking in Fundamental Interactions'} GRK 1581. W.S. is supported by the ERC Advanced Grant SPLE under contract ERC-2012-ADG-20120216-320421, by the grant FPA2012-32828 from the MINECO, and the grant SEV-2012-0249 of the ``Centro de Excelencia Severo Ochoa" Programme.

%\clearpage

%%%%%%%%%%%%%%%%%%%%%%%%%%%%%%%%%%%%%%%%%%%%%%%%%%%%%%%%%%%%%%%%%%%%%%%%%%%%%%%%%%%
%%%%%%%%%%%%%%%%%%%%%%%%%%%%%%%%%%%%%%%%%%%%%%%%%%%%%%%%%%%%%%%%%%%%%%%%%%%%%%%%%%%
%%%%%%%%%%%%%%%%%%%%%%%%%%%%%%%%%%%%%%%%%%%%%%%%%%%%%%%%%%%%%%%%%%%%%%%%%%%%%%%%%%%
%%%%%%%%%%%%%%%%%%%%%%%%%%%%%%%%%%%%%%%%%%%%%%%%%%%%%%%%%%%%%%%%%%%%%%%%%%%%%%%%%%%
\appendix

%%%%%%%%%%%%%%%%%%%%%%%%%%%%%%%%%%%%%%%%%%%%%%%%%%%%%%%%%%%%%%%%%%%%%%%%%%%%%%%%%%%%%%%%%%%%%%%%%%%%%%%%%%%%%%%%%%%%%%%%%%%%%%%%%%%%%%%%%%%%%%%%%%%%%%%%%%%%%%%%%%%%%%%%
\section{Localisation of matter states via Chan-Paton labels}\label{A:ChanPatonMethod}

In this appendix, we briefly summarise the method of employing Chan-Paton labels to determine the localisation of matter states presented first in~\cite{Gmeiner:2008xq} and appendix B.1 of~\cite{Forste:2010gw}
for fractional D6-branes on  orbifolds with some $\Z_2$ or $\Z_2 \times \Z_2$ subsymmetry. We discuss here for the first time explicitly how to not only include displacements, but also
sign factors arising from discrete Wilson lines in the analysis. 
To keep the presentation brief and focussed, we concentrate on the MSSM-like D-brane configuration in table~\ref{tab:5stackMSSMaAAPrototypeI} with the complete massless matter spectrum displayed in table~\ref{tab:5stackMSSMaAAPrototypeISpectrum}.
The discussion can easily be generalised, not only for other rigid D6-brane models but also to T-dual magnetised D9/D5-brane models as in e.g.~\cite{Angelantonj:2009yj,Angelantonj:2011hs}.

All D-brane intersections in $\varrho$-independent global models on $T^6/(\Z_2 \times \Z_6 \times \OR)$ have a vanishing angle along $T^2_{(1)}$, i.e. the angles are $(0,\phi,-\phi)$ or $(0,0,0)$,
and the chiral multiplets at non-vanishing intersections along $T^4_{(1)} \equiv T^2_{(2)} \times T^2_{(3)}$ of the type $x(\omega^k y)$ can be extracted from table~47 of~\cite{Forste:2010gw},
\begin{equation*}
\begin{array}{|c|c|}\hline
\text{state} & (c^{\Z_2^{(1)}},c^{\Z_2^{(2)}},c^{\Z_2^{(3)}}) 
\\\hline\hline
\Psi^3_{(0,\phi,-\phi)} \supset \{\psi^3_{-1/2+\phi} |0 \rangle_{\rm NS}^{({\rm tw})} \, , \, |\tilde{0}\rangle_{\rm R}^{({\rm tw}, 1)} \}
&  (-,-,+)
\\
\Psi^{\ov{2}}_{(0,\phi,-\phi)}  \supset \{ \psi^{\ov{2}}_{-1/2+\phi} |0 \rangle_{\rm NS}^{({\rm tw})} \, , \, \psi^{\mu}_0 \psi^1_0|\tilde{0}\rangle_{\rm R}^{({\rm tw}, 1)}  \}
&  (-,+,-) 
\\\hline
\end{array}
\end{equation*}
where $(c^{\Z_2^{(1)}},c^{\Z_2^{(2)}},c^{\Z_2^{(3)}})$ denotes the $\Z_2 \times \Z_2$ transformation properties of the multiplets if they are localised at some $\Z_2 \times \Z_2$  invariant point {\it in the absence of discrete Wilson lines}.
The missing half of states within $\Psi^i$ not listed explicitly here within the brackets stem from the inverse sectors $(\omega^k y)x$ at angle $(0,-\phi,\phi)$.
The two Weyl fermions within the mulitplets $\Psi^3_{(0,\phi,-\phi)}$ and $\Psi^{\ov{2}}_{(0,\phi,-\phi)}$ have opposite chiralities due to the helicity-flip operator $\psi^{\mu}_0$, 
and the absolute chirality is fixed by the sign of the angle $\phi$.

The common method of Chan-Paton labels in the presence of some $\Z_2$ symmetry(ies)
pioneered in~\cite{Gimon:1996rq}  relies on adding up stacks of $N$ fractional D-branes with opposite $\Z_2$ eigenvalues to arrive at a pure bulk stack of $N$ D-branes 
as first demonstrated in the language of intersecting D-branes on fractional cycles in~\cite{Blumenhagen:2002wn}. 
For simplicity, we choose the following representation of $\gamma$-matrices associated to the $\Z_2$-projections as in~\cite{Forste:2010gw},
\begin{equation}
\begin{aligned}
\gamma_{\Z_2^{(1)} } & = \text{diag}(\unity_{N \times N},\unity_{N \times N},-\unity_{N \times N},-\unity_{N \times N})
, \\
\gamma_{\Z_2^{(2)} } & = \text{diag}(\unity_{N \times N}, -\unity_{N \times N},\unity_{N \times N}, -\unity_{N \times N})
, \\
\gamma_{\Z_2^{(3)} } &= \text{diag}(\unity_{N \times N},- \unity_{N \times N}, -\unity_{N \times N},\unity_{N \times N})
, 
\end{aligned}
\end{equation}
which leads to the decomposition of Chan-Paton labels
\begin{equation}
\lambda_{cd}=
\left(\begin{array}{ccccc}
(\N_c^{1},\ov{\N}_d^{1}) & (\N_c^{1},\ov{\N}_d^{2}) & (\N_c^{1},\ov{\N}_d^{3}) & (\N_c^{1},\ov{\N}_d^{4}) \\
(\N_c^{2},\ov{\N}_d^{1}) & (\N_c^{2},\ov{\N}_d^{2}) & (\N_c^{2},\ov{\N}_d^{3}) & (\N_c^{2},\ov{\N}_d^{4}) \\
(\N_c^{3},\ov{\N}_d^{1}) & (\N_c^{3},\ov{\N}_d^{2}) & (\N_c^{3},\ov{\N}_d^{3}) & (\N_c^{3},\ov{\N}_d^{4}) \\
(\N_c^{4},\ov{\N}_d^{1}) & (\N_c^{4},\ov{\N}_d^{2}) & (\N_c^{4},\ov{\N}_d^{3}) & (\N_c^{4},\ov{\N}_d^{4}) 
\end{array}\right)
\end{equation}
of $\prod_{i=1}^4 U(N^i_c) \times U(N^i_d)$, where we implicitly used the assignment 
\begin{equation*}
\begin{array}{|c|c|c|c|}\hline
\N^1 & \N^2 & \N^3 & \N^4
\\\hline
(+,+,+) & (+,-,-) & (-,+,-) & (-,-,+)
\\\hline
\end{array}
\end{equation*}
of the fractional D-brane label $i$ and its characterisation via the set of $\Z_2 \times \Z_2$ eigenvalues used throughout this article.
Four stacks of $N^i_x$ fractional D-branes thus add up to a stack of $N_x$ bulk D-branes for $N_x=N_x^1=N_x^2=N_x^3=N_x^4$.

The $\Z_2$ projections then act on a given state by
\begin{equation}
\lambda_{cd} |\text{state}\rangle_{(\alpha,\beta)} \stackrel{\Z_2^{(k)}}{\longrightarrow} c_{\text{state}}^{\Z_2^{(k)}} \, \left(\gamma_{\Z_2^{(k)}} \, \lambda_{cd} \gamma_{\Z_2^{(k)}}^{-1}\right) \,  |\text{state}\rangle_{\Z_2^{(k)}(\alpha,\beta)}  
\, ,
\end{equation}
leading to $\lambda_{cd} \stackrel{!}{=}c_{\text{state}}^{\Z_2^{(k)}} \, \left(\gamma_{\Z_2^{(k)}} \, \lambda_{cd} \gamma_{\Z_2^{(k)}}^{-1}\right)$ for states at $\Z_2^{(k)}$-invariant intersections 
\mbox{$(\alpha,\beta)=\Z_2^{(k)}(\alpha,\beta)$}, while for $(\alpha,\beta) \neq \Z_2^{(k)}(\alpha,\beta)$ the matter states are simply spread over the two intersection points paired under 
$\Z_2^{(k)}$.\footnote{The discussion in this appendix is easily adjusted to bulk and fractional D-branes on orbifolds with a single $\Z_2$-subsymmetry such as for $T^6/\Z_{2N}$ with $2N=4,6,6'$.
The computations of vanishing one-loop corrections to K\"ahler metrics for this type of orbifold in~\cite{Berg:2011ij,Berg:2014ama} implicitly use only pure bulk D6-branes by setting prefactors of amplitudes proportional to  $\text{tr}\gamma_{\Z_2}=0$, while all existing phenomenologically appealing models require the use of fractional D6-branes that do not pair up to bulk D6-branes, see e.g.~\cite{Blumenhagen:2002gw,Honecker:2004kb,Honecker:2004np,Bailin:2006zf,Gmeiner:2007we,Gmeiner:2007zz,Bailin:2007va,Gmeiner:2008xq,Bailin:2008xx,Gmeiner:2009fb}.}

Generalising Chan-Paton labels to arbitrary fractional D-branes boils down to selecting the corresponding $\N^i$ while setting all other $\N^{j \neq i}$ to zero, e.g. for the fractional branes $c$ and $d$ in the MSSM-like model of table~\ref{tab:5stackMSSMaAAPrototypeI} we start with:
\begin{equation}
\left(\begin{array}{ccccc}
0 & 0 & 0 & 0  \\
0 & (\Adj_c^{2}) & 0 & 0 \\
0 & 0 & 0 & 0  \\
0 & 0 & 0 & 0  
\end{array}\right)
, \quad
\left(\begin{array}{ccccc}
0 & 0 & 0 & 0  \\
0 & 0 & 0 & 0  \\
0 & 0 & (\Adj_d^{3}) & 0 \\
0 & 0 & 0 & 0  
\end{array}\right)
, \quad
\left(\begin{array}{ccccc}
0 & 0 & 0 & 0  \\
0 & 0 & (\N_c^{2},\ov{\N}_d^{3}) & 0 \\
0 & 0 & 0 & 0  \\
0 & 0 & 0 & 0  
\end{array}\right)
.
\end{equation}
Using the orbifold image wrapping numbers in equation~\eqref{Eq:1-cycle-orbits}, we obtain $I_{x(\omega \, y)}= (-3) \cdot 3$ and $I_{x(\omega^2 y)}=3 \cdot (-3)$ for $x,y \in \{b,c,d\}$, where the signs per two-torus are explicitly shown as a reminder that the angles in these two sectors are exactly opposite. Only one of the nine intersection points is $\Z_2 \times \Z_2$ invariant, and the signs of $\sgn(I^{\Z_2^{(i)},(2\cdot 3)}_{x(\omega \, y)})$
are determined using table~\ref{tab:Z2Z6SignAssignment}.
%%%%%%%%%%%%%%%%%%%%%%%%
\mathtabfix{
\begin{array}{|c||ccc||ccc|ccc|}\hline 
\multicolumn{7}{|c|}{\text{\bf Assignment of prefactors $(-1)^{\tau^{\Z_2^{(i)}}_x}$ \!\!\!\!\! or $(-1)^{\tau^{\Z_2^{(i)}}_x \!\!\!+ \tau^i_x}$ in the counting of states per intersection
}}
\\\hline\hline
& \multicolumn{3}{|c|}{\text{\bf Assignment on }\, T_{(2)}^2 }& \multicolumn{3}{|c|}{\text{\bf Assignment on }\, T_{(3)}^2}\\
\hline \hline
(n^i_x,m^i_x)  & \text{(odd,odd)} \stackrel{\omega}{\to} & \!\!\!\!\text{(odd,even)} \stackrel{\omega}{\to} & \!\!\!\!\text{(even,odd)} & \text{(odd,odd)} \stackrel{\omega}{\to} & \!\!\!\!\text{(even,odd)} \stackrel{\omega}{\to} & \!\!\!\!\text{(odd,even)}\\
\hline
\hline
& (\omega^2 d) & d &  (\omega d) & (\omega b) & (\omega^2 b) & b
\\\cline{2-7}
\sigma^i_x=0 
& 
\left(\begin{array}{c} 1 \\  6\end{array}\right) \to  & \left(\begin{array}{c} 1 \\ 4 \end{array}\right) \to & \left(\begin{array}{c} 1 \\ 5 \end{array}\right)
& \left(\begin{array}{c} 1 \\  6\end{array}\right) \to & \left(\begin{array}{c} 1 \\ 5 \end{array}\right) \to & \left(\begin{array}{c} 1 \\ 4 \end{array}\right)
\\\hline\hline
&  (\omega^2 a), (\omega^2 h) & a,h & (\omega a), (\omega h) &  (\omega a), (\omega h) & (\omega^2 a), (\omega^2 h)  & a,h 
\\\cline{2-7}
&  (\omega^2 b), (\omega^2 c) & b,c & (\omega b), (\omega c) & (\omega c), (\omega d) &(\omega^2 c) ,(\omega^2 d)  & c,d
\\\cline{2-7}
\sigma^i_x=1 
&  \left(\begin{array}{c}  4 \\ 5 \end{array}\right) \to & \left(\begin{array}{c} 5 \\ 6 \end{array}\right)\to   & \left(\begin{array}{c} 6 \\ 4 \end{array}\right)  
&
\left(\begin{array}{c}  4 \\ 5 \end{array}\right) \to & \left(\begin{array}{c} 6 \\ 4 \end{array}\right)\to   & \left(\begin{array}{c} 5 \\ 6 \end{array}\right)  
\\
\hline
\end{array}
}{Z2Z6SignAssignment}{Consistent assignment of the reference point (upper entry)  and the second $\Z_2^{(i)}$ fixed point (lower entry) contributing with  sign factor  
$(-1)^{\tau^{\Z_2^{(i)}}_x}$ and $(-1)^{\tau^{\Z_2^{(i)}}_x+ \tau^i_x}$, respectively, to $\Pi^{\Z_2^{(j),j\neq i}}_x$ and consequently $I_{x(\omega^k y)}^{\Z_2^{(j\neq i)}}$ on the orbifold $T^6/(\Z_2 \times \Z_6 \times \OR)$ with discrete torsion
according to~\cite{Ecker:2014hma}. 
In the rows directly above the fixed points, we list here which D-branes of the MSSM-like model of table~\protect\ref{tab:5stackMSSMaAAPrototypeI} belong to each category.
The same sign factors enter the $\Z_2 \times \Z_2$ projections of the Chan-Paton labels as detailed in the main text of appendix~\protect\ref{A:ChanPatonMethod}.
}
%%%%%%%%%%%%%%%%%%%%%%%%
The same signs enter the $\Z_2 \times \Z_2$ projections of the Chan-Paton labels by effectively shifting
\begin{equation}
(c^{\Z_2^{(1)}},c^{\Z_2^{(2)}},c^{\Z_2^{(3)}}) \longrightarrow (c^{\Z_2^{(1)}}  \!\!\cdot\! (-1)^{\tau^2+\tau^3} ,c^{\Z_2^{(2)}}   \!\!\cdot\! (-1)^{\tau^3},c^{\Z_2^{(3)}}   \!\!\cdot\! (-1)^{\tau^2}) 
\end{equation}
for $\varrho$-independent brane configurations (i.e.  $\tau^1$ here only determines if the mass of some state is shifted away from the massless case).

The nine intersections in the $x(\omega \, y)$ sectors for $x,y \in \{b,c,d\}$ are grouped into one $\Z_2 \times \Z_2$ invariant point, a doublet of points fixed under $\Z_2^{(3)}$, a doublet of points fixed under $\Z_2^{(2)}$
and a quadruplet under $\Z_2 \times \Z_2$. At the $\Z_2 \times \Z_2$ fixed point, all three $\Z_2$ eigenvalues shifted by some discrete Wilson lines have to be taken into account, at each doublet only the respective $\Z_2^{(k)}$
invariance acts as a projection since the other $\Z_2$'s act by exchanging localisations, e.g. for the $\Z_2^{(2)}$ fixed pair in the $d(\omega \, d)$ sector, 
\begin{equation}
\Psi^3 \big|_{(2,5) \text{ on } T^2_{(2)} \times T^2_{(3)}} \stackrel{\Z_2^{(1,3)}}{\longleftrightarrow} \Psi^3 \big|_{(3,5) \text{ on } T^2_{(2)} \times T^2_{(3)}} 
\, ,
\end{equation}
and depending on the relevant $\Z_2$ eigenvalues and discrete Wilson lines the multiplets $\Psi^3$ and $\Psi^{\ov{2}}$ remain in the spectrum ($\Psi^i_{\checkmark}$) or are projected out ($\cancel{\Psi^i}$).
The $\Z_2 \times \Z_2$ symmetry acts on the quadruplet merely by permuting localisations. 
As illustrative examples we compare the localisations of states in the $x(\omega \, y)$ sectors for $x,y \in \{c,d\}$ in table~\ref{Tab:Examples-Localisations}.
%%%%%%%%
\begin{table}
\begin{equation*}
\begin{array}{|c||c|c|c|c|}\hline
\muc{4}{|c|}{\text{\bf Examples of matter localisations in the global MSSM-like model}}
\\\hline\hline
\text{intersection} & c(\omega\, c) & d(\omega\, d) & c(\omega\, d) 
\\\hline\hline
\begin{array}{c} \Z_2 \times \Z_2 \\ \text{fixed} \end{array}
& \begin{array}{c} (6_{\tau^2_c=1} \, , \,  5_{\tau^3_{(\omega\,  c)}=1}) \\ \cancel{ \Psi^{3} }, \cancel{\Psi^{\ov{2}}} \end{array}
& \begin{array}{c} (1 \, , \, 5_{\tau^3_{(\omega\, d)} =1}) \\ \Psi^3_{\checkmark} ,  \cancel{\Psi^{\ov{2}}}  \end{array}
& \begin{array}{c} (5_{\tau^2_{(\omega \, d)}=0} \, , \, 5_{\tau^3_{(\omega \, d)}=1}) \\ \cancel{ \Psi^{3} }, \cancel{\Psi^{\ov{2}}}  \end{array}
\\\cline{2-4}
& \emptyset & ``(-1)'' \times \Adj^3_d &  \emptyset 
\\\hline
\begin{array}{c} \Z_2^{(3)} \\ \text{fixed} \end{array}
& \begin{array}{c} (6_{\tau^2_c=1}  \, , \, 
S_3 \stackrel{\Z_2^{(1,2)}}{\longleftrightarrow}  S_3' 
) \\\cancel{ \Psi^{3} }, \Psi^{\ov{2}}_{\checkmark} \end{array}
& \begin{array}{c} (1 \, , \, 
S_3 \stackrel{\Z_2^{(1,2)}}{\longleftrightarrow}  S_3' 
) \\ \Psi^3_{\checkmark} ,  \cancel{\Psi^{\ov{2}}} \end{array}
& \begin{array}{c} (5_{\tau^2_{(\omega \, d)}=0} \, , \,
S_3 \stackrel{\Z_2^{(1,2)}}{\longleftrightarrow}  S_3' 
) \\  \Psi^3_{\checkmark} ,  \cancel{\Psi^{\ov{2}}}  \end{array}
\\\cline{2-4}
& 1 \times \Adj^2_c &``(-1)'' \times \Adj^3_d & (\ov{\N}^2_c,\N^3_d) 
\\\hline
\begin{array}{c} \Z_2^{(2)} \\ \text{fixed} \end{array}
& \begin{array}{c} ( 
R_2 \stackrel{\Z_2^{(1,3)}}{\longleftrightarrow} R_2'
\, , \,  5_{\tau^3_{(\omega c)}}) \\ \Psi^3_{\checkmark} ,  \cancel{\Psi^{\ov{2}}}  \end{array}
& \begin{array}{c} (2  \stackrel{\Z_2^{(1,3)}}{\longleftrightarrow} 3 \, , \, 5_{\tau^3_{ (\omega \, d)} =1} ) \\  \Psi^3_{\checkmark} ,  \cancel{\Psi^{\ov{2}}}  \end{array}
& \begin{array}{c} 
(S_2 \stackrel{\Z_2^{(1,3)}}{\longleftrightarrow} S_2'
 , \, 5_{\tau^3_{(\omega \, d)}=1}) \\ \cancel{ \Psi^{3} }, \Psi^{\ov{2}}_{\checkmark} \end{array}
\\\cline{2-4}
& 1 \times \Adj^2_c &``(-1)'' \times \Adj^3_d & (\N^2_c,\ov{\N}^3_d) 
\\\hline
\text{quadruplet} & 
\begin{array}{c} 
 \begin{array}{ccc} (R_2,S_3) & \stackrel{\Z_2^{(1)}}{\longleftrightarrow} & (R_2',S_3')  \\ \updownarrow \Z_2^{(2)} & & \updownarrow \\ (R_2,S_3') & \longleftrightarrow&  (R_2',S_3)  \end{array}
\\ \Psi^3_{\checkmark} ,\Psi^{\ov{2}}_{\checkmark} \end{array}
& 
\begin{array}{c} 
\begin{array}{ccc} (2,S_3) & \stackrel{\Z_2^{(1)}}{\longleftrightarrow} & (3,S_3')  \\ \updownarrow \Z_2^{(2)} & & \updownarrow \\ (2,S_3') & \longleftrightarrow&  (3,S_3)  \end{array}
\\ \Psi^3_{\checkmark} ,\Psi^{\ov{2}}_{\checkmark} \end{array}
& 
\begin{array}{c} 
 \begin{array}{ccc} (S_2,S_3) & \stackrel{\Z_2^{(1)}}{\longleftrightarrow} & (S_2',S_3')  \\ \updownarrow \Z_2^{(2)} & & \updownarrow \\ (S_2,S_3') & \longleftrightarrow&  (S_2',S_3)  \end{array}
\\ \Psi^3_{\checkmark} ,\Psi^{\ov{2}}_{\checkmark} \end{array}
\\\cline{2-4}
& |2| \times \Adj^2_c & |2| \times \Adj^3_d & [ (\N^2_c,\ov{\N}^3_d) + c.c.]
\\\hline\hline
\text{total matter} & 4 \times \Adj_c & 5 \times \Adj_d & 2 \times [ (\N_c,\ov{\N}_d) + c.c.]
\\\hline
\end{array}
\end{equation*}
\caption{Examples of matter localisations in dependence of the displacement and Wilson line parameters $(\vec{\sigma})$ and $(\vec{\tau})$ for the global MSSM-like D-brane configuration of table~\protect\ref{tab:5stackMSSMaAAPrototypeI}.
In each box, we list the intersection points, e.g. $(6,5)$ on $T^2_{(2)} \times T^2_{(3)}$, on the first line with the relevant Wilson line(s) according to table~\protect\ref{tab:Z2Z6SignAssignment} as lower index,
and the projections on the massless multiplets $\Psi^3$, $\Psi^{\ov{2}}$ in the second line. 
The following line shows the corresponding matter representation and multiplicity. Since all three sectors $c(\omega \, c)$, $c(\omega \, d)$ and $d(\omega \, d)$ 
have the same intersection angles, the  Weyl fermion within $\Psi^3$ has the same chirality within each column, and as usual the Weyl fermion within $\Psi^{\ov{2}}$ has always the opposite chirality. 
The total amount of matter matches the corresponding entries in table~\protect\ref{tab:Z2Z65stackMSSMTotalSpectrumI} when taking into account that here we implicitly included the inverse sectors for the adjoints, e.g. $(\omega \, c)c \simeq c(\omega^2 c)$, while for notational consistency in table~\protect\ref{tab:Z2Z65stackMSSMTotalSpectrumI} we displayed the counting of states separately.
}
\label{Tab:Examples-Localisations}
\end{table}
%%%%%%%%%%%%%%%%%%

\clearpage

\section{Five-Stack Left Right Symmetric Models}\label{A:5StackLRS}

In this appendix we present an explicit example of the prototype five-stack left-right symmetric model identified in section~\ref{Ss:LR-models},
for which we already anticipated that only {\it semi-local} realisations exist,
 and we discuss some of its properties. The full D6-brane configuration of the example is given in table~\ref{tab:5stackLRSaAAPrototypeI}, giving rise to the massles ${\cal N}=1$ supersymmetric open string spectrum listed in table~\ref{tab:5stackLRSMaAAPrototypeISpectrum}. As we show below, some K-theory constraints are violated while all RR tadpole cancellation conditions are satisfied.

%%%%%%%%%%%%%%%%%%%%%%%%%%%%%%%%%%%%%%%%%%%%%%%%%%%%%%%%%%%%%%%%%%%%%%%%%%%
\mathtabfix{
\begin{array}{|c||c|c||c|c|c||c|}\hline 
\muc{7}{|c|}{\text{\bf D6-brane configuration of a 5-stack Left-Right Symmetric model on the {aAA} lattice}}
\\\hline \hline
&\text{\bf wrapping numbers} &\frac{\rm Angle}{\pi}&\text{\bf $\Z_2^{(i)}$ eigenvalues}  & (\vec \tau) & (\vec \sigma)& \text{\bf gauge group}\\
\hline \hline
 a&(1,0;1,0;1,0)&(0,0,0)&(--+)&(0,1,1) & (0,1,1)& U(3)\\
 b&(1,0;-1,2;1,-2)&(0,\frac{1}{2},-\frac{1}{2})&(+++)&(0,1,0) & (0,1,0)&USp(2)\\
 c&(1,0;-1,2;1,-2)&(0, \frac{1}{2},-\frac{1}{2})&(-+-)&(0,1,0) & (0,1,0)&USp(2)\\ 
  d&(1,0;-1,2;1,-2)&(0, \frac{1}{2},-\frac{1}{2})&(+--)&(0,0,0) & (0,0,0)& U(1)\\
  \hline
    h&(1,0;1,0;1,0)&(0,0,0)&(+++)&(0,1,1) & (0,1,1)& U(4)\\
 \hline
\end{array}
}{5stackLRSaAAPrototypeI}{D6-brane configuration of a {\it semi-local} 5-stack Left-Right Symmetric model with initial gauge group 
$U(3)_a\times USp(2)_b \times USp(2)_c\times U(1)_d \times U(4)_h$ on the {\bf aAA} lattice of the orientifold $T^6/(\Z_2 \times \Z_6 \times \OR)$ with discrete torsion ($\eta=-1$) and the $\OR\Z_2^{(3)}$-plane as the exotic O6-plane ($\eta_{\OR\Z_2^{(3)}}=-1$).}
%%%%%%%%%%%%%%%%%%%%%%%%%%%%%%%%%%%%%%%%%%%%%%%%%%%%%%%%%%%%%%%%%%%%%%%%%%%

Apart from the desired three generations of quarks and leptons, the visible sector also contains an abundant amount of non-chiral massless states in the $bd$ and $cd$ sectors.
Given the nature of the `hidden' $U(4)$ gauge group and the absence of non-chiral states in the $ch$ and $dh$ sectors, one could also interpret this example as a candidate five-stack Pati-Salam model with the $U(3)_a \times U(1)_d$ stack as the `hidden' gauge group. Yet, the absence of a Pati-Salam GUT Higgs  $(\1,\1,\2, \4)_{(0,0,1)}$ (or its complex conjugate) \cite{Antoniadis:1988cm,King:1997ia} refrains us from doing so, as this would require us to discuss the spontaneous breaking of the $SU(4)\times USp(2)_c$ Pati-Salam gauge group to the $SU(3)_{QCD}\times U(1)_Y$ gauge group in a non-standard way.

It can be shown straightforwardly that the bulk and twisted RR tadpole cancellation conditions are satisfied for the D6-brane configuration in table~\ref{tab:5stackLRSaAAPrototypeI}. As already anticipated in section~\ref{Ss:LR-models}, some of the K-theory constraints are not satisfied for this model, and more explicitly the third block in the sufficient K-theory constraints (\ref{Eq:KTheoryZ2Z6}) contains entries which violate the K-theory condition:
\begin{equation}\label{Eq:ViolatedKtheory}
0 \text{ mod } 2 \stackrel{!}{=} \sum_a N_a\left(\begin{array}{c}
24-\frac{3 P_a}{2}  + \frac{3 \,x_{0,a}^{(1)} + x_{1,a}^{(1)} + x_{2,a}^{(1)} + x_{3,a}^{(1)} }{4} - \frac{x_{2,a}^{(2)} +x_{3,a}^{(2)} }{2} + \frac{x_{2,a}^{(3)} +x_{3,a}^{(3)} }{2}\\
 \frac{3 \, x_{0,a}^{(1)} +  x_{2,a}^{(1)}}{2} \\
 \frac{x_{1,a}^{(1)} +  x_{2,a}^{(1)}}{2}\\
x_{3,a}^{(1)}\\
\frac{x_{1,a}^{(1)} +  x_{3,a}^{(1)}}{2} + x_{2,a}^{(3)} +x_{3,a}^{(3)}\\
24-\frac{3 P_a}{2} + \frac{x_{1,a}^{(1)} +x_{5,a}^{(1)} }{2}- \frac{x_{2,a}^{(2)} +x_{3,a}^{(2)} }{2} - \frac{x_{2,a}^{(3)} +x_{3,a}^{(3)} }{2}
\end{array}\right) = \left( \begin{array}{c|c} 220 & {\color{mygr} \checkmark}   \\ 2 & {\color{mygr} \checkmark} \\  1 & {\color{red}\lightning}  \\ -1 & {\color{red}\lightning} \\ 6 &  {\color{mygr} \checkmark}    \\  214 &  {\color{mygr} \checkmark}  \end{array} \right).
\end{equation}
The violation of the K-theory constraints in the third and fourth row implies that the example presented in table~\ref{tab:5stackLRSaAAPrototypeI} is {\it semi-local} or globally not consistent, a characteristic which was also found to be true for all other 1295 models found within this prototype for the choice of exotic O6-plane $\eta_{\Z_2^{(3)}}=-1$. One finds the same amount of models when choosing the $\OR\Z_2^{(2)}$-plane as the exotic O6-plane.

Let us now turn to the search for Abelian symmetries associated to the D6-brane configuration given in table~\ref{tab:5stackLRSaAAPrototypeI}. First, we compute the {\it necessary} conditions (\ref{Eq:Zn-condition-nec}) for the existence of discrete $\Z_n$ symmetries, which reduce to the following three linearly independent constraints: 
\begin{equation}
\begin{array}{rcl}
 k_d &\stackrel{!}{=}& 0 \text{ mod } n,\\
-6 k_a + 8k_h &\stackrel{!}{=}& 0 \text{ mod } n,\\
6 k_a - 2 k_d + 8k_h &\stackrel{!}{=}& 0 \text{ mod } n.
\end{array}
\end{equation}
These constraints have to be supplemented with the {\it sufficient} conditions (\ref{Eq:Zn-condition-suf}) for the existence of discrete $\Z_n$ symmetries, which lead to at most 
three more linearly independent constraints:
\begin{equation}
\begin{array}{rcl}
3k_a - k_d + 4 k_h &\stackrel{!}{=}& 0 \text{ mod } n,\\
3 k_a + k_d &\stackrel{!}{=}& 0 \text{ mod } n,\\
6 k_a  &\stackrel{!}{=}& 0 \text{ mod } n.\\
\end{array}
\end{equation}
Further reductions lead to only three truly independent constraints, \linebreak \mbox{$k_d, \, 3k_a, \, 4k_h \stackrel{!}{=} 0 \text{ mod } n$}, as expected from initially three $U(1)$ gauge factors.
The first observation we can make is that there does not exist any non-trivial combination $(k_a, k_d, k_h)$ for which the eight constraint equations are exactly satisfied for all $n$, indicating that there does not exist any linear combination of $U(1)$'s which stays massless, in particular no (generalised) $U(1)_{B-L}$ symmetry. Next, we can classify the discrete $\Z_n$ symmetries, which arise from linear combinations $(k_a, k_d, k_h)$ satisfying the eight constraints given above:
\begin{itemize}
\item A discrete $\Z_3$ symmetry homomorphic to the centre of the $SU(3)_a$ gauge group appears for the configuration $(k_a, k_d, k_h)=(1,0,0)$, with the charges of the massless states listed in the second-to-last column of table~\ref{tab:5stackLRSMaAAPrototypeISpectrum}. This symmetry acts - as usual - as a baryon-like discrete symmetry, but does not forbid any cubic or higher order coupling which is not already forbidden by the $SU(3)_a$ gauge symmetry.  
\item The combination $(k_a, k_d, k_h)=(0,0,1)$ corresponds to a $\Z_4$ symmetry homomorphic to the centre of the hidden $SU(4)_h$ gauge group, and thus does not constrain additional couplings beyond the ones already constrained by the non-Abelian gauge symmetry. For completeness, we list the charges under the $\Z_4$ symmetry for the massless open string spectrum in the last column of table~\ref{tab:5stackLRSMaAAPrototypeISpectrum}, from which we can clearly see that only exotic matter charged under the hidden gauge group carries $\Z_4$ charges as expected.
\item A last observation is that the violation of the K-theory constraints forbids the existence of a discrete $\Z_2$ symmetry for the combination $(k_a,k_d,k_h)=(1,1,1)$.
\end{itemize}
Thus, the full gauge group of the {\it semi-local} five-stack left-right symmetric model below the string scale corresponds to $SU(3)_a\times USp(2)_b \times USp(2)_c\times SU(4)_h$, free of any non-trivial discrete $\Z_n$ symmetry.

%%%%%%%%%%%%%%%%%%%%%%%%%%%%%%%
\mathtab{
\begin{array}{|c||c|c||c|c|}
\hline \multicolumn{5}{|c|}{\text{\bf Overview of the Spectrum for 5-stack Left-Right Symm.~on the {aAA} lattice}}\\
\hline \hline
\text{sector}& \text{state} & (SU(3)_a \times USp(2)_b \times USp(2)_c \times SU(4)_h)_{U(1)_a \times U(1)_d \times U(1)_h} & \Z_3 & \Z_4\\
\hline 
ab \equiv ab'& Q_L& 3 \times  (\3,\2,\1, \1)_{(1,0,0)} & 1 & 0  \\
ac \equiv ac'& Q_R &3 \times  (\ov \3,\1,\2, \1)_{(-1,0,0)} &2 &0 \\
ad& &(\3,\1,\1,\1)_{(1,-1,0)} + h.c. &1 & 0 \\
ad'& &2 \times \left[ (\3,\1,\1,\1)_{(1,1,0)} + h.c. \right] & 1||2 & 0 \\
bc \equiv bc'&  (H_u, H_d) &10 \times (\1,\2,\2,\1)_{(0,0,0)} & 0 &0  \\
bd\equiv b'd& L &3 \times (\1,\2,\1,\1)_{(0,-1,0)} & 0&  0 \\
bd \equiv b'd& &3 \times \left[ (\1,\2,\1,\1)_{(0,-1,0)} +h.c. \right] &0 & 0\\
cd \equiv c'd & R &3 \times (\1,\1,\2,\1)_{(0,1,0)} &0 & 0\\
cd \equiv c'd & &3 \times \left[  (\1,\1,\2,\1)_{(0,1,0)} +h.c. \right] & 0 & 0\\
\hline \hline
ah& &2 \times \left[ (\3,\1,\1,\ov \4)_{(1,0,-1)} + h.c. \right] & 1||2 & 3||1 \\
ah'& &(\3,\1,\1, \4)_{(1,0,1)} + h.c. &1||2 & 1 ||3 \\
bh \equiv b'h& &3 \times  (\1,\2,\1,\4)_{(0,0,1)}& 0& 1  \\
ch \equiv c'h& &3 \times (\1,\1,\2,\ov\4)_{(0,0,-1)} & 0& 3 \\
dh& &2 \times \left[ (\1,\1,\1,\ov \4)_{(0,1,-1)} + h.c. \right]  &0 & 3 ||1  \\
dh'& &(\1,\1,\1, \4)_{(0,1,1)} + h.c. & 0& 1 \\
\hline \hline
 aa'& & 2\times[ ({\bf 3_{A}},\1,\1,\1)_{(2,0,0)} + h.c.] &2||1 &0  \\
  bb' \equiv bb & &  5 \times  (\1,\1_{\bf A},\1,\1)_{(0,0,0)} &0 &0  \\
   cc' \equiv cc & &  5 \times  (\1,\1,\1_{\bf A},\1)_{(0,0,0)} & 0&  0\\
   dd& &  4 \times  (\1,\1,\1,\1)_{(0,0,0)} & 0&0 \\   
   hh'& &2 \times [ (\1,\1,\1,{\bf 6_{A}})_{(0,0,2)} + h.c.] &0 & 2  \\
\hline
\end{array}
}{5stackLRSMaAAPrototypeISpectrum}{Chiral and non-chiral massless spectrum for the {\it semi-local} five-stack D6-brane model with initial gauge group $U(3)_a\times USp(2)_b \times USp(2)_c \times U(1)_d \times U(4)_h$ corresponding to the configuration from table~\ref{tab:5stackLRSaAAPrototypeI} with vanishing RR tadpoles but violated K-theory constraints (\ref{Eq:ViolatedKtheory}).
 }
%%%%%%%%%%%%%%%%%%%%%%%%%

\clearpage
%%%%%%%%%%%%%%%%%%%%%%%%%%%%%%%%%%%%%%%%%%%%%%%%%%%%%%%%%%%%%%%%%%%%%%%%%%%%%%%%%%%%%%%%%%%%%%%%%%%%%%%%%%%%%%%%%%%%%%%%%%%%%%%%%%%%%%%%%%%%%%%%%%%%%%%%%%%%%%%%%%%%%%%%
\section{Alternative Prototype II models}\label{A:ProtoIIModels}

In this appendix, we present two variants of the prototype II left-right symmetric model discussed in section~\ref{Ss:LR-models}. A sample D6-brane configuration for the  first variant, prototype IIb, is given in table~\ref{tab:6stackLRSaAAPrototypeIIb}, with the corresponding massless open string spectrum listed in table~\ref{tab:6stackLRSMaAAPrototypeIIbSpectrum}. For the second variant, prototype IIc, we opted for the D6-brane configuration in table~\ref{tab:6stackLRSaAAPrototypeIIc}, with corresponding massless open string spectrum in table~\ref{tab:6stackLRSMaAAPrototypeIIcSpectrum}. 
%%%%%%%%%%%%%%%%%%%%%%%%%%%%%%%%%%%%%%%%%%%%%%%%%%%%%%%%%%%%%%%%%%%%%%%%%%%
\mathtabfix{
\begin{array}{|c||c|c||c|c|c||c|}\hline 
\muc{7}{|c|}{\text{\bf D6-brane configuration for a 6-stack LRS model (prototype IIb)  on the {aAA} lattice}}
\\\hline \hline
&\text{\bf wrapping numbers} &\frac{\rm Angle}{\pi}&\text{\bf $\Z_2^{(i)}$ eigenvalues}  & (\vec \tau) & (\vec \sigma)& \text{\bf gauge group}\\
\hline \hline
 a&(1,0;1,0;1,0)&(0,0,0)&(+++)&(0,1,1) & (1,1,1)& U(3)\\
 b&(1,0;-1,2;1,-2)&(0,\frac{1}{2},-\frac{1}{2})&(+++)&(0,1,0) & (1,1,0)&USp(2)\\
 c&(1,0;-1,2;1,-2)&(0, \frac{1}{2},-\frac{1}{2})&(-+-)&(0,1,0) & (1,1,0)&USp(2)\\ 
  d&(1,0;1,0;1,0)&(0, 0,0)&(+--)&(0,1,1) & (1,1,1)& U(1)\\
  \hline
    h_1&(1,0;-1,2;1,-2)&(0,\frac{1}{2},-\frac{1}{2})&(-+-)&(0,0,0) & (0,0,0)& U(1)\\
        h_2&(1,0;-1,2;1,-2)&(0,\frac{1}{2},-\frac{1}{2})&(--+)&(0,0,0) & (0,0,0)& U(1)\\
 \hline
\end{array}
}{6stackLRSaAAPrototypeIIb}{D6-brane configuration for a six-stack Left-Right Symmetric model (prototype IIb) with gauge group 
$SU(3)_a\times USp(2)_b \times USp(2)_c\times  U(1)_a \times U(1)_d \times U(1)_{h_1} \times U(1)_{h_2}$ on the {\bf aAA} lattice of the orientifold $T^6/(\Z_2 \times \Z_6 \times \OR)$ with discrete torsion ($\eta=-1$) and the $\OR\Z_2^{(3)}$-plane as the exotic O6-plane ($\eta_{\OR\Z_2^{(3)}}=-1$).}
%%%%%%%%%%%%%%%%%%%%%%%%%%%%%%%%%%%%%%%%%%%%%%%%%%%%%%%%%%%%%%%%%%%%%%%%%%%
%
%%%%%%%%%%%%%%%%%%%%%%%%%%%%%%%%%%%%%%%%%%%%%%%%%%%%%%%%%%%%%%%%%%%%%%%%%%%
\mathtabfix{
\begin{array}{|c||c|c||c|c|c||c|}\hline 
\muc{7}{|c|}{\text{\bf D6-brane configuration for a 6-stack LRS model (prototype IIc)  on the {aAA} lattice}}
\\\hline \hline
&\text{\bf wrapping numbers} &\frac{\rm Angle}{\pi}&\text{\bf $\Z_2^{(i)}$ eigenvalues}  & (\vec \tau) & (\vec \sigma)& \text{\bf gauge group}\\
\hline \hline
 a&(1,0;1,0;1,0)&(0,0,0)&(+++)&(0,1,1) & (0,1,1)& U(3)\\
 b&(1,0;-1,2;1,-2)&(0,\frac{1}{2},-\frac{1}{2})&(+++)&(0,1,0) & (0,1,0)&USp(2)\\
 c&(1,0;-1,2;1,-2)&(0, \frac{1}{2},-\frac{1}{2})&(-+-)&(0,1,0) & (0,1,0)&USp(2)\\ 
  d&(1,0;1,0;1,0)&(0, 0,0)&(+--)&(0,1,1) & (0,1,1)& U(1)\\
  \hline
    h_1&(1,0;-1,2;1,-2)&(0,\frac{1}{2},-\frac{1}{2})&(+++)&(0,0,1) & (0,0,1)& U(1)\\
        h_2&(1,0;-1,2;1,-2)&(0,\frac{1}{2},-\frac{1}{2})&(+++)&(0,0,1) & (1,1,1)& U(1)\\
 \hline
\end{array}
}{6stackLRSaAAPrototypeIIc}{D6-brane configuration for a six-stack Left-Right Symmetric model (prototype IIc) with gauge group 
$SU(3)_a\times USp(2)_b \times USp(2)_c\times  U(1)_a \times U(1)_d \times U(1)_{h_1} \times U(1)_{h_2}$ on the {\bf aAA} lattice of the orientifold $T^6/(\Z_2 \times \Z_6 \times \OR)$ with discrete torsion ($\eta=-1$) and the $\OR\Z_2^{(3)}$-plane as the exotic O6-plane ($\eta_{\OR\Z_2^{(3)}}=-1$).}
%%%%%%%%%%%%%%%%%%%%%%%%%%%%%%%%%%%%%%%%%%%%%%%%%%%%%%%%%%%%%%%%%%%%%%%%%%%
%
%%%%%%%%%%%%%%%%%%%%%%%%%%%
\mathtab{
\begin{array}{|c||c|c||c|}
\hline \multicolumn{4}{|c|}{\text{\bf Overview of the Spectrum for prototype IIb LRS Model on the {aAA} lattice}}\\
\hline \hline
\text{sector} & \text{state} & (SU(3)_a \times USp(2)_b \times USp(2)_c )_{U(1)_a \times U(1)_d \times U(1)_{h_1} \times U(1)_{h_2}}& \widetilde{U(1)}_{B-L} \\
\hline 
ab \equiv ab'& Q_L &3 \times  (\3,\2,\1)_{(1,0,0,0)} & 1/3  \\
ac \equiv ac'& Q_R &3 \times  (\ov \3,\1,\2)_{(-1,0,0,0)}& -1/3 \\
ad&  &(\3,\1,\1)_{(1,-1,0,0)} + h.c.  &\pm4/3 \\ 
ad'& &2 \times \left[ (\3,\1,\1)_{(1,1,0,0)} + h.c.\right] & \mp 2/3 \\
bc \equiv bc'& (H_u, H_d) &10 \times (\1,\2,\2)_{(0,0,0,0)} & 0   \\
bd \equiv b'd& L &3 \times (\1,\2,\1)_{(0,-1,0,0)} & 1  \\
cd \equiv c'd & R &3 \times (\1,\1,\2)_{(0,1,0,0)} & -1 \\
\hline \hline
h_1 h_2& &5 \times \left[ (\1,\1,\1)_{(0,0,1,-1)} + h.c. \right]   & \mp 2 \\
h_1 h_2'& &6\times \left[ (\1,\1,\1)_{(0,0,1,1)} + h.c. \right] &0\\
\hline \hline
 aa'& & 2\times[ ({\bf 3_{A}},\1,\1)_{(2,0,0,0)} + h.c.] & \pm 2/3 \\
  bb' \equiv bb & &  5 \times  (\1,\1_{\bf A},\1)_{(0,0,0,0)}& 0 \\
   cc' \equiv cc & &  5 \times  (\1,\1,\1_{\bf A})_{(0,0,0,0)}&0  \\
    h_1h_1& & 4 \times (\1,\1,\1)_{(0,0,0,0)} &0  \\
    h_2h_2& &4 \times (\1,\1,\1)_{(0,0,0,0)}& 0  \\
\hline
\end{array}
}{6stackLRSMaAAPrototypeIIbSpectrum}{Chiral and non-chiral massless matter spectrum for the global five-stack D6-brane model with gauge group $U(3)_a\times USp(2)_b \times USp(2)_c \times U(1)_d \times U(1)_{h_1} \times U(1)_{h_2}$ corresponding to the configuration from table~\ref{tab:6stackLRSaAAPrototypeIIb}.
The $\widetilde{U(1)}_{B-L}$ symmetry acts as a chiral, global symmetry and no longer as a massless Abelian generalised $B-L$ gauge symmetry.}
%%%%%%%%%%%%%%%%%%%%%%%%%%%%%
%
%%%%%%%%%%%%%%%%%%%%%%%%%%%
\mathtab{
\begin{array}{|c||c|c||c|}
\hline \multicolumn{4}{|c|}{\text{\bf Overview of the Spectrum for prototype IIc LRS Model on the {aAA} lattice}}\\
\hline \hline
\text{sector} & \text{state} & (SU(3)_a \times USp(2)_b \times USp(2)_c )_{U(1)_a \times U(1)_d \times U(1)_{h_1} \times U(1)_{h_2}}& \widetilde{U(1)}_{B-L} \\
\hline 
ab \equiv ab'& Q_L &3 \times  (\3,\2,\1)_{(1,0,0,0)} & 1/3  \\
ac \equiv ac'& Q_R &3 \times  (\ov \3,\1,\2)_{(-1,0,0,0)}& -1/3 \\
ad&  &(\3,\1,\1)_{(1,-1,0,0)} + h.c.  &\pm4/3 \\ 
ad'& &2 \times \left[ (\3,\1,\1)_{(1,1,0,0)} + h.c.\right] & \mp 2/3 \\
bc \equiv bc'& (H_u, H_d) &10 \times (\1,\2,\2)_{(0,0,0,0)} & 0   \\
bd \equiv b'd& L &3 \times (\1,\2,\1)_{(0,-1,0,0)} & 1  \\
cd \equiv c'd & R &3 \times (\1,\1,\2)_{(0,1,0,0)} & -1 \\
\hline \hline
ah_1& &3 \times (\ov\3,\1,\1)_{(-1,0,1,0)}& - 4/3 \\
ah_1'& &3 \times (\3,\1,\1)_{(1,0,1,0)}& -2/3 \\
bh_1 \equiv b'h_1& &4 \times \left[ (\1,\2,\1)_{(0,0,-1,0)} +h.c. \right]& \pm 1  \\
ch_1\equiv  c'h_1& &6 \times (\1,\1,\2)_{(0,0,-1,0)} & 1  \\
ch_1 \equiv c'h_1 & &2 \times \left[  (\1,\1,\2)_{(0,0,1,0)} +h.c. \right]& \mp 1  \\
dh_1& &3\times (\1,\1,\1)_{(0,1,-1,0)}  & 0 \\
dh_1'& &3\times (\1,\1,\1)_{(0,-1,-1,0)}  & -2 \\
\hline \hline
 aa'& & 2\times[ ({\bf 3_{A}},\1,\1)_{(2,0,0,0)} + h.c.] & \pm 2/3 \\
  bb' \equiv bb & &  5 \times  (\1,\1_{\bf A},\1)_{(0,0,0,0)}& 0 \\
   cc' \equiv cc & &  5 \times  (\1,\1,\1_{\bf A})_{(0,0,0,0)}&0  \\
    h_1h_1& & 5 \times (\1,\1,\1)_{(0,0,0,0)} &0  \\
    h_2h_2& &5 \times (\1,\1,\1)_{(0,0,0,0)}& 0  \\
\hline
\end{array}
}{6stackLRSMaAAPrototypeIIcSpectrum}{Chiral and non-chiral massless matter spectrum for the global five-stack D6-brane model with gauge group $U(3)_a\times USp(2)_b \times USp(2)_c \times U(1)_d \times U(1)_{h_1} \times U(1)_{h_2}$ corresponding to the configuration from table~\ref{tab:6stackLRSaAAPrototypeIIc}.
The $\widetilde{U(1)}_{B-L}$ symmetry acts as a chiral, global symmetry and no longer as a massless Abelian generalised $B-L$ gauge symmetry.}
%%%%%%%%%%%%%%%%%%%%%%%%%%%%%

Notice that the gauge structure supported by the six-stack models and the massless spectrum in the visible sector are identical to the one of the prototype II model from section~\ref{Ss:LR-models}. The differences between subclasses II, IIb and IIc are situated in the massless open string spectrum with  non-trivial charges under the hidden gauge groups, i.e. the `messenger' and the `hidden' sectors. 
In table~\ref{tab:6stackLRSMaAAPrototypeIIbSpectrum} we can see clearly that prototype IIb does not have chiral or non-chiral matter charged under a visible gauge group and a hidden gauge group, in other words 
there is no charged `messenger' sector. 
Hence, this global six-stack left-right symmetric forms an interesting example where the visible and hidden gauge sectors can only communicate with each other through the closed string sector. Regarding the prototype IIc example, we notice that the model only has massless states which are non-trivially charged under the first hidden gauge group $U(1)_{h_1}$. In this model the second hidden gauge group $U(1)_{h_2}$ only communicates to the other gauge sectors through the closed string sector. This ``decoupling" behaviour of the hidden gauge sector for prototypes IIb and IIc should be contrasted to the massless open string spectrum of prototype~II in table~\ref{tab:6stackLRSMaAAPrototypeIISpectrum}, which contains massless matter states charged both under the visible gauge sector and under each of the hidden gauge groups.   

Another crucial difference between prototype II on the one hand and prototypes IIb and IIc on the other hand concerns the generalised $B-L$ symmetry defined in equation (\ref{Eq:GenerBLSymm}). This $\widetilde{U(1)}_{B-L}$ symmetry acts as a massless Abelian gauge symmetry for prototype II models, but turns into a massive chiral global symmetry by virtue of the St\"uckelberg mechanism for the other two prototype models. This subtle difference among prototype II, IIb and IIc results from solving condition (\ref{Eq:MasslessU(1)}) explicitly for appropriate values of $q_a\in \Q$, using the fractional D6-brane configurations in tables~\ref{tab:6stackLRSaAAPrototypeII},~\ref{tab:6stackLRSaAAPrototypeIIb} and~\ref{tab:6stackLRSaAAPrototypeIIc}, respectively. 

The observation regarding the $\widetilde{U(1)}_{B-L}$ symmetry also has consequences for the identification of the discrete $\Z_n$ symmetries. To determine the discrete $\Z_n$ symmetries for the prototype IIb and IIc models, we have to solve the necessary conditions (\ref{Eq:Zn-condition-nec}) and sufficient conditions (\ref{Eq:Zn-condition-suf}) using the D6-brane configurations given in tables~\ref{tab:6stackLRSaAAPrototypeIIb} and~\ref{tab:6stackLRSaAAPrototypeIIc}. Despite the fact that the necessary and sufficient conditions are different for prototype IIb and IIc models, their respective solutions are identical and the identification of the discrete $\Z_n$ symmetries can be discussed simultaneously:
\begin{itemize}
\item First of all, we can identify the discrete $\Z_2$ symmetry guaranteed by the K-theory constraints, which corresponds to the solution $(k_a,k_d,k_{h_1}, k_{h_2}) = (1,1,1,1)$. Nonetheless, this discrete symmetry offers the same selection rules as the $USp(2)_b$ and $USp(2)_c$ gauge groups. In order to see that, we can use the same argument as the one presented in the discussion for the prototype I models in section~\ref{Ss:LRSDiscrete}. 
\item The discrete $\Z_3$ symmetry, emerging as the solution $(k_a,k_d,k_{h_1}, k_{h_2}) = (1,0,0,0)$ is homomorphic to the centre of the non-Abelian $SU(3)_a$ gauge group and does not provide additional selection rules for cubic and higher order couplings beyond the selection rules of the $SU(3)_a$ gauge group.
\item The third discrete symmetry we can identify is a $\Z_6$ symmetry for the solution $(k_a,k_d,k_{h_1}, k_{h_2}) = (1,3,3,3)$. Notice that the above identified discrete $\Z_2$ and $\Z_3$ symmetries form subgroups of this discrete $\Z_6$ symmetry. Modding out the centre $\Z_3\times \Z_2 \times\Z_2$ of the non-Abelian gauge factor from the full set $\Z_6\times\Z_3\times \Z_2$ of identified discrete symmetries results in a quotient group that is homomorphic to the discrete $\Z_3$ symmetry identified above, analogous to the discussion for the prototype II models in section~\ref{Ss:LRSDiscrete}. Hence, we do not expect additional selection rules associated to the $\Z_6$ symmetry for the cubic and higher order couplings beyond the selection rules of the non-Abelian gauge groups.
\end{itemize} 
Note the contrast with the prototype II model in section~\ref{Ss:LRSDiscrete}, where we were able to shift the charges of the states by virtue of the massless $\widetilde{U(1)}_{B-L}$. The absence of this symmetry for the prototype IIb and IIc models prevents us from doing exactly that.

%\clearpage
%%%%%%%%%%%%%%%%%%%%%%%%%%%%%%%%%%%%%%%%%%%%%%%%%%%%%%%%%%%%%%%%%%%%%%%%%%%%%%%%%%%%%%%%%%%%%%%%%%%%%%%%%%%%%%%%%%%%%%%%%%%%%%%%%%%%%%%%%%%%%%%%%%%%%%%%%%%%%%%%%%%%%%%%%%%%%%%%%%%%%%%%%%%%%%%%%%%%%%%%%%%%%%%%%%%%%%%%%%%%%%%%%%%%%%%%%%%%%%%%%%%%%%%%%%%%%%%%%%%%%%%%%%%%%%%%%%%%

\addcontentsline{toc}{section}{References}
\bibliographystyle{ieeetr}
\bibliography{refs_Z2Z6-Pheno}

\end{document}